\documentclass[12pt,a4paper]{amsart} 
\usepackage{amsaddr}
\usepackage[T1]{fontenc}
\usepackage[margin=1.5cm]{geometry}
\usepackage[pdfusetitle]{hyperref}
\usepackage[normalem]{ulem}
\usepackage{wasysym}

\usepackage{amscd,amsmath,amssymb,amsfonts,xspace,mathrsfs,amsthm}
\usepackage{color, mathtools}
\usepackage[dvipsnames]{xcolor}

\usepackage{tikz,textcomp}
\usetikzlibrary{decorations.pathreplacing,calc,fadings,fit,shapes,arrows,positioning,chains,matrix}

\usepackage{subcaption}

\usepackage{url}

\usepackage{latexsym}
\usepackage{graphicx}
\usepackage{dsfont}
\usepackage{longtable}
\usepackage{enumitem}

\usepackage[latin1]{inputenc}
\usepackage{multirow}

\usepackage{tikz-cd}
\usetikzlibrary{bending}
\usepackage{tikz,textcomp}
\usetikzlibrary{decorations.pathreplacing,calc,fadings,fit,shapes,arrows,positioning,chains,matrix}

\usepackage{pifont}
\newcommand{\cmark}{\ding{51}}
\newcommand{\xmark}{\ding{55}}

\numberwithin{equation}{section}

\def\bea{\begin{eqnarray}}
\def\eea{\end{eqnarray}}
\def\be{\begin{equation}}
\def\ee{\end{equation}}
\def\ba{\begin{align}}
\def\ea{\end{align}}
\def\bse{\begin{subequations}}
\def\ese{\end{subequations}}

\newcommand{\nn}{\nonumber}

\def\det{\,{\rm det}\, }

\def\Im{\,{\rm Im}\,}
\def\Re{\,{\rm Re}\,}

\DeclareMathOperator{\rk}{rk}

\DeclareMathOperator{\Coh}{Coh}

\DeclareMathOperator{\Stab}{Stab}
\DeclareMathOperator{\Pic}{Pic}
\DeclareMathOperator{\Li}{Li}

\def\({\left(}
\def\){\right)}
\def\[{\left[}
\def\]{\right]}
\def\<{\left\langle}
\def\>{\right\rangle}

\renewcommand\v{\mathsf v}
\newcommand\w{\mathsf w}

\newcommand{\de}{\mathrm{d}}

\newcommand{\I}{\mathrm{i}}

\newcommand{\cC}{\mathcal{C}}
\newcommand{\cD}{\mathcal{D}}

\newcommand{\cM}{\mathcal{M}}

\newcommand{\cO}{\mathcal{O}}

\newcommand{\cS}{\mathcal{S}}

\newcommand{\cX}{\mathcal{X}}
\newcommand{\cY}{\mathcal{Y}}

\newcommand{\Z}{{\mathbb Z}}

\newcommand{\IC}{\mathds{C}}
\newcommand{\IZ}{\mathds{Z}}
\newcommand{\IQ}{\mathds{Q}}

\newcommand{\IH}{\mathds{H}}

\newcommand{\IP}{\mathds{P}}

\def\bOm{\overline{\Omega}}

\newcommand{\q}{\mbox{q}}

\newcommand\PT{\operatorname{PT}}

\newcommand\NL{\operatorname{NL}}

\def\GW{{\rm GW}}
\def\GV{{\rm GV}}

\def\hCY{Y}

\def\whf{\widehat{W}}

\def\whS{\widehat{S}}

\def\BM{\begin{matrix}}
\def\EM{\end{matrix}}

\def\Thi#1{\Theta^{(#1)}}
\newcommand{\binomial}[2]{ \Big( {#1 \atop #2} \Big)}

\title[Enumerative geometry and modularity in K3-fibered CY threefolds]{
Enumerative geometry and modularity \\[2mm] in two-modulus K3-fibered Calabi-Yau threefolds
}

\author{Charles Doran}
\address{ Department of Mathematical and Statistical Sciences, \\
632 CAB, University of Alberta, Edmonton, AB, T6G \\
2G1, Canada}
\email{charles.doran@ualberta.ca}

\author{Boris Pioline}
\address{Sorbonne Universit\'e, CNRS, 
Laboratoire de Physique Th\'eorique et Hautes Energies, \\
Campus Pierre et Marie Curie, 4 place Jussieu, F-75005, Paris, France}
\email{pioline@lpthe.jussieu.fr}

\author{Thorsten Schimannek}
\address{
Institute for Theoretical Physics \& Department of Mathematics, Utrecht University,\\
3584 CC Utrecht, The Netherlands
}
\email{t.f.a.schimannek@uu.nl}

\begin{document}

\begin{abstract}
Motivated in part by the modular properties of enumerative invariants of 
K3-fibered Calabi-Yau threefolds, 
we introduce a family of 39 Calabi-Yau mirror pairs 
$(X,Y)$ with $h_{1,1}(X)=h_{2,1}(Y)=2$, labelled by certain integer quadruples $(m,i,j,s)$ with $m\leq 11$. On the A-model side, $X$ arises as a complete intersection in a projective bundle over a Fano fourfold $V_m^{[i,j]}$,
and admits a Tyurin degeneration into a pair of degree $m$ Fano threefolds $F_m^{[i]}\cup F_m^{[j]}$ intersecting on an anticanonical K3 divisor of degree $2m$. 
On the B-model side, $Y$ is fibered by $M_{m}$-polarized K3-surfaces of Picard rank 19, and determined by a branched covering of $\IP^1$, consistent with the Doran-Harder-Thompson mirror conjecture. When $s=0$,
$Y$ itself acquires a Tyurin degeneration, and correspondingly 
$X$ acquires a fibration by degree $2m$ K3 surfaces, such that the two K\"ahler moduli control the size of the K3-fiber and base $\IP^1$. 
While the mirror pairs with $m\leq 4$ can be realized as complete intersections in products of projective spaces or as hypersurfaces in toric varieties, 
the examples 
with $m\geq 5$ are intrinsically non-toric.  
We obtain uniform formulae for the genus 0 and 1 topological free energies
near the Tyurin degeneration (mirror to the large base limit), exhibiting modular properties
under the  Fricke-extended congruence group $\Gamma_0(m)^+$. 
We use these results to compute the vertical Gopakumar-Vafa 
and Noether-Lefschetz invariants and check that their generating functions satisfy the expected modular properties.  We also compute generating series of 
Gopakumar-Vafa invariants with fixed non-zero base degree and exhibit their modular properties.
\end{abstract}

\maketitle

\newpage
\tableofcontents

\setlength{\parskip}{0.2cm}

\section{Introduction, summary and outlook}

Since the seminal work \cite{Candelas:1985en}, 
Calabi-Yau (CY) threefolds have played a central role in string theory and generated profound connections
with mathematics, especially with the advent of mirror symmetry \cite{Lerche:1989uy}. At the classical level, mirror symmetry identifies the complexified K\"ahler moduli $\cM_K(X)$ of a CY threefold $X$ with the complex structure moduli space $\cM_C(\hCY)$ of the mirror family $\hCY$, with respective complex dimensions $h_{1,1}(X)=h_{1,2}(\hCY)$. 
At the level of BPS spectra, homological 
mirror symmetry identifies the derived category $D^b(X)$ of coherent sheaves
on $X$ with the Fukaya category ${\rm Fuk}(\hCY)$ 
of Lagrangians on $\hCY$, graded by the
charge lattices $K(X)\simeq H^3(\hCY,\IZ)$ of rank $2h_{1,1}(X)+2=2h_{1,2}(\hCY)+2$ \cite{MR1403918}.
Explicit studies are usually restricted to threefolds with small Hodge numbers, leading to moduli spaces of manageable dimension (see e.g. \cite{Candelas:1990rm,Candelas:1993dm,Candelas:1994hw,hosono:1993qy}).

	Although complete intersections in products of projective spaces (CICYs)~\cite{Green1987,Candelas1988} and hypersurfaces in toric fourfolds~\cite{Kreuzer:2000xy,Batyrev:1993oya} lead to a large number of examples, some of which with small Hodge numbers, a complete classification of CY threefolds still seems far out of reach. Indeed, it seems likely that the vast majority of CY threefolds is neither of CICY nor of hypersurface type.

	An effective strategy to construct  CY threefolds beyond the CICY and hypersurface types is to focus on geometries that exhibit a fibration structure.
        For example, in~\cite{Knapp:2021vkm} the existence of a genus one fibration with a 5-section was used to construct a large number of CY threefolds that can not be realized as complete intersections in any toric ambient space.
	In~\cite{klemm:2004km} the existence of a K3 fibration has been used to construct CY threefolds with small $h_{1,1}(X)\geq 2$ as complete intersections in toric ambient spaces of codimension less than or equal to two.
	On the other hand, in~\cite{doran2015families,doran2016calabi,doran2020calabi,kooistra2021threefolds} a systematic classification was obtained of CY threefolds that are fibered by $M_m$-polarized K3 surfaces. Here, $M_m$ denotes the signature $(1,18)$ lattice 
$M_m= U \oplus E_8 \oplus E_8  \oplus \langle -2m\rangle$ (where
$U$ is the unimodular hyperbolic indefinite lattice, $E_8$ the standard root lattice with negative definite metric, and $\langle -2m\rangle$
the one dimensional lattice $\IZ e$
with $(e,e)=-2m$.
	The latter construction generally leads to CY threefolds $Y$ with small $h_{2,1}(Y)\ge 2$. 

	Our first goal in this paper is to systematically construct the
 CY threefolds which are mirror to those examples of $M_m$-polarized K3 fibered CY threefolds from~\cite{Doran2019,kooistra2021threefolds} that have the smallest possible $h_{2,1}(Y)=2$.
	To this end we shall apply and verify the Doran-Harder-Thompson (DHT) conjecture~\cite{doran:2016uea} in numerous new examples, including several non-toric cases, building in part on~\cite{doran2019doran}.
	Roughly speaking, the DHT conjecture states that a CY threefold $Y$ with a fibration structure is mirror to a resolution $X$ of a Tyurin degeneration of another CY threefold into normal crossing (quasi) Fano components, such that the intersection of the components is the mirror of the generic fiber of $Y$.
 The fibration structure arises from gluing together Landau-Ginzburg models that are  dual to the individual Fano components of the Tyurin degeneration (see Section \S\ref{sec_DHT} for a review). 
 Applying this general strategy to the case at hand, 
 we shall give a uniform construction of a family of Picard rank~$2$ CY threefolds $X^{[i,j]}_{m,s}$, admitting a Tyurin degeneration
 into a pair of degree $m$  Fano threefolds $F_m^{[i]}\cup F_m^{[j]}$ 
 intersecting over an anticanonical K3 surface, and of the two-parameter mirror families $Y^{[i,j]}_{m,s}$, fibered by $M_m$-polarized K3 surfaces. Here $(m,i,j,s)$ runs over the
integer tuples listed in the first column of Table~\ref{tabmij}, with
$0\leq s\leq i-j$, while the pairs $(m,i), (m,j)$ label the Picard rank~1 Fano threefold in Iskovskih's list, see Table \ref{tabFano}. For $s=0$, 
it turns out that $Y^{[i,j]}_{m,0}$ itself admits a Tyurin degeneration, with the Fano components intersecting in an $M_m$-polarized K3 surface, and
correspondingly 
$X^{[i,j]}_{m,0}$ is fibered by degree $2m$ K3 surfaces.

\begin{table}
  $$\footnotesize
  \begin{array}{|c|r|cccc|cc|c|}
  \hline
   (m,i,j,s)  & \chi_X & \kappa_{111} & \kappa_{112} & \kappa_{122} & \kappa_{222} &  c_{2,1} &  c_{2,2} & \cX_m ^{[i,j]} \\
  \hline\hline
(1, 1, 1, 0)&-252&4&2&0&0&52&24& X_{6,2} \\\hline\hline
(1, 2, 1, 0)&-196&3&2&0&0&42&24& X_6 \\\hline
(1, 2, 1, 1)&-196&3&4&4&4&42&52& X_6 \\\hline\hline
(1, 2, 2, 0)&-140&2&2&0&0&32&24& X_{6,4} \\\hline\hline
(2, 1, 1, 0)&-168&8&4&0&0&56&24& X_{4,2} \\\hline\hline
(2, 2, 1, 0)&-140&6&4&0&0&48&24& X_{4,3} \\\hline
(2, 2, 1, 1)&-140&6&8&8&8&48&56& X_{4,3} \\\hline\hline
(2, 2, 2, 0)&-112&4&4&0&0&40&24& X_{4,4} \\\hline\hline
(2, 4, 1, 0)&-168&5&4&0&0&50&24& X_5 \\\hline
(2, 4, 1, 3)&-168&5&16&48&144&50&144& X_5 \\\hline\hline
(2, 4, 1, 1)&-152&5&8&8&8&50&56& X_5 \\\hline
(2, 4, 1, 2)&-152&5&12&24&48&50&96& X_5 \\\hline\hline
(2, 4, 2, 0)&-140&3&4&0&0&42&24& X_6 \\\hline
(2, 4, 2, 2)&-140&3&8&16&32&42&80& X_6 \\\hline\hline
(2, 4, 2, 1)&-132&3&6&6&6&42&48& X_6 \\\hline\hline
(2, 4, 4, 0)&-168&2&4&0&0&44&24& X_8 \\\hline\hline
(3, 1, 1, 0)&-132&12&6&0&0&60&24& X_{3,2,2} \\\hline\hline
(3, 2, 1, 0)&-120&9&6&0&0&54&24& X_{3,3} \\\hline
(3, 2, 1, 1)&-120&9&12&12&12&54&60& X_{3,3} \\\hline\hline
(3, 2, 2, 0)&-108&6&6&0&0&48&24& X_{4,3} \\\hline\hline
(3, 3, 1, 0)&-140&8&6&0&0&56&24& X_{4,2} \\\hline
(3, 3, 1, 2)&-140&8&18&36&72&56&108& X_{4,2} \\\hline\hline
(3, 3, 1, 1)&-128&8&12&12&12&56&60& X_{4,2} \\\hline\hline
(3, 3, 2, 0)&-128&5&6&0&0&50&24& X_5 \\\hline
(3, 3, 2, 1)&-128&5&9&9&9&50&54& X_5 \\\hline\hline
(3, 3, 3, 0)&-148&4&6&0&0&52&24& X_{6,2} \\\hline\hline
(4, 1, 1, 0)&-112&16&8&0&0&64&24& X_{2,2,2,2} \\\hline\hline
(4, 2, 1, 0)&-112&12&8&0&0&60&24& X_{3,2,2} \\\hline
(4, 2, 1, 1)&-112&12&16&16&16&60&64& X_{3,2,2} \\\hline\hline
(4, 2, 2, 0)&-112&8&8&0&0&56&24& X_{4,2} \\\hline\hline
(5, 1, 1, 0)&-100&20&10&0&0&68&24& X_{\mathcal{O}(1)\oplus\mathcal{O}(2)^{\oplus 2}}^{2,5}\text{~\cite{Batyrev:1998kx,Haghighat:2008ut}}\\\hline\hline
(5, 2, 1, 0)&-110&15&10&0&0&66&24& X_{\mathcal{O}(1)^{\oplus 2}\oplus\mathcal{O}(3)}^{2,5}\text{~\cite{Batyrev:1998kx,Haghighat:2008ut}}\\\hline
(5, 2, 1, 1)&-110&15&20&20&20&66&68& X_{\mathcal{O}(1)^{\oplus 2}\oplus\mathcal{O}(3)}^{2,5}\text{~\cite{Batyrev:1998kx,Haghighat:2008ut}}\\\hline\hline
(5, 2, 2, 0)&-120&10&10&0&0&64&24& \text{d.c. of }B_5=X^{2,5}_{\mathcal{O}(1)^{\oplus 3}}\text{~\cite{Borcea:1996mxz,hofmann:2013,Katz:2022lyl}}\\\hline\hline
(6, 1, 1, 0)&-92&24&12&0&0&72&24& X^{2,5}_{\mathcal{S}^{\vee}(1)\oplus\mathcal{O}(2)}\text{~\cite{Ueda:2016wfa}}\\\hline\hline
(7,1,1,0) &  -88 & 28 & 14 & 0 & 0 & 76 & 24 & X^{2,6}_{\mathcal{O}(1)^{\oplus 4}\oplus\mathcal{O}(2)}\text{~\cite{Batyrev:1998kx,Haghighat:2008ut}}\\\hline\hline
(8, 1, 1, 0)&-84&32&16&0&0&80&24& X^{3,6}_{\bigwedge^2\mathcal{S}^{\vee}\oplus\mathcal{O}(1)^{\oplus 2}\oplus\mathcal{O}(2)}\text{~\cite{Ueda:2016wfa}} \\\hline\hline
(9, 1, 1, 0)&-84&36&18&0&0&84&24& X^{5,7}_{\bigwedge^4\mathcal{S}^{\vee}\oplus\mathcal{O}(1)\oplus\mathcal{O}(2)}\text{~\cite{Ueda:2016wfa}}\\\hline\hline
(11,1,1,0) &   -84 & 44 & 22 & 0 & 0 & 92 & 24 & \text{d.c. of }A_{22}= X^{3,7}_{(\bigwedge^2\mathcal{S}^{\vee})^{\oplus 3}}\text{~\cite{Borcea:1996mxz,Batyrev:2008rp,hofmann:2013}}\\\hline
  \end{array}
  $$
\caption{A family of 39 two-parameter CY threefolds $X_{m,s}^{[i,j]}$ labelled by integers $(m,i,j,s)$, whose mirror $Y_{m,s}^{[i,j]}$ is fibered by $M_m$-polarized K3 surfaces. For the 26 models with $s=0$,
$X_{m}^{[i,j]}:=X_{m,0}^{[i,j]}$ is itself fibered by degree $2m$ K3-surfaces.
Each of these models admits a conifold transition to a one-parameter CY threefold
$\cX_{m}^{[i,j]}$, indicated in the last column. Here 
$X_{\{d_i\}}$ refers to the standard hypergeometric models.
The geometries $X^{k,n}_{E}$ are complete intersections in the Grassmannian $\text{Gr}(k,n)$, associated to sections of the vector bundle $E$. The notation for $E$ is explained in Section~\ref{sec:notations}. In the cases $\mathcal{X}_5^{[2,2]}$ and $\mathcal{X}_{11}^{[1,1]}$, ``d.c. of $V$'' means that $\cX_{m}^{[i,j]}$ is a double cover of the Fano threefold $V$.}
\label{tabmij}
\end{table}

\begin{table}
  $$\footnotesize
  \begin{array}{|c|r|c|c|}
  \hline
   (m,i,j,s)  & \chi_X &  & \text{CICY} \\
  \hline\hline
 (1, 1, 1, 0)&-252&   \IP^4_{1,1,2,2,6}[12] &\\\hline\hline
(2, 1, 1, 0)&-168&          \IP^4_{1,1,2,2,2}[8]=   
  \left(\tiny\begin{array}{c|cc} \IP^4 & 4&1  
  \\ \IP^1 & 0 & 2  \end{array}\right)      
& 7886 ,\, 7888 \\\hline\hline
(2, 4, 1, 0)&-168&  
         \left(\tiny\begin{array}{c|cc}  \IP^4 & 4&1  \\ \IP^1 & 1&1 \end{array}\right)\
& 7885 \\\hline
(2, 4, 4, 0)&-168&
       \left(\tiny\begin{array}{c|cc}  \IP^3 & 4  \\ \IP^1 &2 \end{array}\right) &7887 \\\hline\hline
(3, 1, 1, 0)&-132&
  \left(\tiny\begin{array}{c|cccc}  \IP^6 & 3&2&1&1  \\ \IP^1 & 0& 0 & 1 & 1  \end{array}\right) 
&7867 ,\, 7869 \\\hline\hline
(3, 2, 1, 0)&-120&  \left(\tiny\begin{array}{c|ccc}  \IP^5 & 2&3&1  \\ \IP^1 & 1& 0&1 \end{array}\right)
  &7840 \\\hline
(3, 2, 2, 0)&-108&  \left(\tiny\begin{array}{c|cc}  \IP^4 & 3&2  \\ \IP^1 & 0&2 \end{array}\right) &7806 \\\hline\hline
(3, 3, 1, 0)&-140&
  \left(\tiny\begin{array}{c|ccc}  \IP^5 & 2&3&1  \\ \IP^1 & 0& 1&1 \end{array}\right) &7873 \\\hline
(3, 3, 2, 0)&-128& \left(\tiny\begin{array}{c|cc}  \IP^4 & 3&2  \\ \IP^1 & 1&1 \end{array}\right) &7858 \\\hline
(3, 3, 3, 0)&-148&    \left(\tiny\begin{array}{c|cc}  \IP^4 & 3&2  \\ \IP^1 & 2&0 \end{array}\right) &7882 \\\hline\hline
(4, 1, 1, 0)&-112& \left(\tiny\begin{array}{c|cccc}  \IP^6 & 2&2&2&1  \\ \IP^1 & 0&0&0 & 2  \end{array}\right) &7819 ,\, 7823 \\\hline\hline
(4, 2, 1, 0)&-112& \left(\tiny\begin{array}{c|cccc}  \IP^6& 2&2&2&1  \\ \IP^1 & 0 &0& 1&1 \end{array}\right)&7817 \\\hline
(4, 2, 2, 0)&-112&  \left(\tiny\begin{array}{c|cccc}  \IP^5 & 2&2&2  \\ \IP^1 & 0&1&1  \end{array}\right) &7816 ,\, 7822 \\\hline
  \end{array}
  $$
\caption{Alternative realizations of some of the A-model geometries $X^{[i,j]}_{m,s}$.}
\label{tab:alternative}
\end{table}

Our second goal is to compute the topological string free energies at genus 0 and 1 for the models with $s=0$, and analyze the modular properties of the resulting Gopakumar-Vafa invariants. Our main motivation comes from the relation between vertical Gopakumar-Vafa invariants and Donaldson-Thomas invariants counting vertical D4-D2-D0 bound states in K3-fibered threefolds \cite{Bouchard:2016lfg}. Both are determined in terms of the Noether-Lefschetz invariants of the K3-fibration
\cite{maulik2007gromov}. For $\langle 2m\rangle$-polarized fibrations, the generating series of these invariants must transform as vector-valued modular forms in the Weil representation of $SL(2,\IZ)$ attached to the lattice  $\langle 2m\rangle$,
or equivalently (as we shall explain) as skew-holomorphic Jacobi forms of index $m$. On the other hand, the A-model topological string free energies on a K3-fibered CY threefold with polarization $\langle 2m\rangle$ are 
expected to exhibit $\Gamma_0^+(m)$ modular properties with respect to the complexified K\"ahler parameter $T$ governing the volume of the K3 fiber. On the B-model side, we show that these modularity properties are a consequence of the $\Gamma_0^+(m)$ monodromy in the complex structure moduli space of the central K3 $\Sigma_m$ of the Tyurin degeneration. On the A-model side, they arise 
through a Shintani-Skoruppa-Zagier-type correspondence between skew-holomorphic Jacobi forms of index $m$, weight $w$ and (in general mock) modular forms of weight $2w-2$ under $\Gamma_0^+(m)$, as expected from heterotic/type II duality. In particular, we shall obtain uniform expressions for the leading
terms in the genus 0 and 1 topological free energies in the large base limit
for all pairs $(X^{[i,j]}_{m,0},\hCY^{[i,j]}_{m,0})$ with $s=0$, and verify
the expected modularity properties. 

In the following, we describe our uniform construction of two-parameter, K3-fibered CY mirror pairs and explain our results  for the topological string free energies in these models in more detail.

\subsection*{A family of two-parameter CY mirror pairs}
We shall now summarize the main properties of the 
mirror pairs $(X^{[i,j]}_{m,s},\hCY^{[i,j]}_{m,s})$. For convenience,
we refer to the elements in the pair as the  A-model and B-model side, respectively.

On the B-model side, $Y^{[i,j]}_{m,s}$ is fibered by $M_m$-polarized K3 surfaces $\Sigma_m$ of Picard rank 19, parametrized by the modular curve  $X_0(m)^+$ \cite{dolgachev:1996xw}. 
The integers $(i,j,s)$ specify the ramification profile of the generalized functional invariant map~\footnote{This generalization of Kodaira's functional invariant map from the base of an elliptic surface to the $J$-line was introduced by the first author in~\cite{doran2001algebraic}. In general, one also needs to specify the generalized homological invariant which keeps track of the action of monodromy on algebraic cycles~\cite{doran2015families}.  The generalized functional invariant alone suffices here for $m>1$.  For  $m=1$ the choices for the generalized homological invariant were classified in~\cite{kooistra2021threefolds}.}
\be
\label{Lcover}
\Lambda: \begin{cases} \IP^1_B \to \IP^1\simeq X_0(m)^+ \\
y \mapsto \lambda=\Lambda(y) \end{cases}, \qquad
\Lambda(y) =\frac{1}{z_1}\frac{(1-y)^i (1-z_2/y)^j}{y^s}\,,
\ee
describing the fibration of $\Sigma_m$ over the base
$\IP^1_B$.
More precisely, the Ansatz \eqref{Lcover} arises by assuming that $\Lambda$ has ramification profile $[i-s,j+s]$ over 
$\lambda=\infty\in X_0(m)^+$, and $[i,j]$ over 
$\lambda=0$.
The moduli $z_1,z_2$ determine the complex structure of the total space. We will elaborate on this construction in Section~\ref{sec:gfim}.
In \S\ref{sec:periods}, 
we show how to compute
the fundamental period on $Y^{[i,j]}_{m,s}$ from the fundamental period
$f_m(\lambda)$ of the $M_m$-polarized K3 surface via a simple contour integral,
and hence determine the ideal of Picard-Fuchs differential operators.
 In the special case $s=0$, the covering $\IP^1_B\to\IP^1$
splits in the limit $z_2\rightarrow 0$ into two components that intersect but are not exchanged under monodromy, and $Y^{[i,j]}_{m,s}$ acquires
a Tyurin degeneration into a pair of Fano threefolds intersecting on an anticanonical $M_m$-polarized K3 surface $\Sigma_m$. This is the origin of 
modularity, as we discuss further below.

On the A-model side, $X^{[i,j]}_{m,s}$ carries in general no K3-fibration, but
it admits a Tyurin degeneration into a pair of Fano threefolds $(F_m^{[i]},F_m^{[j]})$
intersecting transversally over a Picard rank one, degree $2m$
K3 surface $\Sigma^m$, which is related to $\Sigma_m$ by mirror symmetry for lattice polarized K3 surfaces~\cite{nikulin1980integral,dolgachev:1996xw}.
The parameter $s$ specifies the resolution of the Tyurin degeneration, 
as explained in more detail in Section \S\ref{sec:bmodel}.
The Hodge numbers, intersection form $\kappa_{abc}=\gamma_a\cup\gamma_b\cup\gamma_c$,
second Chern class $c_{2,a}=\int_{\gamma^a}c_2(TX)$ and horizontal Gopakumar-Vafa invariants 
$\GV^{(0)}_{0,k}$ of $X^{[i,j]}_{m,s}$ are summarized below:
\begin{align}
\label{gentopdata}
\begin{split}
h_{1,1}=& 2\,, \quad 
h_{1,2}=22 +m(i+j)^2 + h_{1,2}(F_m^{[i]}) + h_{1,2}(F_m^{[j]})-2m(i-s)(j+s)\,,\\
    \kappa_{111}=&2m\left(\frac{1}{i}+\frac{1}{j}\right)\,,\quad \kappa_{112}=2m\left(1+\frac{s}{j}\right)\,,\quad \kappa_{122}=s\kappa_{112}\,,\quad \kappa_{222}=s^2\kappa_{112}\,,\\
    c_{2,1}=&2m(i+j)+24\left(\frac{1}{i}+\frac{1}{j}\right)\,,\quad
    c_{2,2}=24+ \left( \frac{24}{j}+2mj \right) s+2ms^2\,,\\
    \GV^{(0)}_{0,1}=&2m(i-s)(j+s)\,, \quad \GV^{(0)}_{0,k>0}=0\,.
    \end{split}
\end{align}
Here we use a basis of divisors $\gamma_a,\,a=1,2$ that form a basis of the K\"ahler cone. The Hodge numbers $h_{1,2}(F_m^{[i]})$ of the Fano threefolds can be
read off from Table \ref{tabFano}. In the special case $s=0$, the intersection numbers $\kappa_{122},\kappa_{222}$ vanish and the second Chern number become $c_{2,2}=24$, so by Oguiso's criterium~\cite{oguiso1993algebraic},  
$X^{[i,j]}_m:=X^{[i,j]}_{m,0}$  acquires a fibration by degree $2m$ K3-surfaces.
This fibration is 
mirror to the Tyurin degeneration in the complex structure moduli space of $\hCY^{[i,j]}_m:=\hCY^{[i,j]}_{m,0}$, consistent with the DHT conjecture.

 We note that several of these CY threefolds have appeared previously in the literature. In particular, the 
models $X_1^{[1,1]}$ and $X_2^{[1,1]}$  coincide with the small resolutions of hypersurfaces in weighted 
projective space $\IP^4_{1,1,2,2,6}[12]$ and $ \IP^4_{1,1,2,2,2}[8]$, studied in detail in 
\cite{Candelas:1993dm,Berglund:1993ax,Aspinwall:1994ay,Haghighat:2009nr}. These models
were revisited from the point of view of heterotic-type II duality in \cite{kachru:1995wm,kachru:1995fv,Kaplunovsky:1995tm,Antoniadis:1995zn,klemm:1995tj},
which provides one of the main motivations for focusing on K3-fibered CY threefolds. 
More generally, the threefolds $X^{[i,j]}_m$ with 
$m\leq 4$ include all two-parameter K3-fibered CY threefolds of CICY type (i.e. which  
can be realized as complete intersections in a product of projective spaces,
see Table \ref{tab:alternative}), 
and all have appeared previously (along with some other, non-K3-fibered models) in \cite[Table 1]{klemm:2004km}. The geometries with $m\geq 5$ can not be realized as complete intersections in toric ambient spaces but arise as complete intersections in Grassmanian bundles.
Their topological invariants were predicted earlier in~\cite{Klawer:2021ltm}.
On the other hand, it should be stressed that our construction does not include all known Picard rank 2 K3-fibered CY threefolds. In particular, Table 1 in \cite{klemm:2004km} includes several examples with $\GV^{(0)}_{0,2}\neq 0$ whose mirror is not K3-fibered, although it still admits a Tyurin degeneration into 
a pair of Fano threefolds intersecting over an $M_m$-polarized K3.
Further examples arise from the Generalized CICY construction \cite{Anderson:2015iia,Berglund:2016yqo} or as complete intersections in more general Grassmanian bundles~\cite{Knapp:2021vkm,Klawer:2021ltm} but their mirrors are not yet understood.

We give a uniform construction of the threefolds $X^{[i,j]}_{m,s}$ in terms of a complete intersection inside a projective bundle over a Fano fourfold $V_m^{[i,j]}$ with Picard rank 1.  
For $m=5,6,8$, we alternatively obtain $X^{[i,j]}_{m,s}$ through a conifold transition from a three-parameter family of torically realized K3-fibered CY threefolds. Similarly, for $m=7$ we obtain $X^{[i,j]}_{7}$ through a conifold transition from a five-parameter model.

An important  property of these A-model geometries is that they all admit a 
conifold transition to a one-parameter CY threefold $\cX_m ^{[i,j]}$ with $h_{1,1}( \cX_m ^{[i,j]})=1$. The transition proceeds by simultaneously shrinking the 
$\GV^{(0)}_{0,1}$ rational curves and smoothing the resulting nodal 
singularities. The list of one-parameter CY threefolds $\cX_m ^{[i,j]}$,
given in the last column of Table~\ref{tabmij}, includes the familiar 13 smooth hypergeometric cases \cite{Doran:2005gu} as well as several non-toric examples realized either as complete intersections in Grassmanians or as double covers of Fano threefolds. Moreover,  for $s\le i-j$ the geometries $X^{[i,j]}_{m,s}$ and $X^{[i,j]}_{m,i-j-s}$ are related by flopping the exceptional curves; as a consequence, they share the same mirror and degeneration.

\subsection*{Enumerative invariants and $\Gamma_0(m)^+$ modularity}

As mentioned earlier, one of our main motivations for introducing and studying the geometries $X^{[i,j]}_{m}$ and their mirrors comes from counting D4-D2-D0 black holes in type II string compactifications. 
It is generally expected that the generating series $h_{p,\mu}(\tau)$ of rank 0 Donaldson-Thomas  (DT) invariants, counting BPS states with vanishing D6-brane charge and fixed D4 and D2-brane charges $(p,\mu)$, should exhibit specific mock modular properties \cite{Alexandrov:2018lgp,Alexandrov:2019rth}. This physical prediction was verified mathematically in \cite{Alexandrov:2023zjb,Alexandrov:2023ltz} in the context of one-parameter hypergeometric CY threefolds to great accuracy, using wall-crossing in a certain family of weak stability conditions, but the mathematical origin of the modular symmetry remains mysterious. In contrast, for K3-fibered CY threefolds, it was shown in
\cite{Bouchard:2016lfg} that the DT invariants counting vertical D4-D2-D0 states (i.e. such that the D4-brane is supported along the K3 fiber) are determined from the Noether-Lefschetz (NL) invariants of the fibration, such that the resulting series  $h_{p,\mu}(\tau)$ is modular\footnote{The modular anomaly in this case vanishes, due to the properties of the cubic intersection form for 
K3-fibered manifolds.} by general results of 
Kudla-Millson and Borcherds \cite{kudla1990intersection,0919.11036}.
The NL invariants can in turn be determined 
from the vertical GV invariants by the GV/NL relation of  \cite{maulik2007gromov}.

To determine these invariants for all models $(m,i,j,s)$ with $s=0$, we
establish a universal formula \eqref{d2WTh} for the genus 0 free energy (or Yukawa couplings) in the large base limit $S\to\I\infty$, and similarly \eqref{F1univ} for the genus-one free energy. Since the monodromy group admits a $\Gamma_0(m)^+$
quotient preserving the large base limit, the resulting formulae exhibit modular behavior under $\Gamma_0(m)^+$, although they are not strictly modular due 
to polynomial ambiguities in $F_0$, and to the non-trivial transformation property
of the base modulus $S$ in $F_1$. Nonetheless, we can extract the vertical genus 0 and genus 1 GV invariants from the Fourier expansion in $T$, determine the resulting NL invariants and verify the 
expected modularity properties.

\subsection*{Heterotic-type II duality and modularity at higher base degree}
Another motivation to study the models with $s=0$ comes from heterotic/type II duality, which is expected to hold whenever the CY threefold admits a 
K3-fibration \cite{kachru:1995wm,klemm:1995tj,Aspinwall:1996mn,Aspinwall:2000fd}. In fact, the relation between GV
and NL invariants was initially motivated by heterotic/type II duality, where the generating series of 
NL invariants is identified with the `new supersymmetric index' counting perturbative BPS states on the heterotic side (see  \cite{Kawai:1996te,Marino:1998pg,klemm:2004km} and more recently \cite{Enoki:2019deb}). This duality is fairly well understood
when the K3 fiber is itself elliptically fibered, such that the duality in $D=4$ follows from heterotic/type II duality in $D=6$ by adiabatic arguments \cite{Vafa:1995gm}. In contrast, when the K3 fiber on the type II side has Picard rank one or zero, the decompactification to 6 dimensions is obstructed, and the heterotic dual is expected to involve a non-trivial elliptic fibration over a K3 surface, with fluxes threading both the K3 and the torus fiber \cite{Melnikov:2012cv}.  The worldsheet construction of such heterotic flux compactifications is in general an open problem, which can be informed by our explicit results for the new supersymmetric index (see \cite{Israel:2023itj,Israel:2023tjw} for some recent progress).

Assuming that such an heterotic description exists, the duality group $\Gamma_0(m)^+$ is interpreted as the subgroup of the standard T-duality group $O(2,2,\IZ)$ of the torus compatible with the fluxes. It must therefore be a symmetry of physical observables at each order in heterotic perturbation theory. In particular, it implies that the one-loop contribution to the prepotential (which sums up  contributions of vertical GV invariants $\GV_{(0,d)}^{(0)}$ on the type II side) must transform as a holomorphic Eichler integral of weight $-4$ under 
$\Gamma_0(m)^+$; this property is indeed ensured by the fact that
the one-loop contribution to the prepotential is related to the 
new supersymmetric index by a 
theta lift \cite{Angelantonj:2015rxa,Enoki:2021vid,Enoki:2022cfc}. 
At the non-perturbative level,
it also implies that instanton contributions of order $e^{2\pi\I k S}$ 
(which sum up contributions of  GV invariants $\GV_{(k,d)}^{(0)}$ with fixed base degree $k$ on the type II side) should also have specific modular properties, elucidated in 
\cite{Henningson:1996jf} for the simplest model with $m=1$. In \S\ref{sec_hidegree} we shall verify this modular invariance at base degree $1$ 
and $2$ for all models in Table \ref{tabmij}. It would be very interesting to understand the mathematical origin of these modular properties on the type IIA side.

\begin{table}
\begin{tabular}{|c|c|c|}
\hline 
$(m,i)$ & $h_{1,2}(F_m^{[i]}) $ & \mbox{Construction of $F_m^{[i]}$}\\
\hline \hline
(1,1)   & 52 & $\IP_{1,1,1,1,3}[6]$ \\ \hline
(1,2)   &21 &  $\IP_{1,1,1,2,3}[6] $ \\ \hline
 \hline
(2,1)   & 30 & $\IP^4[4]$ \\  \hline
(2,2)   & 10 & $\IP_{1,1,1,1,2}[4] $ \\ \hline
(2,4)  & 0 &   $ \IP ^3$ \\  \hline
 \hline
 (3,1)   & 20 &  $ \IP ^5[2,3]$ \\  \hline
(3,2)   & 5 &  $ \IP ^4[3]$ \\  \hline
(3,3)  & 0 & $ \IP ^4[2]$ \\  
 \hline\hline
(4,1)   & 14 & $\IP ^6[2,2,2]$ \\  \hline
(4,2)   & 2 &  $\IP ^5[2,2]$ \\  \hline
 \hline
 (5,1)  & 10 & $X^{2,5}_{\mathcal{O}(1)^{\oplus 2}\oplus\mathcal{O}(2)}$ \\  \hline
(5,2)  & 0 & $B_5=X^{2,5}_{\mathcal{O}(1)^{\oplus 3}}$ \\ 
 \hline \hline
(6,1)  & 7 & $X^{2,5}_{\mathcal{S}^{\vee}(1)\oplus\mathcal{O}(1)}$ \\  \hline
 \hline
 (7,1)  & 5 & $X^{2,6}_{\mathcal{O}(1)^{\oplus 5}}$ \\  \hline
 \hline
 (8,1)  & 3& $X^{3,6}_{\bigwedge^2\mathcal{S}^{\vee}\oplus\mathcal{O}(1)^{\oplus 3}}$ \\  \hline
 \hline
 (9,1)  &2 & $X^{5,7}_{\bigwedge^4\mathcal{S}^{\vee}\oplus\mathcal{O}(1)^{\oplus 2}}$ \\ \hline
 \hline
 (11,1) &0 & $A_{22}=X^{3,7}_{(\bigwedge^2 \mathcal{S}^\vee)^{\oplus 3}}$ \\ 
\hline
\end{tabular}
\vspace*{1cm}
\caption{The 17 Fano threefolds of Picard rank one, as classified in \cite{iskovskih1978fano1,iskovskih1978fano2}. The index $i$ is the ratio between the anticanonical class $-K_{F_m^i}$ and the generator of the Picard lattice. Note that $h_{1,2}(F_m^{[i]})=-10+\frac{120+\delta_m}{i(m+1)}$ where 
$\delta_m=4$ if $m=1$, $-1$ if $m=6$, $-3$ if $m=8$ and 0 otherwise.
The geometries $X^{k,n}_{E}$ are complete intersections in $\text{Gr}(k,n)$ associated to vector bundles $E$ and have been identified in~\cite{DeBiase2021}. The notation for the bundles is explained in Section~\ref{sec:notations}.}
\label{tabFano}
\end{table}

\subsection*{Outlook}
In this work, we have uncovered a rather large class of two-parameter K3-fibered CY threefolds, but have only scratched the surface of their rich enumerative geometry. In particular, while we have found general expressions for the vertical GV invariants, and determined the corresponding Noether-Lefschetz invariants, for non-vanishing base degree we have only determined the modular generating series of genus 0 GV invariants at base degree one, two and three in some cases. The Picard-Fuchs equations of course allow to determine the genus 0 GV invariants at arbitrary base degree, but their modular properties become increasingly complicated, and their mathematical origin (in terms of Noether-Lefschetz theory  or otherwise) remain rather mysterious. It would also be desirable to construct modular invariant generating functions for higher genus GV invariants at higher base degree. With this data at hand, it should be possible to determine the complete set of D4-D2-D0 invariants (beyond the vertical case), check (mock) modularity by using similar wall-crossing arguments as in \cite{Alexandrov:2023zjb,Alexandrov:2023ltz}, and recover the D4-D2-D0 invariants
of the one-parameter model through a conifold transition. Another interesting
task is to incorporate K3-fibered threefolds whose mirror is not K3-fibered, but nonetheless admits a Tyurin degeneration into a product of Fano threefolds intersecting over an $M_m$ polarized anticanonical K3. We hope to return to these challenging questions in future work.

\subsection*{Outline}

The remainder of this work is organized as follows. In \S\ref{sec_mirrorK3}, we recall basic facts about the moduli space $X_0(m)^+$ of $M_m$-polarized K3 surfaces, the Dolgachev-Nikulin mirror symmetry relating them to degree $2m$ K3 surfaces, and introduce relatives $\tilde f_m^{[i]}(\lambda)$ of the fundamental period $f_m(\lambda)$ which play a key role in later sections. We also briefly recall the DHT conjecture in the context of K3-fibered CY threefolds. 
In \S\ref{sec:bmodel}, we give construct smooth $M_m$-polarized K3-fibered threefolds $Y^{[i,j]}_{m,s}$
with two complex structure moduli, by judiciously choosing the generalized functional invariant map from the base $\IP^1$ to the moduli space $X_0(m)^+$.
In \S\ref{sec:amirror}, we give a uniform construction of the mirror
$X^{[i,j]}_{m,s}$. In \S\ref{sec:K3mod}, we restrict to the case $s=0$, where
$X^{[i,j]}_{m,s}$ admits a fibration by degree $2m$ hypersurfaces, and determine
universal formulae for the genus-0 and genus-1 free energies in the large base limit. In \S\ref{sec_enum}, we extract various enumerative invariants from these results, namely the Noether-Lefschetz invariants of the fibration, and the 
vertical Gopakumar-Vafa and Donaldson-Thomas invariants, and check that the
generating series have the expected modular properties. We also determine the generating series of non-vertical genus 0 GV invariants,
with base degree one and two, and investigate their
modular properties. Explicit results are collected in Appendix 
\S\ref{sec_results}, after reviewing useful facts about modular forms for $\Gamma_0(m)^+$
and skew-holomorphic Jacobi forms in \S\ref{sec_Gamma0m} and 
\S\ref{sec_skew}. Mathematica notebooks allowing to reproduce and possibly extend our results are provided in a publicly accessible repository:
\begin{center}
\url{https://github.com/bpioline/TwoParameterK3} 
\label{foogithub}
\end{center}
In the notebook we also include the genus 0 and genus 1 GV invariants for all of the examples up to degree 15.
For the cases $X_1^{[1,1]},X_1^{[2,2]},X_2^{[1,1]},X_2^{[2,2]},X_5^{[1,1]},X_6^{[1,1]}$ we also include the corresponding genus 2 GV invariants that we have obtained using the direct integration method of the holomorphic anomaly equations following~\cite{Bershadsky:1993cx,Huang:2006hq,Alim:2012ss}.
In Appendix~\ref{sec:nonToricFundamentals} we discuss the fundamental periods of $M_m$-polarized K3 surfaces $\Sigma_m$ for $m=5,\ldots,9,11$ and in 
Appendix~\ref{sec_Amodel} we discuss conifold transitions from CY threefolds that are complete intersections in toric ambient spaces to $X_m^{[1,1]}$ for $m=6,7,8$.

\subsection*{Acknowledgements}
We are grateful to Sergey Alexandrov, Per Berglund, Andrew Harder, Ilarion Melnikov, Eric Pichon-Pharabod, Emanuel Scheidegger, Alan Thompson, Don Zagier for useful discussions, and to Taizan Watari and Nils Skoruppa for very helpful correspondence. We also want to thank Emanuel Scheidegger for useful comments on the draft. The research of CD is supported by the Natural Sciences and Engineering Research Council of Canada, the Distinguished Visiting Professorship of Mathematics and Physics at Bard College, and the Center of Mathematical Sciences and Applications at Harvard University.  The research of BP is supported by Agence Nationale de la Recherche under contract number ANR-21-CE31-0021.
The research of TS was supported by Agence Nationale de la Recherche under the same contract during the beginning of the work on this paper. {\it For the purpose of Open Access, a CC-BY public copyright licence has been applied by the authors to the present document and will be applied to all subsequent versions up to the Author Accepted Manuscript arising from this submission.}

\subsection{Notations}
\label{sec:notations}
We denote by $\text{Gr}(k,n)$ the $k(n-k)$-dimensional Grassmanian variety of $k$-planes in a complex $n$-dimensional vector space.
On $\text{Gr}(k,n)$ we denote by $\mathcal{S}$ the tautological bundle of rank $k$, $\mathcal{Q}$ is the universal quotient bundle of rank $n-k$ and we use $\mathcal{O}(1)=\det \mathcal{S}^{\vee}$.

For $X$ a CY threefold, 
we denote by $\{\gamma_a\}$, $a=1\dots b_2(X)=h_{1,1}(X)$ a basis of $H_4(X,\IZ)$, by $\{\gamma^a\}$ the Poincar\'e-dual basis of $H_2(X,\IZ)$ such that  $\gamma_a \cap \gamma^b=\delta_a^b$, by $\kappa_{abc}=\gamma_a\cap\gamma_b\cap\gamma_c$ the triple intersection numbers, and by $c_{2,a}=\int_{\gamma_a} c_2(TX)$ the
second Chern class. We denote by $t^a$ the complexified K\"ahler parameters, such that the complexified K\"ahler form reads $\omega=t^a\gamma_a$
with $\Im t^a>0$.
The mirror $Y$ is another family of CY threefolds such that $h_{1,2}(Y)=h_{1,1}(X), h_{1,1}(Y)=h_{1,2}(X)$ hence the Euler numbers are opposite, $\chi_X=2(h_{1,1}(X)-h_{1,2}(X))=-\chi_Y$.

Near the large volume limit $t^a\to\I\infty$ in K\"ahler moduli space and in the holomorphic limit, the topological string partition function on $X$ takes the form
\be
\label{Ftop}
F_{\rm top}(t^a,\lambda) 
:=\sum_{g\geq 0} \lambda^{2g-2} F^{(g)}(t^a)
= F_{\rm pol}(t^a,\lambda) 
+\frac{1}{(2\pi\I)^3} \sum\limits_{g=0}^\infty
\sum\limits_{d_a\geq 0}^\infty \GW^{(g)}_{d_a} 
e^{2\pi \I d_a t^a}\,,
\ee
where the polynomial part is given by 
\bea
\label{Fpol}
F_{\rm pol}(t^a,\lambda) &=&\frac{1}{\lambda^2}
\left(  \frac16 \kappa_{abc}t^a t^b t^c - \frac{1}{24} c_{2,a} t^a -\frac12 A_{ab} t^a t^b   - \frac{\zeta(3)}{2(2\pi\I)^3} \chi_X \right) +\frac{1}{24} c_{2,a} t^a \,.
\eea
Here, $A_{ab}$ is a symmetric matrix with half-integer entries which depends on the choice of basis.
The remainder is exponentially suppressed as $t^a\to\I\infty$,
and determined by the Gromov-Witten invariants $\GW^{(g)}_{d_a}$ counting (families of) curves of genus $g$ and class $d_a \gamma^a$. The latter are rational numbers, but can be expressed in terms of the integer-valued
Gopakumar-Vafa invariants $\GV^{(g)}_{d_a}$ by the multi-cover formula \cite{Gopakumar:1998ii,Gopakumar:1998jq}
\bea
\label{GWtoGV}
\sum\limits_{g=0}^\infty
\sum\limits_{d_a\geq 0} \GW^{(g)}_{d_a} 
e^{2\pi \I d_a t^a}
&=&
\sum\limits_{g=0}^\infty
\sum\limits_{n=1}^\infty
\sum\limits_{d_a \geq 0}
\frac{\GV^{(g)}_{d_a}}{n}
\left(2\sin\frac{n\lambda}{2}\right)^{2g-2}e^{2\pi \I n d_a t^a}\,.
\eea
At the first few orders in $\lambda$, the right-hand side evaluates
to 
\be
\label{GWtoGVexp}
\sum\limits_{d_a\geq 0} \left[
\frac{1}{\lambda^2} \GV^{(0)}_{d_a} \Li_3(q_d )
+ \left( \GV^{(1)}_{d_a} + \frac1{12} \GV^{(0)}_{d_a} \right)
\Li_1(q_d) 
+ \lambda^2 \left( \GV^{(2)}_{d_a} + \frac1{240} \GV^{(0)}_{d_a} \right)
\Li_{-1}(q_d)
+\dots \right]\,,
\ee
where $q_d=e^{2\pi \I d_a t^a}$ and 
$\Li_k(z)=\sum_{n\geq 1} z^n/n^k$ is the polylogarithm.
The genus zero part $F^{(0)}$ is the prepotential governing the
special K\"ahler metric on the K\"ahler moduli space of $X$, or the complex structure moduli space of $Y$.

\section{Mirror symmetry for K3-fibered CY}
\label{sec_mirrorK3}

\subsection{$M_m$-polarized K3 surfaces\label{sec_MmK3}}

Let $\Sigma$ be a complex algebraic K3 surface, i.e. a smooth projective algebraic complex surface 
with vanishing canonical class and first Betti number. Its second cohomology $H^2(\Sigma,\IZ)$ is isomorphic to the signature $(3,19)$ lattice $L=E_8\oplus E_8 \oplus U \oplus U \oplus U$, where 
$E_8$ is the unique even self-dual negative definite lattice of rank 8 and $U$ is the even self-dual hyperbolic lattice of signature $(1,1)$. The Picard lattice is the sublattice of algebraic cycles in $H^2(\Sigma,\IZ)$, 
\be
\Pic\Sigma = H^{1,1}(X) \cap H^2(\Sigma,\IZ)\,,
\ee
with signature $(1,t-1)$ where $t$ is the Picard rank. Given an even, non-degenerate lattice $M$ of signature $(1,t-1)$, an $M$-polarized K3-surface is a pair $(\Sigma,j)$ where $\Sigma$ is a complex algebraic K3 surface and $j:M \hookrightarrow \Pic\Sigma$ is a primitive lattice embedding. By the period map, the moduli space of $M$-polarized K3-surfaces is isomorphic to the orthogonal Grassmannian 
$\cM_M =O(2,20-t)/O(2)\times O(20-t)/\Gamma_M$ quotiented by an arithmetic subgroup $\Gamma_M\subset 
O(2,20-t)$ which is isomorphic to the automorphism group of the orthogonal lattice $M^\perp\subset L$.~\footnote{In general, for the purposes of compactification of moduli spaces, the proper notion of an {\em $M$-quasipolarized} K3 surface was recently introduced by Alexeev-Engel~\cite{alexeevengel23}. This distinction is irrelevant for our one-dimensional moduli spaces of $M_m$-polarized K3 surfaces which are compactified by adding a finite number of points.} 
In this paper we shall be interested in the following two special cases:
\begin{itemize}
\item Degree $2m$ polarized K3 surfaces are K3-surfaces $\Sigma^m$ polarized with respect to
$M=\langle 2m\rangle$, the one-dimensional lattice $e\IZ$ with generator $e$ such that $(e,e)=2m$; they have Picard rank 1 and 19-dimensional moduli space; 
\item $M_m$-polarized K3 surfaces are K3-surfaces $\Sigma_m$ polarized with respect to 
$M_n:=U \oplus E_8 \oplus E_8 \oplus \langle -2m\rangle$: they have Picard rank 19 and 
1-dimensional moduli space. 
\end{itemize}
For each value of $m \in \Z^+$, the K3 surfaces in the two special cases above are mirror to one another.

For $m\leq 4$, families of degree $2m$ polarized K3 surfaces $\Sigma^m$ can be realized as complete intersections in weighted projective space as follows:
\be
\begin{split}
m=1: & \qquad \IP^3_{1,1,1,3}[6]\,, \quad \IP^4_{1,1,1,2,3}[2,6]\,, \quad \IP^4_{1,1,1,3,3}[3,6] \\
m=2: & \qquad \IP^3[4]\,, \quad \IP^4_{2,1,1,1,1}[2,4]\,, \quad \IP^4_{3,1,1,1,1}[3,4] \\
m=3: & \qquad \IP^4[2,3]\\
m=4: & \qquad \IP^5[2,2,2]
\end{split}
\label{K31234}
\ee
For $m\geq 5$, more complicated constructions are required, see e.g. \cite[\S2.3]{Debarre:2018mmi} for a review. In particular, 
for $m=5,\ldots,9,11$ the surfaces $\Sigma^m$ can then be realized as the zero loci of general sections of a vector bundle in the Grassmannian $\text{Gr}(r,n)$ or $r$-dimensional planes in $\IC^n$ (a projective variety of dimension $r(n-r)$)
as follows:
\begin{align}
    \begin{split}
    m=5: & \qquad X^{2,5}_{\mathcal{O}(1)^{\oplus 3}\oplus\mathcal{O}(2)}\\
    m=6: & \qquad X^{2,5}_{\mathcal{S}^{\vee}(1)\oplus\mathcal{O}(1)^{\oplus 2}}\\
    m=7: & \qquad X^{2,6}_{\mathcal{O}(1)^{\oplus 6}}\\
    m=8: & \qquad X^{3,6}_{\bigwedge^2\mathcal{S}^{\vee}\oplus\mathcal{O}(1)^{\oplus 4}}\\
    m=9: & \qquad X^{5,7}_{\bigwedge^4\mathcal{S}^{\vee}\oplus\mathcal{O}(1)^{\oplus 3}}\\
    m=11: &\qquad X^{3,7}_{(\bigwedge^2\mathcal{S}^{\vee})^{\oplus 3}\oplus\mathcal{O}(1)}
    \end{split}
    \label{eqn:nonToricK3}
\end{align}
Degree $2m$ polarized K3 surfaces will be further discussed in \S\ref{sec_GVNL} in relation to Noether-Lefschetz theory.
In the rest of this subsection we focus on $M_m$-polarized K3 surfaces, and, for reasons which will become clear in \S\ref{sec:bmodel}, we restrict attention to the range $m\leq 11$
and $m\neq 10$.

In all cases, the periods are annihilated by a third order differential operator  equation with respect to the complex parameter $\lambda$ in the one-dimensional family, the Picard-Fuchs (PF) operator given in Table \ref{tabPFK3}. We choose $\lambda=\infty$ as the point of maximal unipotent monodromy, and denote by $f_m(\lambda)$ the fundamental period, which is unique solution which is analytic as $\lambda\to\infty$:
 \be
 \label{deffm}
 f_m(\lambda)=\sum_{d\geq 0} c_m(d) \lambda^{-d}\,.
 \ee 
For $m\leq 4$, the coefficients are simple ratios of factorials, 
 \be
 \label{K3fperiod}
 c_1(d)= \frac{(6d)!}{(d!)^3 (3d)!}\,, \quad
c_2(d)=\frac{(4d)!}{(d!)^4}\,, \quad 
c_3(d)= \frac{(2d)!(3d)!}{(d!)^5}\,, \quad
c_4(d)= \frac{[(2d)!]^3}{(d!)^6}\,,
\ee
leading to the familiar hypergeometric series
\begin{align}
\begin{split}
f_1( \lambda)=& _3F_2\left( \tfrac16,\tfrac12,\tfrac56;1,1; 1728/\lambda \right)
=  \left[ _2F_1\left(\tfrac{1}{12},\tfrac{5}{12};1; 1728/\lambda \right) \right] ^2\,,\\
f_2(\lambda) =& _3F_2\left( \tfrac14,\tfrac12,\tfrac34;1,1; 256/\lambda \right)
 =\left[ _2F_1\left(\tfrac{1}{8},\tfrac{3}{8};1; 256/\lambda \right) \right] ^2\,,\\
f_3(\lambda) =& _3F_2\left( \tfrac13,\tfrac12,\tfrac23;1,1; 108/\lambda \right) 
 =\left[ _2F_1\left(\tfrac{1}{6},\tfrac{1}{3};1; 108/\lambda \right) \right] ^2\,,\\
f_4( \lambda) =& _3F_2\left( \tfrac12,\tfrac12,\tfrac12;1,1; 64/\lambda \right) 
= \left[ _2F_1\left(\tfrac{1}{4},\tfrac{1}{4};1; 64/\lambda \right) \right] ^2 \,.
\end{split}
\end{align}
The fundamental periods for the cases $\Sigma_m,\,m=5,\ldots,9,11$, where the mirror $\Sigma^m$ does not admit a realization as a complete intersection in a toric ambient space, are not hypergeometric and are discussed in Appendix~\ref{sec:nonToricFundamentals}.

In addition to the fundamental period, 
the PF equation admits solutions with single and 
double logarithmic singularities at $\lambda=\infty$, such that
 \bea
 \begin{split}
  \label{deffm12}
 f_m^{(1)}(\lambda) =& f_m(\lambda) \log \lambda + \sum_{m\geq 1} c_m^{(1)}(d) \lambda^{-d}\,,\\
 f_m^{(2)}(\lambda) =& -\frac12 f_m(\lambda)\,  (\log \lambda)^2 + 
 f_m^{(1)} (\lambda) \log \lambda 
  + \sum_{m\geq 2} c_m^{(2)}(d) \lambda^{-d} \,.
  \end{split}
 \eea
The solutions are easily obtained by the Frobenius method. 
For example, 
the coefficients for the single log solution with $m\leq 4$ are given by
\bea
 \label{K3fperiodlog}
 c^{(1)}_1(d) &=& 3 \frac{(6d)!}{(d!)^3 (3d)!} \left[ H_d+H_{3d} - 2 H_{6d} \right]\,, \quad
c^{(1)}_2(d)=4\frac{(4d)!}{(d!)^4} \left[ H_d- H_{4d} \right]\,, \quad  \nn\\
c^{(1)}_3(d) &=& \frac{(2d)!(3d)!}{(d!)^5} \left[ 5H_d-2H_{2d} - 3 H_{3d} \right]\,, \quad
c^{(1)}_4(d)= 6\frac{[(2d)!]^3}{(d!)^6} \left[ H_d- H_{2d} \right] \,,
\eea
where $H_n=\sum_{k=1}^n 1/k$ are the harmonic numbers. It turns out that the double log solution
is determined from the fundamental period and single log solution via the quadratic equation 
 \be
  \left( f_m^{(1)}(\lambda) \right)^2 = 2 f_m(\lambda)\, f_m^{(2)}(\lambda) \,.
 \ee
This follows from the fact that the third order Picard-Fuchs equation is 
 the symmetric square of a second order differential 
 equation~\cite{doran1999picard}.

\begin{table}
    \centering
    \begin{tabular}{||c|l||}
    \hline
        $m$ & Picard-Fuchs Operator  \\
        \hline
        \hline
        $1$ &$\frac{d^3}{d\lambda^3}+\frac{3 \lambda - 2592}{\lambda^{2} - 1728 \lambda}\frac{d^2}{d\lambda^2}+\frac{\lambda - 240}{\lambda^{3} - 1728  \lambda^{2}}\frac{d}{d\lambda}+ \frac{120}{\lambda^{4} - 1728  \lambda^{3}}$\\
        \hline
        $2$ &$\frac{d^3}{d\lambda^3}+\frac{3 \lambda - 384}{\lambda^{2} - 256 \lambda}\frac{d^2}{d\lambda^2}+\frac{\lambda - 48}{\lambda^{3} - 256 \lambda^{2}}\frac{d}{d\lambda}+ \frac{24}{\lambda^{4} - 256 \lambda^{3}}$\\
        \hline
        $3$&$\frac{d^3}{d\lambda^3}+\frac{3 \lambda - 162}{\lambda^{2} - 108 \lambda}\frac{d^2}{d\lambda^2}+ \frac{\lambda - 24}{\lambda^{3} - 108 \lambda^{2}}\frac{d}{d\lambda}+ \frac{12}{\lambda^{4} - 108 \lambda^{3}}$\\
        \hline
        $4$&$\frac{d^3}{d\lambda^3}+\frac{3 \lambda - 96}{\lambda^{2} - 64 \lambda}\frac{d^2}{d\lambda^2}+\frac{\lambda - 16}{\lambda^{3} - 64 \lambda^{2}}\frac{d}{d\lambda}+ \frac{8}{\lambda^{4} - 64 \lambda^{3}}$\\
        \hline
        $5$&$\frac{d^3}{d\lambda^3}+\frac{3 \lambda - 66}{\lambda^{2} - 44 \lambda - 16}\frac{d^2}{d\lambda^2}+\frac{\lambda^{2} - 12 \lambda - 12}{\lambda^{4} - 44 \lambda^{3} - 16 \lambda^{2}}\frac{d}{d\lambda}+\frac{6 \lambda + 12}{\lambda^{5} - 44 \lambda^{4} - 16 \lambda^{3}}$\\
        \hline
        $6$&$\frac{d^3}{d\lambda^3}+\frac{3 \lambda - 51}{\lambda^{2} - 34 \lambda + 1}\frac{d^2}{d\lambda^2}+\frac{\lambda^{2} - 10 \lambda + 1}{\lambda^{4} - 34 \lambda^{3} + \lambda^{2}}\frac{d}{d\lambda}+ \frac{5 \lambda - 1}{\lambda^{5} - 34 \lambda^{4} + \lambda^{3}}$\\
        \hline
        $7$&$\frac{d^3}{d\lambda^3}+\frac{3 \lambda - 39}{\lambda^{2} - 26 \lambda - 27}\frac{d^2}{d\lambda^2}+\frac{\lambda^{2} - 8 \lambda - 24}{\lambda^{4} - 26 \lambda^{3} - 27 \lambda^{2}}\frac{d}{d\lambda}+\frac{4 \lambda + 24}{\lambda^{5} - 26 \lambda^{4} - 27 \lambda^{3}}$\\
        \hline
        $8$&$\frac{d^3}{d\lambda^3}+\frac{3 \lambda - 36}{\lambda^{2} - 24 \lambda + 16}\frac{d^2}{d\lambda^2}+\frac{\lambda^{2} - 8 \lambda + 16}{\lambda^{4} - 24 \lambda^{3} + 16 \lambda^{2}}\frac{d}{d\lambda}+\frac{4 \lambda - 16}{\lambda^{5} - 24 \lambda^{4} + 16 \lambda^{3}}$\\
        \hline
        $9$&$\frac{d^3}{d\lambda^3}+\frac{3 \lambda - 27}{\lambda^{2} - 18 \lambda - 27}\frac{d^2}{d\lambda^2}+\frac{\lambda^{2} - 6 \lambda - 27}{\lambda^{4} - 18 \lambda^{3} - 27 \lambda^{2}}\frac{d}{d\lambda}+\frac{3 \lambda + 27}{\lambda^{5} - 18 \lambda^{4} - 27 \lambda^{3}}$\\
        \hline
        $11$&$\frac{d^3}{d\lambda^3}+\frac{3 \lambda^{3} - 30 \lambda^{2} + 66}{\lambda^{4} - 20 \lambda^{3} + 56 \lambda^{2} - 44 \lambda}\frac{d^2}{d\lambda^2}+\frac{\lambda^{3} - 8 \lambda^{2} + 64 \lambda - 132}{\lambda^{5} - 20 \lambda^{4} + 56 \lambda^{3} - 44 \lambda^{2}}\frac{d}{d\lambda}+ \frac{4 \lambda^{2} - 64 \lambda + 132}{\lambda^{6} - 20 \lambda^{5} + 56 \lambda^{4} - 44 \lambda^{3}}$\\
        \hline
    \end{tabular}
    \vspace*{5mm}
    \caption{Picard-Fuchs operators for $M_m$-lattice polarized K3, from
    \cite[Table 7]{doran2019geometric}.
    }
    \label{tabPFK3}
\end{table}

\begin{table}
\centering
\begin{tabular}{|c|c|c|c|c|}
\hline
{ $m$}
&
{ Orbifold Type}
& 
{ $\lambda_1,\dots, \lambda_r$}
& 
{ $\tau_1,\dots, \tau_r$}
& {$\tau_0$}
\\
\hline
$1$&$(3;2;\infty)  $&$1728$ & $\I$ & $\frac12+\frac{\I \sqrt{3}}{2}$\\
\hline
$2$&$(4;2;\infty)$&$256$  & $\frac{\I}{\sqrt2}$ & $\frac{1+\I}{2}$\\
\hline
$3$&$(6;2;\infty)$&$108$ &  $\frac{\I}{\sqrt3}$& $\frac12+\frac{\I}{2\sqrt3}$\\
\hline
$4$&$(\infty;2;\infty)$&$64$ & $\frac{\I}{2}$ & $\frac12$\\
\hline
$5$&$(2;2,2;\infty)$&$22+10\sqrt{5},22-10\sqrt{5}$ &  $\frac{\I}{\sqrt5},\frac49+\frac{\I}{9\sqrt5}$ & $\frac{\I+3}{5} $\\
\hline
$6$&$(\infty;2,2;\infty)$&$17+12\sqrt{2},17-12\sqrt{2}$ &  $\frac{\I}{\sqrt6},\frac25+\frac{\I}{5\sqrt6}$ & $\frac12$ \\
\hline
$7$&$(3;2,2;\infty)$&$27,-1$ &  $\frac{\I}{\sqrt7},\frac12+\frac{\I}{2\sqrt7}$ & 
$\frac{5+\I\sqrt{3}}{14}$\\
\hline
$8$&$(\infty;2,2;\infty)$&$12+8\sqrt{2},12-8\sqrt{2}$ & $\frac{\I}{\sqrt8},-\frac{2}{11}+\frac{\I}{22\sqrt2}$ & $\frac12$ \\
\hline
$9$&$(\infty;2,2;\infty)$&$9+6\sqrt{3},9-6\sqrt{3}$ & $\frac{\I}{9}, \frac12+\frac{\I}{6}$ & $\frac13$ \\
\hline
$11$&$(2;2,2,2;\infty)$&Roots of $\lambda^3-20\lambda^2+56\lambda-44$ & $\frac{\I}{\sqrt{11}},\frac23+\I \frac{\sqrt{11}}{33}, \frac{22}{25}+\frac{\I\sqrt{11}}{275}$ & $\frac16 + \frac{\I}{6\sqrt{11}}$ \\
\hline
\end{tabular}
\vspace*{5mm}
\caption{Structure of the modular surface $X_0(m)^+$. The singularities lie
at $\lambda \in \{ 0;\lambda_1,\dots,\lambda_{r};\infty\}$, or
$\tau \in \{ \tau_0; \tau_1,\dots, \tau_r; \I\infty\}$, with $\lambda_1,\dots,\lambda_r$ being $\IZ_2$-orbifold points.
\label{TableX0m}}
\end{table}

\subsection{Mirror symmetry for $M_m$-polarized K3 surfaces}
\label{sec:K3mirrorsymmetry}

In general, Dolgachev-Nikulin mirror symmetry \cite{dolgachev:1996xw} exchanges families of 
$M$-polarized K3 surfaces $\Sigma$, of Picard rank $t$, with families of 
$M^\vee$-polarized K3 surfaces, of Picard rank $20-t$, where 
$M^\vee$ is such that the orthogonal complement
$M^\perp$ in $H_2(\Sigma,\IZ)$ is isomorphic to $U(k) \oplus M^\vee$ for 
some $k\geq 1$.
For $t=1,k=1$, the case of interest in this paper, 
mirror symmetries exchanges families of 
$M_m$-polarized K3 surfaces with degree $2m$ polarized K3 surfaces. On the $A$-model side,
the K\"{a}hler moduli space is parametrized by
$\tau=\int_{e}(B+\I J)$ in the upper half-plane $\IH$,  where $e$ is the generator of $\Pic\Sigma^m$, $J$ is the K\"ahler form
and $B$ the Kalb-Ramond field.  The global structure is manifest 
on the B-model side, where the complex structure moduli space 
is isomorphic to the modular curve
\be
\label{modcurve}
\cM_{M_m} \simeq X_0(m)^+=\IH/\Gamma_0(m)^+\,,
\ee
where $\Gamma_0(m)$ is the congruence subgroup of $SL(2,\IZ)$ matrices 
${\scriptsize \begin{pmatrix} a & b \\ c & d \end{pmatrix}}$ with 
$c \equiv 0 \mod m$, acting by $\tau\mapsto \frac{a\tau+b}{c\tau+d}$ on $\IH$, 
and $\Gamma_0(m)^+$ denotes its extension by the Fricke involution acting as $\tau\mapsto -1/(m\tau)$ (see Appendix \ref{sec_Gamma0m} for more details). 
For $m=1$, $\Gamma_0(1)^+$ is simply the modular group $SL(2,\IZ)$. 
More generally, for our cases of interest $m=1,\ldots,9,11$, the modular curve $X_0(m)^+$ has genus 0, one cusp of width 1 at $\lambda=\infty$, $r$ $\IZ_2$-orbifold points at $\lambda\in\{\lambda_1,\ldots,\lambda_r\}$ and an additional singular point at $\lambda=0$ which is either a $\IZ_2$-orbifold point for
$m=5,11$, a $\IZ_3$-orbifold point for $m=1,7$, a $\IZ_6$-orbifold point
for $m=6$ or a cusp for $m=4,6,8,9$; (see Table \ref{TableX0m}).
In either case, we refer to the singularity at $\lambda=0$ as the $a$-orbifold
point, with $a\in\{2,3,4,6,\infty\}$. 

The modular parameter $\tau$ is related to the parameter $\lambda$ by the 
mirror map \cite{Lian:1994zv}
\be
\label{mirrormapK3}
\tau=\frac{f_m^{(1)}(\lambda)}{f_m(\lambda)}\ ,\quad 
\lambda = J_m^+ (\tau)\,,
\ee
where $J_m^+(\tau)$ is a weak modular form of weight 0 under $\Gamma_0(m)^+$ with a simple pole at $q=e^{2\pi\I\tau}=0$, and specific constant term, known as the Hauptmodul. For example, for $m=1$, 
\be
\lambda  = J(\tau) =  \frac{E_4^3}{\eta^{24}} = \frac{1}{q}+744 + 196884 q + 21493760 q^2 + 864299970 q^3+\dots \,.
\ee
It is remarkable that the $q$-expansion of 
$\lambda(\tau)$ has integer coefficients, even though the coefficients in the $1/\lambda$-expansion of $f_m^{(1)}(\lambda)$  are rational numbers with unbounded denominators \cite{Lian:1994zv}. 
The fundamental period $f_m(\lambda(\tau))$ itself is a 
modular form of weight 2 under $\Gamma_0(m)$, odd
under the Fricke involution (for $m=1$, its square is an ordinary modular form of weight 2, $f_1(\lambda(\tau))= \sqrt{E_4(\tau)}$). 
As pointed out in \cite[(3.2)]{golyshev2016proof},
for all $m\leq 9$ and for $m=11$, the Hauptmodul is expressed in terms of the fundamental period
and suitable powers of the Dedekind eta function via
\be
\label{ZagierGoly}
J_m^+(\tau) = \left[ \frac{f_m(J_m^+(\tau))}{\eta^2(\tau) \eta^2(m\tau)}\right]^{\frac{12}{m+1}}\,,
\ee
which in particular determines the constant term in $J_m^+$. 
We further note the identity 
\be
\label{dJsq}
    (\partial_\tau J_m^+(\tau))^2=\frac{f_m(J_m^+(\tau))^2}{(2\pi\I)^2}
    \left[J_m^+(\tau)\right]^{2-r} 
    \prod\limits_{k=1}^r ( J_m^+(\tau) - \lambda_k)\,,
\ee
where $\lambda_i$ runs over the 2-orbifold points listed in the middle column of Table  \ref{TableX0m}.
Explicit formulae for the fundamental period and Hauptmodul are collected in Appendix \ref{sec_Gamma0m}.

In preparation for the computation of Yukawa couplings in \S\ref{sec_Yukawa}, it will be 
useful to introduce cousins of the mirror map in \eqref{mirrormapK3}, namely the ratios (for $i$ integer)
\bea
\label{defhtilde}
H_m^{[i]}(\lambda):=\frac{\tilde{f}_m^{[i]}(\lambda)}{f_m(\lambda)}\,,
\eea
where $\tilde{f}_m^{[i]}(\lambda)$ is obtained by inserting the 
harmonic number $H_{id}$ in the Laurent expansion of the fundamental period \eqref{deffm}, 
\be
\label{defftilde}
\tilde{f}_m^{[i]}(\lambda) :=\sum\limits_{d\ge 0} c_m(d) H_{id}\lambda^{-d}\ .
\ee
While the modular properties of 
$H_m^{[i]}(\tau):=H_m^{[i]}(\lambda(\tau))$ are obscure in general, 
by combining \eqref{d2WTh} and \eqref{delf0} below, we shall see
that, 
for the values of $(m,i)$ appearing in Table \ref{tabFano}, the third derivative 
\be
\partial_\tau^3 \left[ H_m^{[i]}(\tau) + \frac13 \sum_{k=1}^r \log(J_m^+(\tau)-\lambda_k)
+\frac13 \left( \frac{1}{i}-r \right) \log J_m^+(\tau)
\right]\,,
\ee
is a meromorphic modular form of weight $6$ under $\Gamma_0(m)^+$, vanishing at $\tau=\I\infty$ with a second
order pole at the 2-orbifold points $\tau_1,\dots,\tau_r$ and (for $m\leq 5$) a 
third order pole at $\tau=\tau_0$ where $J_m^+(\tau_0)=0$. 
For now,  we observe that,  as a consequence 
of \eqref{K3fperiodlog}, 
suitable linear combinations for $m\leq 4$
have a simple expression in terms of $\tau$ and $J_m^+(\tau)$, 
\bea
3 \left( H_{1}^{[1]} + H_{1}^{[3]} -2H_{1}^{[6]} \right) = -\tau - \log J\,,\nn\\
4 \left( H_{2}^{[1]} - H_{2}^{[4]} \right) = -\tau - \log J_2^+ \,,\nn\\
\left( 5H_{3}^{[1]} -2 H_{3}^{[2]} -3 H_{3}^{[3]} \right) = -\tau - \log J_3^+\,,\nn\\
6 \left( H_{4}^{[1]} -  H_{4}^{[2]} \right) = -\tau - \log J_4^+ \,,
\eea
consistent with assigning a vanishing modular weight to $H_m^{[i]}(\tau)$.  
We note that similar functions with insertions of higher harmonic numbers $H^k_{id}=\sum_{m=1}^{id} m^{-k}$ have appeared in \cite{golyshev2016proof}.

\subsection{Doran-Harder-Thompson mirror conjecture\label{sec_DHT}}

While mirror symmetry was originally formulated for CY threefolds, 
 it was generalized to Fano varieties (i.e., varieties with ample anticanonical hypersurfaces) in 
 \cite{eguchi1997gravitational}. The mirror to a Fano variety $X$ is a Landau-Ginzburg model $(X^\vee,W)$ such that $X^\vee$ is a K\"ahler manifold and $W : X^\vee \rightarrow \mathbb{C}$ is a proper map, called the superpotential.

Mirror symmetry for Landau-Ginzburg (LG) models has been generalized further to quasi-Fano varieties in~\cite{harderthesis2016}. We say that a smooth variety X is quasi-Fano if its anticanonical linear system contains a smooth CY member and $H^i(X,\mathcal{O}_X) = 0$ for all $i > 0$. 

A degeneration of CY varieties over the unit disk $\mathcal{V} \rightarrow \Delta \subset \mathbb{C}$  is called a Tyurin degeneration if the total space $\mathcal{V}$ is smooth and the central fiber consists of two quasi-Fano varieties $X_1$ and $X_2$ which meet normally along a common smooth anticanonical (Calabi-Yau) divisor $X_0$.

Doran-Harder-Thompson \cite{doran:2016uea} formulated a conjecture relating mirror symmetry for a CY variety $X$ arising as a smooth fiber of a Tyurin degeneration $\mathcal{V} \rightarrow \Delta$, and mirror symmetry for the quasi-Fano varieties $X_1$ and $X_2$ in the central fiber of $\mathcal{V}$. Roughly speaking, the Doran-Harder-Thompson (DHT) conjecture states that one should be able to ``glue'' the LG models $W_i : X_i^\vee \rightarrow \mathbb{C}$ of the pair $(X_i, X_0)$ for $i = 1, 2$ to form a CY variety $X^\vee$ which is mirror to $X$, with the fibers of the superpotentials combining to form a fibration $W : X^\vee \rightarrow {\mathbb{P}}^1$, and moreover that the compact fibers of the LG models consist of CY manifolds mirror to the common anticanonical divisor $X_0$. 

The gluing in the DHT conjecture can be performed in several different categories, depending on how we match the A/B-models with the fibrations and degenerations. 

First checks of the conjecture in \cite{doran:2016uea} focused on the Tyurin degenerations on the B-model side and fibrations on the A-model side, confirming that the basic topological features (e.g., Euler characteristic and monodromies) and lattice-polarizations in the K3 surface case are consistent with mirror symmetry~\footnote{In the context of mirror symmetry for lattice polarized K3 surfaces, a precise form of the DHT conjecture can be found in~\cite{giovthom2024}.}.  

Flipping sides, in \cite{doran2019doran,doran2021degenerations} the Tyurin degeneration takes place on the A-model, involving complete intersections in toric varieties; the gluing is performed in the B-model and expressed in terms of the generalized functional invariant maps of the two LG models and the generalized functional invariant map of the 
K3 fibration over ${\mathbb{P}}^1$. This is the main setting for DHT mirror symmetry as it arises in this paper.

It was observed in \cite{doran:2016uea} that, as predicted by the DHT conjecture, the classification of CY threefolds fibered by $M_m$-polarized K3 surfaces in~\cite{Doran2016,Doran2019,kooistra2021threefolds} mirrors Iskovskih's classification \cite{iskovskih1978fano1,iskovskih1978fano2} of pairs of smooth Fano threefolds of rank 1   with a common anticanonical K3. Mirror symmetry at the level of Hodge numbers was confirmed, with explicit formulas reflecting geometric sources of the Hodge numbers of the CY threefolds on both sides.

Enumerative consequences of the DHT mirror construction involving genus zero Gromov-Witten invariants, both absolute and relative, were established in~\cite{doran2019doran,doran2021degenerations}.  These include a new B-model motivated degeneration formula for Tyurin degenerations of CY threefolds.

The DHT mirror conjecture itself, and especially its Hodge-theoretic implications, were vastly generalized in~\cite{doranthompMCSseq} as consequences of a four-term long exact sequence DHT mirror to the Clemens-Schmid sequence describing the cohomology of a degeneration.  The role of the CY fibration is now played by any smooth variety admitting a projective morphism to a projective base.

\section{B-model geometries}
\label{sec:bmodel}

In this section we use the results from~\cite{Doran2016,Doran2019,kooistra2021threefolds} to explicitly construct a family of $M_m$-polarized K3 fibered CY threefolds $Y^{[i,j]}_{m,s}$ with $h_{2,1}(Y^{[i,j]}_{m,s})=2$.

\subsection{Generalized functional invariant map}
\label{sec:gfim}

A family of $M_m$-polarized K3 surfaces $\pi:\,Y\rightarrow U$ determines a so-called \textit{generalized functional invariant map} $\Lambda:\,U\rightarrow \mathcal{M}_{M_m}=X_0(m)^+$.
For $m\ge 2$ it was proven in~\cite{Doran2016,Doran2019} that the isomorphism class of a non-isotrivial $M_m$-polarized family of K3 surfaces $\pi:\,Y\rightarrow U$ over a quasi-projective curve $U$ is uniquely determined by the associated generalized functional invariant map.
Concretely, the geometry $Y$ is a crepant resolution of the pullback of the modular families $\overline{\mathcal{X}}_m\rightarrow X_0(m)^+$ listed in Table~\ref{tab:modularFamilies} along $\Lambda$.
The case $m=1$ was subsequently treated in~\cite{kooistra2021threefolds} with the result that an $M_1$-polarized family is uniquely determined by the generalized functional invariant map together with a so called \textit{generalized homological invariant}, determining the monodromies around singular fibers.
When $m\ge 2$, the generalized homological invariant is fixed completely in terms of the generalized functional invariant map while for $m=1$ there is an additional choice of sign for each singular fiber that has to be specified in order to determine the fibration.
\begin{table}[ht!]
\centering
\begin{align*}
\begin{array}{|c|c|c|}\hline
m&\text{Ambient space}&\text{Modular family }\overline{\mathcal{X}}_m\\\hline
2&\mathbb{P}^3[x,y,z,t]&t^4+\lambda xyz(x+y+z-t)=0\\
3&\mathbb{P}^1[r,s]\times\mathbb{P}^2[x,y,z]&s^2z^3+\lambda r(r-s)xy(z-x-y)=0\\
4&\prod_{i=1}^3\mathbb{P}^1[r_i,s_i]&s_1^2s_2^2s_3^2-\lambda r_1(s_1-r_1)r_2(s_2-r_2)r_3(s_3-r_3)=0\\\hline
5&\mathbb{P}^1[r,s]\times\mathbb{P}^2[x,y,z]&(x+y+z)^2(xr+yr+xs)(r+s)-\lambda xyzrs=0\\
6&\mathbb{P}^1[r,s]\times\mathbb{P}^2[x,y,z]&(y+z)(x+y+z)(xr+yr+xs)(r+s)-\lambda xyzrs=0\\
7&\prod_{i=1}^3\mathbb{P}^1[r_i,s_i]&\left(r_1 r_2 r_3-r_1 r_2 s_3+r_1 s_2 s_3-s_1 s_2 s_3\right)\prod_{i=1}^3\left(r_i-s_i\right)+\lambda  r_1 r_2 r_3 s_1 s_2 s_3=0\\
8&\prod_{i=1}^3\mathbb{P}^1[r_i,s_i]&\left(r_1-s_1\right) \left(r_1 \left(r_2-s_2\right){}^2 \left(r_3-s_3\right){}^2-r_2^2r_3^2 s_1\right)+\lambda  r_1 r_2 r_3 s_1 s_2 s_3=0\\\hline
9&\mathbb{P}^3[x,y,z,t]&(x + y + z)(xt^2 + yt^2 + zt^2 + xyt + xzt + yzt + xyz) - \lambda xyzt = 0\\
11&\mathbb{P}^3[x,y,z,t]&(z + t)(x + y + t)(xy + zt) + x^2y^2 + xyz^2 + 3xyzt - \lambda xyzt = 0\\\hline
\end{array}
\end{align*}
\caption{Modular families $\overline{\mathcal{X}}_m$ of $M_m$-polarized K3 surfaces. The cases $m=2,\ldots,4$ as well as $m=9,11$ are reproduced from~\cite[Table 3.1]{doran2020calabi}. The families for $m=5,\ldots,8$ 
are alternatives to the ones listed in~\cite{doran2020calabi} and are, to the best of our knowledge, new.
They are obtained in Section~\ref{sec:Amodelm5} and Appendix~\ref{sec_Amodel}.
}
\label{tab:modularFamilies}
\end{table}
\footnote{We are grateful to Eric Pichon-Pharabod for confirming the generic lattice-polarization prediction for the new cases in Table~\ref{tab:modularFamilies} using his semi-numerical algorithm implemented in Sagemath~\cite{pichonpharabod2024}}

We are interested in families over $U=\mathbb{P}^1$ that are smooth CY threefolds.
The corresponding maps $\Lambda:\,\mathbb{P}^1\rightarrow X_0(m)^+$ have been classified for $m\ge 2$ in~\cite{Doran2019} and the $m=1$ case was treated in~\cite{kooistra2021threefolds}.
The only possible values for $m$ are
\begin{align}
	m\in \{1,\,2,\,3,\,4,\,5,\,6,\,7,\,8,\,9,\,11\}\,.
\end{align}

\begin{figure}[ht!]
	\begin{tikzpicture}[remember picture,overlay,node distance=4mm]
		\node[align=center] at (3.4,5.4) {$[2,1]$};
		\node[align=center] at (1.4,5.4) {$[1,1,1]$};
	\end{tikzpicture}
	\centering
	\includegraphics[width=.4\linewidth]{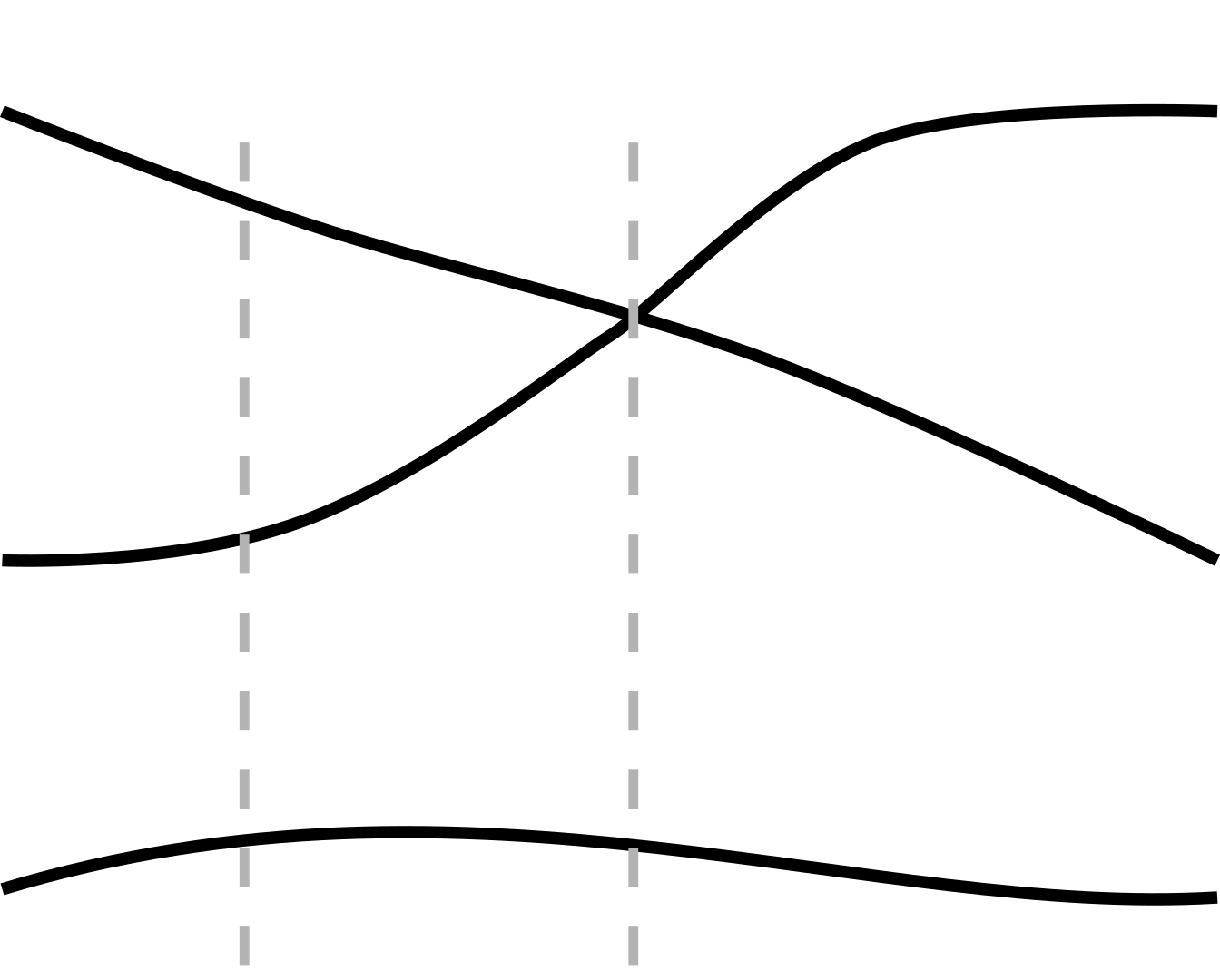}
	\caption{Ramification profiles for a degree three covering of $\mathbb{P}^1$.}
	\label{fig:ramificationProfile}
\end{figure}
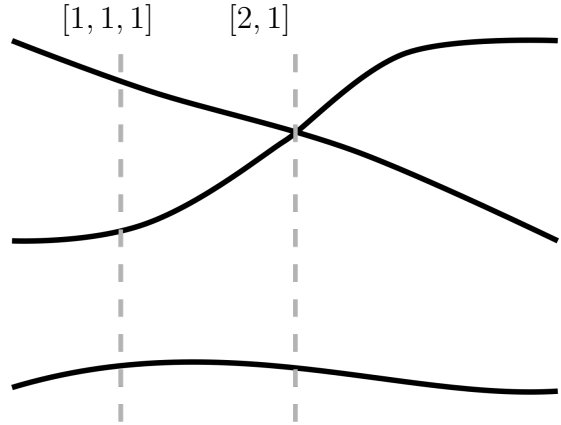

The map $\Lambda$ can then be described in terms of its degree $d$ and the ramification profile over the singular points $\lambda\in\{0,\lambda_1,\ldots,\lambda_r, \infty\}$.
Over any point $p\in \mathcal{M}_{M_m}$, the ramification profile corresponds to a partition of $d$ into positive integers $[d_1,\ldots, d_s]$ such that $\Lambda^{-1}(p)$ consists of $s$ points at which $d_i,\,i=1,\ldots,s$ of the sheets of the covering respectively come together, see Figure~\ref{fig:ramificationProfile}.

We denote the ramification profiles over $\lambda=\infty$, $\lambda=0$ and $\lambda_1,\ldots,\lambda_r$ respectively by $[x_1,\ldots, x_k]$, $[y_1,\ldots,y_l]$ and $[z_{1,1},\ldots,z_{m_1,1}],\ldots,[z_{1,r},\ldots,z_{m_r,r}]$.
The Hurwitz formula specializes to
\begin{align}
	k+l+m_1+\ldots+m_r-rd-\rho-2=0\,.
 \label{eqn:hurwitz}
\end{align}
where the so-called excess ramification $\rho$ is defined as
\begin{align}
	\rho:=\sum\limits_{p \in \mathbb{P}^1_{\rm reg}}(e_p-1)\,,
 \label{eqn:excess}
\end{align}
in terms of the ramification index $e_p$ at points $p$ in the regular locus $\mathbb{P}^1_{\rm reg} = \mathbb{P}^1\backslash\{0,\infty,\lambda_1,\ldots,\lambda_r\}$ (note that $e_p\neq 1$ for only a finite number of points).
We will assume that the generalized functional invariant map is chosen to be sufficiently generic in the sense that it is unramified over the 2-orbifold points $\lambda_1,\ldots,\lambda_r$.
The Hurwitz formula~\eqref{eqn:hurwitz} then takes the simple form $\rho=k+l-2$, see also~\cite[Remark 4.6]{Doran2019}.

\begin{table}[ht!]
\begin{align*}
	\begin{array}{|c|l|}\hline
		m&\text{Partitions}\\\hline
            1& [1,1],\,[2,1],\,[2,2]\\
		2& [1,1],\,[2,1],\,[2,2],\,[4,1],\,[4,2],\,[4,4]\\
		3& [1,1],\,[2,1],\,[2,2],\,[3,1],\,[3,2],\,[3,3]\\
		4,5& [1,1],\,[2,1],\,[2,2]\\
		6,8,9,11& [1,1]\\\hline
	\end{array}
\end{align*}
\caption{Values $m,[i,j]$ for which we consider the mirror families of CY threefolds $Y_{m,s}^{[i,j]},X_{m,s}^{[i,j]}$ with $s=0,\ldots,i-j$.}
\label{tab:mijvalues}
\end{table}

For the geometries $Y_{m,s}^{[i,j]}$ we choose the generalized functional invariant map $\lambda=\Lambda^{[i,j]}_s$ to be
\begin{align}
    \Lambda^{[i,j]}_s(y;z_1,z_2):=\frac{1}{z_1}\frac{(1-y)^i(1-z_2/y)^j}{y^s}\,,
    \label{eqn:gfimGen}
\end{align}
where $y$ is an affine coordinate on $\mathbb{P}^1$ and $z_1,z_2\in\mathbb{C}$ parametrize the complex structure.
For generic choices of $z_1,z_2$, the ramification profiles over $\lambda=\infty$ and $\lambda=0$ are then respectively given by $[x_1,x_2]=[i-s,j+s]$ and $[y_1,y_2]=[i,j]$ while the map is unramified over the two-orbifold points and the excess ramification is $\rho=2$.
We consider the values $(m,i,j)$ contained in Table~\ref{tab:mijvalues}, such that in particular $i\ge j$, while $s=0,\ldots,i-j$.
For the cases with $m=1$ one also has to specify the generalized homological invariant or, equivalently, the types of singular fibers, and we use the choices from Table~\ref{tab:m1fibers}.
For geometries with $s=0$ we also define the abbreviated notations $Y_{m}^{[i,j]}:=Y_{m,0}^{[i,j]}$ and $\Lambda^{[i,j]}:=\Lambda^{[i,j]}_0$.

For $2\le m \le 9$, the result~\cite[Theorem 4.1]{Doran2019} then implies that $Y_{m,s}^{[i,j]}$ is a smooth CY threefold and it follows from~\cite[Proposition 4.4]{Doran2019} that $h_{2,1}(Y_{m,s}^{[i,j]})=2$.
In the case $m=1$, the properties of the different fiber types $\text{I}_d,\text{I}_d^*,\text{II},\text{II}^*,\text{III},\text{III}^*,\text{IV},\text{IV}^*$ are listed in~\cite[Table 4.1]{kooistra2021threefolds}~\footnote{Note that these are singular surfaces appearing as special fibers in threefolds whose generic fiber is a smooth $M_1$-polarized K3 surface, by analogy with the Kodaira naming scheme for the singular curve fibers appearing in elliptic surfaces.}.
Since the generalized functional invariant map is unramified over the 2-orbifold point, such that there are $i+j$ fibers of type $\text{III}$ over $\lambda=1728$, the fibration is smooth~\cite[Proposition 5.3]{kooistra2021threefolds}.
The fact that $Y_{m,s}^{[i,j]}$ is a CY threefold with $h_{1,2}=2$ then respectively follows from~\cite[Proposition 5.5]{kooistra2021threefolds} and~\cite[Proposition 5.9]{kooistra2021threefolds}~\footnote{In principle we could also replace each starred/unstarred fiber with the corresponding unstarred/starred one in Table~\ref{tab:m1fibers}. However, the corresponding geometries will not be relevant for us.}.

\begin{table}[ht!]
    \begin{tabular}{|c|ccc|}\hline
    $[i,j]$&$\lambda=\infty$&$\lambda=1728$&$\lambda=0$\\\hline
    $[1,1]$&$2\times$ I$_1$& $2\times$III& $2\times$II$^*$\\
    $[2,1]$&I$_2$, I$_1$& $3\times$III& IV$^*$, II$^*$\\
    $[2,2]$&$2\times$ I$_2$& $4\times$III& $2\times$IV$^*$\\\hline
    \end{tabular}
    \caption{Types of singular fibers in $Y^{[i,j]}_{1,s}$.}
    \label{tab:m1fibers}
\end{table}

Combining~\cite[Proposition 4.3]{Doran2019} and~\cite[Proposition 5.7]{kooistra2021threefolds}, we obtain the Euler characteristic
\begin{align}
\label{EulerY}
     \chi(Y_{m,s}^{[i,j]})=2\left(20+h_{1,2}(F_m^{[i]})+h_{1,2}(F_m^{[j]})+m(i+j)^2\right)-4m(i-s)(j+s)\,,
\end{align}
where $h_{1,2}(F_m^{[i]})$ is the number of complex structure deformations of the Fano threefold $F_m^{[i]}$ tabulated in Table~\ref{tabFano}.

The case $m=11$ is special in that it is not covered by the smoothness results from~\cite{Doran2019}.
Nevertheless, in \S\ref{sec:amodel} we will be able to extract sensible topological invariants and enumerative invariants from the period, and also construct a smooth A-model geometry. Thus it is plausible that $Y_{11,0}^{[1,1]}$ may also be a smooth CY threefold with $h_{2,1}=2$.

\subsection{Periods}
\label{sec:periods}

Generalizing~\cite{doran2019doran}, we can calculate the holomorphic period of the  unique (up to scale) holomorphic 3-form $\Omega(z_1,z_2)$ on $Y_{m,s}^{[i,j]}$ by taking the contour integral of the pullback of $$\frac{1}{(1-y)(1-z_2/y)}f_m(\lambda)\,,$$ by the generalized functional invariant $\lambda = \Lambda^{[i,j]}_s(y;z_1,z_2)$. Taking the integration contour to circle   around $y=z_2$ counterclockwise (or equivalently, clockwise around $y=1$), we find
the fundamental period
\begin{align}
    \begin{split}
    \varpi_{m,s}^{[i,j]}(z_1,z_2)=&\frac{1}{2\pi \I}\oint \frac{\de y}{y}\frac{1}{(1-y)(1-z_2/y)}f_m\left(\lambda=\Lambda^{[i,j]}_s(y;z_1,z_2)\right)\\
   =&\frac{1}{2\pi \I}\sum\limits_{d\ge 0}c_m(d)z_1^d\oint \frac{\de y}{y^{1-ds}}\frac{1}{(1-y)^{1+di}(1-z_2/y)^{1+dj}}\\
    =&\frac{1}{2\pi \I}\sum\limits_{d\ge 0}c_m(d)z_1^d\oint\frac{\de y}{y}\sum\limits_{k,l=0}^\infty \frac{(k+di)!(l+dj)!}{(di)!(dj)!k!l!}y^{k+ds-l}z_2^l\\
    =&\sum\limits_{d,k\ge 0}c_m(d)\frac{(k+di)!(k+ds+dj)!}{(di)!(dj)!k!(k+ds)!}z_1^dz_2^{k+sd}\,,\\
    =&\sum\limits_{d,k\ge 0}c_m(d)\frac{(k+di-ds)!(k+dj)!}{(di)!(dj)!(k-ds)!k!}z_1^dz_2^{k}\,.
    \end{split}
    \label{eqn:fundamentalPeriod}
\end{align}
As usual, in the special case $s=0$ we abbreviate  $\varpi_{m}^{[i,j]}(z_1,z_2):=\varpi_{m,0}^{[i,j]}(z_1,z_2)$. We note that in the limit $z_2\to 1$
keeping $z:=z_1/(1-z_2)^{i+j}$ fixed, a change of variable $y=(1- z_2)t+z_2$ under the integral leads to 
\begin{align}
    \begin{split}
    \varpi_{m,s}^{[i,j]}(z_1,z_2)=&\frac{1}{2\pi \I}\oint \frac{\de t}{(1-z_2)t (1-t)}   f_m\left( \frac{(1-t)^i t^j}{z (z_2+(1-z_2)t)^{j+s}}\right)\\
    \stackrel{z_2\to 1}{\to} & \frac{1}{2\pi \I (1-z_2) }\oint \frac{\de t}{t (1-t)}   f_m\left( \frac{(1-t)^i t^j}{z}\right) 
    = \sum_{d\geq 0} c_m(d) \frac{[d(i+j)]!}{(di)! (dj)!} z^d\,,
    \label{eqn:fundamentalPeriod1}
    \end{split}
\end{align}
which is recognized, up to a factor of $(1-z_2)$, as the fundamental period of the one-parameter model $\cY_m^{i,j]}$ mirror to $\cX_m^{[i,j]}$ in the last column of Table \ref{tab:mijvalues}. 

Returning to the fundamental period \eqref{eqn:fundamentalPeriod} of the two-parameter model, one can reconstruct a set of Picard-Fuchs operators that must annihilate all periods of the holomorphic 3-form.
Since the complex structure moduli space of $Y_{m,s}^{[i,j]}$ is two-dimensional, the Picard-Fuchs ideal is generated by two operators $\mathcal{D}_2,\mathcal{D}_3$ of respective orders two and three.
While the order two operator annihilating~\eqref{eqn:fundamentalPeriod} takes the universal form
\begin{align}
    \mathcal{D}_2=\theta_2(\theta_2-s\theta_1)-z_2(1+(i-s)\theta_1+\theta_2)(1+j\theta_1+\theta_2)\,,
\end{align}
in terms of the logarithmic derivatives $\theta_{a}=z_a\partial_{z_a},\,a=1,2$, the third order operator $\mathcal{D}_3$ is more complicated and is provided
in the accompanying Mathematica worksheet.

A complete basis of periods then takes the form
\begin{align}
    \begin{split}
        \varpi_0(z_1,z_2)=&\varpi_{m,s}^{[i,j]}(z_1,z_2)\,,\\
        \varpi_{1,a}(z_1,z_2)=&\varpi_{m,s}^{[i,j]}(z_1,z_2)\log(z_a)+\mathcal{O}(z_1,z_2)\,,\quad a=1,2\,,\\
        \varpi_{2,a}(z_1,z_2)=&\varpi_{m,s}^{[i,j]}(z_1,z_2)\sum\limits_{b,c=1}^2\kappa_{abc}\log(z_b)\log(z_c)+\mathcal{O}(z_1,z_2)\,,\quad a=1,2\\
        \varpi_{3}(z_1,z_2)=&\varpi_{m,s}^{[i,j]}(z_1,z_2)\sum\limits_{a,b,c=1}^2\kappa_{abc}\log(z_a)\log(z_b)\log(z_c)+\mathcal{O}(z_1,z_2)\,,\\
    \end{split}
\end{align}
where we have left the dependence on $(m,i,j,s)$ implicit. The
coefficients $\kappa_{abc}$ entering in the double and triple logarithmic period are found to be given by the universal formulae in \eqref{gentopdata}.
One  easily checks that the limit $z_1,z_2\rightarrow 0$ indeed corresponds to a point of maximally unipotent monodromy.
As a result, the coefficients $\kappa_{abc}$ can be identified -- at least up to an overall normalization -- under mirror symmetry with the intersection numbers $\kappa_{abc}=\gamma_a\cap\gamma_b\cap\gamma_c$ for a suitable choice of basis of divisors $\gamma_1,\gamma_2$ on the mirror CY threefold $X_{m,s}^{[i,j]}$. 
We shall choose a basis of divisors $\gamma_1,\gamma_2\in H_4(X_{m,s}^{[i,j]},\mathbb{Z})$ such that, upon expanding the complexified K\"ahler form as $\omega=t^1\gamma_1+t^2\gamma_2$, the leading behaviour of the mirror map takes the form
\begin{align}
    t^1=\frac{1}{2\pi \I}\log(z_1)+\mathcal{O}(z_1,z_2)\,,\quad t^2=\frac{1}{2\pi \I}\log(z_2)+\mathcal{O}(z_1,z_2)\,.
\end{align}
In the Section~\ref{sec:amodel} we shall  construct the mirror CY threefolds explicitly and verify that the intersection numbers and other topological invariants are indeed given by the universal formulae in~\eqref{gentopdata}.

\subsection{Moduli space}
\label{sec:bmodelmoduli}

We will now discuss the complex structure moduli space and the components of its boundary.
First, we observe that the involution
\begin{align}
    \sigma:\quad z_1\mapsto (-1)^{i-j}z_1/z_2^{j}\,,\quad z_2\mapsto 1/z_2\,,\quad y\mapsto 1/y\,,
    \label{eqn:vwinvolution}
\end{align}
acting on the generalized functional invariant map~\eqref{eqn:gfimGen}, relates $Y_{m,s}^{[i,j]}$ and $\widetilde{Y}_{m,s}^{[i,j]}:=Y_{m,i-j-s}^{[i,j]}$.
For cases with $i-j=2s$ one has $\widetilde{Y}_{m,s}^{[i,j]}=Y_{m,s}^{[i,j]}$ such that $\sigma$ is a symmetry and the complex structure moduli space is globally a quotient.

Given $i,j$ and a choice of complex structure $z_1,z_2$, the excess ramification points $\lambda_\pm$ of~\eqref{eqn:gfimGen} can be obtained as the non-zero roots in $\lambda$ of the discriminant
\begin{align}
    \widetilde{\Delta}_{s}^{[i,j]}(z_1,z_2,\lambda):=\text{Disc}_y\left[(y-z_2)^j(1-y)^i-z_1y^{s+j}\lambda\right]\,.
\end{align}

As we have observed in the previous subsection, the combined limit $z_1,z_2\rightarrow 0$ corresponds to a point of maximally unipotent monodromy.
We will show in Section~\ref{sec:amodel} that it is mirror to the large volume limit of a smooth CY threefold $X_{m,s}^{[i,j]}$.
Using the involution~\eqref{eqn:vwinvolution} it then also follows that the limit $z_1/z_2^j\rightarrow 0,\,1/z_2\rightarrow0$ corresponds to a point of maximally unipotent monodromy that is mirror to the large volume limit of $X_{m,i-j-s}^{[i,j]}$.
In Section~\ref{sec:amodel} we will explicitly show that $X_{m,s}^{[i,j]}$ and $X_{m,i-j-s}^{[i,j]}$ are related by a flop transition.

One can further distinguish three types of codimension-one boundaries of the complex structure moduli space, corresponding to different degenerations of the generalized functional invariant map~\eqref{eqn:gfimGen}:

\begin{figure*}[t!]
    \centering
    \begin{subfigure}[t]{0.5\textwidth}
        \centering
        \includegraphics[width=.9\linewidth]{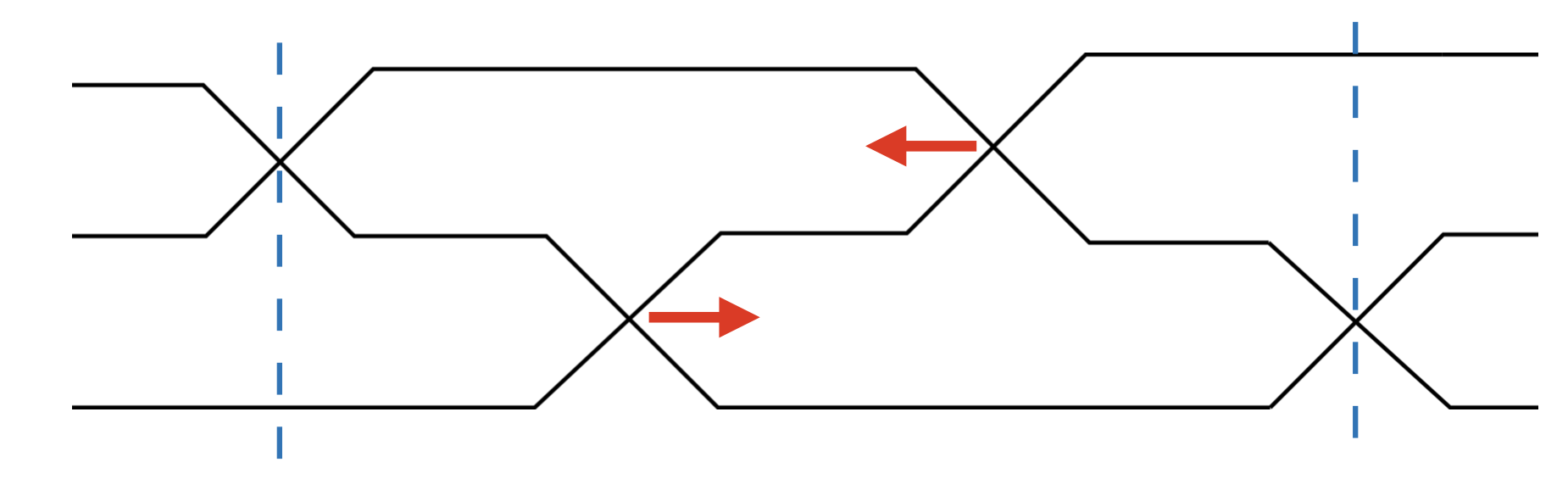}\\[.5cm]
        \includegraphics[width=.9\linewidth]{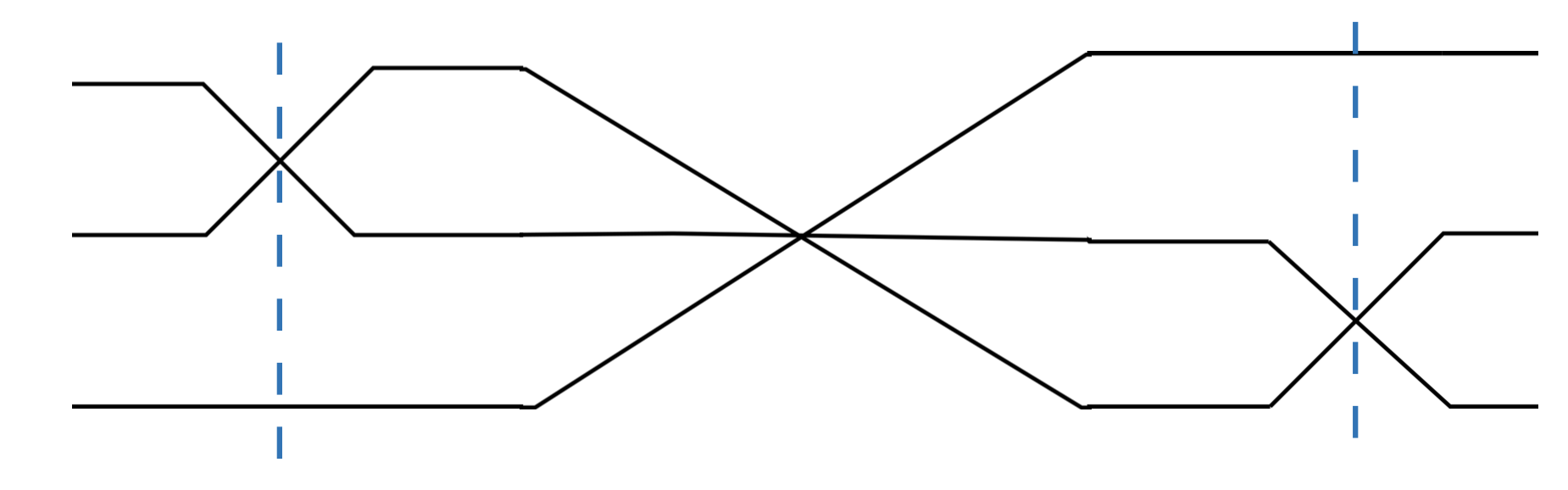}\\[0cm]
\begin{tikzpicture}[remember picture,overlay,node distance=4mm]
\begin{scope}[xshift=-4.1cm,yshift=3.8cm]
            \node[] at (1.5,-3.5) {$\lambda=0$};
            \node[] at (7.05,-3.5) {$\lambda=\infty$};
            \draw[->,dashed,line width=.3mm] (4.2,-.1)   -- (4.2,-1.0);
            \node[] at (5.2,-.5) {$z_2\rightarrow 0$};
            \end{scope}
        \end{tikzpicture}
        \caption{Example $(i,j,s)=(2,1,1)$}
    \end{subfigure}%
    ~ 
    \begin{subfigure}[t]{0.5\textwidth}
        \centering
        \includegraphics[width=.9\linewidth]{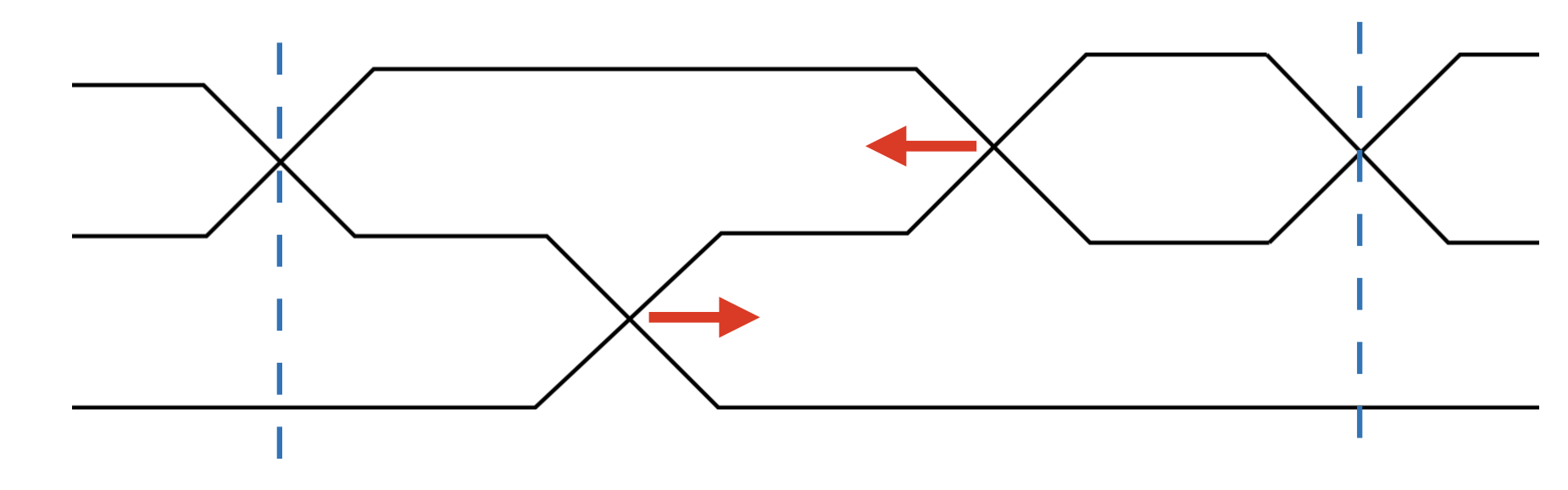}\\[.5cm]
        \includegraphics[width=.9\linewidth]{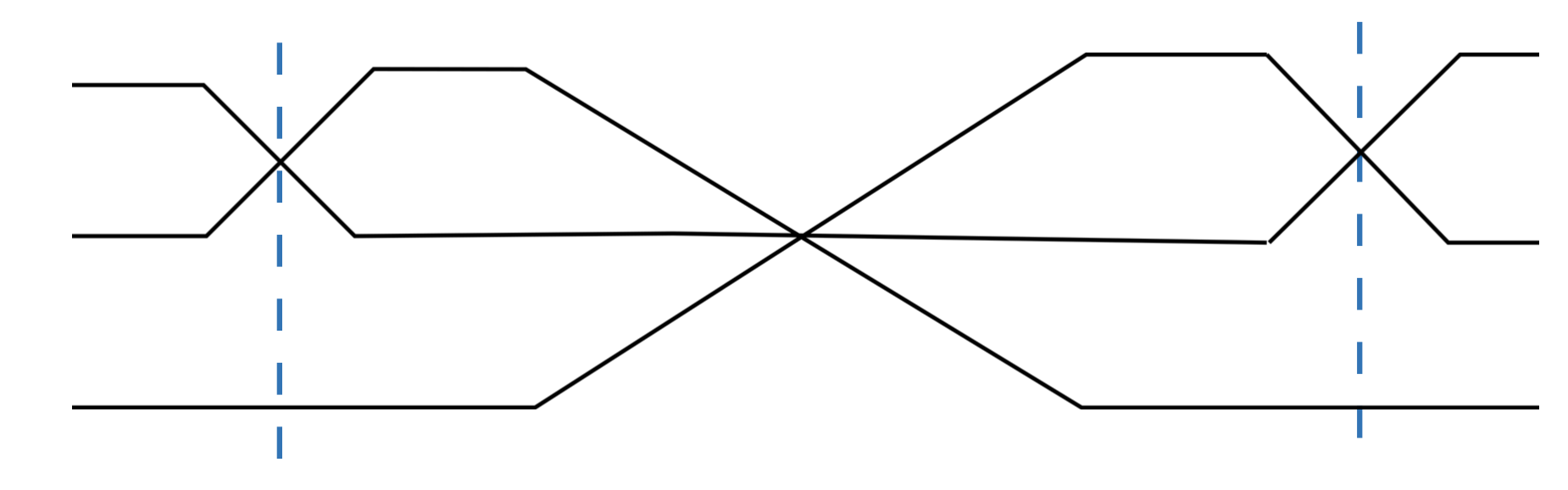}\\[.0cm]
\begin{tikzpicture}[remember picture,overlay,node distance=4mm]
\begin{scope}[xshift=-4.1cm,yshift=3.8cm]
            \node[] at (1.5,-3.5) {$\lambda=0$};
            \node[] at (7.05,-3.5) {$\lambda=\infty$};
            \draw[->,dashed,line width=.3mm] (4.2,-.1)   -- (4.2,-1.0);
            \node[] at (5.2,-.5) {$z_2\rightarrow 0$};
            \end{scope}
        \end{tikzpicture}
        \caption{Example $(i,j,s)=(2,1,0)$}
    \end{subfigure}
    \caption{Degenerations of the generalized functional invariant map in the limit $z_2\rightarrow 0$. The images depict the ramification over a circular slice along the $\mathbb{P}^1$ base of the K3-fibration $Y_{m,s}^{[i,j]}$ and are periodic in the horizontal direction.}
    \label{fig:degz2to0}
\end{figure*}
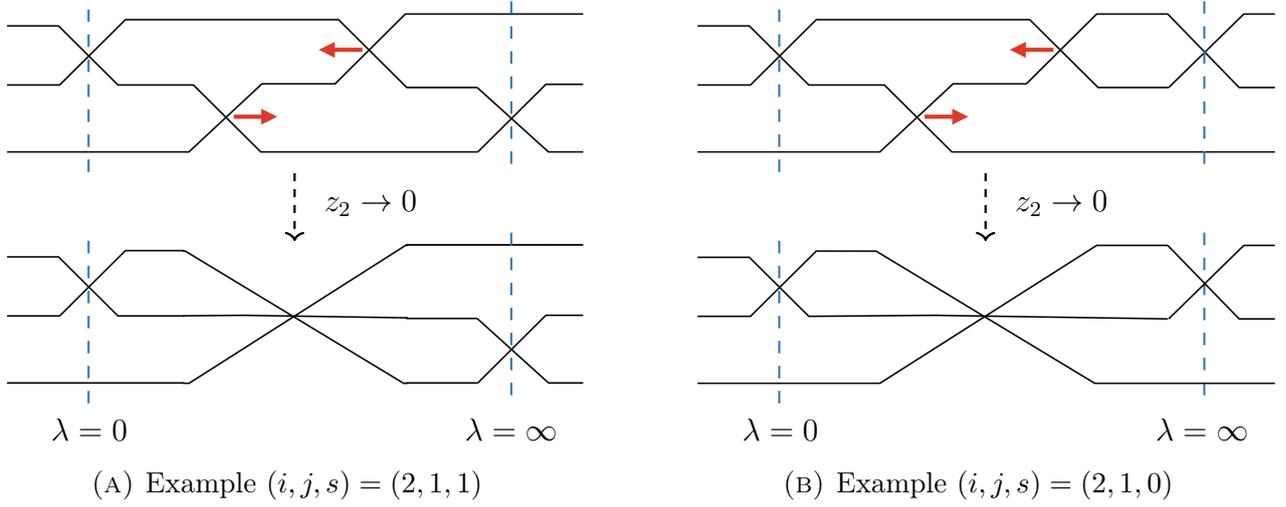
\begin{itemize}
\setlength\itemsep{.5em}
\item \textbf{Ramification points collide with each other.}
The corresponding part of the boundary is reducible and exhibits itself three different components:\vspace{.2cm}
\begin{enumerate}\setlength\itemsep{.5em}
\item In the limit $z_2\rightarrow 0$ the two excess ramification points collide and one has
\begin{align}
    \lambda_+=\lambda_-=\frac{(-1)^s}{z_1}\frac{(i+j)^{i+j}}{(i-s)^{i-s}(j+s)^{j+s}}\,,
    \label{eqn:lambdavtozero}
\end{align}
over $y=-(j+s)/(i-s)$.
This is illustrated in Figure~\ref{fig:degz2to0}.
The combined limit $z_1,z_2\rightarrow 0$ corresponds to a boundary point of maximally unipotent monodromy that is mirror to the large volume limit of $X_{m,s}^{[i,j]}$.

For $s=0$ the covering splits in the limit $z_2\rightarrow 0$ and one obtains a Tyurin degeneration where a component corresponding to the ramification profile $[y_1]=[i],\,[x_1]=[i]$ intersects another component with ramification profile $[y_2]=[j],\,[x_2]=[j]$.
The two components intersect in an $M_m$-polarized K3 surface over $\lambda=\lambda_+=\lambda_-$.
The DHT conjecture then implies that the mirror CY threefold $X_{m}^{[i,j]}$ is fibered by degree $2m$ K3-surfaces, and the limit $z_2\rightarrow 0$ is mirror to the limit of large base volume for $X_{m}^{[i,j]}$.

\item In the limit $z_2\rightarrow 1$, one of the excess ramification points moves to $\lambda=0$ such that the ramification profile degenerates to $[y_1]=[i+j],\,[x_1,x_2]=[i-s,j+s]$. 
In the mirror $X_{m,s}^{[i,j]}$ this limit will correspond to moving to a boundary of the K\"ahler cone where $N=2m(i-s)(j+s)$ rational curves shrink simultaneously and lead to nodal singularities. By smoothing the nodes, and thus performing the conifold transition, one arrives at the one-parameter model $\mathcal{X}_m^{[i,j]}$ listed in the last column of Table \ref{tabmij}.

\item The limit $z_2\rightarrow \infty$ of $\Lambda^{[i,j]}_s(y;z_1,z_2)$ keeping $z_1z_2^j$ finite is equivalent to the limit $z_2\rightarrow 0$ of the functional invariant map $\Lambda^{[i,j]}_{i-j-s}(y;z_1,z_2)$ obtained after the involution~\eqref{eqn:vwinvolution}.
The combined limit $z_1z_2^j\rightarrow0$, $z_2\rightarrow \infty$ therefore corresponds to a boundary point of maximally unipotent monodromy that is mirror to the large volume limit of $X_{m,i-j-s}^{[i,j]}$.

\end{enumerate}
\item \textbf{Ramification points collide with $2$-orbifold points.}
This happens at the vanishing locus of the product of discriminants
\begin{align}
    \widetilde{\Delta}_{m,s}^{[i,j]}(z_1,z_2)=\prod\limits_{k=1}^q\widetilde{\Delta}_{s}^{[i,j]}(z_1,z_2,\lambda_k)\,.
\end{align}
One can check that $\widetilde{\Delta}_{m,s}^{[i,j]}(z_1,z_2)\in\mathbb{Q}[z_1,z_2]$.
We also define $\Delta_{m,s}^{[i,j]}(z_1,z_2):=c z_1^az_2^b\widetilde{\Delta}_{m,s}^{[i,j]}(z_1,z_2)$, where $c\in\mathbb{Q}$ and $a,b\in\mathbb{Z}$ are chosen such that $\Delta_{m,s}^{[i,j]}(z_1,z_2)=1+\mathcal{O}(z_1,z_2)$.
From the perspective of the mirror $X_{m,s}^{[i,j]}$, this corresponds to the locus where the 6-brane becomes massless and is sometimes referred to as the principal component of the discriminant locus.

\item \textbf{The map becomes constant.} This happens in the limits $z_1\rightarrow 0,\infty$. The limit $z_1\rightarrow 0$ is mirror to a large volume limit of $X_{m,s}^{[i,j]}$ and in the cases with $s=0$ corresponds to the limit of large fiber volume of the K3-fibration when $\vert z_2\vert \ll 1$.
On the other hand, the nature of the limit $z_1\rightarrow \infty$ depends on the choice of $m$.
\end{itemize}

In order to get a smooth CY, we can therefore let $z_1,z_2$ take values in
\begin{align}
    \widetilde{\mathcal{M}}_{\text{cpx.}}=(\mathbb{C}^\times)^2\backslash(\{\Delta_{m,s}^{[i,j]}(z_1,z_2)=0\}\cup\{z_2=1\})\,.
\end{align}
Taking into account the potential quotient due to~\eqref{eqn:vwinvolution}, this is, in general, a double cover of a dense open subset of the complex structure moduli space of $Y_{m,s}^{[i,j]}$.

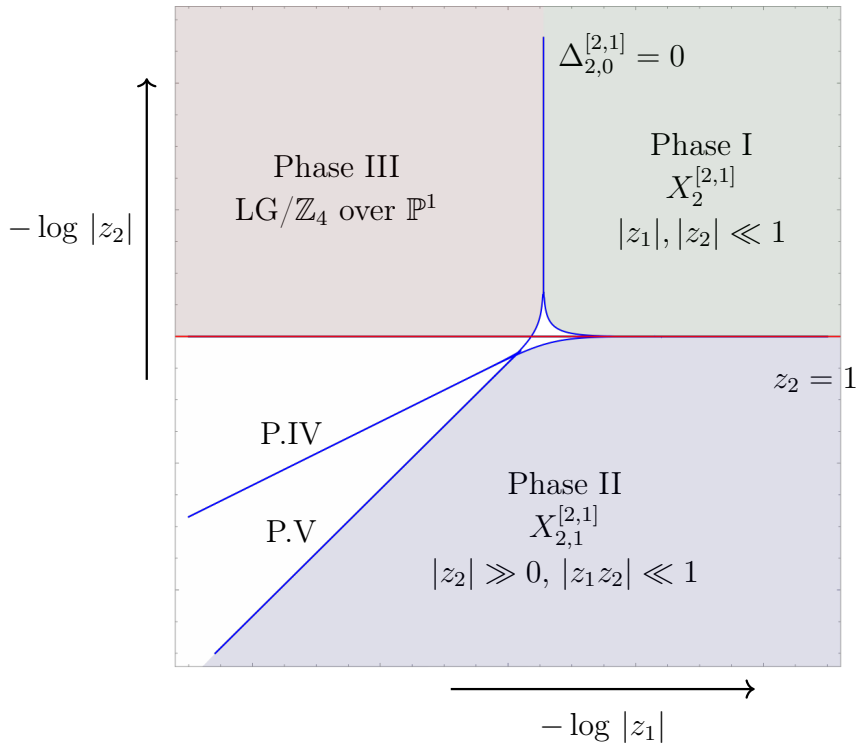
\begin{figure}[ht!]
	\begin{tikzpicture}[remember picture,overlay,node distance=4mm]
		\node[align=center] at (7.3,6.5) {Phase I\\[.2em]$X_2^{[2,1]}$\\[.2em]$\vert z_1\vert ,\vert z_2\vert \ll 1$};
		\node[align=center] at (5.5,2) {Phase II\\[.2em]$X_{2,1}^{[2,1]}$\\[.2em]$\vert z_2\vert\gg0,\,\vert z_1z_2\vert \ll 1$};
		\node[align=center] at (2.5,6.5) {Phase III\\[.2em]LG/$\mathbb{Z}_4$ over $\mathbb{P}^1$};
		\node[align=center] at (1.9,3.3) {P.IV};
		\node[align=center] at (1.9,2.0) {P.V};
		\node[align=center] at (8.8,4) {$z_2=1$};
		\node[align=center] at (6.25,8.3) {$\Delta_{2,0}^{[2,1]}=0$};
            \draw[->,line width=.3mm]        (4,-.1)   -- (8,-.1);
            \draw[->,line width=.3mm]        (0,4)   -- (0,8);
            \node[align=center] at (-1,6) {$-\log\,\vert z_2\vert $};
            \node[align=center] at (6,-.6) {$-\log\,\vert z_1\vert$};
	\end{tikzpicture}
	\centering
	\includegraphics[width=.5\linewidth]{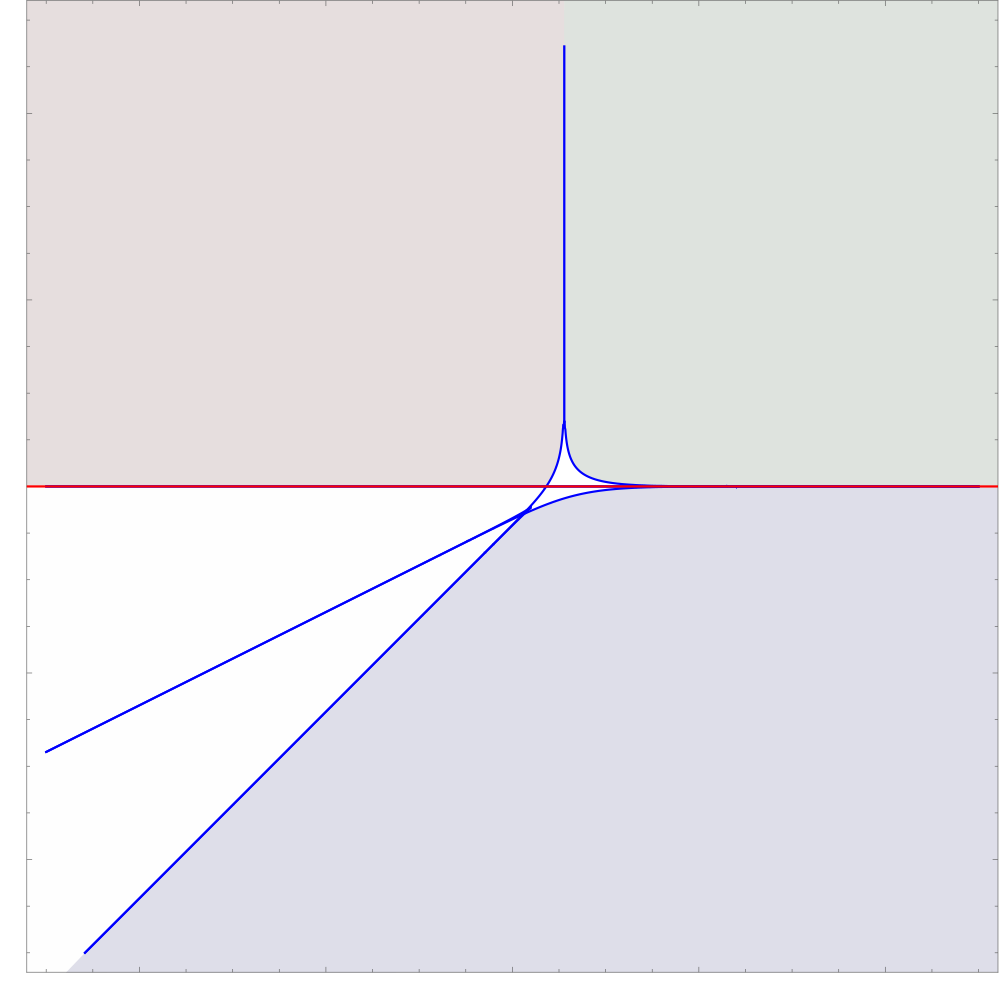}\vspace{.5cm}
	\caption{The phase structure of the stringy K\"ahler moduli space of $X_2^{[2,1]}$. The discriminant component $\Delta_{2,0}^{[2,1]}(z_1,z_2)=0$ is depicted in blue while $z_2=1$ is shown in red.}
	\label{fig:phasesmij212}
\end{figure}

The structure of the moduli space is illustrated for the examples $X_2^{[2,1]}$ and $X_{2,1}^{[2,1]}$ in Figure~\ref{fig:phasesmij212}.

\section{A-model geometries}
\label{sec:amodel}

We shall now construct the CY threefolds $X_{m,s}^{[i,j]}$ that are mirror to $Y_{m,s}^{[i,j]}$, and check that
they satisfy the properties stated in the introduction.

\subsection{Toric cases $m=2$}
\label{sec:Amodelm2}
We first discuss the case $m=2$ in detail, as 
the essential features of the construction will apply to the other cases as well.

Using the results from~\cite[Section 3]{hosono:1994ax}, We first 
recognize that the period~\eqref{eqn:fundamentalPeriod} can formally be obtained as the mirror fundamental period of the complete intersection
\begin{align}
    X_{2,s}^{[i,j]}=&\IP\left(\begin{array}{cccccccc}
    i&j&1&1&1&1&-s&0\\[.2em]
    0&0&0&0&0&0& 1&1
    \end{array}\right)\left[\begin{array}{ccc}
    4&i-s&j\\[.2em]
    0&1&1
    \end{array}\right]\,.
    \label{eqn:X2sij}
\end{align}

Let $P:=\mathbb{P}^5_{i,j,1,1,1,1}$ such that the $\mathbb{P}^1$-bundle $V:=\mathbb{P}(\mathcal{O}_{P}(-s)\oplus\mathcal{O}_P)$ is the toric ambient space in~\eqref{eqn:X2sij}~\footnote{Strictly speaking, we have to restrict to the complement of $S=\{\,u_3=\ldots=u_6=0\,\}\subset P$ since $\mathcal{O}_{P}(-s)$ is not a line bundle if $s$ is not divisible by $i$ and $j$, as happens for the cases with $s\ne 0$. It is easy to check that the Calabi-Yau does not intersect this locus and we leave the restriction implicit.}.
We denote the homogeneous coordinates on $P$ by $[u_1:\ldots:u_6]$ and on the $\mathbb{P}^1$-fibers of $V$ by $[v_1:v_2]$.
Now, consider a generic homogeneous polynomial $P^{(4)}(u_1,\ldots,u_6)$ of weighted degree $4$ and 
\begin{align}
    \begin{split}
    Q_1:=&p_1^{(i)}(u_1,\ldots, u_6)v_1+p_2^{(i-s)}(u_1,\ldots, u_6)v_2\,,\\
    Q_2:=&p_3^{(j+s)}(u_1,\ldots, u_6)v_1+p_4^{(j)}(u_1,\ldots, u_6)v_2\,,
    \label{eqn:m2defP2P3}
    \end{split}
\end{align}
where $p_a^{(b)}$ is a generic homogeneous polynomial of degree $b$ for $a=1,\ldots,4$, such that $X_{2,s}^{[i,j]}$ is the complete intersection
\begin{align}
	X_{2,s}^{[i,j]}=\{\,P^{(4)}=Q_1=Q_2=0\,\}\subset V\,.
\end{align}
For $s=0$ the ambient space of $X_{2}^{[i,j]}=X_{2,0}^{[i,j]}$ takes the form $V=P\times\mathbb{P}^1$ and the induced projection
\begin{align}
    \pi:\,X_{2}^{[i,j]}\rightarrow\mathbb{P}^1\,,
\end{align}
has as generic fiber the $\langle 4\rangle$-polarized K3 surface $\Sigma^2=\mathbb{P}^3[4]$.

\textbf{Conifold transition to $\mathcal{X}_{2}^{[i,j]}$.}
In order to understand the geometry of $X_{2,s}^{[i,j]}$, let us briefly recall some standard results on small resolutions of ordinary double point  singularities (ODP or conifold for brevity).
Locally, the  ODP takes the form
\begin{align}
	W=\{\,uv-zw=0\,\}\subset\mathbb{C}^4\,.
\end{align}
Introducing divisors $D_+=\{\,u=z=0\,\}$ and $D_-=\{\,u=w=0\,\}$ we obtain two inequivalent small resolutions, $\widehat{W}=\text{Bl}_{D_+}W$ and $\widehat{W}'=\text{Bl}_{D_-}W$, that are both isomorphic to $\mathcal{O}_{\mathbb{P}^1}(-1)\oplus \mathcal{O}_{\mathbb{P}^1}(-1)$ and related by a flop.
The blow-ups can explicitly be constructed as complete intersections
\begin{align}
	\begin{split}
		\widehat{W}=&\{\,uy_1-zy_2=wy_1-vy_2=0\,\}\subset \mathbb{C}^4\times\mathbb{P}^1\,,\\
		\widehat{W}'=&\{\,uy_1-wy_2=zy_1-vy_2=0\,\}\subset \mathbb{C}^4\times\mathbb{P}^1\,.
	\end{split}
	\label{eqn:odp}
\end{align}

To relate this to $X_{2,s}^{[i,j]}$ we first define
\begin{align}
	\widetilde{X}:=\{\,P^{(4)}=\widetilde{Q}=0\,\}\subset P\,,\quad \widetilde{Q}=p_1^{(i)}p_4^{(j)}-p_2^{(i-s)}p_3^{(j+s)}\,.
\end{align}
Since $\widetilde{Q}$ is the resultant of $(Q_1,Q_2)$, we find that the projection from $V$ to $P$ induces a projection
\begin{align}
	\tilde{\pi}:\,X_{2,s}^{[i,j]}\rightarrow\widetilde{X}\,.
\end{align}
Comparing with~\eqref{eqn:m2defP2P3} we see that $X_{2,s}^{[i,j]}$ is the blow-up of $\widetilde{X}$ along $\{\,p_1^{(i)}=p_2^{(i-s)}=0\,\}\subset\widetilde{X}$ and therefore a simultaneous small resolution of the $N=4(i-s)(j+s)$ ordinary double points
\begin{align}
	Z=\{\,p_1^{(i)}=p_2^{(i-s)}=p_3^{(j+s)}=p_4^{(j)}=0\,\}\subset \widetilde{X}\,.
\end{align}
Since $\widetilde{X}$ is smooth away from $Z\subset\widetilde{X}$ we then also conclude that $X_{2,s}^{[i,j]}$ is smooth.
The smooth deformation of $\widetilde{X}$ is the CY threefold $\mathcal{X}_{2}^{[i,j]}=\mathbb{P}^5_{i,j,1,1,1,1}[4,i+j]$ with $h_{1,1}(\mathcal{X}_{2}^{[i,j]})=1$.

\textbf{Birationality of $X_{2,s}^{[i,j]}$  and $X_{2,i-j-s}^{[i,j]}$.}
It is now easy to show that $X_{2,s}^{[i,j]}$ and $X_{2,i-j-s}^{[i,j]}$ are birational to each other.
To this end we first denote the toric ambient space of $X_{2,i-j-s}^{[i,j]}$ by $\widetilde{V}:=\mathbb{P}(\mathcal{O}_{P}(j+s-i)\oplus\mathcal{O}_P)$ and write
\begin{align}
\begin{split}
    \tilde{Q}_1=&\tilde{p}_1^{(i)}(u_1,\ldots, u_6)v_1+\tilde{p}_2^{(j+s)}(u_1,\ldots, u_6)v_2\,,\\
    \tilde{Q}_2=&\tilde{p}_3^{(i-s)}(u_1,\ldots, u_6)v_1+\tilde{p}_4^{(j)}(u_1,\ldots, u_6)v_2\,.
    \end{split}
\end{align}
After choosing
\begin{align}
        \tilde{p}_1^{(i)}=p_1^{(i)}\,,\quad \tilde{p}_2^{(j+s)}=p_3^{(j+s)}\,,\quad \tilde{p}_3^{(i-s)}=p_2^{(i-s)}\,,\quad \tilde{p}_4^{(j)}=p_4^{(j)}\,,
\end{align}
we see that $X_{2,i-j-s}^{[i,j]}$ is the blow-up of $\widetilde{X}$ along the divisor $\{\,p_1^{(i)}=p_3^{(j+s)}=0\,\}\subset \tilde{X}$.
Therefore $X_{2,s}^{[i,j]}$  and $X_{2,i-j-s}^{[i,j]}$ are related by flopping each of the exceptional curves. More generally, one can check that $X_{m,s}^{[i,j]}$  and $X_{m,i-j-s}^{[i,j]}$ are related by flopping the $N=2m(i-s)(j+s)$ exceptional curves\footnote{The flop transition relating the models $(m,i,j,s)=(3,2,1,0)$ and $(3,2,1,1)$ was discussed in \cite[Example 4.3]{Brodie:2021toe}.}.

At the level of the free energies the intersection numbers and second Chern classes from~\eqref{gentopdata} behave appropriately under $(s,T,S)\mapsto (i-j-s,T-i S,-S)$, such that the polynomial part of the prepotential and one-loop free energy transform into 
\bea
F_0^{\rm pol}(T,S) \mapsto  F_0^{\rm pol}(T-i S,-S) -\frac{N}{6} B_3(-S) - \frac{N}{4} S^2,\nn\\
F_1^{\rm pol}(T,S) \mapsto F_1^{\rm pol}(T-i S,-S) + \frac{N}{12} B_1(-S) - \frac{N}{24}
\eea
where $B_n(x)$ are the Bernoulli polynomials, originating from Jonqui\`ere's formula 
\be
\Li_n(e^{2\pi\I S})+(-1)^n \Li_n(e^{-2\pi\I S})=-\frac{(2\pi\I)^n}{n!}B_n(S) .
\ee

\textbf{Tyurin degeneration and Hodge numbers.}
Let us now construct the Tyurin degeneration on the A-model side.
To this end we consider the family of singular CY threefolds
\begin{align}
    \widetilde{X}(t):=\{\,P^{(4)}=p_1^{(i)}p_4^{(j)}-t\,p_2^{(i-s)}p_3^{(j+s)}=0\,\}\subset P\,,
    \label{eqn:tyurin}
\end{align}
such that $\widetilde{X}=\widetilde{X}(1)$.
In the limit $t\rightarrow 0$, the fiber degenerates into the union of Fano threefolds
\begin{align}
	\tilde{X}(0)=P[4,i]\cup_{\tilde{X}(0)} P[4,j]\simeq F_2^{[i]}\cup_{\tilde{X}(0)} F_2^{[j]}\,,
\end{align}
with $F_2^{[a]}=\mathbb{P}^4_{1,1,1,1,a}[4]$.
We can further identify $F_2^{[1]}=\mathbb{P}^4[4],F_2^{[2]}=\mathbb{P}^4_{1,1,1,1,2}[4]$ and $F_2^{[4]}=\mathbb{P}^3$ and the intersection of the two Fano components is the degree $4$ polarized K3 surface $\Sigma^2=F_2^{[i]}\times_{\tilde{X}(0)} F_2^{[j]}=\mathbb{P}^3[4]$. 

However,~\eqref{eqn:tyurin} is not yet a Tyurin degeneration because the total space of the family is not smooth.
As discussed above, we can blow up each of the fibers in the family along the divisor $\{\,p_1^{(i)}=p_2^{(i-s)}=0\,\}\subset \widetilde{X}(t)$, such that for $t\ne 0$ the resulting blow-up of the fiber is smooth.
As discussed in~\cite[Section 5.3]{doran:2016uea}, we still need to blow up along $\{\,p_1^{(i)}=p_3^{(j+s)}=0\,\}\subset \widetilde{X}(t)$ in order for the total space of the family~\eqref{eqn:tyurin} to be smooth as well. This only affects the central fiber.
Note that the Tyurin degeneration associated to the flopped geometry $X_{2,i-j-s}^{[i,j]}$ is obtained by changing the order of the blow ups.

We denote by $\mathcal{O}_{\Sigma^2}(1)$ the restriction of the hyperplane bundle on $\mathbb{P}^3$ to $\Sigma^2$, the result of the two blow ups of $\widetilde{X}(0)$ takes the form $\tilde{F}\cup_{\widetilde{X}}F_2^{[j]}$, where $\tilde{F}$ is the blow-up of $F_2^{[i]}$ first along the zero locus of a section of $\mathcal{O}_{\Sigma^2}(i-s)$ and then of $\mathcal{O}_{\Sigma^2}(j+s)$.
We can now use~\cite[Proposition 5.6]{doran:2016uea} to find the Hodge numbers
\begin{align}
\label{HodgeXs}
	h_{1,1}(X_{2,s}^{[i,j]})=2\,,\quad h_{1,2}(X_{2,s}^{[i,j]})=2+2\frac{i+j}{i j}(20+i^2 j+i j^2)-4(i-s)(j+s)\, ,
\end{align}
such that the Euler characteristic is indeed opposite to \eqref{EulerY}.

\textbf{Batyrev-Borisov mirror.}
The realization of $X_{2,s}^{[i,j]}$ as a complete intersection $V$ does not in general correspond to a nef-partition of the anti-canonical class on $V$.
Nevertheless it turns out that the Batyrev-Borisov mirror construction~\cite{Batyrev:1994pg} still works.~\footnote{The reason is that in some cases one of the component divisors of the ``nef-partition'' and/or the canonical divisor itself fail to be Cartier, which can happen if the ambient space is singular. This is a well known phenomenon and arises for example also in the case $X_{6,4}=\mathbb{P}^5_{3,2,2,1,1,1}[6,4]$. The problem can usually be solved by a suitable blowup of the toric ambient space that does not intersect the Calabi-Yau complete intersection itself.}

Let us first recall the mirror construction in the context of actual nef-partitions.
A mirror pair of nef-partitions with $r$ parts then corresponds to a pair of $d$-dimensional reflexive polytopes $\Delta,\nabla$ and lattice polytopes $\Delta_i,\nabla_j,\,i,j=1,\ldots,r$ such that $\Delta=\Delta_1+\ldots+\Delta_r$, $\Delta^\circ=\text{Conv}(\nabla_1,\ldots,\nabla_r)$, $\nabla=\nabla_1+\ldots+\nabla_r$ and $\nabla^\circ=\text{Conv}(\Delta_1,\ldots,\Delta_r)$, while for each pair of points $\vec{m}\in\Delta_i$, $\vec{n}\in\nabla_j$ one has $\langle \vec{m}, \vec{n}\rangle\ge -\delta_{i,j}$.
Denoting by $P_{\Delta},P_{\nabla}$ the toric varieties that are respectively associated to (refinements of) the face fans of $\Delta^\circ$ and $\nabla^\circ$, one then has nef divisors $D_{\Delta_i},D_{\nabla_j},\,i,j=1,\ldots,r$ on $P_{\Delta}$ and $P_{\nabla}$ and a corresponding mirror pair of  complete intersection CY $(d-r)$-folds $X_{\Delta_1,\ldots,\Delta_r}\subset P_{\Delta},\,X_{\nabla_1,\ldots,\nabla_r}\subset P_{\nabla}$.
Denoting the torus invariant divisors on $P_{\Delta},P_{\nabla}$ that are associated to points $\vec{n}\in\partial\Delta^\circ,\,\vec{m}\in\partial\nabla^\circ$ by $D_{\vec{n}},D_{\vec{m}}$, one has
\begin{align}
    D_{\Delta_i}=\sum\limits_{\vec{n}\in \nabla_i-\vec{0}}D_{\vec{n}}\,,\quad D_{\nabla_j}=\sum\limits_{\vec{m}\in \Delta_j-\vec{0}}D_{\vec{m}}\,,\quad i,j=1,\ldots,r\,.
\end{align}
On the torus $(\mathbb{C}^*)^d\subset P_{\nabla}$, one can choose coordinates $x_1,\ldots,x_d$ and the threefold $X_{\nabla_1,\ldots,\nabla_r}$ is the intersection of the vanishing loci of the polynomials
\begin{align}
    p_i=\sum\limits_{\vec{n}\in \nabla_i}c_{i,\vec{n}}\vec{x}^{\,\vec{n}}\,,\quad i=1,\ldots,r\,,
    \label{eqn:nefequations}
\end{align}
for some choice of coefficients $c_{i,\vec{n}}\in\mathbb{C}$, with analogous expressions for the polynomials defining $X_{\Delta_1,\ldots,\Delta_r}$ on $(\mathbb{C}^*)^d\subset P_{\Delta}$.

\begin{table}[ht!]
\begin{align*}
	\begin{array}{c}\\u_1\\u_2\\u_3\\u_4\\u_5\\u_6\\v_1\\v_2\\\vspace{.3cm}\\ \\ \\\end{array}
	\left[\begin{array}{cccccc|cc|c}
	\multicolumn{6}{c|}{\vec{p}\in\mathbb{Z}^n}& l^{(1)} & l^{(2)} & \nabla_i\\\hline
	 1 & 0 & 0 & 0 & 0 & 0 & i & 0 & 2 \\
	 0 & 1 & 0 & 0 & 0 & 0 & j & 0 & 3 \\
	 0 & 0 & 1 & 0 & 0 & 0 & 1 & 0 & 1 \\
	 0 & 0 & 0 & 1 & 0 & 0 & 1 & 0 & 1 \\
	 0 & 0 & 0 & 0 & 1 & 0 & 1 & 0 & 1 \\
	 -i & -j & -1 & -1 & -1 & s & 1 & 0 & 1 \\
	 0 & 0 & 0 & 0 & 0 & 1 & -s & 1 & 2 \\
	 0 & 0 & 0 & 0 & 0 & -1 & 0 & 1 & 3 \\\hline\hline
	 0 & 0 & 0 & 0 & 0 & 0 & -4 & 0 & 1 \\
	 0 & 0 & 0 & 0 & 0 & 0 & s-i & -1 & 2 \\
	 0 & 0 & 0 & 0 & 0 & 0 & -j & -1 & 3 \\
	\end{array}\right]
\end{align*}
	\caption{The toric data associated to $X_{2,s}^{[i,j]}$. In the first six columns we list the coordinates of the relevant points in $\Delta^\circ=\text{Conv}(\nabla_1,\nabla_2,\nabla_3)$, with the associated homogeneous coordinate indicated on the left, while the sevenths and eights column contain the coefficients of the points in the two independent linear relations. The last column indicates whether the point belongs to $\nabla_1$, $\nabla_2$ or $\nabla_3$. In each case there is a unique fine regular star triangulation of the listed points and the corresponding fan defines the toric variety $P_{\Delta}=P$.}
	\label{tab:tdatam2}
\end{table}
The toric data associated to $X_{2,s}^{[i,j]}$ is summarized in Table~\ref{tab:tdatam2}.
Let us stress that in general $\Delta^\circ$ is not reflexive and should not be interpreted as the polar dual of a lattice polytope $\Delta$.
Nonetheless we can apply the Batyrev-Borisov construction to obtain the mirror at least on a dense open subset.
To this end we choose coordinates $(x_1,\ldots,x_6)$ on $(\mathbb{C}^{\times})^6$ and write down the defining polynomials~\eqref{eqn:nefequations}.
We can use the freedom to rescale the coordinates and the defining polynomials to obtain
\begin{align}
	\begin{split}
		p_1=&1+x_3+x_4+x_5+(-1)^{i+j+s}z_1\frac{x_6^s}{x_1^ix_2^jx_3x_4x_5}\,,\\
		p_2=&1+x_1+x_6\,,\quad p_3=1+x_2+\frac{z_2}{x_6}\,,
	\end{split}
\end{align}
in terms of the invariant complex structure coordinates $z_1,z_2\in\mathbb{C}$.
After suitable restriction to a dense open subset $V\subset (\mathbb{C}^{*})^6$ we can eliminate $x_1,x_2$ and rewrite this as $\{\,p=0\,\}\subset V$ with
\begin{align}
    \begin{split}
        p=&1 + \frac{1}{ z_1}\frac{(1-y)^i(1-z_2/y)^j}{y^s}u_1 u_2 u_3(u_1+u_2+u_3-1)\,,
    \end{split}
\end{align}
where we use $u_1=-x_3$, $u_2=-x_4$, $u_3=-x_5$ and $y=-x_6$.
Comparing this to the fundamental modular family of $M_2$-polarized K3 surfaces from Table~\ref{tab:modularFamilies}, we find that the CY threefold mirror to $X_{2,s}^{[i,j]}$ is indeed the $M_2$-polarized fibration $Y_{2,s}^{[i,j]}$  associated to the generalized functional invariant map $\Lambda^{[i,j]}_s(y;z_1,z_2)$ in~\eqref{eqn:gfimGen}.

Using the representation of the fundamental period from~\cite{hosono:1994ax}, we obtain 
\begin{align}
    \varpi_0(z_1,z_2)=\int_\Gamma \frac{1}{p_1p_2p_3}\prod\limits_{a=1}^6\frac{\de x_a}{x_a}=\oint\frac{dy}{y(1-y)(1-z_2/y)}\int_{\Gamma_u}
    \frac{ \de u_1 \de u_2 \de u_3}{p}\,,
\end{align}
with the contour $\Gamma\subset (\mathbb{C}^\times)^6$ corresponding to $|x_1|=\ldots=|x_6|=\epsilon\to 0$, and $\Gamma_u$ to $|u_1|=|u_2|=|u_3|=\epsilon\to 0$. Expanding $1/p$  in powers of $z_1$ and using 
\begin{align}
\label{Laurent2}
\int_{\tilde{\Gamma}_0}
\frac{\prod_{i=1}^3du_i\,}{\left[ u_1u_2u_3(1-u_1-u_2-u_3) \right]^{d+1}}
=\frac{(4d)!}{(d!)^4} = c_2(d)\,,
\end{align}
we get 
\begin{align}
\begin{split}
    \varpi_0(z_1,z_2)=&\frac{1}{2\pi \I}\oint \frac{\de y}{y}\frac{1}{(1-y)(1-z_2/y)}f_2\left(\lambda=\Lambda^{[i,j]}_s(y;z_1,z_2)\right)\\
    =&\sum\limits_{d,k\ge 0}c_2(d)\frac{(k+di-ds)!(k+dj)!}{(di)!(dj)!(k-ds)!k!}z_1^dz_2^{k}\,,
    \end{split}
    \label{eqn:bbmirrorperiod}
\end{align}
which precisely matches the expression~\eqref{eqn:fundamentalPeriod}.

\subsection{Toric cases $m=1,3,4$}
\label{sec:Amodelm134}
The other toric cases $m=1,3,4$ can be treated in the same way as the $m=2$ case in the previous subection. 
In those cases, using again the results from~\cite[Section 3]{hosono:1994ax}, we find that the periods~\eqref{eqn:fundamentalPeriod} can be obtained as the mirror fundamental periods of the complete intersections
\begin{align}
    \begin{split}
   X_{1,s}^{[i,j]}=&\IP\left(\begin{array}{cccccccc}
    i&j&1&1&1&3&-s&0\\[.2em]
    0&0&0&0&0&0& 1&1
    \end{array}\right)\left[\begin{array}{ccc}
    6&i-s&j\\[.2em]
    0&1&1
    \end{array}\right]\,,\\
    X_{3,s}^{[i,j]}=&\IP\left(\begin{array}{ccccccccc}
    i&j&1&1&1&1&1&-s&0\\[.2em]
    0&0&0&0&0&0&0& 1&1
    \end{array}\right)\left[\begin{array}{cccc}
    2&3&i-s&j\\[.2em]
    0&0&1&1
    \end{array}\right]\,,\\
    X_{4,s}^{[i,j]}=&\IP\left(\begin{array}{cccccccccc}
    i&j&1&1&1&1&1&1&-s&0\\[.2em]
    0&0&0&0&0&0&0&0& 1&1
    \end{array}\right)\left[\begin{array}{ccccc}
    2&2&2&i-s&j\\[.2em]
    0&0&0&1&1
    \end{array}\right]\,.
    \end{split}
    \label{eqn:cicy1234}
\end{align}
In particular, the generalized functional invariant map~\eqref{eqn:gfimGen} can in each case be reproduced by applying the Batyrev-Borisov construction as in Section~\ref{sec:Amodelm2}.
While the generalized homological invariant in the cases with $m=1$ is harder to determine, it is fixed by comparing the Hodge numbers of the B-model geometries with those calculated in~\cite{kooistra2021threefolds}.

\subsection{Non-toric cases $m=5$}
\label{sec:Amodelm5}

We now come to non-toric cases, starting with $m=5$.

As discussed e.g. in~\cite{Debarre:2018mmi}, the generic degree 10 polarized K3 surface $\Sigma^5$ can be realized as the intersection of the Pl\"ucker embedding of $\text{Gr}(2,5)$ in $\mathbb{P}^9$ with a quadric and a linear subspace of codimension three, i.e.
\begin{align}
    \Sigma^5=X^{2,5}_{\mathcal{O}(1)^{\oplus 3}\oplus\mathcal{O}(2)}\,.
\end{align}
Using the ``Trick with the Factorials'' from~\cite{Batyrev:1998kx}, one finds that the coefficients \eqref{eqn:fundamental_c5} of the fundamental period $f_5(\lambda)$ of the mirror $\Sigma_5$ can be rewritten as
\begin{align}
    c_5(d)=\sum\limits_{l_1,l_2\ge 0}\binom{l_1}{l_2}\binom{d}{l_1}^2\binom{d}{l_2}\binom{2d}{d}=\sum\limits_{l=0}^d\binom{2d}{d}\binom{d}{l}^2\binom{d+l}{l}\,,
\end{align}
where we have used the identity
\begin{align}
    \binom{d+l}{l}=\sum\limits_{n=0}^d\binom{d}{n}\binom{l}{n}\,.
\end{align}
We can then write
\begin{align}
    \begin{split}
    \varpi_{5,s}^{[i,j]}(z_1,z_2)=&\sum\limits_{d,k\ge 0}c_5(d)\frac{(k+di-ds)!(k+dj)!}{(di)!(dj)!(k-ds)!k!}z_1^dz_2^{k}=\varpi_0'(z_1,z_2,1,1)\,,\\
    \end{split}
\end{align}
in terms of
\begin{align}
    \varpi_0'(z_1,z_2,z_3,z_4)=\sum\limits_{d,k,l_1,l_2\ge 0}\binom{l_1}{l_2}\binom{d}{l_1}^2\binom{d}{l_2}\binom{2d}{d}\frac{(k+di-ds)!(k+dj)!}{(di)!(dj)!(k-ds)!k!}z_1^dz_2^{k}z_3^{l_1}z_4^{l_2}\,.
\label{eqn:m5periodRelation}
\end{align}
This is the mirror fundamental period of a CY threefold $\widetilde{X}_{5,s}^{[i,j]}$, with $h_{1,1}(\widetilde{X}_{5,s}^{[i,j]})=4$, that is a codimension six complete intersection in a nine-dimensional toric ambient space whose toric data is given in Table~\ref{tab:tdatam5}.

\begin{table}[ht!]
\begin{align*}
	\begin{array}{c}\\w_1\\w_2\\u_1\\u_2\\u_3\\u_4\\u_5\\u_6\\u_7\\u_{8}\\u_{9}\\v_{1}\\v_{2}\\\vspace{.3cm}\\ \\ \\ \\ \\ \\ \end{array}
	\left[
	\begin{array}{ccccccccc|cccc|c}
		\multicolumn{9}{c|}{\vec{p}\in\mathbb{Z}^n}& l^{(1)} & l^{(2)} & l^{(3)} & l^{(4)} & \Delta_i\\\hline
	 1 & 0 & 0 & 0 & 0 & 0 & 0 & 0 & 0 & i & 0 & 0 & 0 & 5 \\
	 0 & 1 & 0 & 0 & 0 & 0 & 0 & 0 & 0 & j & 0 & 0 & 0 & 6 \\
	 0 & 0 & 1 & 0 & 0 & 0 & 0 & 0 & 0 & 1 & 0 & 0 & 0 & 4 \\
	 0 & 0 & 0 & 1 & 0 & 0 & 0 & 0 & 0 & 1 & 0 & 0 & 0 & 3 \\
	 0 & 0 & 0 & 0 & 1 & 0 & 0 & 0 & 0 & 1 & 0 & -1 & 0 & 1 \\
	 0 & 0 & 0 & 0 & 0 & 1 & 0 & 0 & 0 & 1 & 0 & -1 & 0 & 2 \\
	 -i & -j & -1 & -1 & -1 & -1 & 0 & 0 & s & 1 & 0 & 0 & -1 & 2 \\
	 0 & 0 & 0 & 0 & 0 & 0 & 1 & 0 & 0 & 0 & 0 & 1 & 0 & 2 \\
	 0 & 0 & 0 & 0 & 1 & 1 & -1 & 0 & 0 & 0 & 0 & 1 & -1 & 1 \\
	 0 & 0 & 0 & 0 & 0 & 0 & 0 & 1 & 0 & 0 & 0 & 0 & 1 & 1 \\
	 -i & -j & -1 & -1 & 0 & 0 & -1 & -1 & s & 0 & 0 & 0 & 1 & 2 \\
	 0 & 0 & 0 & 0 & 0 & 0 & 0 & 0 & 1 & -s & 1 & 0 & 0 & 5 \\
	 0 & 0 & 0 & 0 & 0 & 0 & 0 & 0 & -1 & 0 & 1 & 0 & 0 & 6 \\\hline\hline
	 0 & 0 & 0 & 0 & 0 & 0 & 0 & 0 & 0 & -1 & 0 & 0 & 0 & 1 \\
	 0 & 0 & 0 & 0 & 0 & 0 & 0 & 0 & 0 & -2 & 0 & 0 & 0 & 2 \\
	 0 & 0 & 0 & 0 & 0 & 0 & 0 & 0 & 0 & -1 & 0 & 0 & 0 & 3 \\
	 0 & 0 & 0 & 0 & 0 & 0 & 0 & 0 & 0 & -1 & 0 & 0 & 0 & 4 \\
	 0 & 0 & 0 & 0 & 0 & 0 & 0 & 0 & 0 & s-i & -1 & 0 & 0 & 5 \\
	 0 & 0 & 0 & 0 & 0 & 0 & 0 & 0 & 0 & -j & -1 & 0 & 0 & 6 
	\end{array}
	\right]
\end{align*}
	\caption{The toric data associated to the Calabi-Yau $\widetilde{X}_{5,s}^{[i,j]}$ that is connected to $X_{5,s}^{[i,j]}$ via a conifold transition.
 The format is explained in the description of Table~\ref{tab:tdatam2}.
 In each case there are five different fine regular star triangulation of the listed points and we choose the one such that the linear relations among the points correspond to a basis of the Mori cone on the associated toric variety $P_{\Delta}=P$.}
	\label{tab:tdatam5}
\end{table}

The relationship~\eqref{eqn:m5periodRelation} among the periods suggests that $\widetilde{X}_{5,s}^{[i,j]}$ is connected to $X_{5,s}^{[i,j]}$ via a conifold transition and we will show that this is indeed the case.
To this end let us first discuss the mirror.
Using affine coordinates $(x_1',\ldots,x_9')\in(\mathbb{C}^\times)^9$, we can write the defining polynomials of the complete intersection $\widetilde{Y}_{5,s}^{[i,j]}$ mirror to  $\widetilde{X}_{5,s}^{[i,j]}$ on this open subset as
\begin{align}
	\begin{split}
		p_1=&1+x'_5+x'_8+z_3\frac{x'_5x'_6}{x'_7}\,,\quad p_3=1+x'_4\,,\quad p_4=1+x'_3\,,\quad p_5=1+x'_1+x'_9\,,\\
		p_2=&1+x'_6+x'_7-(-1)^{i+j+s}\left(\frac{z_1}{x'_5x'_6}+\frac{z_1z_3z_4}{x'_7x'_8}\right)\frac{(x'_9)^s}{(x'_1)^i(x'_2)^jx'_3x'_4}\,,\quad \quad p_6=1+x'_2+\frac{z_2}{x'_9}\,.
	\end{split}
\end{align}
After setting $z_3=z_4=1$, using $p_1,p_3,p_4,p_5,p_6$ to eliminate $x'_1,\ldots,x'_5$ and introducing
\begin{align}
	x_1=x_6'\,,\quad x_2=x_7'\,,\quad y_1=x_8'\,,\quad y=-x_9'\,,
\end{align}
we can reduce this to the hypersurface $\{\,p'=0\,\}\subset(\mathbb{C}^{\times})^{4}$ with
\begin{align}
	p=x_2y_1-x_1+\frac{1}{z_1}\frac{(1-y)^i(1-z_2/y)^j}{y^s}x_1x_2(1+x_1+x_2)y_1(1+y_2)\,,
	\label{eqn:pm5}
\end{align}
We see that this can be compactified to a fibration of
\begin{align}
	\left\{\,(x_1+x_2+x_3)^2(y_1+y_2)(x_1y_1+x_2y_1+x_1y_2)-\lambda x_1x_2x_3y_1y_2=0\,\right\}\subset \mathbb{P}^2\times\mathbb{P}^1\,,
	\label{eqn:universalm5}
\end{align}
where~\eqref{eqn:pm5} is recovered after identifying $\lambda=\Lambda_s^{[i,j]}(y;z_1,z_2)$ and setting
\begin{align}
	x_3\rightarrow -x_1-x_2-1,\,\quad y_2\rightarrow -y_1-1\,.
\end{align}
The hypersurfaces \eqref{eqn:universalm5} parametrized by $\lambda$ in fact define a universal family 
of $M_5$-polarized K3-surfaces, whose equation differs from the 
standard one in the literature, see Table~\ref{tab:modularFamilies}.
After specializing the complex structure of $\widetilde{Y}_{5,s}^{[i,j]}$ setting $z_3=z_4=1$, and performing a small resolution of all nodal singularities, we therefore obtain indeed $Y_{5,s}^{[i,j]}$.

Let us now return to the A-model geometries.
To understand the conifold transition from $\widetilde{X}_{5,s}^{[i,j]}$ to $X_{5,s}^{[i,j]}$ we distinguish two cases.

\textbf{Case I:} $(i,j,s)\in\{(1,1,0),\,(2,1,0),\,(2,1,1)\}$. 
		We first use the equations associated to $\Delta_i$ and $\Delta_3$ to eliminate the homogeneous coordinates $w_1$ and $w_2$.
		This allows us to construct $\widetilde{X}_{5,s}^{[i,1]}$ as a codimension four complete intersection in a seven-dimensional toric ambient space $P_{\Delta}'$.
		The corresponding toric data is obtained from that in Table~\ref{tab:tdatam5} by projecting the points onto the last seven coordinates and dropping the two rows that are respectively associated to the origin in $\Delta_i$ and $\Delta_3$.

		The homogeneous coordinates $[u_1:\ldots:u_9]$ then parametrize a toric variety $\widehat{P}(2,5)$ which is a small resolution of the toric degeneration $P(2,5)$ of $\text{Gr}(2,5)$ discussed in~\cite{sturmfels1996grobner,Batyrev:1998kx}.
            The conifold transition from $\widetilde{Y}_{5,s}^{[i,j]}$ to $Y_{5,s}^{[i,j]}$ discussed above is dual on the A-model side to the transition from $\widetilde{X}_{5,s}^{[i,j]}$ to $X_{5,s}^{[i,j]}$ that is induced by the conifold transition on the ambient space from $\widehat{P}(2,5)$ via $P(2,5)$ to $\text{Gr}(2,5)$.

		The projection $\pi:\,P_{\Delta}'\rightarrow \widehat{P}(2,5)$ gives $P_{\Delta}'$ the structure of a $\mathbb{P}^1$-bundle on $\widehat{P}(2,5)$.
	The 10 independent sections of $\mathcal{O}_{\widehat{P}(2,5)}(1):=\mathcal{O}_{\widehat{P}(2,5)}(D_{u_1}')$, with $D_{u_1}'=\{\,u_1=0\,\}\subset\widehat{P}(2,5)$, are given by
	\begin{align}
	    \{\,u_1,\,u_2,\,u_3u_6,\,u_4u_6,\,u_3u_7u_8,\,u_3u_7u_9,\,u_4u_7u_8,\,u_4u_7u_9,\,u_5u_8,\,u_5u_9\,\}\subset\Gamma\left(\mathcal{O}_{\widehat{P}(2,5)}(D_{u_1}')\right)\,,
	\end{align}
	and define an embedding $\iota:\,\widehat{P}(2,5)\rightarrow\mathbb{P}^9$ such that $\mathcal{O}_{\widehat{P}(2,5)}(1)=\iota^*\mathcal{O}_{\mathbb{P}^9}(1)$.
	We then see that
	\begin{align}
		P_{\Delta}'=\mathbb{P}\left(\mathcal{O}_{\widehat{P}(2,5)}(-s)\oplus\mathcal{O}_{\widehat{P}(2,5)}\right)\,,
	\end{align}
	and we denote the relative hyperplane bundle on $P_{\Delta}'$ by $\zeta$.

	The four defining polynomials of $\widetilde{X}_{5,s}^{[i,1]}$ in $P'_{\Delta}$ are respectively sections of $\mathcal{L}'_1=\pi^*\mathcal{O}_{\widehat{P}(2,5)}(3-i)$, $\mathcal{L}'_2=\pi^*\mathcal{O}_{\widehat{P}(2,5)}(1)$, $\mathcal{L}'_3=\pi^*\mathcal{O}_{\widehat{P}(2,5)}(i-s)\otimes\zeta$ and $\mathcal{L}'_4=\pi^*\mathcal{O}_{\widehat{P}(2,5)}(1)\otimes\zeta$.
	They take the form
	\begin{align}
		\begin{split}
			P_1=&q_1^{(3-i)}(u_1,u_2,u_3u_6,\ldots,u_7u_{11})\,,\quad P_2=q_2^{(1)}(u_1,u_2,u_3u_6,\ldots,u_7u_{11})\,,\\
			Q_1=&p_1^{(i)}(u_1,u_2,u_3u_6,\ldots,u_7u_{11})v_1+p_2^{(i-s)}(u_1,u_2,u_3u_6,\ldots,u_7u_{11})v_2\,,\\
			Q_2=&p_3^{(1+s)}(u_1,u_2,u_3u_6,\ldots,u_7u_{11})v_1+p_4^{(1)}(u_1,u_2,u_3u_6,\ldots,u_7u_{11})v_2\,,
		\end{split}
	\end{align}
	where $q_i^{(d},\,i=1,2$ and $p_j^{(d)},\,j=1,\ldots,4$ are homogeneous polynomials in ten variables of degree $d$.

	After following the conifold transition from $\widehat{P}(2,5)\rightarrow\text{Gr}(2,5)$ on the ambient space, we can just replace $\mathcal{O}_{\widehat{P}(2,5)}(1)$ with the pullback $\mathcal{O}_{\text{Gr}(2,5)}(1)=\tilde{\iota}^*\mathcal{O}_{\mathbb{P}^9}(1)$ along the Pl\"ucker embedding $\tilde{\iota}:\text{Gr}(2,5)\rightarrow \mathbb{P}^9$.
	We denote the homogeneous coordinates on $P=\mathbb{P}\left(\mathcal{O}_{\mathbb{P}^9}(-s)\oplus\mathcal{O}_{\mathbb{P}^9}\right)$ by $[x_1:\ldots:x_{10}:v_1:v_2]$, with projection $\pi_P:\,P\rightarrow \mathbb{P}^9$, and obtain
	\begin{align}
		\begin{split}
		X_{5,s}^{[i,1]}=\{\,&q_1^{(3-i)}(x_1,\ldots,x_{10})=q_2^{(1)}(x_1,\ldots,x_{10})=p_1^{(i)}(x_1,\ldots,x_{10})v_1+p_2^{(i-s)}(x_1,\ldots,x_{10})v_2\\
			=&p_3^{(1+s)}(x_1,\ldots,x_{10})v_1+p_4^{(1)}(x_1,\ldots,x_{10})v_2=0\,\}\cap \pi_P^{-1}\tilde{\iota}\left(\text{Gr}(2,5)\right)\subset P\,.
		\end{split}
	\end{align}

	An analogous discussion to the one identifying $X_{2,s}^{[i,j]}$ as a small resolution of a nodal degeneration of $\mathcal{X}_{2}^{[i,j]}$ in Section~\ref{sec:Amodelm2} then shows the following:
	\begin{itemize}
        \setlength\itemsep{1em}
		\item $X_{5}^{[1,1]}=X_{5,0}^{[1,1]}$ is a small resolution of a degeneration of the Calabi-Yau threefold $\mathcal{X}_5^{[1,1]}=X_{1,2,2}\subset\text{Gr}(2,5)$ that has been studied e.g. in~\cite{Batyrev:1998kx,Haghighat:2008ut}.
			The degeneration has 10 nodes  $\{\,q_1^{(2)}=q_2^{(1)}=p_1^{(1)}=p_2^{(1)}=p_3^{(1)}=p_4^{(1)}=0\,\}\subset\text{Gr}(2,5)$.
		\item $X_{5}^{[2,1]}=X_{5,0}^{[2,1]}$ is a small resolution of a degeneration of the Calabi-Yau threefold $\mathcal{X}_5^{[2,1]}=X_{1,1,3}\subset\text{Gr}(2,5)$ that has also been studied in~\cite{Batyrev:1998kx,Haghighat:2008ut}.
			The degeneration has 20 nodes  $\{\,q_1^{(1)}=q_2^{(1)}=p_1^{(2)}=p_2^{(2)}=p_3^{(1)}=p_4^{(1)}=0\,\}\subset\text{Gr}(2,5)$.
		\item $X_{5,1}^{[2,1]}$ is a small resolution of the same degeneration of $\mathcal{X}_5^{[2,1]}=X_{1,1,3}\subset\text{Gr}(2,5)$ and obtained from $X_{5}^{[2,1]}$ by flopping all of the exceptional curves.
	\end{itemize}

\textbf{Case II}: $(i,j,s)=(2,2,0)$. To understand this case we first use the equation associated to $\Delta_2$ to eliminate $w_2$ and without loss of generality we can then write the defining polynomials of $\widetilde{X}_{5,0}^{[2,2]}$ as
	\begin{align}
		\begin{split}
			P_1=&q_1^{(1)}(u_1,\ldots,u_7u_{11})\,,\quad P_2=q_2^{(1)}(u_1,\ldots,u_7u_{11})\,,\quad P_3=q_3^{(1)}(u_1,\ldots,u_7u_{11})\,,\\
			Q_1=&\left[w_1+p_1^{(2)}(u_1,\ldots,u_7u_{11})\right]v_1+p_2^{(2)}(u_1,\ldots,u_7u_{11})v_2\,,\\
			Q_2=&p_3^{(2)}(u_1,\ldots,u_7u_{11})v_1+\left[w_1-p_1^{(2)}(u_1,\ldots,u_7u_{11})\right]v_2\,.
		\end{split}
	\end{align}
        After following again the conifold transition from $\widehat{P}(2,5)\rightarrow\text{Gr}(2,5)$ we then obtain
	\begin{align}
		\begin{split}
		X_{5,0}^{[2,2]}=\{\,&q_1^{(1)}(x_1,\ldots,x_{10})=q_2^{(1)}(x_1,\ldots,x_{10})=q_3^{(1)}(x_1,\ldots,x_{10})\\
                =&\left[w_1+p_2^{(2)}(x_1,\ldots,x_{10})\right]v_1-p_1^{(2)}(x_1,\ldots,x_{10})v_2\\
			=&p_4^{(2)}(x_1,\ldots,x_{10})v_1+\left[w_1-p_2^{(2)}(x_1,\ldots,x_{10})\right]v_2=0\,\}\cap \pi_P^{-1}\tilde{\iota}\left(\text{Gr}(2,5)\right)\subset P\,,
		\end{split}
	\end{align}
        where $P:=U\times\mathbb{P}^1$, with $[w_1:x_1:\ldots:x_{10}]$ homogeneous coordinates on $W:=\mathbb{P}^{10}_{2,1,\ldots,1}$ and $U:=W\backslash\{\,x_1=\ldots=x_{10}=0\,\}$ while $\pi_P$ is the projection from $P$ to $\mathbb{P}^9$ that factors through $U$.

        We denote by $B_5$ the del Pezzo threefold of degree $5$ obtained as the intersection of the image of $\text{Gr}(2,5)$ under the Pl\"ucker embedding with a linear subspace of codimension three in $\mathbb{P}^9$, i.e. $B_5=X^{2,5}_{\mathcal{O}(1)^{\oplus 3}}$.
        Then $X_{5}^{[2,2]}=X_{5,0}^{[2,2]}$ is a small resolution of
        \begin{align}
            \widetilde{X}:=\{\,w_1^2+p_1^{(2)}p_4^{(2)}-\left(p_2^{(2)}\right)^2=0\,\}\cap V_5^{[2,2]}\subset W\,,
        \end{align}
        where we use $V_5^{[2,2]}$ to denote the closure in $\mathbb{P}^{10}_{2,1,\ldots,1}$ of the preimage of the Pl\"ucker embedding of $B_5$ in $\mathbb{P}^9$ under the projection $\pi_U:\,U\rightarrow\mathbb{P}^9$, i.e.
        \begin{align}
            V_5^{[2,2]}:=\overline{\pi_U^{-1}\left(B_5\right)}\subset W\,.
            \label{eqn:v522}
        \end{align}
        The Calabi-Yau threefold $\widetilde{X}$ is a double cover of $B_5$ with ramification locus
        \begin{align}
            R=\{\,p_1^{(2)}p_4^{(2)}-\left(p_2^{(2)}\right)^2=0\,\}\cap B_5\subset\mathbb{P}^9\,.
        \end{align}
        Assuming that the polynomials $p_i^{(2)},\,i=1,2,4$ are generic, we can use them as local coordinates around the 40 points $\{\,p_1^{(2)}=p_2^{(2)}=p_4^{(2)}=0\,\}\cap B_5$ and see that these are isolated nodes of $R$.
        The double cover $\widetilde{X}$ then also has nodes over the $40$ nodes of $R$ and is smooth everywhere else.
        The smooth deformation $\mathcal{X}_5^{[2,2]}$ of $\widetilde{X}$ is a generic Calabi-Yau double cover of $B_5$.

\textbf{Tyurin degenerations.} In each case the Tyurin degeneration is obtained by replacing $p_2^{(\bullet)}$ with $t p_2^{(\bullet)}$ and can be discussed analogously to $X_{2,s}^{[i,j]}$.

\subsection{Uniform construction}
\label{sec:amirror}

\begin{table}[ht!]
    \centering
    \begin{align*}
        \begin{array}{|c|c|c|c|c|c|}\hline
		(m,i,j)&\text{index }r&\text{degree }d&\text{variety }V_m^{[i,j]}&&\text{S}\\\hline
		(2,4,4)&8&1/4&\mathbb{P}^4_{1^4,4}&&\text{\xmark}\\\hline
		(2,4,2)&6&1/2&\mathbb{P}^4_{1^4,2}&&\text{\xmark}\\\hline
		(3,3,3)&6&2/3&\mathbb{P}^5_{1^5,3}[2]&&\text{\xmark}\\\hline
		(2,4,1)&5&1&\mathbb{P}^4&&\text{\cmark} \\\hline
		(3,3,2)&5&1&\mathbb{P}^4&&\text{\cmark}\\\hline
		(1,2,2)&4&1/2&\mathbb{P}^5_{1^3,2^2,3}[6]&&\text{\xmark}\\
		(2,2,2)&4&1&\mathbb{P}^5_{1^4,2^2}[4]&&\text{\xmark}\\
		(3,3,1)&4&2&\mathbb{P}^5[2]&Q^4&\text{\cmark}\\
		(4,2,2)&4&2&\mathbb{P}^5[2]&Q^4&\text{\cmark}\\
		(3,2,2)&4&3/2&\mathbb{P}^5_{1^5,2}[3]&&\text{\xmark}\\
		(5,2,2)&4&5/2&\text{  see~\eqref{eqn:v522}}&&\text{\xmark}\\\hline
		(1,2,1)&3&1&\mathbb{P}^5_{1^4,2,3}[6]&\text{FI}_1^4&\text{\cmark}\\
		(2,2,1)&3&2&\mathbb{P}^5_{1^5,2}[4]&\text{FI}_2^4&\text{\cmark}\\
		(3,2,1)&3&3&\mathbb{P}^5[3]&\text{FI}_3^4&\text{\cmark}\\
		(4,2,1)&3&4&\mathbb{P}^6[2,2]&\text{FI}_4^4&\text{\cmark}\\
		(5,2,1)&3&5&X^{2,5}_{\mathcal{O}(1)^{\oplus 2}}&\text{FI}_5^4&\text{\cmark}\\\hline
		(1,1,1)&2& 2&\mathbb{P}^5_{1^5,3}[6]&V^4_{ 2}&\text{\cmark}\\
		(2,1,1)&2& 4&\mathbb{P}^5[4]&V^4_{ 4}&\text{\cmark}\\
		(3,1,1)&2& 6&\mathbb{P}^6[2,3]&V^4_{ 6}&\text{\cmark}\\
		(4,1,1)&2& 8&\mathbb{P}^7[2,2,2]&V^4_{ 8}&\text{\cmark}\\
		(5,1,1)&2&10&X^{2,5}_{\mathcal{O}(1)\oplus\mathcal{O}(2)}&V^4_{10}&\text{\cmark}\\
		(6,1,1)&2&12&X^{2,5}_{\mathcal{E}(1)}&V^4_{12}&\text{\cmark}\\
		(7,1,1)&2&14&X^{2,6}_{\mathcal{O}(1)^{\oplus 4}}&V^4_{14}&\text{\cmark}\\
		(8,1,1)&2&16&X^{3,6}_{\mathcal{O}(1)^{\oplus 2}\oplus\bigwedge^2\mathcal{E}}&V^4_{16}&\text{\cmark}\\
		(9,1,1)&2&18&X^{5,7}_{\mathcal{O}(1)\oplus\bigwedge^4\mathcal{E}}&V^4_{18}&\text{\cmark}\\
		(11,1,1)&2&22&\text{ see~\eqref{eqn:V1111}}&&\text{\xmark}\\\hline
        \end{array}
    \end{align*}
    \caption{Fano varieties $V_m^{[i,j]}$ with Picard rank $1$. The smoothness is indicated in the last column and in the smooth cases we also provide the name of the Fano manifold, following the conventions from~\cite{Coates2018}, in the second to last column.}
    \label{tab:fano4folds}
\end{table}
So far the construction of the geometries $X_{m,s}^{[i,j]}$ for $m=1,\ldots,5$ has always led to the same pattern that we will now use to describe a uniform construction for all $m=1,\ldots,9,11$.

Recall that a variety $V$, which in the following we assume to be of complex dimension $4$, is called Fano if the anti-canonical bundle $-K_V$ is ample.
The index of $V$ is the largest integer $r$ such that $-K_V=rH$ for some Weil divisor $H$ on $V$ and then the degree of $V$ is $d=H^4$~\cite{Parshin2010-pn}~\footnote{Note that some authors define the degree of a Fano fourfold instead as $(-K_V)^4=r^4d$.}.
We list in Table~\ref{tab:fano4folds} the four-dimensional Fano manifolds of Picard rank $1$ and index $r=2,\ldots,5$ from~\cite{Coates2018} together with singular Fano varieties of index $4,6$ and $8$.

All but two of the geometries can be realized directly as complete intersections in weighted projective spaces or Grassmanians.
The construction of the singular Fano fourfold $V_5^{[2,2]}$ of index $r=4$ has been described in Section~\ref{sec:Amodelm5} while that of $V_{11}^{[1,1]}$ with index $r=2$ will be described at the end of the section.

We denote by $V:=V_m^{[i,j]}$ the Fano fourfold of index $r=i+j$ and degree $d=2m/(ij)$ from Table~\ref{tab:fano4folds} and write $\mathcal{O}_V(1):=\mathcal{O}_V(H)$ with $H$ as above.
We denote by $V[d_1,\ldots,d_k]$ the generic complete intersection associated to $\mathcal{O}_{V}(d_1)\oplus\ldots\oplus\mathcal{O}_{V}(d_k)$.
Then $V[i,j]$ is the degree $2m$ K3-surface $\Sigma^m$ and we define $\mathcal{X}_m^{[i,j]}:=V[i+j]$.
For each $(m,i,j)$ in Table~\ref{tab:fano4folds} we see that $\mathcal{X}:=\mathcal{X}_m^{[i,j]}$ is a smooth Calabi-Yau threefold with $\text{\normalfont Pic}(\mathcal{X})\simeq\mathbb{Z}$.
As discussed in~\ref{sec:Amodelm5}, $\mathcal{X}_5^{[2,2]}$ is a double cover of $B_5$ and similarly, the construction of $V_{11}^{[1,1]}$ given below  will show that $\mathcal{X}_{11}^{[1,1]}$ is a double cover of the Fano threefold $A_{22}$.

Given generic sections $p^{(i)}_1,p^{(i-s)}_2,p^{(j+s)}_3,p^{(j)}_4$ with $p_a^{(b)}\in\Gamma\left(V,\mathcal{O}_V(b)\right)$ we define
\begin{align}
    \widetilde{X}:=\{\,p^{(i)}_1p^{(j)}_4-p^{(i-s)}_2p^{(j+s)}_3=0\,\}\subset V\,,
\end{align}
which has $2m(i-s)(j+s)$ isolated nodal singularities at
\begin{align}
    Z:=\{\,p^{(i)}_1=p^{(i-s)}_2=p^{(j+s)}_3=p^{(j)}_4=0\,\}\subset V\,.
\end{align}
For $s=0,\ldots,i-j$ we denote by $X_{m,s}^{[i,j]}$ the blow-up of $\widetilde{X}$ along $p_1^{(i)}=p_2^{(i-s)}=0$ and define $X_m^{[i,j]}:=X_{m,0}^{[i,j]}$.
Note that for $m=2,\ldots,5$ this reproduces the construction from the previous sections.
Again $X_{m}^{[i,j]}$ and $X_{m,i-j}^{[i,j]}$ are related by flopping the exceptional curves.

We can again explicitly construct $X_{m,s}^{[i,j]}$ as a complete intersection in a projective bundle on $V$.
To this end we define the projective bundle $P:=\mathbb{P}\left(\mathcal{O}_V(-s)\oplus\mathcal{O}_V\right)$ with projection $\pi:\,P\rightarrow V$, denote by $\zeta_P$ the relative hyperplane bundle on $P$ and write $\mathcal{O}_P(a,b):=\pi^*\mathcal{O}_V(a)\otimes\zeta_P^{\otimes b}$~\footnote{Again we leave the restriction to the locus where $\mathcal{O}_V(-s)$ is actually a line bundle implicit.}.
Then $X_{m,s}^{[i,j]}$ is the complete intersection associated to the bundle
\begin{align}
    E=\mathcal{O}_P(i-s,1)\oplus\mathcal{O}_P(j,1)\,.
\end{align}
For $m=2$ this coincides with the definition in Section~\ref{sec:Amodelm2} and for $m=5$ we recover the geometries from Section~\ref{sec:Amodelm5}.

We then define the family
\begin{align}
    \widetilde{X}(t)=\{\,p^{(i)}_1p^{(j)}_4-tp^{(i-s)}_2p^{(j+s)}_3=0\,\}\subset V\,,
\end{align}
such that $\widetilde{X}=\widetilde{X}(1)$.
The central fiber takes the form
\begin{align}
    \widetilde{X}(0)=F_m^{[i]}\cup_{\widetilde{X}(0)}F_m^{[j]}\,,
\end{align}
where $F_m^{[i]}$ and $F_m^{[j]}$ are respectively the smooth Fano threefolds that are obtained as the vanishing loci of generic sections of $\mathcal{O}_V(j)$ and $\mathcal{O}_V(i)$.
They intersect in the K3 surface $\Sigma^m=F_m^{[i]}\times_{\widetilde{X}(0)}F_m^{[j]}$.
In order to obtain a smooth total space we first blow up each fiber of the family $\widetilde{X}(t)$ along $p_1^{[i]}=p_2^{[i-s]}=0$ and then along $p_3^{[j+s]}=p_4^{[j]}=0$.
The resulting family $\widehat{X}(t)$ is a Tyurin degeneration.
We denote the restriction of $\mathcal{O}_{F_m^{[i]}}(1)$ to $\Sigma^m$ by $\mathcal{O}_{\Sigma^m}(1)$.
Then the generic fiber of $\widehat{X}(t)$ is $X_{m,s}^{[i,j]}$ while the central fiber takes the form
\begin{align}
    \widehat{X}(0)=\widetilde{F}\cup_{\widehat{X}(0)}F_m^{[j]}\,,
\end{align}
with $\widetilde{F}$ being the blow-up of $F_m^{[i]}$ first along the zero locus of a section of $\mathcal{O}_{\Sigma^m}(i-s)$ and then along the zero locus of a section of $\mathcal{O}_{\Sigma^m}(j+s)$.

\subsection*{The construction of $V_{11}^{[1,1]}$}

The construction of the Fano variety $V_{11}^{[1,1]}$ of index $2$ is similar to that of $V_5^{[2,2]}$ described in Section~\ref{sec:Amodelm5}. In this case we let $W:=\mathbb{P}^{21}$, denote the homogeneous coordinates by $[w:x_1:\ldots:x_{21}]$ and consider $U=W\backslash\{\,x_1=\ldots=x_{21}=0\,\}$ with projection $\pi_U:\,U\rightarrow \mathbb{P}^{20}$.
Then $A_{11}^{[1,1]}$ is the closure
\begin{align}
    V_{11}^{[1,1]}:=\overline{\pi_U^{-1}\left(A_{22}\right)}\subset W\,,
    \label{eqn:V1111}
\end{align}
where we implicitly use the Pl\"ucker embedding of $A_{22}:=X^{3,7}_{(\bigwedge^2\mathcal{S}^{\vee})^{\oplus 3}}\subset\text{Gr}(3,7)\subset\mathbb{P}^{20}$.
This is the projective cone over the smooth Fano threefold $A_{22}$ of genus $\frac12(-K_V)^3+1=12$.

A generic Calabi-Yau double cover of $A_{22}$ then takes the form
\begin{align}
    \mathcal{X}_{11}^{[1,1]}:=\{\,w^2-Q(x_1,\ldots,x_{21})=0\,\}\cap V_{11}^{[1,1]}\subset W\,,
\end{align}
where $Q(x_1,\ldots,x_{21})$ is a generic homogeneous polynomial of degree two.
Note that $\mathcal{X}_{11}^{[1,1]}$ does not intersect the point $\{\,x_1=\ldots=x_{21}=0\,\}\subset W$.
We can then consider the nodal degeneration
\begin{align}
    \widetilde{X}:=\{\,w^2-p_1^{(1)}p_4^{(1)}+p_2^{(1)}p_3^{(1)}=0\,\}\cap V_{11}^{[1,1]}\subset W\,,
\end{align}
of $\mathcal{X}_{11}^{[1,1]}$ where $p_i^{(1)},\,i=1,\ldots,4$ are homogeneous polynomial of degree one in $x_1,\ldots,x_{21}$.
This has 22 nodes at $Z:=\{\,p_1^{(1)}=\ldots=p_4^{(1)}=0\,\}\subset V_{11}^{[1,1]}$.
The small resolution and corresponding Tyurin degeneration can then be constructed as discussed above.

\subsection*{Quantum periods and the quantum Lefschetz principle}

The so-called quantum period of a Fano manifold $V$ with index $r$ was 
introduced in~\cite{Coates}. It takes the form
\begin{align}
    G_V(z)=\sum\limits_{d\ge 0}c_V(d) z^{rd}\,,
\end{align}
in terms of certain genus zero Gromov-Witten invariants $c_V(d)\in\mathbb{Q}$ of $V$.
The quantum period is expected to uniquely determine the Fano manifold and was therefore proposed as a basis for a classification program, see e.g.~\cite{Coates_2022} and references therein.
If the complete intersection associated to a bundle $\mathcal{O}_V(a_1)\oplus\ldots\oplus\mathcal{O}_V(a_k)$ is Calabi-Yau, then the quantum Lefschetz theorem~\cite{Coates2007} implies that
\begin{align}
    f(z)=\sum\limits_{d\ge 0}(a_1 d)!\cdot\ldots\cdot(a_k d)! c_V(d)z^d\,,
\end{align}
is the fundamental period of the mirror.

Let us now consider the case $V=V_m^{[i,j]}$.
Since the complete intersection associated to $\mathcal{O}_V(i)\oplus\mathcal{O}_V(j)$ is the degree $2m$ K3 surface $\Sigma^m$, we can use the quantum Lefschetz theorem~ to identify
\begin{align}
    c_V(d)=\frac{c_m(d)}{(id)!(jd)!}\,,
\end{align}
in terms of the coefficients $c_m(d)$ of the fundamental period of the $M_m$ polarized K3  $\Sigma_m$ mirror to $\Sigma^m$.
We also recall the quantum period of $\mathbb{P}^1$, given by
\begin{align}
    G_{\mathbb{P}^1}(z)=\sum\limits_{k\ge 0}c_{\mathbb{P}^1}(k)z^{2k}\,,\quad\text{with}\quad c_{\mathbb{P}^1}(k)=\frac{1}{(k!)^2}\,.
\end{align}
For the cases with $s=0$ we can then use~\cite[Proposition E.3]{Coates2016} and again the quantum Lefschetz theorem to conclude that the mirror fundamental period of $X_{m}^{[i,j]}=X_{m,0}^{[i,j]}$ takes the form
\begin{align}
    \varpi_m^{[i,j]}(z_1,z_2)=\sum\limits_{d,k\ge 0}c_V(d)c_{\mathbb{P}^1}(k)(k+id)!(k+jd)! \, z_1^dz_2^k\,,
\end{align}
in agreement with~\eqref{eqn:fundamentalPeriod}.

Note that so far we have assumed that $V$ is a Fano manifold and therefore smooth.
This assumption applies in particular to the examples with $m=6,\ldots,9$.
However, the results agree with those from Sections~\ref{sec:Amodelm2},~\ref{sec:Amodelm134} and~\ref{sec:Amodelm5} even in cases where $V_m^{[i,j]}$ is singular, such as $(m,i,j,s)=(5,2,2,0)$.
We therefore expect that the argument extends to $(m,i,j,s)=(11,1,1,0)$ as well.

\subsection{Other geometric realisations}
Many of the geometries that we consider have alternative representations as complete intersections, or even hypersurfaces, in different toric ambient spaces.
Consider the example $(m,i,j,s)=(2,4,1,0)$.
We can construct $X:=X_{2,0}^{[4,1]}$ as a complete intersection $X=\{\,Q_1=Q_2=0\,\}\subset\mathbb{P}^4\times\mathbb{P}^1$ with
\begin{align}
    \begin{split}
    Q_1:=&p_1^{(4)}(u_1,\ldots,u_5)v_1+p_2^{(4)}(u_1,\ldots, u_5)v_2\,,\\
    Q_2:=&p_3^{(1)}(u_1,\ldots, u_5)v_1+p_4^{(1)}(u_1,\ldots, u_5)v_2\,,
    \label{eqn:m2defP2P31}
    \end{split}
\end{align}
where $[u_1:\ldots:u_5]$ and $[v_1:v_2]$ are now respectively homogeneous coordinates on $\mathbb{P}^4$ and $\mathbb{P}^1$.
Compared to the discussion in the previous section, we have so far just performed the trivial modification of eliminating the weight $4$ coordinate on $P$ using the equation $P^{(4)}=0$. This model has a conifold transition to the quintic, which was analyzed in \cite{Greene:1995hu} (see also \cite{klemm:2004km,Chialva:2007sv,Grimm:2008ed}).

However, we can also represent $X$ as a hypersurface in a toric fourfold.
To see this we first note that by changing coordinates on $\mathbb{P}^4$, we can  assume that $p_3^{(1)}(u_1,\ldots, u_5)=u_4$ and $p_4^{(1)}(u_1,\ldots, u_5)=u_5$.
Following the discussion in the previous section, we then observe that $X$ is the blow-up of the nodal quintic $\widetilde{X}:=\{\,u_5p_1^{(4)}(u_1,\ldots,u_5)-u_4p_2^{(4)}(u_1,\ldots, u_5)\,\}\subset \mathbb{P}^4$ along the divisor $\{\,u_4=u_5=0\,\}\subset \widetilde{X}$.
But this can equally be realized by performing a toric blow-up of $\mathbb{P}^4$ along $\{\,u_4=u_5=0\,\}\subset\mathbb{P}^4$ and taking the proper transform of $\widetilde{X}$.
The aforementioned blow-up of $\mathbb{P}^4$ is the toric variety
\begin{align}
    P''=\IP\left(\tiny\begin{array}{cccccc}
            \BM 0&0&0&1&1&-1\\
            1&1&1&0&0&1\EM
            \end{array}\right)=\mathbb{P}\left(\mathcal{O}_{\mathbb{P}^1}^{\oplus 3}\oplus\mathcal{O}_{\mathbb{P}^1}(-1)\right)\,,
\end{align}
and the proper transform of $\tilde{X}$ along $\pi:\,P''\rightarrow\mathbb{P}^3$ is the anti-canonical hypersurface
\begin{align}
    X=\{\,u_5p_1^{(4)}(u_1,u_2,u_3,eu_4,eu_5)-u_4p_2^{(4)}(u_1,u_1,u_3,eu_4,e u_5)=0\,\}\subset P''\,,
    \label{eqn:Xhypersurface}
\end{align}
where we use homogeneous coordinates $[u_1:u_2:u_3:u_4:u_5:e]$ on $P''$.

Let us now compare the mirror fundamental periods.
From the realisation as a complete intersection in $P'$ we obtain the expression~\eqref{eqn:bbmirrorperiod}
\begin{align}
    \begin{split}
    \varpi_0(z_1,z_2)=&\sum\limits_{d,k\ge 0}\frac{(d+k)!(4d+k)!}{(d!)^5(k!)^2}z_1^dz_2^k=\sum\limits_{d\ge 0}\frac{(4d)!}{(d!)^4}{_2}F_1(1+d,1+4d;1;z_2)z_1^d\,.
    \end{split}
\end{align}
On the other hand, the realisation as a toric hypersurface~\eqref{eqn:Xhypersurface} leads to the mirror fundamental period
\begin{align}
\label{X52mirror}
    \varpi_0'(w_1,w_2)=&\sum\limits_{d,k\ge 0}\frac{(4d+k)!}{(d!)^3(d-k)!(k!)^2}w_1^dw_2^k=\sum\limits_{d\ge 0}\frac{(4d)!}{(d!)^4}{_2}F_1(1+4d,-d;1;-w_2)w_1^d\,.
\end{align}
Using the Pfaff transform
\begin{align}
    {_2}F_1(a,b;c;z)=(1-z)^{-b}{_2}F_1\left(b,c-a;c;\frac{z}{z-1}\right)\,,
\end{align}
we find that the two fundamental periods are related by 
\begin{align}
    \varpi_0(z_1,z_2)=\frac{1}{1-z_2}\varpi_0'\left(\frac{z_1}{(1-z_2)^{4}},\frac{z_2}{1-z_2}\right)\,.
\end{align}
Since the overall factor can be absorbed by a K\"ahler transformation, or a rescaling of the holomorphic 3-form, we find that the two constructions 
are related by a birational transformation of the complex structure moduli, 
\begin{align}
\label{ztow}
    w_1=\frac{z_1}{(1-z_2)^{4}}\,,\quad w_2=\frac{z_2}{1-z_2}\,.
\end{align}
Moreover, defining $z=w_1 w_2$ and shifting the summation variable $d\mapsto d+k$ in \eqref{X52mirror}, we get
\begin{align}
    \varpi_0'(w_1,w_2)=&\sum\limits_{d,k\ge 0}\frac{(4d+5k)!}{ d! (k!)^2 ((d+k)!)^3} w_1^d z^k\ ,
\end{align}
which reduces to the fundamental period of the mirror quintic when $w_1\to 0$ keeping $z$ fixed \cite{Chialva:2007sv}.

Recall that the limit $z_1,z_2\rightarrow 0$ is mirror to the large volume limit of $X_{2,0}^{[4,1]}$ and this just corresponds to the limit $w_1,w_2\rightarrow 0$ in the coordinates \eqref{ztow}.
However, the large volume limit associated to the flopped geometry $X_{2,3}^{[4,1]}$ is mirror to the limit $z_2\rightarrow \infty$ while $z_1z_2\rightarrow 0$.
In terms of $w_1,w_2$ this corresponds to the limit $w_1\rightarrow 0,\,w_2\rightarrow -1$.
From the perspective of the realisation of $X$ as a toric hypersurface, the flopped geometry is thus hidden at a phase boundary.
We therefore find concrete geometric realisations of the so-called ``non-toric'' flops that  have recently been discussed in the context of Calabi-Yau threefolds that are hypersurfaces in toric ambient spaces e.g. in~\cite{Gendler:2022ztv}.
Let us also note that from a GLSM perspective, the phenomenon that large volume limits can be realized both at phase limits and at phase boundaries, depending on the underlying toric realisation, is of course well known and discussed e.g. in~\cite{Greene:1996dh,Addington:2013gpa,Aspinwall:2014vea}.

The geometries $X^{[i,j]}_m$ with $m=1,2,3$ have also appeared in~\cite{klemm:2004km}, albeit as hypersurfaces and/or complete intersections in different toric ambient spaces, see~\cite[Table 1]{klemm:2004km}.
The CY threefolds that admit a realization as a complete intersection in a product of projective spaces (so called CICYs classified in~\cite{Candelas:1987kf,Green:1987cr}) are listed in Table~\ref{tab:alternative}.

\section{Modularity of $\langle 2m\rangle$-polarized K3 fibrations}
\label{sec:K3mod}

In the previous sections we have constructed mirror pairs $Y_{m,s}^{[i,j]},X_{m,s}^{[i,j]}$ of Calabi-Yau threefolds where $Y_{m,s}^{[i,j]}$ exhibits an $M_m$-polarized K3 fibration structure.
When $s=0$ we showed that $X_m^{[i,j]}:=X_{m,0}^{[i,j]}$ itself admits a fibration by $\langle 2m\rangle$-polarized K3 surfaces.

In order to unburden the notation we will assume that $m,[i,j]$ are fixed and write $Y=Y_m^{[i,j]}$ and $X=X_m^{[i,j]}$.
Let us denote again by $\gamma_i,\,i=1,2$ the divisors that form a basis of a K\"ahler cone on $X$ and that have been introduced in~\eqref{gentopdata}.
Then $\gamma_2=[\Sigma]$ is the class of a generic fiber $\Sigma$ and $\gamma_1$ induces the polarization, meaning that the restriction $L_{\Sigma}:=L\big|_\Sigma\in\text{Pic}(\Sigma)$ of the line bundle $L$ associated to $\gamma_1$ generates $\text{Pic}(\Sigma)$ and satisfies $\langle L_{\Sigma},L_{\Sigma}\rangle=2m$.
We have $\kappa_{112}=2m$, $\kappa_{122}=\kappa_{222}=0$ and $c_{2,2}=24$ and for brevity we define
\begin{align}
    \kappa:=\kappa_{111}\,,\quad c_2:=c_{2,1}\,.
\end{align}
Note that any CY3 with a triple intersection and second Chern class of the form above (in some basis) admits a K3-fibration~\cite{oguiso1993algebraic}. 

We also choose a dual basis of curves $\gamma^a,\,a=1,2$ with $\gamma_a\cap \gamma^b=\delta_a^b$.
This implies that $\gamma^1$ is an isolated curve in one of the K3 fibers while $\gamma^2$ is any of $4 ij$ holomorphic sections and therefore an embedding of the $\mathbb{P}^1$ base into the fibration.
In order to make contact with the string theory literature, we introduce
\begin{align}
    T:=t^1\,,\quad S:=t^2\,,
\end{align}
such that $T,S$ are respectively the complexified volumes of $\gamma^1$ and $\gamma^2$.
As predicted by the DHT-conjecture, the large base limit $S\rightarrow {\rm i}\infty$ is mirror to the Tyurin degeneration of $Y_{m}^{[i,j]}=Y_{m,0}^{[i,j]}$ at $z_2\rightarrow 0$ that has been discussed below~\eqref{eqn:lambdavtozero}.
Recall that in this limit $Y_m^{[i,j]}$ degenerates into a union of two Fano threefolds that intersect in an $M_m$-polarized K3 $\Sigma_m$.

Mirror symmetry implies that the stringy K\"ahler moduli space of a $\langle 2m\rangle$-polarized K3 surface is the modular curve $X_0(m)^+=\mathbb{H}/\Gamma_0(m)^+$.
The complex structure modulus of the $M_m$-polarized K3 fibers in $Y$ varies over the base $\mathbb{P}^1$ and is only indirectly related to the complex structure moduli of the Calabi-Yau threefold $Y$.
However, in the limit $z_2\rightarrow 0$ the modulus $z_1$ can be interpreted is the complex structure parameter of the central $\Sigma_m$ where the two Fano components of $Y$ intersect.
On the other hand, the complexified K\"ahler modulus of the generic $\langle 2m\rangle$-polarized K3 fiber in $X$ directly corresponds to a complexified K\"ahler modulus of $X$  itself.
As a consequence, the modular curve $X_0(m)^+$ can be embedded close to the boundary in the large base limit of the stringy K\"ahler modulus of $X$.
The $\Gamma_0(m)^+$ monodromies in the stringy K\"ahler moduli space of the fiber lift to monodromies in the moduli space of the fibration and lead to the well known modular properties of the topological string A-model on $X$.

\subsection{Integral periods and topological B-branes}
We first construct an integral symplectic period basis by using homological mirror symmetry, following~\cite{Gerhardus:2016iot,Cota:2019cjx}.

The periods of the holomorphic 3-form $\Omega$ on $Y$,
\begin{align}
    \Pi(L)=\int_L\Omega\,,\quad L\in H_3(Y,\mathbb{Z})\,,
    \label{eqn:periodsBmodel}
\end{align}
can be interpreted as central charges of topological A-branes that are supported on $L$ and are identified under homological mirror symmetry with central charges of topological B-branes on $X$.
The topological B-branes can be represented by complexes of coherent sheaves
\begin{align}
    \mathcal{F}^\bullet=\left(\ldots\rightarrow \mathcal{F}_{-1}\rightarrow \mathcal{F}_0\rightarrow\mathcal{F}_1\rightarrow \ldots\right)\in D^b(X)\,,
\end{align}
in the bounded derived category $D^b(X)$.
The central charges are multi-valued functions of the complexified K\"ahler moduli and the asymptotic behaviour near the large volume limit $T,S\rightarrow i\infty$ is determined 
by the Gamma class formula~\cite{iritani2009integral} (see \cite{Iritani:2023ngp} for a recent review)
 \begin{align}
     \Pi(\mathcal{F}^\bullet)=\int_X e^{\omega}\Gamma_{\mathbb{C}}(X)\left(\text{ch}\,\mathcal{F}^\bullet\right)^\vee+\mathcal{O}(e^{2\pi i T},e^{2\pi i S})\,,
 \end{align}
 with the terms in the Gamma class that are relevant for Calabi-Yau threefolds given by
 \begin{align}
    \Gamma_{\mathbb{C}}(X)=1+\frac{1}{24}c_2(TX)+\frac{\zeta(3)}{(2\pi i )^3}c_3(TX)+\ldots\,.
 \end{align}
The open string index, that in the B-model can be identified with the intersection form on 3-cycles, is given by
\begin{align}
\label{openindex}
    \chi(\mathcal{F}^\bullet,\mathcal{E}^\bullet):=\int_X\text{Td}(X)\text{ch}(\mathcal{E}^\bullet)^{\vee}\text{ch}(\mathcal{F}^\bullet)\,.
\end{align}
 
 A set of branes that generates the charge lattice is given by
 \begin{align}
  \begin{split}
    \mathcal{F}_{(6)}^\bullet=&0\rightarrow(\mathcal{O}_X)_0\rightarrow 0\,,\\
    \mathcal{F}_{(4),a}^\bullet=&0\rightarrow \mathcal{O}_X(-\gamma_a)_{-1}\rightarrow (\mathcal{O}_X)_0\rightarrow 0\,,\quad a=1,2\,,\\
    \mathcal{F}_{(2),a}^\bullet=&\iota_{!}\mathcal{O}_{\gamma^a}(K_{\gamma^a}^{1/2})\,,\quad a=1,2\,,\\
    \mathcal{F}_{(0)}=&0\rightarrow (\mathcal{O}_p)_0\rightarrow 0\,,
  \end{split}
 \end{align}
 where $\iota:\,\gamma^a\rightarrow X$ is the inclusion and $\mathcal{O}_p$ is the skyscraper sheaf supported on a generic point $p\in X$.
Note that $\mathcal{F}_{(6)}^\bullet$ is a D6-brane that wraps the Calabi-Yau while $\mathcal{F}_{(4),a}^\bullet$ are D4-branes supported on the divisors $\gamma_a,\,a=1,2$, $\mathcal{F}_{(2),a}^\bullet$ are D2-branes supported on the curves $\gamma^a,\,a=1,2$ and $\mathcal{F}_{(0)}$ is a D0-brane supported on the point $p$.

Using $\mathcal{F}_{i=1,\ldots,6}=\left([\mathcal{F}_{(0)}^\bullet],[\mathcal{F}_{(2),1}^\bullet],[\mathcal{F}_{(2),2}^\bullet],[\mathcal{F}_{(4),1}^\bullet],[\mathcal{F}_{(4),2}^\bullet],[\mathcal{F}_{(6)}^\bullet]\right)$, where $[\mathcal{F}^\bullet]\in K(X)$ denotes the K-theory class associated to a brane $\mathcal{F}^\bullet\in D^b(X)$, the dual Chern characters are given by
\begin{align}
\begin{split}
    \text{ch}(\mathcal{F}_{i=1,\ldots,6})^\vee=\left(-\gamma_2\gamma^2,\,\gamma^1,\,\gamma^2,\,-\gamma_1-\frac12(\gamma_1)^2-\frac16(\gamma_1)^3,\,-\gamma_2,\,1\right)\,,
    \end{split}
\end{align}
where we also use $\gamma_a,\gamma^b$ to refer to the Poincar\'{e} dual cohomology classes.
In order to obtain a symplectic intersection form we will actually choose a slightly different basis
\begin{align}
    \widetilde{\mathcal{F}}_{i=1,\ldots,6}=\left(-\mathcal{F}_1,\,\mathcal{F}_2,\,\mathcal{F}_3,\,\rho\mathcal{F}_1-\kappa\mathcal{F}_2-m\mathcal{F}_3-\mathcal{F}_4,\,\mathcal{F}_1-\mathcal{F}_5,\,\mathcal{F}_3-\mathcal{F}_6\right)\,,
\end{align}
with $\rho:=\frac{c_2+2\kappa}{12}\in\mathbb{N}$.
The normalized central charges in this basis take the form
\begin{align}
\label{vecPiLeading}
    \vec{\Pi}(T,S):=\frac{\Pi(\widetilde{\mathcal{F}}_{i=1,\ldots,6})}{\Pi(\widetilde{\mathcal{F}}_1)}=\left(\begin{array}{c}
        1\\[.2em]
        T\\[.2em]
        S\\[.2em]
        \frac12\kappa T^2+2mTS-\frac{\kappa}{2}T-\frac{c_2}{24}+\mathcal{O}(e^{2\pi i T},e^{2\pi i S})\\[.2em]
        mT^2+\mathcal{O}(e^{2\pi i T},e^{2\pi i S})\\[.2em]
        -\frac16\kappa T^3-mT^2S-\frac{c_2}{24}T-\frac{\zeta(3)}{(2\pi i )^3}\chi+\mathcal{O}(e^{2\pi i T},e^{2\pi i S})
    \end{array}\right)\,.
\end{align}
The subleading terms can be obtained by solving the Picard-Fuchs system associated to the fundamental period~\eqref{eqn:fundamentalPeriod}.

Recalling that for a CY threefold, $\text{Td}(X)=1+\frac{1}{12}c_2(TX)$,
the open string index \eqref{openindex} in this basis is 
\begin{align}
    \chi(\widetilde{\mathcal{F}}_i,\widetilde{\mathcal{F}}_j)=\left(\begin{array}{cccccc}
        0&0&0&0&0&1\\
        0&0&0&1&0&0\\
        0&0&0&0&1&0\\
        0&-1&0&0&0&0\\
        0&0&-1&0&0&0\\
        -1&0&0&0&0&0
    \end{array}\right)\,.
\end{align}

\subsection{Monodromies and Fourier-Mukai transforms}
The periods~\eqref{eqn:periodsBmodel} experience monodromies along loops in the complex structure moduli space of $Y$.
Under homological mirror symmetry these lift to autoequivalences of the category of topological B-branes on $X$ that are associated to loops in the corresponding stringy K\"ahler moduli space.
By a result from Orlov~\cite{Orlov1997} every autoequivalence of $D^b(X)$ can be represented by a Fourier-Mukai transform
\begin{align}
    \Phi_{\mathcal{E}^\bullet}:\,D^b(X)\rightarrow D^b(X)\,,\quad \mathcal{F}^{\bullet}\mapsto {\rm R}p_{1*}\left(\mathcal{E}^\bullet \otimes_{\rm L} p_2^*\mathcal{F}^\bullet\right)\,,
    \label{eqn:fmtrafo}
\end{align}
where $p_1,p_2$ are the two projections from $X\times X$ to $X$ and $\mathcal{E}^\bullet\in D^b(X\times X)$ is the so-called Fourier-Mukai kernel that determines the map.

For example, the large volume monodromies around $T\rightarrow T+1$ and $S\rightarrow S+1$ respectively correspond to the Fourier-Mukai kernels $j_*\mathcal{O}_X(-\gamma_1)$ and $j_*\mathcal{O}_X(-\gamma_2)$, where $j:\,\Delta\rightarrow X\times X$ is the embedding of the diagonal $\Delta\simeq X$.
The corresponding image of a brane $\mathcal{F}^{\bullet}\in D^b(X)$ is respectively given by $\mathcal{F}^{\bullet}\otimes \mathcal{O}_X(-\gamma_i)$, $i=1,2$ and the action on the central charges~\eqref{vecPiLeading} takes the form
\begin{align}
    \vec{\Pi}(T+1,S)=M_T\,\vec{\Pi}(T,S)\,,\quad \vec{\Pi}(T,S+1)=M_S\,\vec{\Pi}(T,S)\,,
\end{align}
with the large volume monodromy matrices $M_T,M_S$ defined as
\begin{align}
\label{MTMS}
    M_T=\left(
\begin{array}{cccccc}
 1 & 0 & 0 & 0 & 0 & 0 \\
 1 & 1 & 0 & 0 & 0 & 0 \\
 0 & 0 & 1 & 0 & 0 & 0 \\
 0 & \kappa  & 2 m & 1 & 0 & 0 \\
 m & 2 m & 0 & 0 & 1 & 0 \\
 -\rho & -\kappa  & -m & -1 & 0 & 1 \\
\end{array}
\right)\,,\quad
    M_S=\left(
\begin{array}{cccccc}
 1 & 0 & 0 & 0 & 0 & 0 \\
 0 & 1 & 0 & 0 & 0 & 0 \\
 1 & 0 & 1 & 0 & 0 & 0 \\
 0 & 2 m & 0 & 1 & 0 & 0 \\
 0 & 0 & 0 & 0 & 1 & 0 \\
 0 & 0 & 0 & 0 & -1 & 1 \\
\end{array}
\right)\,.
\end{align}

The second type of monodromy that is generically present in the stringy K\"ahler moduli space of a CY threefold corresponds to a loop around a generic point on the principal component of the discriminant $\{\,\Delta_m=0\,\}$ where the 6-brane $\mathcal{O}_X$ becomes massless.
Mathematically, this amounts to a spherical twist, also called Seidel-Thomas twist, corresponding to the spherical object $\mathcal{O}_X$.
In the string theory literature this is usually referred to as the conifold monodromy.
The corresponding Fourier-Mukai kernel is the ideal sheaf $\mathcal{I}_{\Delta}\in D^b(X\times X)$ of $\Delta$.
It acts on the brane charges as
\begin{align}
    [\mathcal{F}^\bullet]\mapsto [\mathcal{F}^\bullet]- \chi\left([\mathcal{F}_{(6)}],[\mathcal{F}^\bullet]\right)\,[\mathcal{F}_{(6)}]\,,
\end{align}
and the action on the central charges~\eqref{vecPiLeading} takes the form $\vec{\Pi}\mapsto M_C\vec{\Pi}$ with the matrix $M_C$ given by
\begin{align}
    M_C=\left(\begin{array}{cccccc}
     1 & 0 & 1 & 0 & 0 & -1 \\
 0 & 1 & 0 & 0 & 0 & 0 \\
 0 & 0 & 1 & 0 & 0 & 0 \\
 0 & 0 & 0 & 1 & 0 & 0 \\
 0 & 0 & 1 & 0 & 1 & -1 \\
 0 & 0 & 0 & 0 & 0 & 1 \\
        \end{array}\right)\,.
\end{align}

Using the K3 fibration $\pi:\,X\rightarrow B$ over $B=\mathbb{P}^1$ we can also consider the ideal sheaf $\mathcal{I}_{\Delta_B}$ of the relative diagonal $\Delta_B\subset X\times_B X$.
Although we are not aware of a proof of this fact, it was observed in~\cite[Section 3]{Cota:2019cjx} that the Fourier-Mukai transform $\Phi_{\mathcal{I}_{\Delta_B}}$ associated to $\mathcal{I}_{\Delta_B}$ generically appears to be an autoequivalence as well and was referred to as the `relative conifold monodromy'.
As was explained in~\cite[Section 3.4]{Cota:2019cjx}, the corresponding matrix $M_U$ acting on~\eqref{vecPiLeading} can be expressed as
\begin{align}
\label{MU}
    M_U=M_S^{-1}M_CM_S^{-1}M_C M_S^2=\left(
\begin{array}{cccccc}
 0 & 0 & 0 & 0 & -1 & 0 \\
 0 & 1 & 0 & 0 & 0 & 0 \\
 -1 & 0 & 0 & 0 & -1 & 1 \\
 0 & 0 & 0 & 1 & 0 & 0 \\
 -1 & 0 & 0 & 0 & 0 & 0 \\
 1 & 0 & 1 & 0 & 1 & 0 \\
\end{array}
\right)\,.
\end{align}
Therefore, at least on the level of brane charges, the action of $\Phi_{\mathcal{I}_{\Delta_B}}$ can be identified with a composition of twists by line bundles and Seidel-Thomas twists.

As we will discuss in Section~\ref{sec_hidegree}, following~\cite{Henningson:1996jf,Berglund:1997eb}, the modular properties of the topological string free energies can be understood in terms of their transformation under the monodromies $M_U$, $M_S$ and $M_T$.

\subsection{Yukawa couplings and genus zero free energies}
\label{sec_Yukawa}

We will now derive the modular properties of the base degree zero contribution to the genus zero free energies from the generic behaviour of the Yukawa couplings.
The algebraic Yukawa couplings in algebraic coordinates $z_1,z_2$ are defined by~\footnote{We use lower indices for the algebraic coordinates $z_a$, but upper indices for the flat coordinates $t^a$.}
\begin{align}
    c_{abc}(z_1,z_2):=\int_Y \Omega\, \frac{\partial^3\Omega}{\partial z_a\partial z_b\partial z_c}\,,\quad a,b,c=1,2\,.
\end{align}
They can be fixed using the Picard-Fuchs system and the technique described e.g. in~\cite[Section 5.3]{Haghighat:2015qdq}.
In the large base limit $z_2\to 0$, we find that their leading behaviour is given by
\begin{align}
	\begin{split}
		c_{111}(z_1,z_2)=&\frac{2m}{z_1^3\Delta_m(1/z_1)}c_{m}^{[i,j]}(1/z_1)+\mathcal{O}(z_2)\,,\quad 
  c_{112}(z_1,z_2)=\frac{2m}{z_1^2z_2\Delta_m(1/z_1)}+\mathcal{O}(z_2^0)\,,
	\end{split}
\end{align}
where in terms of $\lambda=1/z_1$ one has
\begin{align}
\begin{split}
    \Delta_m(\lambda) =& \prod\limits_{k=1}^r(1-\lambda_k/\lambda)\,,\\
    c_{m}^{[i,j]}(\lambda)=&\frac{i+j}{i j}+2\sum\limits_{k=1}^r \frac{\lambda_k}{\lambda-\lambda_k}\,,
\end{split}
\end{align}
with $\lambda_1,\ldots,\lambda_r$ being the 2-orbifold points listed in Table~\ref{TableX0m}.
The components $c_{122}$ and $c_{222}$ will only contribute at order $\mathcal{O}(e^{2\pi i t^2})$ and can be neglected.

In flat coordinates $t^1=T,t^2=S$, the Yukawa couplings take the form
\begin{align}
    C_{ABC}(T,S):=(\varpi_0)^{-2}\frac{\partial z_a}{\partial t^A}\frac{\partial z_b}{\partial t^B}\frac{\partial z_c}{\partial t^C}
    c_{abc}(z_1,z_2)\big|_{z_1\rightarrow z_1(T,S),\,z_2\rightarrow z_2(T,S)}\,,
\end{align}
and are equal to the third derivative of the genus zero topological string free energy $F^{(0)}(T,S)$,
\begin{align}
    C_{ABC}(T,S)=\partial_A\partial_B\partial_C F^{(0)}(T,S)\,.
    \label{eqn:yukawaFreeEnergy}
\end{align}
Applying the Frobenius method to~\eqref{eqn:fundamentalPeriod} one obtains
\begin{align}
    2\pi \I T =\frac{f_m^{(1)}(\lambda)}{f_m(\lambda)}+\mathcal{O}(z_2)\,,\quad 
    2\pi \I S =\log{z_2}+
    \frac{\tilde{f}_m^{[i]} (\lambda)+\tilde{f}_m^{[j]} (\lambda)}
    {f_m(\lambda)}
   +\mathcal{O}(z_2)\,,
    \label{TSv}
\end{align}
where $f_m(\lambda)$ and $f_m^{(1)}(\lambda)$ are the fundamental and logarithmic periods of the K3-fiber defined in~\eqref{deffm} and~\eqref{deffm12},
and $\tilde{f}_m^{[i]}(\lambda)$ is the variant defined in \eqref{defftilde}. 
Setting $H_m^{[i,j]}(\tau) :=H_m^{[i]}(\lambda(\tau))+H_m^{[j]}(\lambda(\tau))$
where $H_m^{[i]}(\lambda)$ was defined in \eqref{defhtilde}, 
the mirror map in the large base limit \eqref{TSv} becomes 
\be
2\pi\I T = \tau+\cO(z_2), \quad 
2\pi \I S = \log{z_2}+ H_m^{[i,j]}(\lambda)+\cO(z_2).
\ee
Moreover, one has 
\be
z_1=J_m^+(T)^{-1}+\mathcal{O}(z_2)\ ,\quad 
\varpi_0(z_1,z_2)=f_m(1/z_1)+\mathcal{O}(z_2)\,,
\ee
and it follows from~\eqref{TSv} that
\begin{align}
    \partial_T z_2=-z_2\, \partial_T H_{m}^{[i,j]}(\lambda(T))+\mathcal{O}(z_2^2)\,.
\end{align}
We also need the relation
\begin{align}
    (\partial_T\log J_m^+(T))^2=f_m(T)^2\Delta_m(J_m^+(T))\,.
\end{align}
consistent with the fact that 2-orbifold points coincide with roots of $\Delta_m(\lambda)$.
Focusing only on the constant term of $C_{TTT}$ in $q_S=e^{2\pi i S}$, one then has
\begin{align}
    \begin{split}
    C_{TTT}=&(\varpi_0)^{-2}\left[\left(\frac{\partial z_1}{\partial T}\right)^3c_{111}
    +3\left(\frac{\partial z_1}{\partial T}\right)^2\frac{\partial z_2}{\partial T}c_{112}+\mathcal{O}(q_S)\right]\\
    =&\frac{2m (J_m^+(T))^3}{f_m^{2}\Delta_m(J_m^+(T))}\left(\frac{\partial z_1}{\partial T}\right)^2\left(\left(\frac{\partial z_1}{\partial T}\right)c_m^{[i,j]}(J_m^+(T)))-\partial_T H_{m}^{[i,j]}(\lambda(T))\right)+\mathcal{O}(q_S)\\
    =&2mJ_m^+(T)\left(\left(\frac{\partial z_1}{\partial T}\right)c_m^{[i,j]}(J_m^+(T))-3\partial_T H_{m}^{[i,j]}(\lambda(T))\right)+\mathcal{O}(q_S)\\
    =&-2m\, c_m^{[i,j]}(J_m^+(T))\, \partial_T\log J_m^
    +(T) -6m\, \partial_T H_{m}^{[i,j]}(\lambda(T))+\mathcal{O}(q_S)\,.
    \end{split}
    \label{CTTT}
\end{align}
Integrating once, we get
\be
\label{d2WTh}
\partial_T^2 F^{(0)} = -2m \left[ \frac{1}{i}+\frac{1}{j} -2r \right] \log J_m^+
-4m \sum_{k=1}^r \log( J_m^+ -\lambda_k) 
- 6m  \, H_m^{[i,j]}
\ee
which determines  the leading behaviour of the genus zero free energy $F^{(0)}$, up to linear and constant terms in $T$. As we shall see in \S\ref{sec_hidegree}, the fifth derivative  $C_{TTTTT}:=\partial_T^5 F^{(0)}$ is a meromorphic modular form of weight 6 under $\Gamma_0(m)^+$. As a consequence of the second term on the r.h.s. of \eqref{d2WTh}, it admits a third order pole at the 2-orbifold points $\tau_k$, with coefficient
\be
C_{TTTTT}  \sim -\frac{8m}{(\tau-\tau_k)^3}\,.
\ee

\subsection{Genus one free energy}
In the so-called holomorphic limit, the genus one topological string free energy on $X_m^{[i,j]}$ takes the form
\begin{align}
    \begin{split}
        F^{(1)}=&-\frac12\left(5-\frac{\chi(X_m^{[i,j]})}{12}\right)\log\,\varpi_0-\frac12\log\text{det}\,\frac{\partial t^b}{\partial z_a}\\
        &-\frac{1}{24}(12+c_2(X_m^{[i,j]}))\log{z_1}-\frac32\log{z_2}-\frac{1}{12}\log\,\Delta_{m}^{[i,j]}-\frac{a_m^{[i,j]}}{12}\log(1-z_2)\,,
    \end{split}
\end{align}
where $a_m^{[i,j]}$ are given in an accompanying Mathematica worksheet. Using the leading behaviour
\begin{align}
    \begin{split}
    \text{det}\,\frac{\partial t^b}{\partial z_a}=&\frac{1}{z_2}\frac{\partial T}{\partial z_1}+\mathcal{O}(z_2^0)=-\frac{1}{z_2}\lambda^2\frac{\partial T}{\partial\lambda}+\mathcal{O}(z_2^0)\,,
    \end{split}
\end{align}
as well as~\eqref{dJsq}, one obtains
\begin{align}
    \begin{split}
        F^{(1)}=&\left(\frac{\chi(X_m^{[i,j]})}{24}-2\right)\log\,f_m+\frac{c_2(X_m^{[i,j]})}{24}\log J_m^+(T)+\frac{1}{12}\log\,\Delta_m(J_m^+(T))+H_m^{[i,j]}-S+\mathcal{O}(q_S)\,.
    \end{split}
    \label{F1univ}
\end{align}
We note that the coefficient of $\log f_m$ is equal to $-\frac{b_{\rm grav}}{24}$ where $b_{\rm grav}=48-\chi_X$ is 
the coefficient of the conformal anomaly  
\cite[(22)]{Kaplunovsky:1995tm}. This ensures that 
$\Re(F^{(1)})- \frac{b_{\rm grav}}{12}\log T_2$ is modular invariant, after expressing
$S$ in terms of the invariant heterotic dilaton, as we discuss in \S\ref{sec_hidegree}. 
Using \eqref{ZagierGoly} we can also express the result as 
\begin{align}
    \begin{split}
        F^{(1)}=&\left(\frac{\chi_X-48}{12}\right)\log[\eta(T)\eta(mT)]
        +\left[\frac{m+1}{12} \left(\frac{\chi_X-48}{24}\right) + 
        \frac{c_2-2r}{24} \right]
        \log J_m^+(T) \\
        &+\frac{1}{12}\log\,\Delta_m(J_+^m(T))+H_m^{[i,j]}-S+\mathcal{O}(q_S)\,.
    \end{split}
    \label{F1univ2}
\end{align}
The resulting genus one free energies and conformal anomaly coefficients are summarized in Table \ref{tabgenus1}. 

\begin{table}
\be
\begin{array}{|c|c|c|l|c|c|}
\hline
m & i & j & F^{(1)} - H_m^{[i,j]} + S & 48-\chi_X  &  \# T  \\ 
\hline
1 & 1 & 1 & \frac1{12} \log(J-1728) -50  \log \eta(\tau) & 300 & 300 \\
1 & 1 & 2 & \frac1{12} \log(J-1728) -\frac{122}{3}  \log \eta(\tau) -\frac{1}{36} \log J  & 244 & 240 \\
1 & 2 & 2 &  \frac1{12} \log(J-1728) -\frac{94}{3}  \log \eta(\tau) -\frac{1}{18} \log J & 188 & 180\\
\hline
2 & 1 & 1 &  \frac1{12} \log(J_2^+-256) -18  \log \eta(\tau)\eta(2\tau) & 216 & 324 \\
2 & 1 & 2 &  \frac1{12} \log(J_2^+-256) -\frac{47}{3}  \log \eta(\tau)\eta(2\tau) -\frac{1}{24}\log J_2^+
& 188 & 276 \\
2 & 2& 2 &  \frac1{12} \log(J_2^+-256) -\frac{40}{3}  \log \eta(\tau)\eta(2\tau) -\frac{1}{12}\log J_2^+
& 160 & 228 
\\
2 & 1& 4 &  \frac1{12} \log(J_2^+-256) -18  \log \eta(\tau)\eta(2\tau) -\frac{1}{4}\log J_2^+
& 216 & 288 \\
2 & 2& 4 &  \frac1{12} \log(J_2^+-256) -\frac{47}{3}  \log \eta(\tau)\eta(2\tau) -\frac{7}{24}\log J_2^+
& 188 & 240 \\
2 & 4 & 4 &  \frac1{12} \log(J_2^+-256) -18  \log \eta(\tau)\eta(2\tau) -\frac{1}{2}\log J_2^+
& 216 & 252 \\
\hline
3 & 1 & 1 &  \frac1{12} \log(J_3^+-108) -15  \log \eta(\tau)\eta(3\tau) -\frac{1}{12}\log J_3^+
& 180 & 348\\
3 & 1 & 2 &  \frac1{12} \log(J_3^+-108) -14  \log \eta(\tau)\eta(3\tau) -\frac{1}{6}\log J_3^+
& 168 & 312 \\
3 & 1 & 3 &  \frac1{12} \log(J_3^+-108) -\frac{47}{3}  \log \eta(\tau)\eta(3\tau) -\frac{13}{36}\log J_3^+
& 188 & 324 \\
3 & 2 & 2 &  \frac1{12} \log(J_3^+-108) -13  \log \eta(\tau)\eta(3\tau) -\frac{1}{4}\log J_3^+
& 156 & 276
\\
3 & 2 & 3 &  \frac1{12} \log(J_3^+-108) -\frac{44}{3} \log \eta(\tau)\eta(3\tau) -\frac{4}{9}\log J_3^+
&176 & 288 \\
3 & 3 & 3 &  \frac1{12} \log(J_3^+-108) -\frac{49}{3}  \log \eta(\tau)\eta(3\tau) -\frac{23}{36}\log J_3^+
&196 & 300 \\
\hline
4 & 1 & 1 &  \frac1{12} \log(J_4^+-64) -\frac{40}{3}  \log \eta(\tau)\eta(4\tau) -\frac{7}{36}\log J_4^+
& 160 & 372 \\
4 & 1 & 2 &  \frac1{12} \log(J_4^+-64) -\frac{40}{3}  \log \eta(\tau)\eta(4\tau) -\frac{13}{36}\log J_4^+
& 160 & 348 \\
4 & 2 & 2 &  \frac1{12} \log(J_4^+-64) -\frac{40}{3}  \log \eta(\tau)\eta(4\tau) -\frac{19}{36}\log J_4^+
& 160 & 324 \\
\hline
5 & 1 & 1 &  \frac1{12} \log(J_5^{+2} -44 J_5^+-16) 
-\frac{37}{3}  \log \eta(\tau)\eta(5\tau) -\frac{5}{12}\log J_5^+
& 148 & 420 \\
5 & 1 & 2 &  \frac1{12} \log(J_5^{+2} -44 J_5^+-16) 
-\frac{79}{6}  \log \eta(\tau)\eta(5\tau) -\frac{17}{24}\log J_5^+
& 158 & 408 \\
5 & 2 & 2 &  \frac1{12} \log(J_5^{+2} -44 J_5^+-16) 
-14  \log \eta(\tau)\eta(5\tau) -\log J_5^+
& 168 & 396 \\
\hline
6 & 1 & 1 &  \frac1{12} \log(J_6^{+2} -34 J_6^+ +1) 
-\frac{35}{3}  \log \eta(\tau)\eta(6\tau) -\frac{41}{72}\log J_6^+
& 140 & 444\\
\hline
7 & 1 & 1 &  \frac1{12} \log(J_7^{+2} -26 J_7^+ -27) 
-\frac{34}{3} \log \eta(\tau)\eta(7\tau) -\frac79\log J_7^+
& 136 & 468\\
\hline
8 & 1 & 1 &  \frac1{12} \log(J_8^{+2} -24 J_8^+ +16) 
-11  \log \eta(\tau)\eta(8\tau) -\frac{23}{24}\log J_8^+
& 132 & 492\\
\hline
9 & 1 & 1 &  \frac1{12} \log(J_9^{+2} -18 J_9^+ -27) 
-11  \log \eta(\tau)\eta(9\tau) -\frac{5}{4}\log J_9^+
& 132 & 516\\
\hline
11 & 1 & 1 &  \frac1{12} \log(J_{11}^{+3}-20J_{11}^{+2} +56 J_{11}^+ -44) 
-11  \log \eta(\tau)\eta(11\tau) -\frac53 \log J_{11}^+
& 132 & 564\\
\hline
\end{array}\nn
\ee
\caption{Genus one free energy for various models. The penultimate column gives the coefficient of the conformal anomaly, while the last colum gives the coefficient of $T/(2\pi\I)$ (conveniently rescaled).\label{tabgenus1}}
\end{table}


\section{Enumerative geometry}
\label{sec_enum}
In this section, we discuss the general properties of enumerative invariants on CY threefolds fibered by Picard rank 1 K3-surfaces, and verify that these properties indeed hold on the family of CY threefolds $X_m^{[i,j]}$ introduced in Section~\ref{sec:amodel}.

\subsection{CY threefolds fibered by Picard rank 1 K3-surfaces}

Let $X$ be a CY threefold with a projection $\pi:X\to B=\IP^1$
such that the generic fiber $\Sigma=\pi^{-1}(x)$  is a quasi-polarized K3-surface of degree $2m$ and of Picard rank 1.  We assume
that there are no reducible fibers and that $b_2(X)=2$.
We use the same notation for the divisors, curves and topological invariants as introduced in Section~\ref{sec:K3mod}.

The prepotential governing the special K\"ahler metric on K\"ahler moduli space is the genus zero part of the topological string partition function  \eqref{Ftop}, \footnote{We omit the term linear term in $S$, which can be absorbed by an integral change of basis.}
\be
\label{FK3fib}
F^{(0)}(T,S)= m ST^2 + \frac{\kappa}{6} T^3 -\frac{\kappa}{4} T^2
-\frac{c_2}{24} T - \frac{\zeta(3)}{2(2\pi\I)^3} \chi_X +\frac{1}{(2\pi\I)^3} \sum_{\ell,k\geq 0}
 \GV^{(0)}_{\ell,k} \Li_3\left(e^{2\pi\I (\ell T+ k S)}\right)\,.
\ee
In the limit $S\to\I\infty$, the base $B$ becomes of infinite size, and one may invoke heterotic/type II duality in 6 dimensions \cite{Hull:1994ys,kachru:1995wm,Sen:1995ff} to relate type IIA strings on the K3 surface $\Sigma$ with heterotic strings on $T^4$. At finite $S$, one expects a similar duality between type IIA strings on $X$ and heterotic strings on $\widehat\Sigma \times T^2$ (or more generally, a $T^2$ fibration over $\widehat\Sigma$ \cite{Melnikov:2012cv}), where $\widehat\Sigma$ is a K3-surface of Picard rank $19$,  $T^2$ a two-torus at a special point $T=U$ in vector multiplet moduli space where the gauge symmetry 
enhances to $E_8\times E_8\times SU(2)$, accompanied by a suitable choice of bundle. Since the modulus $S$ controlling the size of the base $B$ is identified with 
the heterotic axio-dilaton, such a dual description opens the possibility to compute the
`vertical' GV invariants $\GV^{(g)}_{\ell,0}$ by a one-loop computation in perturbative heterotic string theory. The
latter involves a modular integral of the product of a Siegel-Narain theta series (corresponding to the partition function of momentum and winding modes around $T^2$) and of the `new supersymmetric index' counting perturbative BPS 
excitations \cite{deWit:1995dmj,Antoniadis:1995ct}.
Since the latter transforms as a vector-valued weak modular form of weight $-3/2$ (see e.g. \cite{Enoki:2019deb}), this reasoning indicates that vertical GV invariants should be governed by a suitable theta lift of the new supersymmetric index. 

Unfortunately, heterotic/type II duality in the Picard rank one case has so far only been convincingly demonstrated for the so-called ST model
 $X=\IP_{1,1,2,2,6}[12]$ (corresponding to $(m,i,j)=(1,1,1)$
 in our list), where the second Chern class of the gauge bundle decomposes as $4+10+10$ 
 under $E_8\times E_8\times SU(2)$ \cite{kachru:1995fv,Kaplunovsky:1995tm,Antoniadis:1995zn,Melnikov:2012cv}. 
The case $X=\IP_{1,1,2,2,2}[8]$  (corresponding to $(m,i,j)=(2,1,1)$ in our list)  
was studied using similar methods in \cite{Antoniadis:1995cy}, but its heterotic dual remains elusive \cite{Melnikov:2012cv}. Nonetheless, motivated by heterotic/type II duality, one may formulate (and prove) a precise modularity conjecture for the vertical GV invariants, which we now spell out.

\subsection{GV/NL correspondence}
\label{sec_GVNL}

As proposed in \cite{maulik2007gromov}, the vertical GV invariants
of K3-fibered CY threefolds can be computed in terms of the Noether-Lefschetz invariants of the fibration, defined as follows.

The family $X\stackrel{\pi}{\rightarrow}B$
defines a map $\iota_\pi: B\to \cM_{\langle 2m\rangle}$ into the moduli space 
of quasi-polarized K3 surfaces of degree $2m$.
As mentioned in \S\ref{sec_MmK3}, 
$\cM_{\langle 2m\rangle}$ is a quotient of the symmetric domain $O(2,19)/O(2)\times O(19)$ by an arithmetic group $\Gamma_{\langle 2m\rangle}\subset O(2,19)$. This quotient 
 carries a natural set of Heegner 
 divisors $D_{h,d}$, which have support on
quasi-polarized K3 surfaces $\Sigma$ such that there exists a class $\beta\in\Pic\Sigma$ satisfying
\be
\int_\Sigma \beta^2 = 2h-2\,,
\qquad
\int_\Sigma \beta\cdot L=d\,.
\ee
where $L\in\Pic\Sigma$ is the line bundle of degree $2m=\int_\Sigma L^2$ which determines the polarization of $\Sigma$.
The Noether-Lefschetz number $\NL_{h,d}$ is defined as the intersection number of the divisor $D_{h,d}$ with the base $B$,
\be
\NL_{h,d}= \int_B \iota_\pi^* [D_{h,d}]\,.
\ee
It vanishes unless
\be
\Delta(h,d):=d^2-4m h+ 4m\geq 0 \qquad \mbox{i.e.} \qquad h\leq \frac{d^2}{4m}+1\,,
\ee
and is invariant under   \cite[Prop. 2]{maulik2007gromov}
\be
\label{NLflow}
(h,d)\mapsto (h+kd+m k^2, d+2km)\,,
\ee
for $k\geq 0$,
hence it depends only on $\Delta(h,d)$ and $d\mod 2m$. We shall extend the definition of 
$\NL_{h,d}$ to $h<0$ by requiring invariance under \eqref{NLflow}.
We note that $\NL_{1,0}=-2$ since the degree of the Hodge bundle is $-1$.

A key result of Kudla-Millson \cite{kudla1990intersection} (see also \cite{0919.11036}) is that the generating series of divisors 
\be
\Phi_{\rm KM}(\tau) = 
\sum_{d=0}^{2m-1} v_{d} \sum_{h\leq \frac{d^2}{4m}+1}
D_{h,d} \, q^{\frac{\Delta(h,d)}{4m}} \,,
\ee
transforms as a vector-valued modular form of weight $21/2$ in the Weil representation of the metaplectic cover of $SL(2,\IZ)$ attached to the lattice $\Lambda=\langle 2m\rangle$, with values in $H^2(\cM_{\langle 2m\rangle},\IZ)$. Here, the vectors $v_d$ with $d\in\IZ/(2m\IZ)=\Lambda^*/\Lambda$ are the basis vectors of the space $V=\IC[\Lambda^*/\Lambda]$ on which the Weil representation acts.
As a result, its intersection with the base of the fibration
\be
\Phi_{\rm NL} (\tau)
= \sum_{d=0}^{2m-1} v_{d} \sum_{h\leq \frac{d^2}{4m}+1}
\NL_{h,d} \, q^{\frac{\Delta(h,d)}{4m}} \,,
\ee
also defines a vector-valued modular form of the same weight, now 
valued in complex numbers. Conjecturally, the ratio $\Phi_{\rm NL}/\eta^{24}$ coincides with the new supersymmetric index of the
heterotic dual, when such a dual exists. 

As explained in Appendix \ref{sec_skew}, the space of vector-valued modular form of weight $21/2$ in the Weil representation attached to 
$\Lambda=\langle 2m\rangle$ is isomorphic to the space of
skew-holomorphic Jacobi forms of weight 11 and index $m$,
the dimension of which is tabulated in Table \ref{tabskew}.
A generating basis can be constructed by starting with the theta series of weight $1/2$, 
\be
\label{defThi}
\Thi{m}_d(\tau) = \sum_{k\in\IZ+\frac{d}{2m}} q^{mk^2}\,, 
\ee
acting with Serre derivatives \eqref{SerreD} and multiplying by powers of $E_4$ and $E_6$. For $m\geq 5$, one requires further theta series $T_{m_1,m_2}$ as well as acting with Hecke-like operators.  
Once a generating basis has been constructed, it suffices to compute the first few coefficients in the $q$-expansion (for example, the polar coefficients in $\Phi_{\rm NL}/\eta^{24})$ in order to determine the complete series. 

Furthermore, it is conjectured (and proven for genus $g=0,1$ in \cite{maulik2007gromov,Klemm2010})
that the vertical GV invariants are related 
to the NL numbers of the fibration via
\be
\GV_{d,0}^{(g)} = \sum_{h\geq g} r_{g,h} \, \NL_{h,d}\,.
\ee
Here $r_{g,h}$ are the reduced GV invariants for K3, determined by~\cite{katz:1999xq}
\be
\label{defrgh}
\sum_{h,g} r_{g,h}\, (2-y-1/y)^g q^{h} = \prod_{n=1}^{\infty} \frac{1}
{(1-q^n)^{20}(1-y q^n)^2(1-q^n/y)^2}\, ,
\ee
as was proven for primitive curve classes in~\cite{Maulik:2010jw}.
These integers vanish unless $h\geq g$ and satisfy $r_{g,g}=(-1)^g(g+1)$ (see Table \ref{tab_rgh}).

\begin{table}
$$
\begin{array}{|c|cccccccccc|}
\hline
 g\backslash h & 0 & 1 & 2 & 3 & 4 & 5 & 6 & 7 & 8 & 9 \\ \hline
 0 & 1 & 24 & 324 & 3200 & 25650 & 176256 & 1073720 & 5930496 & 30178575 & 143184000 \\
 1 & 0 & -2 & -54 & -800 & -8550 & -73440 & -536860 & -3459456 & -20119050 & -107388000 \\
 2 & 0 & 0 & 3 & 88 & 1401 & 15960 & 145214 & 1118880 & 7568403 & 46046040 \\
 3 & 0 & 0 & 0 & -4 & -126 & -2136 & -25750 & -246720 & -1993854 & -14098560 \\
 4 & 0 & 0 & 0 & 0 & 5 & 168 & 3017 & 38328 & 385380 & 3257008 \\
 5 & 0 & 0 & 0 & 0 & 0 & -6 & -214 & -4056 & -54138 & -569828 \\
 6 & 0 & 0 & 0 & 0 & 0 & 0 & 7 & 264 & 5265 & 73656 \\
 7 & 0 & 0 & 0 & 0 & 0 & 0 & 0 & -8 & -318 & -6656 \\
 8 & 0 & 0 & 0 & 0 & 0 & 0 & 0 & 0 & 9 & 376 \\
 9 & 0 & 0 & 0 & 0 & 0 & 0 & 0 & 0 & 0 & -10 \\ \hline
\end{array}
$$
\caption{Reduced GV invariants $r_{g,h}$}
\label{tab_rgh}
\end{table}

By rewriting  \eqref{defrgh} as follows \cite{Marino:1998pg} 
\be
\sum_{h,g} r_{g,h}\, (2-y-1/y)^g q^{h-1} =
\frac{1}{ \eta^{24}} \left( 
\frac{\sin(\lambda/2)}{\lambda/2}\right)^2
\exp\left( \sum_{k\geq 1} \frac{2\zeta(2k)}{k} \left(\frac{\lambda}{2\pi}\right)^{2k} E_{2k}(\tau) \right)\,,
\ee
with $y=e^{\I\lambda}$, and expanding in powers of $t=2-y-1/y=\lambda^2 -\frac1{12} \lambda^4+ \dots$,
we find  the following generating series
for fixed genus,
\be
\begin{split}
R_0:=\sum_{h\geq 0} r_{0,h} q^{h-1} =& \frac{1}{\eta^{24}} \\
R_1:=\sum_{h\geq 0} r_{1,h} q^{h-1} =& \frac{E_2-1}{12\eta^{24}}  \\
R_2:=\sum_{h\geq 0} r_{2,h} q^{h-1} =& \frac{E_4+5E_2^2-6}{1440\eta^{24}}  \\
R_3:=\sum_{h\geq 0} r_{3,h} q^{h-1} =& \frac{4E_6+21 E_2E_4+35E_2^3+105E_2^2-186}{362880\eta^{24}}, \ \mbox{etc}
\end{split}
\ee
While the series $R_g$ for $g\geq 1$ 
have no nice modular properties, the linear combinations $\tilde R_g$ defined by 
\be
\sum_{g\geq 0} t^g\, R_g  = 
\left(\frac{\sin(\lambda/2)}{\lambda/2}\right)^2 
\sum_{g\geq 0}  \lambda^{2g} \,\tilde R_g\,,
\ee
are quasi-modular forms of depth $g$ and weight $2g-12$:
\be
\tilde R_1= R_1+\frac{1}{12}R_0 = \frac{E_2}{12\eta^{24}}\ ,\quad
\tilde R_2 = R_2+\frac{1}{240}R_0 = \frac{E_4+5E_2^2}{1440\eta^{24}},  \ \mbox{etc}
\ee
The relevance of the 
generating series $R_g$ is that  the 
invariant $\GV_{d,0}^{(g)}$ appears as the 
coefficient of $q^{\frac{d^2}{4m}}$ in the product $R_g \Phi$,
\be
\GV_{d,0}^{(g)} =  C_{R_g\Phi}(n^2,n)\,,
\ee
where $C_{R_g\Phi}(\Delta,r)$ denotes the Fourier coefficients of the skew-Jacobi form $R_g\Phi$, see Appendix \ref{sec_skew}. 
Similarly, the combination of invariants $\GV_{d,0}^{(g'\leq g)}$
occurring in the genus $g$ topological amplitude 
appears as the coefficient of $q^{\frac{d^2}{4m}}$ in the product $\tilde R_g \Phi$, for example (compare with \eqref{GWtoGVexp})
\bea
\GV_{d,0}^{(1)} + \frac{1}{12}\GV_{d,0}^{(0)} &=&  C_{\tilde R_1\Phi}(n^2,n)\,, \nn\\
\GV_{d,0}^{(2)} + \frac{1}{240}\GV_{d,0}^{(0)} &=&  C_{\tilde R_2\Phi}(n^2,n) \,.
\eea

Conversely, one can compute NL invariants using the knowledge of the vertical GV invariants, and check whether they are indeed consistent with modular invariance. We have carried out this check for all manifolds $X_m^{[i,j]}$ and found a unique skew-holomorphic Jacobi form which reproduces the GV invariants up to the degree and genus which we could compute them, see \S\ref{sec_results}. For $m\leq 4$, this reproduces the generating series found in \cite[\S6.4]{maulik2007gromov}, at least for some choice of $(i,j)$ \footnote{Note that the authors of \cite{maulik2007gromov} used a different basis of vector-valued modular form, constructed as Rankin-Cohen brackets of Eisenstein series with the theta series \eqref{defThi}. They further assumed that NL invariants vanish when $0<\Delta(h,d)<4m$, which is not true in general.}. 

In general, the GV/NL relation implies that the contribution of vertical GW invariants to the topological string amplitude can be
written as 
\be
\label{WgShimura}
W^{(g)}(T) := \sum_{d\geq 0} \GW_{d,0}^{(g)} \, e^{2\pi\I d T}\
=
\sum_{n\geq 1} C_{\tilde R_g\Phi}(n^2,n) \Li_{3-2g}
\left( e^{2\pi\I n T}\right)\,.
\ee
This is recognized as a Shimura-type lift \label{eqShimura} of 
the vector-valued quasi-modular form $\tilde R_g \Phi$,
in agreement with general expectations from heterotic/type II duality. If $\tilde R_g \Phi$ was a skew-holomorphic Jacobi cusp form of weight $2g-1$, its Shimura lift $W^{(g)}(T)$ would be a holomorphic cusp form of weight $4g-4$ under $\Gamma_0(m)^+$, as reviewed in \S\ref{sec_skew}. However, $\tilde R_g \Phi$ has a pole at the cusp $\tau=\I\infty$
for any $g$, and is skew-holomorphic only for $g=0$. The resulting
analytic and modular properties of $W^{(g)}(T)$ are therefore more complicated.

For $g=0$, \eqref{WgShimura} encodes the contribution of genus 0 GV invariants 
to the prepotential $F^{(0)}$ in \eqref{FK3fib}, 
\be
W^{(0)}(T) = \sum_{n\geq 0} \GV^{(0)}_{n,0} 
\Li_{3}\left( e^{2\pi\I n T}\right) 
= \sum_{n\geq 1} C_{\Phi/\eta^{24}}(n^2,n) \Li_{3}\left( e^{2\pi\I n T}\right)\,.
\ee
The latter is known to transform as a holomorphic Eichler integral of weight $-4$ under $\Gamma_0(m)^+$ \cite{Antoniadis:1995ct,Harvey:1995fq,Angelantonj:2015rxa,Enoki:2019deb,Enoki:2021vid,Enoki:2022cfc}. 
Equivalently, it
can be viewed as the holomorphic part of a local harmonic Maass form of weight $-4$. Acting with Bol's operator $D^5=\partial_T^5$, we find a meromorphic modular form of weight $6$ under $\Gamma_0(m)^+$ \cite{Antoniadis:1995cy}:
\be
\partial_T^5 W^{(0)}(T) = \sum_{n\geq 1} n^5 C_{\Phi/\eta^{24}}(n^2,n) \Li_{-2}\left( e^{2\pi\I n T} \right)
\ee

For $g=1$, \eqref{WgShimura} instead encodes the contribution of genus 0 and 1 GV invariants to the genus one topological free energy
$F^{(1)}(T,S)=S + \frac{c_2}{24} T + W^{(1)}(T)$
with
\be
W^{(1)}(T) = \sum_{n\geq 0} \left[ \GV^{(1)}_{n,0} + \frac{1}{12}
\GV^{(0)}_{n,0} \right]
\Li_{1}\left( e^{2\pi\I n T}\right) \,.
\ee
We can indeed write $W^{(1)}(T) - H_m^{[i,j]}$
as a combination of 
$\log(J_m^+-\lambda_i)$ and of the logarithm of a certain eta product, see Table \ref{tabgenus1}, which is formally of weight 0.

\subsection{NL/DT correspondence}
\label{sec_NLDT}
In this subsection, we briefly review the relation between
NL invariants and vertical Donaldson-Thomas invariants \cite{Bouchard:2016lfg}, which is one of the prime motivations 
for this work.

Recall that BPS states in type IIA string theory compactified on a CY threefold $X$ can be understood mathematically as semi-stable objects in the bounded derived category of coherent sheaves 
$\cC=D^b\Coh(X)$. This category is graded by the Grothendieck group $K(\cC)$, which is identified with the lattice of electromagnetic charges.  
The notion of semi-stability refers to the choice of a Bridgeland stability condition $\sigma\in\Stab\cC$, determined locally by the K\"ahler moduli of $X$. For fixed electromagnetic charge vector $\gamma\in K(\cC)$ and stability condition $\sigma$, one considers the moduli stack of $\sigma$-semi-stable objects in $\cC$, and associates to it a number $\bOm_\sigma(\gamma)\in\IQ$, known as the (rational) Donaldson-Thomas invariant. For generic $\sigma$, the combination $\Omega_\sigma(\gamma):=\sum_{d|\gamma} \frac{1}{d^2} \bOm_\sigma(\gamma/d)$ is conjecturally integer, and correspond to the physical index counting BPS bound states of D6-D4-D2-D0 branes. 

For charge $\gamma=[-1,0,\beta,-n]\in H^0(X)\oplus H^2(X)\oplus H^4(X)\oplus H^6(X)$ (corresponding
to one unit of anti-D6 brane charge, $n$ units of D0-brane charge and D2-brane charge $\beta$), the index $\Omega_\sigma(\gamma)$  in a suitable large volume limit agrees with the Pandharipande-Thomas (PT) invariant $\PT(\beta,n)$ counting stable pairs. The latter is related to the Gopakumar-Vafa invariants $\GV(\beta',n')$ by the GV/PT correspondence \cite{gw-dt,gw-dt2}. Restricting to vertical D2-brane charges and using the GV/NL correspondence, the index $\Omega_\sigma(\gamma)$ is therefore computable in terms of the NL invariants. On the other hand, by results of \cite{Gholampour:2013mka}, the same index 
is related by wall-crossing to the D4-D2-D0 indices $\Omega(0,r[\Sigma],\mu,-n)$ counting 
rank $r$
Gieseker semi-stable sheaves supported on the K3-fiber $\Sigma$, with first Chern class $\mu$ and second Chern class $-n$. Combining these two relations, the authors of \cite{Bouchard:2016lfg} show that the generating series of D4-D2-D0 indices
\be
\label{defhmuk3}
h_{r,\mu} = \sum_{n\geq 0} \bOm(0,r[\Sigma],\mu,-n)\, \q^{n+\frac{\mu^2}{4mr}+\frac{r\mu}{2}- \frac{\chi(r[\Sigma])}{24}}\,,
\ee
is obtained from the series of NL invariants $\Phi_{\rm NL}$ by 
\be
\label{hmuNL}
h_{r,\mu} = V_{r,\mu} \left( \frac{\Phi_{\rm NL}}{\eta^{24}} \right) \,,
\ee
where $V_{r,\mu}$ is a certain Hecke-like operator, which maps the
Weil representation of $\IC[\IZ/(2m \IZ)]$ to the Weil representation attached to the rescaled lattice $\IC[\IZ/(2m r \IZ)]$, preserving the modular weight. As a result, $h_{r,\mu}$ transforms as a vector-valued modular form of weight $-3/2$
in the Weil representation attached to $\IC[\IZ/(2m r \IZ)]$.

To see why this is in precise agreement with constraints from S-duality in string theory \cite{Alexandrov:2018lgp,Alexandrov:2019rth}, recall that 
a general CY 3-fold and
a fixed divisor class $[\cD]$,
the generating series of D4-D2-D0 invariants are in general defined by
\be
\label{defhmu}
h_{p^a,\mu_a} = \sum_{q_0} \bOm_*(0,p^a,q_a,q_0)\, \q^{-q_0+\frac12 q_a \kappa^{ab} q_b}\,,
\ee
where $p^a$ are the components of the divisor class $[\cD]$
on a basis of $\Lambda=H^4(X)$, $\kappa^{ab}$ is the inverse of the 
quadratic form $\kappa_{ab}=\kappa_{abc} p^c$
on $\Lambda^*=H_4(X,\IZ)$,
the D2-brane charges $q_a$ are such that $q_a=\mu_a+\frac12 \kappa_{ab} (p^b +\epsilon^b)$ for $\epsilon^a\in H_2(X,\IZ)$, and
the DT invariant is evaluated in the large volume attractor chamber $t^a = \kappa^{ab} q_b + \I \lambda p^a$, $\lambda\to+ \infty$. 
By spectral flow invariance, the series \eqref{defhmu} is invariant under $\mu_a\mapsto \mu_a+\kappa_{ab} \epsilon^b$, and defines a vector in $\IC[\Lambda^*/\Lambda]$. 
The general prediction from \cite{Alexandrov:2018lgp,Alexandrov:2019rth} is that 
$h_{p,\mu}$ should transform as  a vector-valued mock modular form of weight $-1-\frac12 \rk\Lambda$, associated to the Weil representation on $\IC[\Lambda^*/\Lambda]$, 
with depth $r-1$ where $r$
is the maximal number of irreducible components in which the divisor class $[\cD]$ can be decomposed.

When $X$ is fibered by K3-surfaces of degree $2m$, the quadratic form for divisor class $[\cD]=r_1 [D]+ r_2 [\Sigma]$ is given by 
\be
\kappa_{ab}=\begin{pmatrix} \kappa r_1+2m r_2  & 2m r_1 \\ 2m r_1 & 0 \end{pmatrix}\,.
\label{kappagen}
\ee
For a non-vertical divisor ($r_1\neq 0$), the quadratic form
\eqref{kappagen} is invertible, with inverse 
\be
\kappa^{ab}= \begin{pmatrix}0  & \frac{1}{2m r_1} \\
\frac{1}{2m r_1} & -\frac{\kappa}{4m^2 r_1} -\frac{r_2}{2m r_1^2}\end{pmatrix}\,,
\ee
and $h_{p,\mu}$ is expected to transform as a vector-valued mock modular form of weight $-2$ and depth $r=\gcd(r_1,r_2)-1$. 

For a vertical divisor however, 
$r_1=0$  and the quadratic form \eqref{kappagen} has a one-dimensional null space which should be divided out (similar to the analysis of $X=K3\times T^2$ in \cite{Alexandrov:2020qpb}). This  leads to a lattice $\Lambda$ of rank 1, hence a  vector-valued mock modular form of weight $-\frac32$. The power of $q$ in 
\eqref{defhmu} can also be matched to the power in \eqref{defhmuk3} by using the standard formulae for the Chern character for push forward of torsion free sheaves on $\Sigma$ to 
coherent sheaves on $X$ \cite{Diaconescu:2007bf}. 
Moreover, the fact that $(p^1,p^2)=(0,r)$ lies in the kernel of $\kappa_{ab}$ ensures the absence of wall-crossing and the
disappearance of the modular anomaly, leading to an ordinary vector-valued modular form. \footnote{We are grateful to Sergey Alexandrov for suggesting this line of argument.}

Applying \eqref{hmuNL} and using the modular forms $\Phi_{\rm NL}$ from Appendix \ref{sec_results}, we can readily compute 
the vertical D4-D2-D0 rational indices $\bOm(0,r[\Sigma],\mu,-n)$ for arbitrary $r$. A non-trivial fact is that 
the resulting `integer invariants' $\Omega(0,r[\Sigma],\mu,-n)$ indeed turn out to be integer,
even when the charge vector is not primitive. An outstanding
problem is to compute D4-D2-D0 indices for non-vertical divisor class, and identify the corresponding mock modular forms
of higher weight.

\subsection{Higher base degree}
\label{sec_hidegree}
As discussed in \S\ref{sec_GVNL}, the generating series $W^{(0)}$ of vertical genus 0 GW invariants transforms as a holomorphic Eichler integral of weight $-4$ under $\Gamma_0(m)^+$, consistent with the predictions of heterotic-type II symmetry. In fact, it was shown in \cite{Henningson:1996jf,Berglund:1997eb} that 
this modular behavior follows from the transformation properties of the period vector $\vec\Pi$ under the monodromies $M_S,M_T,M_U$ in the large base limit $S\mapsto \I\infty$, and that the same transformation properties also determine the modular properties of the generating series of genus 0 GW invariants of higher base degree. Here we review the reasoning in \cite{Henningson:1996jf,Berglund:1997eb}, and find generating series of genus zero, base degree one or two GW invariants for the models $X^{[i,j]}_m$ with the appropriate modular behaviour. Generating series of higher base degree and higher genus GW invariants will be discussed elsewhere.

Separating the genus-zero free energy as $F^{(0)}=m S T^2 + W(T,S)$, such that the period vector \eqref{vecPiLeading} becomes
\be
\label{defPi}
\vec\Pi = (1,T,S,  2mST+\partial_T W, mT^2+\partial_S W,  -mST^2+(2-S\partial_S-T\partial_T) W)^t\,,
\ee
it is straightforward to check that the actions of the monodromies $M_S$ and $M_T$ \eqref{MTMS} imply the following periodicity conditions on $f(S,T)$,
\bea
\begin{split}
W(T+1,S) =& W(T,S) + \frac{\kappa}{2} T^2 - \frac{c_2+2\kappa}{24}\,,
 \\
W(T,S+1) =& W(T,S)\,.
\end{split}
\eea
This is solved by 
\be
W(T,S) = \frac{\kappa}{6}T^3 - \frac{\kappa}{4} T^2 - \frac{c_2}{24} T + \frac{1}{(2\pi\I)^3} \sum_{k\geq 0,\ell\geq 0} 
\GW^{(0)}_{(\ell,k)}\, e^{2\pi\I (\ell T+kS)} \,,
\ee
where $\GW^{(0)}_{(\ell,k)}$ are the genus 0, base degree $k$ GW invariants. We denote the Fourier coefficients of $W(S,T)$ with respect to $S$ by $W_k(T)$, such that 
\be
W_0(T) = \frac{\kappa}{6}T^3 - \frac{\kappa}{4} T^2 - \frac{c_2}{24} T +\frac{W^{(0)}}{(2\pi\I)^3} , \quad
W_{k\geq 1}(T) = \frac{1}{(2\pi\I)^3} \sum_{\ell\geq 0} 
\GW^{(0)}_{(\ell,k)}\, e^{2\pi\I \ell T} \,.
\ee

The modular properties of $W_k(T)$ will come by requiring consistency with  the relative conifold monodromy $M_U$ in \eqref{MU}, acting as follows:
\be
\begin{split}
\label{MUPi}
T\mapsto& \ -\frac{T}{mT^2+\partial_S W} \\
S \mapsto & \ S+1 + \frac{T\partial_T W-2W+1}{mT^2+\partial_S W} \\
 W \mapsto&\ 
 \frac{mT^2(2W-T\partial_T W-1)}{ \left(m T^2+\partial_S W\right)^3}
 +\frac{ T\partial_T W-W-mT^2+\frac12}{\left(m T^2+\partial_S W\right)^3} - \frac12
 \\
\partial_ T W \mapsto &\  \frac{2mT}{mT^2+\partial_S W}   +
  \frac{2m(T\partial_T W-2W+1)}{(mT^2+\partial_S W)^2} \\
\partial_S W \mapsto &\  \frac{\partial_S W }{(mT^2+\partial_S W)^2}
\end{split}
\ee
It is cumbersome but straightforward to check that the transformation properties of $\partial_T W$ and $\partial_S W$
follow from the first three equations by the chain rule. 

In the limit $S\to\I\infty$, $\partial_S W$ is exponentially suppressed and at leading order, one may replace $W$ by its
zero-mode $W_0$ in the Fourier expansion $W=\sum_{k\geq 0} W_k(T) q_S^k$ where $q_S:=e^{2\pi\I S}$. As a result, the first equation in \eqref{MUPi} reduces to $T\mapsto -1/(mT)$, and the third equation becomes
\be
\label{f0mod}
W_0(-1/mT) = \frac{W_0(T)}{m^2 T^4} - \frac{1}{2} \left( 1+ \frac{1}{mT^2} \right)^2 \,.
\ee
This implies that $W^{(0)}$ should transform as a holomorphic Eichler integral of weight $-4$, as advertized previously.
To determine the modular properties of $W_k(T)$, we note that
it follows from \eqref{f0mod} that the  second and fifth derivatives  of $f_0(T)$ transforms as 
\bea
\label{delf0}
\begin{split}
\partial_T^2 W_0(-1/mT)  =& \partial_T^2 W_0(T) -2m -\frac{6\partial_T W_0(T)}{m}+\frac{12}{T^2} (W_0(T)-\tfrac12) \,,\\
\partial_T^5 W_0(-1/mT)  =& m^3 T^6 \, \partial_T^5 W_0(T)\,.
\end{split}
\eea
In particular, $\partial_T^5 W_0(T)$ transforms as a (meromorphic) modular form of weight $6$ under $\Gamma_0(m)^+$. 

Using the first equation in \eqref{delf0} and the second equation in \eqref{MUPi}, one easily checks that the combination, sometimes called `heterotic dilaton' in the context of heterotic/type II duality,
\be
\label{hetdil}
\widehat{S} := S +\frac{\partial_T^2 W_0}{6m} \,,
\ee
transforms as\footnote{In \cite{Berglund:1997eb}, the authors define an `improved heterotic dilaton' which transforms as \eqref{MSTUwS} with no 
$\cO(q_S)$ corrections, as well as an `improved modulus' $\widetilde{T}$ transforming as $\widetilde{T}\mapsto -1/(m \widetilde{T})$ without $\cO(q_S)$ corrections. These constructions are not needed for the present purposes. } 
\be
\label{MSTUwS}
M_S:\widehat{S}\mapsto \widehat{S} +1+\cO(q_S)\,, \quad 
M_T: \widehat{S}\mapsto \widehat{S} +\frac{\kappa}{6m}+\cO(q_S)\,, \quad 
M_U:\widehat{S}\mapsto \widehat{S}+\frac23+\cO(q_S)\,.
\ee
This motivates the definition 
\be
\label{deffhat}
\whf_k(T) := e^{-\frac{\I\pi k}{3m} \partial_T^2 W_0} W_k(T)\,,
\ee
such that the Fourier expansion of $W$ becomes 
\be
\label{Fourierfh}
W(T,S) = W_0(T)+ \sum_{k\geq 1} \whf_k(T) \, q_{\whS}^k\,, \quad q_{\whS}:=e^{2\pi\I \whS}\,.
\ee
As explained in \cite{Henningson:1996jf}, one can now obtain
the modular behavior of $\whf_k(T)$ by substituting \eqref{Fourierfh} in the transformation property \eqref{MUPi} of $\partial_S W$ and expanding in powers of $q_{\whS}$. At leading order in $q_{\whS}$, one immediately finds that $h_1(T):=\whf_1(T)$
transforms as a modular form of weight $-4$ under $\Gamma_0(m)^+$,
\be
\whf_1(-1/(mT)) = e^{2\I\pi/3} \whf_1(\tau)/(m^2 T^4)\,,\quad 
\whf_1(T+1) = e^{-\frac{\I\pi\kappa}{3m}} \whf_1(T) \,.
\ee
At next-to-leading order, one finds a rather complicated transformation property for $\whf_2(T)$, which becomes much simpler in terms of\footnote{To avoid cluttering we omit factors of $2\pi\I$ multiplying $\partial_T$.} 
\begin{align}
h_2(T) :=& \whf_2(T) - \frac{1}{6m}  \whf_1 \partial_T^2 \whf_1
- \frac{1}{24m^2} \partial_T^4W_0\,  (\whf_1)^2 \nn\\
=& e^{-\frac{2\I\pi}{3m} \partial_T^2W_0} \left[
W_2-\frac{1}{6m} W_1\, \partial_T^2 W_1 +\frac{1}{18m^2} \partial_T^3W_0\, \partial_T W_1 
-\frac{1}{72m^2 } \partial_T^4W_0\, (W_1)^2 + \frac{1}{216m^3} ( \partial_T^3W_0)^2   (W_1 )^2\right]\,, \nn\\
\label{h2W2h}
\end{align}
namely
\be
h_2(-1/(mT)) = e^{4\I\pi/3} h_2(T)/(m^2 T^4)\, ,\quad 
h_2(T+1) = e^{-\frac{2\I\pi\kappa}{3m}} h_2(T) \,.
\ee
At next-to-next-to-leading order, one finds a rather complicated transformation property for $\whf_3(T)$, which becomes much simpler in terms of 
\begin{align}
    h_3(T)  :=&\whf_3(T) -\frac{h_2 \partial_T^2 h_1}{3m}
    -\frac{h_1 \partial_T^2 h_2}{6m} 
-\frac{h_1 h_2 \partial_T^4 W_0}{6m^2}
-\frac{h_1 (\partial_T^2 h_1)^2}{24m^2}
- \frac{h_1^2 \partial_T^4 h_1}{24m^2}
\nn\\
& -\frac{h_1^2 \partial_T^2 h_1  \partial_T^4 W_0}{36m^3}
+  \frac{h_1^3  \partial_T^6 W_0}{144m^3} 
- \frac{h_1^3  \partial_T^4 W_0}{288m^4}
\end{align}
namely
\be
h_3(-1/(mT)) =  h_3(T)/(m^2 T^4)\, ,\quad 
h_3(T+1) = e^{-\frac{\I\pi\kappa}{m}} h_3(T) \,.
\ee
More generally, the authors of \cite{Henningson:1996jf} construct 
a combination $h_k(T):=\whf_k(T)+\dots$ where the dots include monomials in $\whf_{k_i}$ and derivatives thereof with $\sum_k k_i=k$, such that for any $k\geq 1$, $h_k(T)$ transforms as a modular form of weight $-4$ under $\Gamma_0(m)^+$,
\be
\label{modhk}
h_k(-1/(mT)) = e^{2\I\pi k/3} h_k(\tau)/(m^2 T^4)\ ,\quad 
h_k(T+1) = e^{-\frac{\I\pi k \kappa}{3m}} h_k(T) \,.
\ee
Further requiring that $f_k$ behaves as $f_k\sim(T-T_i)^{2-4k}$
for $T\to T_i$ (as required by the field theory analysis \cite[(2.7)]{kachru:1995fv}), and using the behavior
of the second derivative of $W_0$, 
\be
\partial^2_T W_0(T)  \sim \begin{cases} \kappa T &  \mbox{as}\  T\to\I\infty 
\\
 -\frac{8m}{2\pi\I} \log(T-T_i) & \mbox{as}\ T \to T_i
 \end{cases}\,,
 \ee
imposes the following boundary conditions on $h_k(T)$ \cite{Henningson:1996jf}
\be
\label{boundhk}
h_k(T) \stackrel{T\to\infty} \sim q^{-\frac{\kappa k}{6m}}, \quad 
h_k(T) \stackrel{T\to T_i} \sim (T-T_i)^{2-\frac83 k}\,.
\ee
Combining \eqref{modhk} and \eqref{boundhk}, one can in principle determine the modular form $h_k(T)$ explicitly, from the knowledge of the first base degree $k$ GW invariants. 

In practice, we find it more convenient to rescale $h_k(T)$ and define
\be
\tilde h_k(T) :=  h_k(T) \times (J_m^+)^{-\frac{k\kappa}{6m}} \prod_{i=1}^r \left( 
\frac{J_m^+}{J_m^+-\lambda_i}\right)^{2k/3} \,.
\ee
Using \eqref{d2WTh}, we see that $\tilde h_k(T)$ is expressed in terms of the Fourier coefficients $W_k(T)$ via 
\be
\tilde h_k(T)  = e^{kH_m^{[i,j]}} \left[ W_k(T) + \dots \right]\,,
\ee
where $H_m^{[i,j]}$ was introduced in \eqref{defhtilde} and the dots are polynomials in $W_{k'}(T)$ with $k'<k$ and derivatives thereof,
see e.g. the second equation in \eqref{h2W2h} for $k=2$. The advantage of this normalization is 
that the modular properties and boundary conditions take the simpler form
\be
\tilde h_k(-1/(mT)) = \tilde h_k(T)/(m^2 T^4)\ ,\quad  
\tilde h_k(T+1) = \tilde h_k(T) \,,
\ee
\be
\tilde h_k(T) \stackrel{T\to\infty}\sim q^{0}, \quad 
\tilde h_k(T) \stackrel{T\to T_i} \sim (T-T_i)^{2-4k}\,,
\ee
where we recall that $J_m^+-\lambda_i \sim (T-T_i)^2$.
Recalling that $J_m^+(T)-\lambda_i$ has a double zero at $T=T_i$,
we can thus use the Ansatz
\be
\label{htkansatz}
\tilde h_k(T) = \frac{P_w(T)}{\left(\Xi_m(T)\right)^{(2k-1)r} \prod_{i=1}^r (J_m^+(T)-\lambda_i)^{2k-1}}\,,
\ee
where $\Xi_m(\tau)$ is a $\Gamma_0(m)$ cusp form of lowest possible weight 
(except for $m=7$, where we take  the cusp form of next-to-lowest weight, $m=9$
where we take the square of the lowest weight cusp form, and 
$m=11$, where we take the cube of the lowest weight cusp form). The numerator 
$P_w(\tau)$, to be determined, is then a  modular form of weight $w=(2k-1)r w_0 -4$ under $\Gamma_0(m)^+$, where $w_0$ is the modular weight of $\Xi_m(\tau)$.
More explicitly, for $m=1,2,3,5$ we choose 
\be
\Xi_m(\tau) = [\eta(\tau)\eta(m\tau)]^{\frac{24}{m+1}}\,, \quad w_0=\frac{24}{m+1}\,,
\ee
while for the remaining values we take 
\be
\begin{cases}
\Xi_4(\tau) = \eta(2\tau)^{12}, & w_0=6 \\
\Xi_6(\tau) = [\eta(\tau) \eta(2\tau) \eta(3\tau) \eta(6\tau) ]^4, &w_0=8  \\
\Xi_7(\tau) = \eta(\tau)^6 \eta(7\tau)^6, & w_0=6 \\
\Xi_8(\tau) = [\eta(2\tau)\eta(4\tau)]^{8},  &w_0=8 \\
\Xi_9(\tau) = \eta(3\tau)^{16},  &w_0=8 \\
\Xi_{11}(\tau) = [\eta(\tau) \eta(11\tau) ]^6,  &w_0=6 
\end{cases}\,.
\ee
In Appendix \ref{sec_results}, we give the modular 
form $\tilde h_1, \tilde h_2$ which encodes the genus 0 
GW invariants  of base degree one and two
for each of the models $X_m^{[i,j]}$. 
Interestingly,
$\tilde h_1$ turns out to be universal for fixed $m$, up to an overall factor of $ij$, while the generating series $W_1$ is not, due to the non-trivial $(i,j)$ dependence of  $H_m^{[i,j]}$. 

\appendix

\section{Fundamental periods of $\Sigma_m$ for $m=5,\ldots,9$}
\label{sec:nonToricFundamentals}
In this section we collect various representations
of the coefficients $c_m(d)$ of the fundamental period $f_m(\lambda)=\sum_{d\ge 0}c_m(d)\lambda^{-d}$ of a $M_{m}$-polarized K3 surface $\Sigma_m$ with $m=5,\ldots,9$. The coefficients for $m\leq 4$ are standard and given in \eqref{K3fperiod}.

\begin{itemize}
\item $m=5$:
\be
\begin{split}
        c_5(d)=&\binom{2d}{d} \sum\limits_{l_1,l_2\ge 0}\binom{l_1}{l_2}\binom{d}{l_1}^2\binom{d}{l_2}
        \\
        =&\binom{2d}{d}\sum\limits_{l=0}^d \binom{d}{l}^2\binom{d+l}{l}
        \\
        =& (-1)^d \binom{2d}{d} \sum_{l=0}^d
        \binom{d}{l}^5\left(1+5l\left[H_{d-l}-H_{l}\right]\right) \\
        =&\{1, 6, 114, 2940, 87570, 2835756,\dots \}
        \label{eqn:fundamental_c5}
 \end{split}       
\ee
The first expression follows directly from \cite{Batyrev:1998kx,Batyrev2000}. 
The second equality is obtained by using 
Vandermonde's identity
\begin{align}
\label{vandermonde}
    \sum\limits_{l_2\ge 0}\binom{l_1}{l_2}\binom{d}{l_2}=\sum\limits_{l_2\ge 0}\binom{l_1}{l_1-l_2}\binom{d}{l_2}=\binom{d+l_1}{l_1}\,.
\end{align}
The last equality follows from~\cite[\S F.12]{Coates2016}, upon identifying the the holomorphic period of $\Sigma_5$ with the regularized quantum period of the Fano fourfold $V^4_{10}$. The equality between these two representations is shown in~\cite[Eq (5)]{Paule2003}. After multiplication with  $\binomial{2d}{d}$, one reproduces the holomorphic solution of the CY operator \#25 in the
AESZ list~\cite{almkvist2005tables}.

\medskip

\item $m=6$:
\be
\begin{split}
        c_6(d)=&\sum\limits_{l_1,l_2,l_3\ge 0}\binom{l_1}{l_2}\binom{l_1}{l_3}\binom{d}{l_1}^2\binom{d}{l_2}\binom{d}{l_3}
        \\
        =&\sum\limits_{l_1,l_2,l_3\ge 0}\binom{l_1}{l_2}\binom{l_1}{l_3}\binom{d}{l_1}^2\binom{d}{d-l_2}\binom{d}{d-l_3}
        \\
        =&\sum\limits_{l=0}^d\binom{d}{l}^2\binom{d+l}{l}^2
        \\
        =& (-1)^d  \sum_{l=0}^d
        \binom{d}{l}^4\binom{d+l}{l}\binom{2d-l}{d}\left(1+l\left[5H_{d-l}-5H_{l}+H_{d+l}-H_{2d-l}\right]\right)
        \\
        =&\{ 1, 5, 73, 1445, 33001, 819005, \dots\}
        \label{eqn:fundamental_c6}
\end{split}
\ee
The equality between the first three representations follows from the Vandermonde identity \eqref{vandermonde}.
The last representation follows from the regularized quantum period of the Fano fourfold $V^4_{12}$~\cite[\S F.13]{Coates2016}.
After multiplication with  $\binomial{2d}{d}$, one recovers the holomorphic solution of the CY operator \#29 in the AESZ list.

\medskip

\item $m=7$:
\be
\begin{split}
        c_7(d)=&\sum\limits_{l_1,l_2,l_3\ge 0}\binom{l_1}{l_2}\binom{l_2}{l_3}\binom{d}{l_1}^2\binom{d}{l_2}\binom{d}{l_3}
        \\
        =&\sum\limits_{l_1,l_2,l_3\ge 0}\binom{l_1}{l_2}\binom{l_1}{l_3}\binom{d}{l_1}^2\binom{d}{d-l_2}\binom{l_1}{d-l_3}
        \\
        =&\sum\limits_{l=0}^d\binom{d}{l}^2\binom{2l}{d}\binom{d+l}{l}
        \\
        =& (-1)^d \sum_{l=0}^d \binom{d}{l}^6\left(1+6l\left[H_{d-l}-H_{l}\right]\right)
        \\
        =& \{1, 4, 48, 760, 13840, 273504,\dots \}
        \label{eqn:fundamental_c7}
\end{split}
\ee
The first expression follows directly from \cite{Batyrev:1998kx,Batyrev2000}, and the third from
the Vandermonde identity. The equality between the first and the third expressions can be reduced to an analogue of
the Strehl identity
\begin{align}
    \sum\limits_{l=0}^d\binom{d}{l}^3\frac{\binom{d+l}{l}}{\binom{d+d}{l}}=\sum\limits_{l=0}^d\binom{d}{l}^2\binom{2l}{d}\frac{\binom{d+l}{l}}{\binom{d+d}{d}}\,,
\end{align}
which we have verified to hold at least for $d\le 1000$.
The last representation follows from the regularized quantum period of the Fano 
fourfold $V^4_{14}$~\cite[\S F.14]{Coates2016}.
The equality between with the last representation can be established directly using the same method as  in~\cite{Paule2003}.
After multiplication with  $\binomial{2d}{d}$, one recovers the holomorphic solution of the CY operator \#26 in the AESZ list. 

\medskip

\item $m=8$:
\be
\begin{split}
        c_8(d)=&\sum\limits_{l_1,l_2,l_3\ge 0}\binom{d}{l_1}^2 \binom{l_1}{l_2} \binom{l_1}{l_3} \binom{l_1}{d-l_2}\binom{l_1}{d-l_3}\\
        =&\sum\limits_{l= 0}^d\binom{d}{l}^2\binom{2l}{d}^2\\
        =& \{1, 4, 40, 544, 8536, 145504,\dots\}
        \label{eqn:fundamental_c8}
\end{split}
\ee
This is checked to agree with the regularized quantum period of the Fano fourfold $V^4_{16}$~\cite[\S F.15]{Coates2016} for high values of $d$, although the latter expression is too complex to display. After multiplication with  $\binomial{2d}{d}$, one recovers the holomorphic solution of the CY operator \#42 in the AESZ list.

\medskip

\item $m=9$:
\be
\begin{split}
        c_9(d)=&\sum\limits_{l_1,l_2\ge 0}\binom{d}{l_1}^2 \binom{d}{l_2} \binom{l_1}{l_2}\binom{l_1+l_2}{d}
        \\
        =& \sum\limits_{l_1,l_2,l_3\ge 0}\binom{l_1}{l_2}\binom{l_1}{l_3}\binom{d}{l_1}^2\binom{d}{d-l_2}\binom{l_2}{d-l_3}
        \\
        =&\{1, 3, 27, 309, 4059, 57753,\dots\}
        \label{eqn:fundamental_c9}
\end{split}
\ee
This is checked to agree with the regularized quantum period of the Fano fourfold $V^4_{18}$~\cite[\S F.16]{Coates2016} for high values of $d$, although the latter expression is too complex to display. After multiplication with  $\binomial{2d}{d}$, one recovers the holomorphic solution of the CY operator \#185 in the AESZ list.

\noindent

\item $m=10$ 
\be
\begin{split}
c_{10}(d)=&
 \sum_{j=0}^d \sum_{\ell=0}^{j}  \binomial{j}{\ell}^4 \binomial{j}{2j-d} \\
 =& \{1, 2, 20, 200, 2320, 28984, 381128, \dots\}
 \end{split}
 \label{eqn:fundamental_c10}
 \ee

\item $m=11$:
The coefficient are not known in any simple closed form, but can be determined to any order by the recurrence relation in \cite[Thm 4.7]{cooper2015hypergeometric}, equivalent to the Picard-Fuchs equation in Table \ref{tabPFK3}, 
\be
c_{11}(d)=\{1, 4, 28, 268, 3004, 36784, 476476, 6418192, 88986172, 1261473136, \dots\}
 \label{eqn:fundamental_c11}
\ee
The numbers $c_{11}(d)$ can also be obtained as the coefficient of $(xyz)^d$ in
\begin{align}
    (z+(1+y(1+z(1+x)))(1+x(1+z(1+y))))^d\,.
\end{align}
\end{itemize}

\section{Conifold transitions for $m=6,7,8$}
\label{sec_Amodel}

\subsection{The non-toric case $m=6$}
Let us now study $X_6^{[1,1]}=X_{6,0}^{[1,1]}$.
As discussed in Appendix~\ref{sec:nonToricFundamentals}, the coefficients $c_6(d)$ of the fundamental period $f_6(\lambda)$ of $\Sigma_6$ can be expressed as
\begin{align}
    c_6(d)=\sum\limits_{l_1,l_2,l_3\ge 0}\binom{l_1}{l_2}\binom{l_1}{l_3}\binom{d}{l_1}^2\binom{d}{l_2}\binom{d}{l_3}\,.
\end{align}
This leads us to consider the expression
\begin{align*}
    \varpi_0'(z_1,z_2,w_1,w_2,w_3)=\sum\limits_{d,k,l_1,l_2,l_3\ge 0}\binom{l_1}{l_2}\binom{l_1}{l_3}\binom{d}{l_1}^2\binom{d}{l_2}\binom{d}{l_3}\binom{d+k}{d}^2z_1^dz_2^kw_1^{l_1}w_2^{l_2}w_3^{l_3}\,,
    \label{eqn:m6period5par}
\end{align*}
which satisfies
\begin{align*}
    \varpi_6^{[1,1]}(z_1,z_2)=\varpi_0'(z_1,z_2,1,1,1)\,.
\end{align*}
By interpreting the coefficients that appear in the factorials in~\eqref{eqn:m6period5par} as coefficients of linear relations among points we can reconstruct the toric data in Table~\ref{tab:tdatam6} associated to a CY threefold $\widetilde{X}_{6}^{[1,1]}$ such that~\eqref{eqn:m6period5par} is the fundamental period of the mirror.
\begin{table}[ht!]
\begin{align*}
\begin{array}{c}\\u_1\\u_2\\u_3\\u_4\\u_5\\u_6\\u_7\\u_{8}\\u_{9}\\u_{10}\\v_{1}\\v_{2}\\\vspace{.25cm}\\ \\ \\ \\ \end{array}
\left[\begin{array}{ccccccc|ccccc|c}
\multicolumn{7}{c|}{\vec{p}\in\mathbb{Z}^n}& l^{(1)} & l^{(2)} & l^{(3)} & l^{(4)} & l^{(5)} & \Delta_i\\\hline
 1 & 0 & 0 & 0 & 0 & 0 & 0 & 1 & 0 & -1 & 0 & 0 & 1 \\
 0 & 1 & 0 & 0 & 0 & 0 & 0 & 1 & 0 & -1 & 0 & 0 & 2 \\
 0 & 0 & 1 & 0 & 0 & 0 & 0 & 1 & 0 & 0 & -1 & 0 & 3 \\
 -1 & -1 & -1 & 0 & 0 & 0 & 0 & 1 & 0 & 0 & 0 & -1 & 4 \\
 0 & 0 & 0 & 1 & 0 & 0 & 0 & 0 & 0 & 1 & -1 & 0 & 1 \\
 1 & 1 & 0 & -1 & 0 & 0 & 0 & 0 & 0 & 1 & 0 & -1 & 2 \\
 0 & 0 & 0 & 0 & 1 & 0 & 0 & 0 & 0 & 0 & 1 & 0 & 1 \\
 0 & 0 & 1 & 1 & -1 & 0 & 0 & 0 & 0 & 0 & 1 & 0 & 3 \\
 0 & 0 & 0 & 0 & 0 & 1 & 0 & 0 & 0 & 0 & 0 & 1 & 2 \\
 0 & 0 & -1 & -1 & 0 & -1 & 0 & 0 & 0 & 0 & 0 & 1 & 4 \\
 0 & 0 & 0 & 0 & 0 & 0 & 1 & 0 & 1 & 0 & 0 & 0 & 3 \\
 0 & 0 & 0 & 0 & 0 & 0 & -1 & 0 & 1 & 0 & 0 & 0 & 4 \\\hline\hline
  0 & 0 & 0 & 0 & 0 & 0 & 0 & -1 & 0 & 0 & 0 & 0 & 1 \\
 0 & 0 & 0 & 0 & 0 & 0 & 0 & -1 & 0 & 0 & 0 & 0 & 2 \\
 0 & 0 & 0 & 0 & 0 & 0 & 0 & -1 & -1 & 0 & 0 & 0 & 3 \\
 0 & 0 & 0 & 0 & 0 & 0 & 0 & -1 & -1 & 0 & 0 & 0 & 4
\end{array}\right]
\end{align*}
        \caption{The toric data associated to the Calabi-Yau $\widetilde{X}_{6,0}^{[1,1]}$ that is connected to $X_{6,0}^{[1,1]}$ via a conifold transition.
 The format is explained in the description of Table~\ref{tab:tdatam2}.}
	\label{tab:tdatam6}
\end{table}

Using affine coordinates $x_1',\ldots,x_7'$ on $(\mathbb{C}^\times)^7$ we obtain a dense open subset of the Batyrev-Borisov mirror as the complete intersection of the vanishing loci of the polynomials
\begin{align}
    \begin{split}
        p_1=&1+x_1'+x_4'+x_5'\,,\quad p_2=1+x_2'+x_6'+w_1\frac{x_1'x_2'}{x_4'}\,,\\
        p_3=&1+x_3'+x_7'+w_2\frac{x_3'x_4'}{x_5'}\,,\quad p_4=1+\frac{z_1}{x_3'}\left(\frac{1}{x_1'x_2'}+\frac{w_1w_3}{x_4'x_6'}\right)+\frac{z_2}{x_7'}\,.
    \end{split}
\end{align}
To follow the conifold transition we set $w_1=w_2=w_3=1$. Using $p_i$ to eliminate $x_i'$ for $i=1,2,3$ and introducing
\begin{align}
    x_1=x_4'\,,\quad x_2=x_5'\,,\quad y_1=x_6'\,,\quad y=-x_7'\,,
\end{align}
we then obtain
\begin{align}
    p=(x_1+x_2)(1+x_1+x_2+x_1y_1)-\frac{1}{z_1}(1-y)(1-z_2/y)x_1x_2(1+x_1+x_2)y_1(1+y_1)\,.
    \label{eqn:pm6}
\end{align}
This can be compactified to a fibration of
\begin{align}
    \{\,(x_1+x_2)(x_1+x_2+x_3)(x_1y_1+x_3y_1+x_3y_2)(y_1+y_2)-\lambda x_1x_2x_3y_1y_2=0\,\}\subset\mathbb{P}^2\times\mathbb{P}^1\,,
\end{align}
and we recover~\eqref{eqn:pm6} after identifying $\lambda=\Lambda_0^{[1,1]}(y;z_1,z_2)$ and setting $x_3\rightarrow -1-x_1-x_2$, $y_2\rightarrow -1-y_1$.

\subsection{The non-toric case $m=7$}
\label{eqn:nontoric7}
To construct $X_7^{[1,1]}=X_{7,0}^{[1,1]}$ we again consider a codimension six complete intersection CY threefold $\widetilde{X}_{7}^{[1,1]}$ in a toric ambient space and then perform a conifold transition.
The toric data associated to $\widetilde{X}_{7}^{[1,1]}$ is given in Table~\ref{tab:tdatam7}.
The toric ambient space takes the form $P_{\Delta}=\widehat{P}(2,6)\times\mathbb{P}^1$, where $\widehat{P}(2,6)$ is a small resolution of the toric degeneration $P(2,6)$ of $\text{Gr}(2,6)$ that has also been discussed in~\cite{sturmfels1996grobner,Batyrev:1998kx}.
Following the conifold transition from $\widehat{P}(2,6)$ to $\text{Gr}(2,6)$ we then obtain $X_7^{[1,1]}$ as defined in Section~\ref{eqn:nontoric7}.

\begin{table}[ht!]
\begin{align*}
	\begin{array}{c}\\u_1\\u_2\\u_3\\u_4\\u_5\\u_6\\u_7\\u_{8}\\u_{9}\\u_{10}\\u_{11}\\u_{12}\\v_{1}\\v_{2}\\\vspace{.3cm}\\ \\ \\ \\ \\ \\ \end{array}
	\left[
	\begin{array}{ccccccccc|ccccc|c}
	\multicolumn{9}{c|}{\vec{p}\in\mathbb{Z}^n}& l^{(1)} & l^{(2)} & l^{(3)} & l^{(4)} & l^{(5)} & \Delta_i\\\hline
	 1 & 0 & 0 & 0 & 0 & 0 & 0 & 0 & 0 & 1 & 0 & 0 & 0 & 0 & 5 \\
	 0 & 1 & 0 & 0 & 0 & 0 & 0 & 0 & 0 & 1 & 0 & 0 & 0 & 0 & 6 \\
	 0 & 0 & 1 & 0 & 0 & 0 & 0 & 0 & 0 & 1 & 0 & 0 & 0 & -1 & 1 \\
	 0 & 0 & 0 & 1 & 0 & 0 & 0 & 0 & 0 & 1 & 0 & 0 & -1 & 0 & 2 \\
	 0 & 0 & 0 & 0 & 1 & 0 & 0 & 0 & 0 & 1 & 0 & -1 & 0 & 0 & 3 \\
	 -1 & -1 & -1 & -1 & -1 & 0 & 0 & 0 & 0 & 1 & 0 & -1 & 0 & 0 & 4 \\
	 0 & 0 & 0 & 0 & 0 & 1 & 0 & 0 & 0 & 0 & 0 & 0 & 0 & 1 & 1 \\
	 0 & 0 & 0 & 0 & 0 & 0 & 1 & 0 & 0 & 0 & 0 & 0 & 1 & 0 & 2 \\
	 0 & 0 & 0 & 0 & 0 & 0 & 0 & 1 & 0 & 0 & 0 & 1 & 0 & 0 & 3 \\
	 -1 & -1 & 0 & 0 & 0 & -1 & -1 & -1 & 0 & 0 & 0 & 0 & 0 & 1 & 4 \\
	 -1 & -1 & -1 & 0 & 0 & 0 & -1 & -1 & 0 & 0 & 0 & 0 & 1 & -1 & 4 \\
	 -1 & -1 & -1 & -1 & 0 & 0 & 0 & -1 & 0 & 0 & 0 & 1 & -1 & 0 & 4 \\
	 0 & 0 & 0 & 0 & 0 & 0 & 0 & 0 & 1 & 0 & 1 & 0 & 0 & 0 & 5 \\
	 0 & 0 & 0 & 0 & 0 & 0 & 0 & 0 & -1 & 0 & 1 & 0 & 0 & 0 & 6 \\\hline\hline
	 0 & 0 & 0 & 0 & 0 & 0 & 0 & 0 & 0 & -1 & 0 & 0 & 0 & 0 & 1 \\
	 0 & 0 & 0 & 0 & 0 & 0 & 0 & 0 & 0 & -1 & 0 & 0 & 0 & 0 & 2 \\
	 0 & 0 & 0 & 0 & 0 & 0 & 0 & 0 & 0 & -1 & 0 & 0 & 0 & 0 & 3 \\
	 0 & 0 & 0 & 0 & 0 & 0 & 0 & 0 & 0 & -1 & 0 & 0 & 0 & 0 & 4 \\
	 0 & 0 & 0 & 0 & 0 & 0 & 0 & 0 & 0 & -1 & -1 & 0 & 0 & 0 & 5 \\
	 0 & 0 & 0 & 0 & 0 & 0 & 0 & 0 & 0 & -1 & -1 & 0 & 0 & 0 & 6 \\
	\end{array}
	\right]
\end{align*}
        \caption{The toric data associated to the Calabi-Yau $\widetilde{X}_{7}^{[1,1]}$ that is connected to $X_{7}^{[1,1]}$ via a conifold transition.
 The format is explained in the description of Table~\ref{tab:tdatam2}.}
	\label{tab:tdatam7}
\end{table}

The mirror fundamental period associated to $\widetilde{X}_{7,0}^{[1,1]}$ takes the form
\begin{align}
\begin{split}
    \varpi_0'(z_1,z_2,w_1,w_2,w_3)
    =&\sum\limits_{d,k,l_1,l_2,l_3\ge 0}\binom{l_1}{l_2}\binom{l_2}{l_3}\binom{d}{l_1}^2\binom{d}{l_2}\binom{d}{l_3}\binom{d+k}{d}^2z_1^dz_2^kw_1^{l_1}w_2^{l_2}w_3^{l_3}\,.
    \end{split}
\end{align}
As discussed in Appendix~\ref{sec:nonToricFundamentals} we can identify
\begin{align}
    c_7(d)=\sum\limits_{l_1,l_2,l_3\ge 0}\binom{l_1}{l_2}\binom{l_2}{l_3}\binom{d}{l_1}^2\binom{d}{l_2}\binom{d}{l_3}\,,
\end{align}
as the coefficients of the fundamental period of the mirror $\Sigma_7$ and obtain
\begin{align}
    \varpi_7^{[1,1]}(z_1,z_2)=\varpi_0'(z_1,z_2,1,1,1)\,.
\end{align}

Using affine coordinates $x_1',\ldots,x_9'$ on $(\mathbb{C}^\times)^9$, a dense open subset of the Batyrev-Borisov mirror is the complete intersection of the vanishing loci of the polynomials
\begin{align}
	\begin{split}
		p_1=&1+x'_3+x'_6\,,\quad p_2=1+x'_4+x'_7\,,\quad p_3=1+x'_5+x'_8\,,\quad p_5=1+x'_1+x'_9\,,\\
		p_4=&1+\frac{z_1}{x'_1 x'_2}\left(\frac{1}{x'_3 x'_4 x'_5}+\frac{w_1}{x'_3 x'_4 x'_8}+\frac{w_1 w_2}{x'_3 x'_7 x'_8}+\frac{w_1 w_2
		w_3}{x'_6 x'_7 x'_8}\right)\,,\quad p_6=1+x'_2+\frac{z_2}{x'_9}\,.
	\end{split}
\end{align}
We then set $w_1=w_2=w_3=1$, use $p_1,p_2,p_3,p_5,p_6$ to eliminate $x_i,\,i=1,\ldots,5$ and define
\begin{align}
	r_1=x_6'\,,\quad r_2=x_7'\,,\quad r_3=x_8'\,,\quad y=-x_9'\,,
\end{align}
to identify this with the hypersurface $\{\,p'=0\,\}\subset(\mathbb{C}^{\times})^4$ where
\begin{align}
	p'=1+r_2+r_3+r_1r_2+r_2r_3+\frac{1}{z_1}(1-y)\left(1-\frac{z_2}{y}\right)r_1(1+r_1)r_2(1+r_2)r_3(1+r_3)\,.
	\label{eqn:pm7}
\end{align}
This can be compactified to a fibration of
\begin{align}
	\left\{\, \left(r_1 r_2 r_3-r_1 r_2 s_3+r_1 s_2 s_3-s_1 s_2 s_3\right)\prod_{i=1}^3\left(r_i-s_i\right)+\lambda  r_1 r_2 r_3 s_1 s_2 s_3=0\,\right\}\subset\left(\mathbb{P}^1\right)^3\,,
\end{align}
and we recover~\eqref{eqn:pm7} after identifying $\lambda=\Lambda_0^{[1,1]}(y;z_1,z_2)$ and setting $s_i\rightarrow r_i+1,\,i=1,2,3$.

\subsection{The non-toric case $m=8$}
\label{eqn:nontoric8}
\begin{table}[ht!]
\begin{align*}
\begin{array}{c}\\u_1\\u_2\\u_3\\u_4\\u_5\\u_6\\u_7\\u_{8}\\v_{1}\\v_{2}\\\vspace{.21cm}\\ \\ \\ \\ \end{array}
\left[\begin{array}{ccccccc|ccc|c}
\multicolumn{7}{c|}{\vec{p}\in\mathbb{Z}^n}& l^{(1)} & l^{(2)} & l^{(3)} & \Delta_i\\\hline
 1 & 0 & 0 & 0 & 0 & 0 & 0 & -1 & 0 & 2 & 2 \\
 0 & 1 & 0 & 0 & 0 & 0 & 0 & -1 & 0 & 2 & 1 \\
 0 & 0 & 1 & 0 & 0 & 0 & 0 & 0 & 0 & 1 & 4 \\
 0 & 0 & 0 & 1 & 0 & 0 & 0 & 0 & 0 & 1 & 3 \\
 0 & 0 & 0 & 0 & 1 & 0 & 0 & 1 & 0 & -1 & 4 \\
 2 & 2 & 1 & 1 & -1 & 0 & 0 & 1 & 0 & -1 & 3 \\
 0 & 0 & 0 & 0 & 0 & 1 & 0 & 1 & 0 & 0 & 2 \\
 -1 & -1 & -1 & -1 & 0 & -1 & 0 & 1 & 0 & 0 & 1 \\
 0 & 0 & 0 & 0 & 0 & 0 & 1 & 0 & 1 & 0 & 4 \\
 0 & 0 & 0 & 0 & 0 & 0 & -1 & 0 & 1 & 0 & 3 \\\hline\hline
  0 & 0 & 0 & 0 & 0 & 0 & 0 & 0 & 0 & -2 & 1 \\
 0 & 0 & 0 & 0 & 0 & 0 & 0 & 0 & 0 & -2 & 2 \\
 0 & 0 & 0 & 0 & 0 & 0 & 0 & -1 & -1 & 0 & 3 \\
 0 & 0 & 0 & 0 & 0 & 0 & 0 & -1 & -1 & 0 & 4 \\
\end{array}\right]
\end{align*}
        \caption{The toric data associated to the Calabi-Yau $\widetilde{X}_{8}^{[1,1]}$ that is connected to $X_{8}^{[1,1]}$ via a conifold transition.
 The format is explained in the description of Table~\ref{tab:tdatam2}.}
	\label{tab:tdatam8}
\end{table}

To construct the conifold transition to $X_8^{[1,1]}$ we first recall the coefficients of the fundamental period of an $M_8$-polarized K3 surface $\Sigma^8$ from~\eqref{eqn:fundamental_c8},
\begin{align}
 c_8(d)=&\sum\limits_{l= 0}^d\binom{d}{l}^2\binom{2l}{d}^2\,,
\end{align}
and write down the expansion
\begin{align*}
    \varpi_0'(z_1,z_2,w_1)=\sum\limits_{d,k,l\ge 0}\binom{d}{l}^2\binom{2l}{d}^2\binom{d+k}{d}^2z_1^dz_2^kw_1^{l}\,,
    \label{eqn:m8period3par}
\end{align*}
which satisfies
\begin{align*}
    \varpi_8^{[1,1]}(z_1,z_2)=\varpi_0'(z_1,z_2,1)\,.
\end{align*}
We can identify~\eqref{eqn:m8period3par} as the mirror fundamental period of a CY threefold $\widetilde{X}_8^{[1,1]}$ that is a complete intersection in a toric ambient space with the toric data summarized in Table~\ref{tab:tdatam8}.
Using affine coordinates $x_1',\ldots,x_7'$ on $(\mathbb{C}^\times)^7$ we obtain a dense open subset of the Batyrev-Borisov mirror as the complete intersection of the vanishing loci of the polynomials
\begin{align}
    \begin{split}
          p_1=&1+x'_2+\frac{w_1}{x'_1 x'_2 x'_3 x'_4 x'_6}\,,\quad p_2=1+x'_1+z_1 x'_6\,,\\
          p_3=&1+x'_4+\frac{(x'_1 x'_2)^2 x'_3 x'_4}{w_1 x'_5}+\frac{1}{x'_7}\,,\quad 1+x'_3+x'_5+z_2 x'_7\,.     \end{split}
\end{align}
The conifold transition is implemented by setting $w_1=1$. Using $p_i$ to eliminate $x_i'$ for $i=1,2,4$, introducing
\begin{align}
    r_1=\frac{x_5'}{1+z_2x_7'}\,,\quad r_2=x_1'\,,\quad r_3=x_2'\,,\quad y=-1/x_7'\,,
\end{align}
and dropping an overall factor, we then obtain 
$\{\,p=0\,\}\subset(\mathbb{C}^\times)^4$ with
\begin{align}
    p=\left(1+r_1\right) r_2^2 r_3^2-r_1+\frac{1}{z_1}(1-y)\left(1-\frac{z_2}{y}\right) r_1 \left(1+r_1\right) r_2 \left(1+r_2\right) r_3\left(1+r_3\right)\,.
    \label{eqn:pm8}
\end{align}
This can be compactified to a fibration of
\begin{align}
    \{\,\left(r_1-s_1\right) \left(r_1 \left(r_2-s_2\right){}^2 \left(r_3-s_3\right)^2-r_2^2r_3^2 s_1\right)+\lambda  r_1 r_2 r_3 s_1 s_2 s_3=0\,\}\subset\left(\mathbb{P}^1\right)^3\,,
\end{align}
and we recover~\eqref{eqn:pm8} after identifying $\lambda=\Lambda_0^{[1,1]}(y;z_1,z_2)$ and setting $s_i\rightarrow r_i+1,\,i=1,2,3$.

\section{Modular forms for $\Gamma_0(m)^+$}
\label{sec_Gamma0m}

In this section, we collect basic facts about modular forms of $\Gamma_0(m)$ and its Fricke extension $\Gamma_0(m)^+$. Recall
that $\Gamma_0(m)$ is the group integer matrices of matrices $g=\begin{pmatrix} a & b \\ c & d \end{pmatrix}$ with $ad-bc=1$ and $c=0\mod m$. It is a  congruence subgroup of $SL(2,\IZ)$, of index
$m+1$ with $m$ is prime, or more generally of index
$m \prod_{p|m} \left(1+\tfrac{1}{p}\right)$, generated by the
matrices $T=\begin{pmatrix} 1 & 1 \\ 0 & 1 \end{pmatrix}$,
$S=\begin{pmatrix} 1 & 0 \\ m & 1 \end{pmatrix}$ and 
$P=\begin{pmatrix} -1 & 0 \\ 0 & -1 \end{pmatrix}$. $\Gamma_0(m)$
acts on the upper half-plane $\IH$ by $\tau\mapsto \frac{a\tau+b}{c\tau+d}$, with $P$ acting trivially. The quotient $\IH/\Gamma_0(m)$ can be compactified by adding $h=\sum_{d|m}\varphi((d,m/d))$ cusps (where $\varphi(n)$ is the totient function), corresponding to the orbits of the points
$\tau=b/d$ with $d|m$, $(b,d)=1$ and $0\leq b< (d,m/d)$. This includes the cusp at infinity $\tau=\I\infty$ and $\tau=0$;
these are the only cusps when $m$ is not prime. The modular curve
$X_0(m)$ is the corresponding compactification of $\IH/\Gamma_0(m)$. For $m\leq 10$ or $m\in\{12,13,16,18,25\}$,
$X_0(m)$ is a rational curve with $h$ marked points.

For any $g\in \Gamma_0(m)$ and integer $k$, we define the Peterson slash operator $\vert_k g$ acting on
functions $f(\tau)$ on $\IH$ by
\be
(f\vert_k g)(\tau) = (c\tau+d)^{-k}\, 
f\left( \frac{a\tau+b}{c\tau+d}
\right)
\ee
Modular forms of weight $k$ (and level $m$) are holomorphic functions $f(\tau)$ on $\IH$, bounded at all cusps, such that $(f\vert_k g)(\tau)=f(\tau)$ for all $g\in\Gamma_0(m)$. In particular, $f(\tau)$ vanishes unless $k$ is even, is constant unless $k\geq 2$, and is invariant under $\tau\mapsto\tau+1$,
hence admits a convergent Fourier expansion $f(\tau)=\sum_{n\geq 0} f_n q^n$ with $q=e^{2\pi\I\tau}$.
Cusp forms are modular forms which vanish at all cusps (in particular, $f_0=0$). 
Weak modular forms are holomorphic functions on $\IH$, invariant under $\vert_k g$,  with poles of finite order at the cusps. 
We denote by $M_k(\Gamma_0(m))$  the space of modular forms of weight $k$.

The  space $M_k(\Gamma_0(m))$ contains $M_k(\Gamma_0(d))$ for any $d|m$, and injects into $M_k(\Gamma_0(Nm))$ for any $N>1$ by mapping $f(\tau)\mapsto f_N(\tau)=f(N\tau)$. The image of these maps spans the so-called `old forms'. In particular, the ring
$M_\star(\Gamma_0(m)):=\oplus_{k\geq 0} M_k(\Gamma_0(m))$  
is a module  over $M_\star(SL(2,\IZ))$.
The latter is the polynomial ring generated by the Eisenstein series 
\be
E_4(\tau):=1+240\sum_{n\geq 1} \frac{n^3 q^n}{1-q^n},
\quad E_6(\tau) := 1-504 \sum_{n\geq 1} \frac{n^5 q^n}{1-q^n}
\ee
In particular, $M_\star(\Gamma_0(m))$ contains $E_4(\tau),E_4(m\tau)$ in weight 4 and $E_6(\tau),E_6(m\tau)$ in weight 6. In addition, the linear combination
\be
\label{defXm}
E_2^{(m)}(\tau):= E_2(\tau) - m E_2(m\tau)\ ,\quad 
E_2(\tau):= 1-24 \sum_{n\geq 1} \frac{n q^n}{1-q^n}
\ee
defines a weight 2, level $m$ modular form.\footnote{In contrast, the Eisenstein series
$E_2(\tau)$ is quasi-modular, with $\widehat E_2(\tau):=E_2(\tau)-\frac{3}{\pi\tau_2}$ transforming as a (non-holomorphic) modular form of weight $2$. The ring of quasi-modular forms of level $m$ is defined as $M_\star(\Gamma_0(m))[E_2]$.}
Since $M_\star(\Gamma_0(m))$ is a finitely generated module of rank $h$ over $M_\star(SL(2,\IZ))$, the generating series of dimensions is a rational function of the form
\be
\label{eqHilbert}
\sum_{k\geq 0} \dim M_k(\Gamma_0(m)) t^{k} = \frac{P_m(t)}{(1-t^4)(1-t^6)}
\ee
where $P_m(t)$ is a polynomial with $P_m(1)=h$.  
As shown in \cite{skoruppa1987jacobi}, the dimensions are given for any $k\geq 2, m\geq 1$ by 
\be
\label{dimMk}
\dim M_k(\Gamma_0(m)) = \sum_{m'|m \atop m/m' \mbox{\tiny square-free}}
\left[ \delta( m'(k-1)) + \frac12 a' \right]
\ee
where $\delta(n)=\frac{n}{12}-\frac13 \chi_3(n)-\frac14\chi_4(n)$, with $\chi_3$ and $\chi_4$ the primitive Dirichlet characters of order 3 and 4, and $a'$ is the unique positive integer such that $m'=a'^2 b'$ with $b'$ square-free.

When $m$ is such that 
$X_0(m)$ has genus zero, there exists a weak modular modular form,
$J_m(\tau)$ of weight 0, unique up to Moebius transformations,
which maps $X_0(m)$ to $\IP_1$ isomorphically and generates the field of weight 0 weak modular forms. This weak modular form,
known as Hauptmodul, is determined uniquely by requiring the behavior $J_m(\tau)=\frac{1}{q}+\cO(q)$
at $\tau\to\I\infty$. For $m=1$, this is
\be
J_1 = J-744 =\frac{E_4^3}{\eta^{24}} -744 = \frac{1}{q}+ 196884 q + 21493760 q^2 + 864299970 q^3+\dots  
\ee
where $\eta(\tau):=q^{1/24}\prod_{n\geq 1}(1-q^n)$ is the Dedekind eta function. As a manifestation of Monstruous Moonshine, $J_m(\tau)$ is equal, up to an additive constant, to the McKay-Thompson series attached to a suitable conjugacy class of the Monster \cite{conway1979monstrous}.

The group $\Gamma_0(m)^+$ is defined as $\Gamma_0(m)\cup \Gamma_0(m) W_m \subset SL(2,\IQ)$, where  $W_m=\tfrac{1}{\sqrt{m}}\begin{pmatrix} 0 & -1 \\ m & 0 \end{pmatrix}$ is the Fricke involution. $W_m$ acts on $\IH$ as $\tau\mapsto -\frac{1}{m\tau}$, and identifies the cusps of $\Gamma_0(m)$ pairwise (except for $m=4$, where the cusp at $1/2$ is invariant). It
acts on $M_k(\Gamma_0(m))$ via
\be
f(\tau) \mapsto (f\vert_{W_m})(\tau) = \tau^{-k} m^{-k/2}\, 
f\left(-\frac{1}{m\tau} \right)
\ee
Modular forms of $\Gamma_0(m)^+$ are modular forms of $\Gamma_0(m)$ which are invariant under the Fricke involution.
This includes in particular the even linear combinations
$E_{2n}^+$ defined for $n\geq 2$ by
\be
E_{2n}^\pm (\tau):= E_{2n}(\tau) \pm m^{n} E_{2n}(m\tau)
\ee
but not $E_2^{(m)}$ in \eqref{defXm}, which is odd under Fricke. The modular curve $X_0(m)^+$ is the quotient of $X_0(m)$ under the Fricke involution; unlike $X_0(m)$, it is rational for all $m\leq 20$
as well as $m=24,26,28,30,31,\dots$ (see \cite{conway1979monstrous}). In those cases, there again exists a weak modular modular form, determined uniquely 
by requiring the behavior $J^+_m(\tau)=\frac{1}{q}+c_m+\cO(q)$
at $\tau\to\I\infty$, which maps
$X^+_0(m)$ to $\IP_1$ isomorphically and generates the field of
weak modular modular forms of weight 0 under $\Gamma_0(m)^+$.
It also coincides with the McKay-Thompson series attached to a suitable conjugacy class of the Monster, up to an additive constant.
Unlike $J_m$, we shall not
require that the constant $c_m$ in $J^+_m(\tau)$ vanishes, but rather fix it to a particular value which is natural in the context of mirror symmetry for $M_n$-polarized K3 surfaces. When $X_0(m)$ itself
has genus 0, one has $J_m^+=J_m + J_m\vert_{W_m} +\gamma_m$ where
$J_m\vert_{W_m}=\frac{\alpha_m J_m}{J_m + \beta_m}$ for some real constants 
$\alpha_m,\beta_m,\gamma_m$. For $m=1$, we set 
$J_1^+=E_4^3/{\eta^{24}}=J$. With this definition, the fundamental period for $M_2$-polarized K3 surfaces is given by
\bea
f_1(J_1^+(\tau)) &=& \sum_{d\geq 1} \frac{(6d)!}{(d!)^3 (3d)!}
(J_1^+(\tau))^{-d} = \sqrt{E_4(\tau)} \nn\\
&=&1 + 120 q - 6120 q^2 + 737760 q^3 - 107249640 q^4+\dots 
\eea

In the following, we record the generators of the ring of modular forms and the Hauptmoduls of $\Gamma_0(m)$ and $\Gamma_0(m)^+$ for $m\leq 11$.

\subsection{$m=2$}
The congruence subgroup $\Gamma_0(2)=\Gamma_1(2)$ has index 3 in $SL(2,\IZ)$ and 2 cusps at $\tau=\I\infty$ and $0$. The ring of modular forms is generated by $E_2^{(2)}(\tau)$ and 
by $E_4^{(2,-)}=E_4(\tau)-4E_4(2\tau)=2E_4(\tau)-5(E_2^{(2)})^2$, which are
both under the Fricke involution. 
The  dimensions of $M_k(\Gamma_0(2))$ are given by 
\bea
\sum_{k\geq 0} \dim M _k(\Gamma_0(2))\, x^k &=& 
\frac{1+x^2+x^4}{(1-x^4)(1-x^6)} = \frac{1}{(1-x^2)(1-x^4)}
\nn\\
&=& 1+ (0+1)x^2+(1+1)x^4+(1+1)x^6+(2+1)x^8
+(1+2)x^{10} \nn\\&&  +(2+2)x^{12}
+(2+2)x^{14}+(3+2)x^{16}+(2+3)x^{18}
+(3+3)x^{20}+\dots\nn\\&&
\eea 
where we indicated the dimensions of the Fricke-even and Fricke-odd subspaces. The first cusp form occurs at weight 8 and is even under Fricke, 
\be
\begin{split}
\Xi_2(\tau)=& [\eta(\tau)\eta(2\tau)]^8 
=  -\frac{1}{144} (E_2^{(2)})^4 + \frac{5}{576} (E_2^{(2)})^2 E_4 - \frac{1}{576} E_4^2 \\
= & q - 8 q^2+12 q^3+64 q^4-210 q^5+\dots
\end{split}
\ee
The Hauptmodul for $\Gamma_0(2)$ is the McKay-Thompson series associated to the conjugacy class $2B$ of the Monster,
\be
J_2(\tau) = \frac{\Delta(\tau)}{\Delta(2\tau)}+24 = \frac{192 (E_2^{(2)})^2}{E_4-(E_2^{(2)})^2}-40 = 
q^{-1} + 276 q-2048 q^2 + 11202 q^3 + \dots
\ee
The Hauptmodul of $SL(2,\IZ)$ is given in terms of $J_2$ by
\be
J(\tau) = \frac{(J_2-232)^3}{(J_2-24)^2}, \quad 
J(2\tau)=\frac{(J_2-8)^3}{J_2-24}\ .
\ee
The Hauptmodul for $\Gamma_0(2)^+$ is instead the McKay-Thompson series of
class $2A$,
\be
\begin{split}
J_2^+(\tau)=& 
 \frac{(\theta_3^4+\theta_4^4)^4}{16(\eta(\tau)\eta(2\tau))^8}
= J_2 + \frac{4096}{J_2-24} + 104
\\ 
= & q^{-1}+104+4372 q+96256 q^2+1240002q^3+\dots
\end{split}
\ee
With this definition, the fundamental period for $M_2$-polarized K3 surfaces is given by
\bea
f_2(J_2^+(\tau)) &=& \sum_{d\geq 0}\frac{(4d)!}{(d!)^4}
(J_2^+(\tau))^{-d} = -(E_2^{(2)}) \nn\\
&=& 1 + 24 q + 24 q^2 + 96 q^3 + 24 q^4 +\dots 
\eea

\subsection{$m=3$}

The congruence subgroup $\Gamma_0(3)$ has index 4 in $SL(2,\IZ)$ and 2 cusps at $\tau=\I\infty$ and $0$.
The ring of modular forms of $\Gamma_0(3)$
is generated by the Fricke-odd weight 2 form $E_2^{(3)}(\tau)$, the Fricke-odd weight 4 form $E_4^{(3,-)}=E_4(\tau)-9E_4(3\tau)$ and the unique cusp form of weight 6, odd
under Fricke,
\be
\begin{split}
\Xi_3(\tau) = & [\eta(\tau) \eta(3\tau)]^6 =  \frac{1}{384}(E_2^{(3)})^3-\frac{7}{864}(E_2^{(3)}) E_4+\frac{1}{216} E_6\\
 = & q - 6 q^2+9 q^3+4 q^4+6 q^5-54 q^6+\dots
\end{split}
\ee 
The dimensions of $M_k(\Gamma_0(3))$ are given by 
\bea
\sum_{k\geq 0} \dim M _k(\Gamma_0(3))\, x^k &=& 
\frac{1+x^2+x^4+x^6}{(1-x^4)(1-x^6)} = \frac{1+x^4}{(1-x^2)(1-x^6)}\nn\\
&=& 1+x^2+2x^4+3 x^6+3 x^8 + 4 x^{10}+
5 x^{12} + 5 x^{14} + 6 x^{16} + 7 x^{18} + 7 x^{20}+\dots
\eea

The Hauptmodul of $\Gamma_0(3)$ is the McKay-Thompson series of class $3B$
\be
\begin{split}
J_3(\tau)=&\left(\frac{\eta(\tau)}{\eta(3\tau)}\right)^{12}+12 = 
\left(-\frac{23}{128} (E_2^{(3)})^3 + \frac{53}{288} (E_2^{(3)}) E_4 - \frac{5}{72} E_6\right)/\Xi_6^{(3)} \\
=&q^{-1}+54 q-76 q^2-243 q^3+1188 q^4 + \dots
\end{split}
\ee
The Hauptmodul of $SL(2,\IZ)$ is given in terms of $J_3$ by
\be
J(\tau) =  \frac{(J_3+15)(J_3+231)^3}{(J_3-12)^3}, \quad 
J(3\tau) = \frac{(J_3-9)^3(J_3+15)}{J_3-12}  
\ee
The Hauptmodul of $\Gamma_0(3)^+$ is the McKay-Thompson series of class $3A$,
\be
J_3^+ = J_3 +  \frac{729}{J_3-12} +42= q^{-1} +42 + 783 q + 8672 q^2 + 65367 q^3 + \dots\ ,\quad 
\ee
With this definition, the fundamental period for $M_3$-polarized K3 surfaces is given by
\bea
f_3(J_3^+(\tau)) &=& \sum_{d\geq 0} \frac{(2d)!(3d)!}{(d!)^5}
(J_3^+(\tau))^{-d} =-\frac12 E_2^{(3)}
\nn\\ &=& 
1+ 12 q + 36 q^2 + 12 q^3 + 84 q^4+\dots 
\eea

\subsection{$m=4$}
The congruence subgroup $\Gamma_0(4)$ has index 6 in $SL(2,\IZ)$ and 3 cusps at $\tau=\I\infty,0$ and $1/2$. 
The ring of modular forms is freely generated by
the weight 2 elements 
\be
E_2^{(4,+)}(\tau)=E_2^{(2)}(2\tau)+\frac32 [\theta_2(2\tau)]^4
\ ,\quad 
E_2^{(4,-)}(\tau)=E_2^{(2)}(2\tau)-\frac12 [\theta_2(2\tau)]^4
\ee
which are even and odd under the Fricke involution, respectively.
The dimensions of $M_k(\Gamma_0(4))$ are given by 
\bea
\sum_{k\geq 0} \dim M _k(\Gamma_0(4))\, x^k &=& 
\frac{1+2x^2+2x^4+x^6}{(1-x^4)(1-x^6)}=\frac{1}{(1-x^2)^2} \nn\\
&=& 1+2x^2+3x^4+4x^6+5x^8+\dots
\eea
hence the dimension is $\frac{k}{2}+1$, with
$\lceil \frac{k+2}{4} \rceil$ even forms and 
$\lfloor \frac{k+2}{4} \rfloor$ odd forms.
The first cusp form occurs in weight $6$ and is odd under Fricke,
\be
\begin{split}
\Xi_4(\tau) =& \eta(2\tau)^{12} = \frac{1}{16} E_2^{(2)}(2\tau)^2 \theta_2(2\tau)^4 - \frac{1}{64}\theta_2(2\tau)^{12} \\
=&q-12 q^3+54 q^5-88 q^7 - 99 q^9 + \dots
\end{split}
\ee
The Hauptmodul of $\Gamma_0(4)$ is the McKay-Thompson series of class 4C,
\be
\begin{split}
J_4(\tau) = & 16 \frac{\theta_3^4(2\tau)}{\theta_2^4(2\tau)} -8 
= \left(\frac{\eta(\tau)}{\eta(4\tau)}\right)^8 + 8
= \frac{E_2^{(2)}(2\tau) \theta_2^8(2\tau) - 4E_2^{(2)}(2\tau) ^3}{4\Xi_6^{(4)}} \\
=&q^{-1} + 20 q - 62 q^3 +216 q^5 + \dots
\end{split}
\ee
The Hauptmodul of $SL(2,\IZ)$ and $\Gamma_0(2)$ are expressed in terms of $J_4$ as
\be
J(\tau)=\frac{(J_4^2+240J_4+2112)^3}{(J_4-8)^4 (J_4+8)},\quad
J(4\tau) = \frac{(J_4^2-48)^3}{(J_4-8)^2(J_4+8)^2},
\ee
\be
J_2(\tau) = J_4 + \frac{256}{J_4+8}
\ee
The Hauptmodul of $\Gamma_0(4)^+$ is instead 
the McKay-Thompson series of class 4A,
\be
J_4^+ = J_4 + \frac{256}{J_4-8}+24=\frac{1}{q}+24
+276 q + 2048 q^2 + 11202 q^3 + 49152 q^4+\dots
\ee
With this definition, the fundamental period for $M_4$-polarized K3 surfaces is given by
\bea
f_4(J_4^+(\tau)) &=& \sum_{d\geq 0}  \frac{[(2d)!]^3}{(d!)^6} 
(J_4^+(\tau))^{-d} =-\frac13 X_4
\nn\\ &=& 
1 + 8 q + 24 q^2 + 32 q^3 + 24 q^4 \dots 
\eea

\subsection{$m=5$}
The congruence subgroup $\Gamma_0(5)$ has index 6 in $SL(2,\IZ)$ and 2 cusps at $\tau=\I\infty$ and $0$. 
The ring of modular forms is generated by the odd weight 2 form, $(E_2^{(5)})(\tau)$, the odd weight 4 form $E_4^{(5,-)}(\tau)=E_4(\tau) - 25 E_4(5\tau)$
and  the unique cusp form of 
weight 4, even under Fricke,
\be
\begin{split}
\Xi_5 = & [ \eta(\tau)\, \eta(5\tau)]^4 =
 \frac{13}{576}(E_2^{(5)})^2-\frac{1}{72}E_4-\frac{25}{72} E_4(5\tau) \\
= & q - 4 q^2+2 q^3+8 q^4-5q^5-8 q^6+\dots
\end{split}
\ee 
The dimensions of $M_k(\Gamma_0(5))$ are given by 
\bea
\sum_{k\geq 0} \dim M_k(\Gamma_0(5))\, x^k &=& 
\frac{1+x^2+2x^4+x^6+x^8}{(1-x^4)(1-x^6)} 
=\frac{1+x^4}{(1-x^2)(1-x^4)}\nn\\
&=& 1+x^2+3x^4+3 x^6+5 x^8 + 5 x^{10} \\
&&+7 x^{12}+7 x^{14}+9 x^{16}+9 x^{18}+11
   x^{20}+\dots \nn
\eea
The Hauptmodul of $\Gamma_0(5)$ is the McKay-Thompson series of class $5B$
\be
\begin{split}
J_5(\tau)=& \left(\frac{\eta(\tau)}{\eta(5\tau)}\right)^6+6 = \frac{28 E_4+1300 E_4(5\tau)-47 (E_2^{(5)})^2}{576
\Xi_4^{(5)}}\\
&= q^{-1}+9q+10q^2-30q^3+6 q^4-25 q^5+\dots
\end{split}
\ee
such that
\be
J(\tau)=\frac{(J_5^2+238  J_5+166)^3}{(J_5-6)^5},\quad
J(5\tau)=   \frac{\left(J_5^2-2
   J_5-19\right)^3}{J_5-6}
   \ee
The Hauptmodul of $\Gamma_0(5)^+$ is instead the McKay-Thompson series of class 5A,
\be
J_5^+ = J_5 + \frac{125}{J_5-6}+16 =
\frac{1}{q} + 16 + 134 q + 760 q^2 + 3345 q^3 + 12256 q^4 +\dots\ .
\ee
With this definition, the fundamental period for $M_{5}$-polarized K3 surfaces is given by
\bea
f_5(J_5^+(\tau)) &=& \sum_{d\geq 0}  c_5(d)
(J_5^+(\tau))^{-d} =-\frac14 E_2^{(5)}
=
1 + 6 q + 18 q^2 + 24 q^3 + 42 q^4 + \dots 
\eea
where the coefficients $c_5(d)$ are given in \eqref{eqn:fundamental_c5}.

\subsection{$m=6$}
The congruence subgroup $\Gamma_0(6)$ has index 12 in $SL(2,\IZ)$ and 4 cusps at at $\tau=\I\infty,0,1/2,1/3$. The ring of modular forms under is generated by the weight 2 forms
$E_2^{(6)}$ and 
\bea
E_2^{(6,-)}(\tau)&:=&2E_2^{(3)}(2\tau)-3E_2^{(2)}(3\tau)\nn\\
E_2^{(6,+)}(\tau)&:=&E_2^{(6)}(\tau)-3E_2^{(2)}(3\tau)-2E_2^{(3)}(2\tau)
\eea
which are odd, odd and even under the Fricke involution, respectively. 
The dimensions of $M_k(\Gamma_0(6))$ are given by 
\bea
\sum_{k\geq 0} \dim M_k(\Gamma_0(6))\, x^k &=& 
\frac{1+3x^2+4x^4+3x^6+x^8}{(1-x^4)(1-x^6)} =\frac{1+x^2}{(1-x^2)^2} \nn\\
&=&  1+3x^2+5x^4+7 x^6+9 x^8 + 11 x^{10}+\dots
\eea
The first cusp form occurs in weight 4, and is even under Fricke,
\be
\Xi_6(\tau) = 
\eta^2(\tau)\eta^2(2\tau)\eta^2(3\tau)\eta^2(6\tau)\\
= q - 2 q^2-3 q^3+4 q^4+6 q^5 - 16 q^7 + \dots
\ee
 The Hauptmodul is the McKay-Thompson series of class 6E,
\be
J_6(\tau) = \left(\frac{\eta(2\tau)\eta^3(3\tau)}{\eta(\tau) \eta^3(6\tau)}\right)^3 -3  =
q^{-1} + 6q+ 4 q^2 - 3 q^3 -12 q^4 + \dots
\ee
such that
\bea
J(\tau)&=&\frac{(J_6+7)^3 \left(J_6^3+237 J_6^2+1443
   J_6+2287\right)^3}{(J_6-5)^6 (J_6+3)^2
   (J_6+4)^3},\quad \nn\\
J(6\tau) &=&   
   \frac{(J_6+1)^3 \left(J_6^3+3 J_6^2-21
   J_6-71\right)^3}{(J_6-5) (J_6+3)^3 (J_6+4)^2}
   \eea
The Hauptmodul of $\Gamma_0(6)^+$ is the
McKay-Thompson series of class 6B
\be
\begin{split} 
J_6^+ = & \left( \frac{\eta(2\tau) \eta(3\tau)}{\eta(\tau)\, \eta(6\tau)}\right)^{12}=
J_6+\frac{72}{J_6-5}+12\\
= &q^{-1} + 12+78q+364 q^2+ 1365 q^3  + 4380 q^4 + \dots \ .
\end{split}
\ee
With this definition, the fundamental period for $M_{6}$-polarized K3 surfaces is given by
\bea
f_6(J_6^+(\tau)) &=& \sum_{d\geq 0}  c_6(d)
(J_6^+(\tau))^{-d} 
=-\frac{5}{24} E_2^{(6)}+\frac1{12}E_2^{(3)}(2\tau)+\frac18 E_2^{(2)}(3\tau) \nn\\
&=& \frac{\eta^7(2\tau)\eta^7(3\tau)^7}{\eta^5(\tau)\eta^5(6\tau)} =
1 + 5 q + 13 q^2 + 23 q^3 + 29 q^4 + 30 q^5 + \dots 
\eea
where the coefficients $c_6(d)$ are given in \eqref{eqn:fundamental_c6}.

\subsection{$m=7$}
The congruence subgroup $\Gamma_0(7)$ has index 8 in $SL(2,\IZ)$ and 2 cusps at $\tau=\I\infty$ and $0$. The ring of modular forms is generated by 
$E_2^{(7)}(\tau)$ and the Eisenstein series 
\be
E_4^{(7,\pm)} =E_4(\tau)\pm 49 E_4(7\tau) \ ,\quad 
E_6^{(7,\pm)} = E_6(\tau) \pm 749 E_6(7\tau) 
\ee 
The dimensions of $M_k(\Gamma_0(7))$ are given by
\bea
\sum_{k\geq 0} \dim M_k(\Gamma_0(7))\, x^k &=& 
\frac{1+x^2+2x^4+3x^6+x^8 }{(1-x^4)(1-x^6)}=\frac{1+2x^4+x^6}{(1-x^2)(1-x^6)}\nn\\
&=& 1 + x^2 + 3x^4 + 5x^6 + 5x^8 + 7x^{10} + 9x^{12} \nn\\&& +9 x^{14}+11 x^{16}+13 x^{18}+13
   x^{20}+\dots \nn
\eea
The first cusp form occurs in weight 4 and is even under Fricke,
\bea
\Delta_4^{(7)}(\tau)&=&
\frac{\eta(\tau)^5 \eta(7\tau)^5}{\eta(2\tau)   \eta(14\tau)}
+4 \eta(\tau)^2
   \eta(2\tau)^2 \eta(14\tau)^2
   \eta(7\tau)^2 \\
&=&2\frac{\eta(\tau)^5 \eta(7\tau)^5}{\eta(\tau/2)   \eta(7\tau/2)}
+ \eta(\tau)^2
   \eta(2\tau)^2 \eta(\tau/2)^2
   \eta(7\tau/2)^2 \nn\\   
   &=& -\frac{1}{160} E_4^{(7,+)} + \frac{5}{576} (E_2^{(7)})^2 \nn\\
   &=&q-q^2-2 q^3-7 q^4+16 q^5+2 q^6-7 q^7+\dots \nn
\eea
The unique cusp form of weight 6 is odd under Fricke,
\be
\Xi_7(\tau) := \eta(\tau)^6 \eta(7\tau)^6 = 
-\frac{19}{15552} (E_2^{(7)})^3 + \frac{25}{54} E_2^{(7)} \Delta_4^{(7)} + \frac{1}{1296} E_6^{(7,-)}
\ee
The Hauptmodul of $\Gamma_0(7)$ is the
McKay-Thompson series of class 7B,
\be
J_7(\tau) =\frac{\eta^4(\tau)}{\eta^4(7\tau)}+4= \frac{1}{q}+2 q+8 q^2-5 q^3-4 q^4-10 q^5+12 q^6-7 q^7+8 q^8+46 q^9+\dots
\ee
such that
\be
J(\tau)=\frac{\left(J_7^2+5 J_7+13\right) \left(J_7^2+237
   J_7+1437\right)^3}{(J_7-4)^7},
   \quad 
   J(7\tau)=\frac{\left(J_7^2-3 J_7-3\right)^3
   \left(J_7^2+5 J_7+13\right)}{J_7-4}
   \ee
The Hauptmodul of $\Gamma_0(7)^+$ is the
McKay-Thompson series of class 7A
\be
J_7^+ (\tau) =J_7 + \frac{49}{J_7-4} +9  
=\frac{1}{q} +9 + 51 q + 204 q^2 + 681 q^3 + 1956 q^4+\dots
\ee
With this definition, the fundamental period for $M_{7}$-polarized K3 surfaces is given by
\bea
f_7(J_7^+(\tau)) &=& \sum_{d\geq 0}  c_7(d)
(J_7^+(\tau))^{-d} =
-\frac16 E_2^{(7)} =1 + 4 q + 12 q^2 + 16 q^3 + 28 q^4+\dots \dots 
\eea
where the coefficients $c_7(d)$ are given in \eqref{eqn:fundamental_c7}.

\subsection{$m=8$}
The congruence subgroup $\Gamma_0(8)$ has index 12 in $SL(2,\IZ)$ and 4 cusps at $(0,1/2,1/4,\infty)$. The ring of modular forms is generated by the weight 2 forms $E_2^{(8)}(\tau)$ and
\bea
E_2^{(8,-)}(\tau) &:=& E_2^{(4)}(2\tau)-2 E_2^{(2)}(4\tau) \nn\\
E_2^{(8,+)}(\tau) &:=& E_2^{(8)}(\tau)-4 E_2^{(2)}(4\tau)-2 X_4(2\tau)
\eea
which are odd, odd and even under Fricke, respectively. 
The  dimensions of $M_k(\Gamma_0(8))$ are given by
\bea
\sum_{k\geq 0} \dim M_k(\Gamma_0(8))\, x^k &=& 
\frac{1+3x^2+4x^4+3x^6+x^8}{(1-x^4)(1-x^6)}
=\frac{1+x^2}{(1-x^2)^2} 
\\
&=& \sum_{n\geq 0} (2n+1) x^{2n} = 1 + 3 x^2+5 x^4+7 x^6+\dots
\nn
\eea
The first cusp form occurs in weight 4 and is Fricke invariant,
\be
\Xi_8(\tau)=\left[ \eta(2\tau) \eta(4\tau)\right]^4 = q-4q^3-2q^5+24 q^7-11 q^9-44 q^{11}+\dots
\ee
The Hauptmodul of $\Gamma_0(8)$ is the McKay-Thompson series of class 8E,
\be
J_8(\tau) = \frac{\eta(4\tau)^{12}}{\eta(2\tau)^4 \eta(8\tau)^8}=\frac{1}{q}+4q+2q^3
-8 q^5 - q^7 + 20 q^9 - 2 q^{11} - 40 q^{13} + 3 q^{15}+\dots,\quad  \nn\\
\ee
such that
\bea
J(\tau)&=&\frac{\left(J_8^4+240 J_8^3+2144 J_8^2+3840
   J_8+256\right)^3}{(J_8-4)^8 J_8
   (J_8+4)^2},\quad
J(8\tau)=  \frac{\left(J_8^4-16 J_8^2+16\right)^3}{J_8^2
   \left(J_8^2-16\right)}
\eea
The Hauptmodul of $\Gamma_0(8)^+$ is the McKay-Thompson series of class 8A,
\be
J_8^+ = J_8 + \frac{32}{J_8-4} +8 = 
\frac{1}{q}+8 +36 q + 128 q^2 + 386 q^3 + 1024 q^4 + 2488 q^5 + 5632 q^6 + 
 12031 q^7 + \dots
\ee
With this definition, the fundamental period for $M_{8}$-polarized K3 surfaces is given by
\bea
f_8(J_8^+(\tau)) &=& \sum_{d\geq 0}  c_8(d)
(J_8^+(\tau))^{-d} =\frac{(\eta(2\tau)\eta(4\tau))^6}{(\eta(\tau)\eta(8\tau))^4}=
\theta^2(2\tau) \theta^2(4\tau) \nn\\ &=& 
-\frac16 (E_2^{(8)} -2 E_2^{(2)}(4\tau)- X_4(2\tau))=
1 + 4 q + 8 q^2 + 16 q^3 + 24 q^4+\dots 
\eea
where the coefficients $c_8(d)$ are given in \eqref{eqn:fundamental_c8}.

\subsection{$m=9$}
The congruence subgroup $\Gamma_0(8)$ has index 12 in $SL(2,\IZ)$, and 4 cusps $(0,1/3,2/3,\infty)$.
The ring of modular forms is generated by the weight 2 forms $E_2^{(9)}(\tau)$ and  the even weight 2 form
\bea
E_2^{(9,+)}(\tau) &:=& E_2^{(3)}(3\tau)-\frac16 E_2^{(9)}(\tau) \nn \\
E_2^{(9,-)}(\tau) &:=& \frac{\eta^3(\tau)\eta^3(9\tau)}{\eta^2(3\tau)}=q-3 q^2+7 q^4-6 q^5+8 q^7-15 q^8+18 q^{10}-12 q^{11}+14 q^{13}+\dots \nn\\
\eea
which are odd, even and odd under the Fricke involution, respectively.
The  dimensions of $M_k(\Gamma_0(9))$ are given by
\bea
\sum_{k\geq 0} \dim M_k(\Gamma_0(9))\, x^k &=& 
\frac{1+3x^2+4x^4+3x^6+x^8}{(1-x^4)(1-x^6)} =
\frac{1+x^2}{(1-x^2)^2} \\
&=& \sum_{n\geq 0} (2n+1) x^{2n} = 1 + 3 x^2+5 x^4+7 x^6+\dots
\nn
\eea
The first cusp form occurs in weight 4 and is Fricke even,
\be
\Xi_9(\tau)=\eta(3\tau)^8 = q-8q^4+20q^7-70 q^{13}+ 64 q^{16} +\dots
\ee
The Hauptmodul is given by the McKay-Thompson series of class 9B,
\be
J_9 = \frac{\eta^3(\tau)}{\eta^3(9\tau)}+3 = \frac{1}{q}+5 q^2-7 q^5+3 q^8+15 q^{11}-32 q^{14}+9 q^{17}+58 q^{20}
+\dots
\ee
such that
\be
J(\tau)= \frac{(J_9+6)^3 \left(J_9^3+234 J_9^2+756 J_9+2160\right)^3}{(J_9-3)^9 \left(J_9^2+3
   J_9+9\right)} \ ,\quad
   J(9\tau)=
\frac{J_{9}^3 \left(J_{9}^3-24\right)^3}{J_{9}^3-27}  
\ee
The Hauptmodul of $\Gamma_0(9)^+$ is given by the McKay-Thompson series of class 9A,
\be
J_9^+ = J_9 + \frac{27}{J_9-3}+6 = \frac{1}{q} +6+ 27 q + 86 q^2 + 243 q^3 + 594 q^4 + 1370 q^5 + 2916 q^6 + 
 5967 q^7 
 +\dots
\ee
With this definition, the fundamental period for $M_{9}$-polarized K3 surfaces is given by
\bea
f_9(J_9^+(\tau)) &=& \sum_{d\geq 0}  c_9(d)
(J_9^+(\tau))^{-d} =-\frac{1}{8} E_2^{(9)} =
\frac{\eta^{10}(3\tau)}{\eta^3(\tau)\eta^3(9\tau)}\nn\\
 &=& 
1 + 3 q + 9 q^2 + 12 q^3 + 21 q^4 +\dots
\eea
where the coefficients $c_9(d)$ are given in \eqref{eqn:fundamental_c9}.

\subsection{$m=10$}
The congruence subgroup $\Gamma_0(10)$ has index 18 in $SL(2,\IZ)$, and 4 cusps $(0,1/2,1/5,\infty)$.
The dimensions  of $M_k(\Gamma_0(10))$ are given  by
\bea
\sum_{k\geq 0} \dim M_k(\Gamma_0(10))\, x^k &=& 
\frac{1+3x^2+6x^4+5x^6+3x^8}{(1-x^4)(1-x^6)} =
\frac{1+2x^2+3x^4}{(1-x^2)(1-x^4)} \nn\\
&=&  1+3 x^2+7 x^4+9 x^6+13 x^8+15 x^{10}+19 x^{12} \\ 
&&+21 x^{14}+25 x^{16}+27 x^{18} 
 +31 x^{20}+33 x^{22}+37 x^{24}+\dots \nn
\eea
In weight 2, we have the Eisenstein series 
$E_2^{(10)}(\tau)$ and 
\bea
E_2^{(10,-)}(\tau) &:=& 2E_2^{(5)}(2\tau)-5 E_2^{(2)}(5\tau) \nn\\
E_2^{(10,+)}(\tau) &:=& E_2^{(10)}(\tau)-5 E_2^{(2)}(5\tau)-2 E_2^{(5)}(2\tau)
\eea
which are odd, odd and even under  the Fricke involution,
respectively. 
The first cusp forms arise in weight 4, consisting of  
the two old forms $\Xi_4^{(5)}(\tau)$ and 
$4\Xi_5(2\tau)$, exchanged by the Fricke involution,
and the new form 
\be
\Xi_{10}(\tau)=
\frac{\eta^5(2\tau)\eta^5(10\tau)}
{\eta(\tau)\eta(10\tau)}=q+q^2-3 q^3-2 q^4-8 q^6+q^7+20 q^8+7 q^9+5 q^{10}+\dots
\ee
which is even under Fricke.
The Hauptmodul is given by the McKay-Thompson series of class 10E,
\be
J_{10}(\tau) = \frac{\eta^5(5\tau)\eta(2\tau)}{\eta^5(10\tau) \eta(\tau)}-1=
\frac{1}{q}+ q + 2 q^2 + 2 q^3 - 2 q^4 - q^5 - 4 q^7 - 2 q^8 + 5 q^9 +\dots
\ee
such that
\bea
J(\tau)&=& \frac{\left(J_{10}^6+242 J_{10}^5+2635 J_{10}^4+10060 J_{10}^3+20615 J_{10}^2+20642
   J_{10}+7949\right)^3}{(J_{10}-3)^{10} (J_{10}+1)^2 (J_{10}+2)^5},\nn\\
J(10\tau)  &=&  
   \frac{\left(J_{10}^6+2 J_{10}^5-5
   J_{10}^4-20 J_{10}^3-25 J_{10}^2+2 J_{10}+29\right)^3}{(J_{10}-3) (J_{10}+1)^5 (J_{10}+2)^2}
\eea
The Hauptmodul of $\Gamma_0(10)^+$ is the McKay-Thompson series of class 10D, given by 
\be
J^+_{10} = \frac{(\eta(5\tau)\eta(2\tau))^6}{(\eta(\tau)\eta(10\tau))^6}-1 = 
J_{10} + \frac{20}{J_{10}-3} +5= 
\frac{1}{q} + 5+ 21 q + 62 q^2 + 162 q^3 + 378 q^4 
+\dots
\ee
With this definition, the fundamental period for $M_{10}$-polarized K3 surfaces is given by 
\bea
f_{10}(J^{+}_{10}(\tau)) &=& \sum_{d\geq 0}  c_{10}(d)
(J^{+}_{10}(\tau))^{-d} =-\frac{1}{12} (E_2^{(10)}-5E_2^{(2)}(5\tau)+2E_2^{(5)}(2\tau)) =\nn\\
 &=& 
1 + 2 q + 10 q^2 + 8 q^3 + 26 q^4 + 2 q^5  +\dots
\eea
where the coefficients $c_{10}(d)$ are given in \eqref{eqn:fundamental_c10}.

\subsection{$m=11$}
The congruence subgroup $\Gamma_0(11)$ has index 12 in $SL(2,\IZ)$ and 2 cusps $(0,\infty)$. The ring of modular forms is generated by $E_2^{(11)}$, 
the weight 2 cusp form 
\be
E_{2}^{(11,-)}(\tau) =\eta(\tau)^2 \eta(11\tau)^2 = q - 2 q^2 - q^3 + 2 q^4 + q^5 + 2 q^6 - 2 q^7 - 2 q^9 +\dots
\ee
and the weight 4 Eisenstein series $E_4^{(11,-)}(\tau)=E_4(\tau)-121 E_4(11\tau)$,
all of which are odd under the Fricke involution.
The dimensions  of $M_k(\Gamma_0(11))$ are given  by the generating series
\bea
\sum_{k\geq 0} \dim M_k(\Gamma_0(11))\, x^k &=& 
\frac{1+2x^2+3x^4+3x^6+2x^8+x^{10}}{(1-x^4)(1-x^6)} =
\frac{1+x^4}{(1-x^2)^2}= \sum_{n\geq 0} (2n) x^{2n} \nn\\
&=& 1 + 2 x^2+4 x^4+6 x^6+\dots
\eea
While the modular curve $X_{0}(11)$ has genus 1,
its quotient $X_{0}(11)^+$ by the Fricke involution has
genus 0. The Hauptmodul of $\Gamma_0(11)^+$
is the McKay-Thompson series of class 11A,
\be
J_{11}^+(\tau) = J_{11} + \frac{16}{J_{11}-2} + \frac{16}{(J_{11}-2)^2} + 6 
= \frac{1}{q}+6 + 17 q + 46 q^2 + 116 q^3 + 252 q^4+ \dots 
\ee
where 
\be
J_{11}(\tau)=\frac{\eta^2(\tau) \eta^2(11\tau)}{\eta^2(2\tau)\eta^2(22\tau)}+2=
\frac{1}{q}+q-2 q^2+4 q^3-4 q^4+5 q^5-6 q^6+9 q^7
+\dots
\ee
The $J$ invariant can be expressed in terms of $J_{11}^+$ by\footnote{This is obtained by setting $t=J_{11}^+-6$ in \cite{vanHoeij}.}
\bea
J(\tau) &=& P_{11}(J_{11}^+) + P_9(J_{11}^+) \sqrt{ J_{11}^+ \left[ (J_{11}^+)^3-20 (J_{11}^+)^2+56 J_{11}^+-44 \right]}
\nn\\
J(11\tau) &=& P_{11}(J_{11}^+(\tau) ) - P_9(J_{11}^+) 
\sqrt{ J_{11}^+ \left[ (J_{11}^+)^3-20 (J_{11}^+)^2+56 J_{11}^+-44 \right]} \nn\\
\eea
where
\bea
P_9(x) &=& \frac{1}{2} (x-16) (x-7) (x-4) (x-2) (x-1) \left(x^2-14 x+4\right) \left(x^2-12 x+16\right)\nn\\
P_{11}(x) &=& \frac{x^{11}}{2}-33 x^{10}+\frac{1793 x^9}{2}-13024 x^8+110528 x^7-566335 x^6+1767920 x^5 \nn\\ &&
-3341536 x^4 +3709376 x^3-2230272 x^2+591872   x-32768
\eea
With this definition, the fundamental period for $M_{11}$-polarized K3 surfaces is given by
\bea
f_{11}(J_{11}^+(\tau)) &=& \sum_{d\geq 0}  c_{11}(d)
(J_{11}^+(\tau))^{-d} =-\frac{1}{10}E_2^{(11)}+\frac85 \eta^2(\tau)\eta^2(11\tau)\nn\\
 &=& 
1 + 4 q + 4 q^2 + 8 q^3 + 20 q^4 + 16 q^5  +\dots
\eea
where the coefficients $c_{11}(d)$  are given in \eqref{eqn:fundamental_c5}.

\section{Skew-holomorphic Jacobi forms}
\label{sec_skew}
Skew-holomorphic Jacobi forms of weight $w$ and index $m$ are functions $\phi(\tau,z)$ which 
are smooth in $\tau\in\IH$, holomorphic in $z\in \IC$, periodic in each variable with period 1, satisfying 
the functional equation (see e.g. See e.g. \cite{skoruppa1989developments})
\be
\label{skewprop}
\phi\left(-\frac{1}{\tau},\frac{z}{\tau}\right) \, e^{-2\pi \I m z^2/\tau} = \bar\tau^{w-1}\, |\tau|\, \phi(\tau,z) 
\ee
and has Fourier expansion
\be
\label{skewFourier}
\phi(\tau,z) = \sum\limits_{\Delta,r\in \IZ, \Delta\geq 0 \atop \Delta=r^2 \mod 4m}
C_\phi(\Delta,r)\, e^{2\pi\I\left( \frac{r^2-\Delta}{4m} \tau_1 + \I \frac{r^2+|\Delta|}{4m} \tau_2 + r z\right)}
\ee
where the coefficients $C_\phi(\Delta,r)$ depend only on $r$ modulo $2m$. Equivalently, 
$\phi(\tau,z)$ has the theta decomposition
\be
\label{skewdec}
\phi(\tau,z) = \sum_{r=0}^{2m-1} \overline{ \phi_r(\tau)} \, 
\sum\limits_{k\in \IZ\atop k=r \mod 2m} e^{2\pi\I \left( \frac{k^2}{4m}\tau + k z \right)}
\ee
such that 
\be
\label{skewvec}
\phi_r(\tau) =  \sum\limits_{\Delta\geq 0 \atop \Delta=r^2 \mod 4m} \overline{C_\phi(\Delta,r)}\, q^{\frac{\Delta}{4m}}
\ee
transforms like a vector-valued modular form of weight $w-\frac12$ under the Weil representation
associated to the lattice $\langle 2m\rangle$.  We denote by $J^+_{w,m}$ the space of skew-holomorphic 
Jacobi forms of weight $w$ and index $m$, and by $S^+_{w,m}$ the subspace of skew-holomorphic cusp
Jacobi forms, such that $C_\phi(\Delta,r)=0$ unless $\Delta>0$.  Note that the space $J^-_{w,m}$ of ordinary holomorphic Jacobi forms is defined similarly, but with  
a factor of $\tau^w$ on the r.h.s. of \eqref{skewprop}, Fourier expansion restricted to $\Delta\leq 0$ in \eqref{skewFourier} and \eqref{skewdec}, and $\tau$ replaced by $-\bar \tau$ in \eqref{skewvec},
such that \eqref{skewdec} is holomorphic both in $\tau$ and $z$. In the following we shall restrict
to the case where $w\geq 1$ is odd.

Prime examples of skew-holomorphic Jacobi forms are
the signature $(1,1)$ theta series 
\be
\label{defT}
T_{m_1,m_2} (\tau,z)= \sum_{s,t\in\IZ^2}  q^{st} e^{-\frac{\pi \tau_2 (m_1 s-m_2 t)^2}{m_1 m_2}} e^{2\pi\I z(m_1 s+m_2 t)}
\ee
which belong to $J^+_{w,m}$ with $w=1, m=m_1 m_2$ (as follows from the Poisson resummation formula).
In fact, the space $J^+_{1,m}$ for any $m,\geq 1$ is spanned by the theta series with \eqref{defT} with $(m_1,m_2)$ running
over all factorizations $m=m_1 m_2$, up to the trivial exchange of $m_1$ and $m_2$.
The Fourier coefficients of \eqref{defT} are given by
\be
\label{FourierT}
C_{m_1,m_2}(\Delta,r) =  \#\{ (s,t) \in \IZ^2: (m_1 s-m_2 t)^2=\Delta,  m_1s+m_2 t=r\}
\ee
For $(m_1,m_2)=(m,1)$, one can eliminate $t=r-ms$ such that $C_{m_1,m_2}(\Delta,r)=\sum_{s\in\IZ} 
\delta_{\Delta,(r-2ms)^2}$. The corresponding vector-valued modular form is therefore the standard theta series
\be
T_{m,1;r}(\tau) = \sum_{s\in\IZ}  q^{\frac{(r-2ms)^2}{4m}} = 
 \sum_{k\in \IZ+\frac{r}{2m}}  q^{mk^2}  = \Thi{m}_r
\ee
For $(m_1,m_2)=(2,m)$, we find that the components of $T_{2,m}$ can written in terms of 
$\Thi{m}_r$ by applying \cite[Prop. 2]{Alexandrov:2022pgd}, e.g. 
\bea
T_{2,2;r} &=& \delta^{(2)}_{r}(\Thi{4}_r+\Thi{4}_{r+4}) \nn \\ 
T_{2,3;r} &=& \Thi{6}_r - \delta^{(2)}_{r+1}(\Thi{6}_r-\Thi{6}_{r+6}) \nn\\
T_{2,4;r} &=& \delta^{(2)}_{r}(\Thi{8}_r+\Thi{8}_{r+8}) \\
T_{2,5;r} &=& \Thi{10}_r -  \delta^{(2)}_{r+1}(\Thi{10}_r-\Thi{10}_{r+10}) \nn\\
T_{2,6;r} &=& \delta^{(2)}_{r}(\Thi{12}_r+\Thi{12}_{r+12}) \nn\\
T_{3,3;r} &=& \Thi{9}_r+\delta^{(3)}_{r}(\Thi{9}_r+\Thi{9}_{r+6}+\Thi{9}_{r+12})
\eea

Since multiplication $\phi_{r}(\tau)\mapsto f(\tau)\phi_{r}(\tau)$ by any holomorphic modular form $f(\tau)$
of weight $k$ (or equivalently, $\phi(\tau,z)\mapsto \overline{f(-\bar\tau)} \phi(\tau,z)$) maps $J^+_{w,m}\mapsto J^+_{w+k,m}$, the total sum $J^+_{*,m}=\oplus_{w\geq 0} J^+_{w,m}$
forms a graded module over the ring $M_*(SL(2,\IZ))=\IC[E_4,E_6]$ of holomorphic modular forms. Since all
elements of $M_*$ have even weight, the odd weight part $J^{+}_{{\rm odd},m}=\oplus_{n\geq 0} J^+_{2n+1,m}$ also forms
a graded module, which can be shown to be free of rank $m+1$ over $M_*$. This implies that the Hilbert series generating the graded dimensions of $J^{+}_{{\rm odd},m}$ takes the form 
\be
\label{eqHilbertJ}
\sum_{n\geq 0} \dim J^+_{2n+1} t^{n+1} = \frac{Q_m(t)}{(1-t^4)(1-t^6)}
\ee
where $Q_m(t)$ is an odd polynomial in $t$ with positive integer coefficients, whose leading $\cO(t)$ term is given by the number of unordered factorizations $m=m_1 m_2$ and such that $P_m(1)=m+1$.

The dimension of the space $J^+_{w,m}$ can be computed along the lines of \cite[\S A]{Alexandrov:2022pgd}. For odd $w$ in the range $1\leq w \leq 11$, such that $\phi(\tau,z)/\eta^{24}$ has negative weight, the dimension is given by 
\bea
\dim J^+_{w,m} &=& \frac{(w-\frac{27}{2})(m+1)}{12} + \frac{3m+5}{2} 
+\frac{1}{8\sqrt{2m}} \Re\left[ \frac{e^{\I\pi w/2}}{\I} G(2,4m) \right] \\
&& + \frac{1}{6\sqrt{6m} }  \Re\left[ \frac{e^{\I\pi w/3}}{\sqrt{\I}}  \left( G(1,4m)
+ G(-3,4m) \right) \right] -\sum_{k=0}^m \{ k^2/4m \} -1 \nn
\eea
where $\{x\}=x-\lfloor x \rfloor$ denotes the fractional part and $G(n,m)$ is the Gauss sum
\be
G(n,m) = \sum_{k=0}^{m-1}e^{2\pi\I nk^2/m}
\ee
For $w=11$, the formula reduces to the result for $\dim\Pic(\cM_{2m})^{\rm NL}\otimes\IQ$
in \cite[p45]{maulik2007gromov}, except for replacing the constant $C$ (counting isotropic vectors $k$ such that $k^2=0\mod 4m$) by 1. 
The dimensions for $m\leq 12, w\leq 11$ are summarized in Table \ref{tabskew}, along with the polynomial
$P_m(t)$ appearing in the numerator of \eqref{eqHilbert}, as well as the number of polar terms in 
$\phi(\tau,z)/\eta^{24}$ for $w=11$ and number isotropic vectors. The last column indicates the dimension of the space of generating series of Noether-Lefschetz invariants for K3-fibrations of degree $2m$.

\begin{table}
\be
\nn
\begin{array}{|c|l||c|c|c|c|c|c||c|c||c|}
\hline
 m & Q_m(t) & w= 1\  & \ 3\  & \ 5\  &\  7\  &\  9\  &\ 11\  & \# {\rm pol} & \# {\rm iso} & \# {\rm NL} \\
 \hline
1& t+t^3 & 1 & 1 & 1 & 2 & 2 & 2 & 2 & 0 & 2\\
2& t+t^3+t^5 & 1 & 1 & 2 & 2 & 3 & 3 & 3 & 0 & 3 \\
3& t+t^3+t^5+t^7 &1 & 1 & 2 & 3 & 3 & 4 & 4 & 0 & 4  \\
4& 2t+2t^3+t^5 & 2 & 2 & 3 & 4 & 5 & 5 & 5 & 1 & 4 \\
5& t+2t^3+2t^5+t^7 & 1 & 2 & 3 & 4 & 5 & 6 & 6 & 0 & 6 \\
6& 2t+2t^3+2t^5+t^7 & 2 & 2 & 4 & 5 & 6 & 7 & 7 & 0 & 7 \\
7& t+2t^3+2t^5+2t^7+t^9 &1 & 2 & 3 & 5 & 6 & 7 & 8 & 0 & 7 \\
8& 2t+3t^3+3t^5+t^7     & 2 & 3 & 5 & 6 & 8 & 9 & 9 & 1 & 8\\
9& 2t+3t^3+3t^5+2t^7 &2 & 3 & 5 & 7 & 8 & 10 & 10 & 1 & 9 \\
10& 2t+3t^3+3t^5+2t^7+t^9 & 2 & 3 & 5 & 7 & 9 & 10 & 11 & 0 & 10\\
11 &t+3t^3+4t^5+4t^7+t^9 & 1 & 3 & 5 & 7 & 9 & 11 & 12 & 0 & 11 \\
12 &3t+4t^3+4t^5+2t^7 & 3 & 4 & 7 & 9 & 11 & 13 & 13 & 1 & 12\\
\hline
\end{array}
\ee
\caption{Dimension of the space of skew-Jacobi forms $J^+_{m,w}$ for odd $1\leq w\leq 11$, $m\leq 12$. 
The last three columns indicate the number of polar coefficients
in $\phi(\tau,z)/\eta^{24}$, number of isotropic vectors, and the dimension of the space of generating series of NL invariants for $K3$-fibrations of degree $2m$. 
\label{tabskew}} 
\end{table}

It is straightforward to explicitly construct a basis of $J^+_{m,w}$ for the relevant values of $(m,w)$ in 
Table \ref{tabskew}. First, the theta series $\Thi{m}=T_{m,1}$ provides one element in $J^+_{1,m}$ for any $m$, to be supplemented by $T_{2,m/2}$ for $m=4,8,10,12$, $T_{3,m/3}$ for $m=6,9,12$. Second, given any element  $\phi\in J^+_{w,m}$, one obtains an element $D\phi\in J^+_{w+2,m}$
 by applying the Serre derivative
\be
\label{SerreD}
\phi_{r}\mapsto
(D\phi)_r(\tau) = \left(q\partial_q - \frac{w-\frac12}{12} E_2\right)\phi_r
\ee
where $E_2(\tau)=1-24 \sum_{n\geq 1} n q^n/(1-q^n) $ is the quasi-modular Eisenstein series of weight 2. Elements of higher weight are obtained by iterating the Serre derivative and multiplying by arbitrary powers of $E_4$ and $E_6$. In general however, there are relations between the resulting elements.  
For example, for $m=1$ the second Serre derivative of  $T_{1,1}$ is proportional to $E_4 T_{1,1}$,
in agreement with the fact that $\dim J^+_{5,1}=1$. We record this relation and similar relations for 
$m=2,3,4$ below, which can be checked by comparing the first few Fourier coefficients,
\bea
0 &=& D^2 \Thi{1}- \frac{5}{576} E_4\, \Thi{1} \nn\\
0  &=& D^3 \Thi{2}  - \frac{23}{576} E_4\, D\Thi{2} + \frac{11}{6912} E_6 \Thi{2}  \\
0 &=& D^4 \Thi{3}  - \frac{29}{288} E_4\, D^2\Thi{3} + \frac{11}{864} E_6 \Thi{3}
-\frac{119}{331776} E_4^2 \Thi{3} \nn\\
0 &=&\left(D^3 -\frac{5}{576} E_4 D + \frac{5}{1728} E_6 \right)T_{2,2} \nn\\
0 &=& D^3 \Thi{4} + E_4 \left( 
-\frac{155}{2304} D\Thi{4} + \frac{15}{512} DT_{2,2} \right) 
+E_6 \left( \frac{25}{55296} \Thi{4} + \frac{5}{4096} T_{2,2} \right)\nn
 \eea

Beyond this differential graded ring structure, there are also 
two Hecke-like operators 
$U_N: J^+_{w,m}\to J^+_{w,m N^2}$ and $V_N: J^+_{w,m}\to J^+_{w,m N^2}$, whose action on 
Fourier coefficients is given by 
\bea
\label{FourierUV}
C_{U_N \phi}(\Delta,r) &=& 
\begin{cases} C_\phi(\Delta/N^2, r/N) & \mbox{if}\  r=0\mod N \\
0 & \mbox{otherwise}
\end{cases}
\\
C_{V_N \phi}(\Delta,r) &=& 
\begin{cases} \sum_{d| \left( N, r, \frac{r^2-\Delta}{4mN} \right)} d^{w-1} \, C_\phi(\Delta/d^2, r/d) 
& \mbox{if} \Delta=r^2 \mod 4mN \\
0 & \mbox{otherwise}
\end{cases}
\eea
as well as Hecke operators $T_N:J^+_{w,m}\to J^+_{w,m}$, 
acting as
\be
\label{FourierTN}
C_{T_N \phi}(\Delta,r)= \sum_{d|N}
  \left( {\Delta \over d} \right)\ C_\phi\left( \frac{N^2 D}{d^2}, \frac{N}{d} r \right)
\ee
when $\Delta$ is a fundamental discriminant, i.e. $\Delta=1\mod 4$ square-free or $\Delta=4m$ with $m=2\, \mbox{or} \mod 4$ with $m$ square-free, and $\left( {\Delta \over d} \right)$ is the Legendre symbol (the action for general $\Delta$ is more complicated, see e.g. \cite[(3.3.6)]{Cheng:2016klu}).

In particular, in weight $w=1$ we find 
\bea
\label{Vw1}
V_2 \Thi{1}&=& 2 \Thi{2} , \quad  \nn\\
V_3 \Thi{1} &=& 2 \Thi{3}, \quad  \nn \\
V_4 \Thi{1} &=& 2 \Thi{4}+T_{2,2}, \quad V_2\Thi{2} = \Thi{4}+T_{2,2}, \quad 
     U_2 \Thi{1} = T_{2,2} \nn \\
V_5 \Thi{1} &=& 2 \Thi{5} \nn \\
V_6 \Thi{1} &=& 2 V_3 \Thi{2} = 2 V_2 \Thi{3} = 2 \Thi{1} + 2 T_{2,3} \nn\\
V_7 \Thi{1} &=& 2 \Thi{7} \nn \\
V_8 \Thi{1} &=& 2 V_2 \Thi{4} = 2 \Thi{8} + 2 T_{2,4}, \quad
   V_4\Thi{2} = \Thi{8} + 2 T_{2,4}, \quad 
   U_2 \Thi{2} = T_{2,4} \nn \\
V_9 \Thi{1} &=& 2 \Thi{9} + T_{3,3}, \quad
   V_3\Thi{3} = \Thi{9} + T_{3,3}, \quad 
   U_3 \Thi{1} = T_{3,3} \nn \\ 
V_{10} \Thi{1}   &=& 2 V_5 \Thi{2} = 2 V_2 \Thi{5} = \Thi{10} + T_{2,5} \nn\\
V_{11} \Thi{1} &=& 2 \Thi{11} \nn\\
V_{12} \Thi{1}   &=& 2 V_4 \Thi{3} = 2\Thi{12}+2T_{2,6} + 2T_{3,4}\ ,\quad 
   V_3 \Thi{4} = \Thi{12}+T_{3,4}, \quad  \nn\\ 
   V_6 \Thi{2}  &=& \Thi{12}+2T_{2,6} + T_{3,4}\ ,\quad
   U_2 \Thi{3} =  T_{2,6}
\eea
Moving on to weight 3, there are  similar relations to \eqref{Vw1}
\bea
\label{Vw3}
 V_2 D\Thi{1} &=& 5 D\Thi{2} \ ,\quad\nn\\
 V_3 D\Thi{1} &=& D\Thi{3} \ ,\quad \nn\\
 V_4 D\Thi{1} &=& 20 D\Thi{4}+DT_{2,2}, \quad 
 V_2 D\Thi{2} = 4D\Thi{4}+DT_{2,2}, \quad 
 U_2 D\Thi{1} = DT_{2,2}  \nn
\eea
In contrast, for $m=5$,   $V_5 D\Thi{1}$ turns out to be linearly independent from $D\Thi{5}$, providing a basis of 
$J^+_{3,5}$. For $m=6$, $J^+_{3,6}$ is also two-dimensional, but $D\Thi{6}$ and $DT_{2,3}$ turn out to coincide; the second generator can be chosen to be $V_6 D\Thi{1}=5 V_3 D\Thi{2} = 10 V_2 D\Thi{3}$.
Similarly, $V_7 D\Thi{1}$ and $D\Thi{7}$ span $J^+_{3,7}$. For $8\leq m \leq 10$, $J^+_{3,m}$ is three-dimensional and generators can be constructing using the same ideas. The same holds for $m=12$, where
$\dim J^+_{3,m}=4$. 
For $m=11$ however, $V_{11} D\Thi{1}$ and $D\Thi{11}$ only span a codimension-one space of $J^+_{3,11}$, and it is unclear at present how to construct the missing third generator. 
Fortunately, for weight $w=11$, the case relevant for generating series of NL invariants, we can construct sufficiently many elements to span $J^+_{11,m}$ by applying Serre derivatives and Hecke-like operators without having to resolve this issue. 

An important property of the space $S^+_{w,m}$ of skew-holomorphic
Jacobi cusp forms is the existence of Hecke-equivariant maps, known as Shimura-Skoruppa-Zagier lifts \cite{skoruppa1987jacobi,skoruppa1988explicit,skoruppa1991heegner},
\be
\cS_{D,r}: S^+_{w,m} \rightarrow M_{2w-2}(\Gamma_0(m))
\ee
for any $D=r^2\mod 4m$ and $m,w\geq 1$, acting by
\be
\phi \mapsto \sum_{n\geq 0} C_{T_n\phi}(D, r)\, q^n
\ee 
where the zero-th Fourier coefficient vanishes for $w>2$. 
Moreover, the image is an eigenfunction of the Fricke involution with eigenvalue $(-1)^{w+1}$. 
For $D=r=1$ and $w>2$, using \eqref{FourierT} this simplifies to 
\be
\label{S11phiF}
\cS_{1,1}: \phi \mapsto \sum_{n=1}^{\infty} C_{\phi}(n^2, n)\, \Li_{2-w}(q^n)
\ee
where $\Li_s(x):=\sum_{k\geq 1}x^n/n^s$. In general, $\cS_{D,r}$ can be written as a theta lift,
\be
\cS_{D,r} \phi = \int_{\IH/SL(2,\IZ)} \sum_{s \mod 2m} \Theta_{D,r,s}(\tau,\tau') 
\overline{\phi_{s}(\tau')} (\Im\tau')^{w-\frac52} \de\tau'\de\bar\tau'
\ee
where $\Theta_{D,r,s}(\tau,\tau')$ is a suitable Siegel theta 
which is modular of weight $2w-2$ under $\Gamma_0(2m)$
with respect to $\tau$ and vector-valued of weight $w-\frac12$ under $SL(2,\IZ)$ acting on $\tau'$ (see \cite{skoruppa1991heegner,Cheng:2016klu} for details). More relevant for our purposes, there exists a similar 
Shimura-type lift between skew-holomorphic weak Jacobi forms of weight $3-w$ and harmonic Maass modular forms of weight $4-2w$,  of the form \cite{zbMATH06251204,bringmann2014locally}
\be
\cS_{D,r}^\star \phi = {\rm Reg} \int_{\IH/SL(2,\IZ)} \sum_{s \mod 2m} \Theta^\star_{D,r,s}(\tau,\tau') 
\overline{\phi_{s}(\tau')} (\Im\tau')^{w-\frac52} \de\tau'\de\bar\tau'
\ee
where $\Theta_{D,r,s}^\star(\tau,\tau')$ is now modular of weight $4-2w$ with respect to $\tau$ and weight $\frac52-w$ with respect to $\tau'$. Due to the poles of $\phi_{s}$ at $\tau=\I\infty$, the integral must be regularized as in \cite{0919.11036}.
This automorphic form is in general a harmonic Maass form
along the codimension-one loci $a|\tau|^2+b\Re\tau+c=0$ with 
$b^2-4ac=D$. Its holomorphic part is given by the same formula as \eqref{S11phi} with $w$ replaced by $3-w$, up to a polynomial of degree at most $2w-4$:
\be
\label{S11phi}
\cS_{1,1}^{\star,{\rm hol}}: \phi \mapsto P_{2w-4}(\tau) + \sum_{n=1}^{\infty} C_{\phi}(n^2, n)\, \Li_{w-1}(q^n)
\ee
Applying this lift to the ratio $\Phi_{\rm NL}/\eta^{24}$, which is a weak skew-holomorphic Jacobi form of weight $-1$ (hence $w=4$), we obtain in this way the contribution of vertical GV invariants to the prepotential, transforming as an holomorphic Eichler integral of weight $-4$ under $\Gamma_0(m)^+$.  

\section{Explicit results}
\label{sec_results}
In this section, for each model 
$X_m^{[i,j]}$ we collect explicit results for the Yukawa coupling $C_{TTT}$, generating series of NL invariants $\Phi_m^{[i,j]}$, generating series of base degree 1, genus 0 GW invariants $\tilde h_1$, and generating series of base degree 2, genus 0 GW invariants $\tilde h_2$. 
For brevity, we display the first few Fourier coefficients of 
$\Phi_{\rm NL}$ as a flattened list $\sum_{d=0}^m \Phi_{m,d}^{[i,j]}$. Moreover $\Delta_\chi$ denotes the difference of Euler numbers $\chi(X^{[i,j]}_m)-\chi(X^{[1,1]}_m)$.

\subsection{$m=1$}
Using \eqref{CTTT} we get 
\bea
(i,j)=(1,1): && C_{TTT}= 4 + 2496 q + 1792512 q^2 + 1043215104 q^3 + 582416191488 q^4 + \dots
\nn\\
(i,j)=(2,1): && C_{TTT}= 3 + 2880 q + 1672704 q^2 + 1073721600 q^3 + 575027490816 q^4+\dots
\nn\\
(i,j)=(2,2): && C_{TTT}=2 + 3264 q + 1552896 q^2 + 1104228096 q^3 + 567638790144 q^4+ \dots
\nn\\
\eea
Taking two further derivatives, we get a meromorphic modular form of weight 6  (generalizing \cite[(5)]{Kaplunovsky:1995tm} \cite[(3.8)]{Antoniadis:1995zn} for $(i,j)=(1,1)$)
\be
C_{TTTTT} = \left[  \frac{192  (13 J+8640)}{(J-1728)^2}  + \frac{48\Delta_\chi}{7J(\tau)}  \right] E_6
\ee
where $\Delta_\chi=-2(i^2+j^2-2)-124(\tfrac{1}{i}+\tfrac{1}{j}-2)$, which equals $-112(\tfrac{1}{i}+\tfrac{1}{j}-2)$ for the relevant values of $(i,j)$.
By the GV/NL correspondence, this implies 
\be
\Phi_1^{[i,j]} = -\frac53 E_4 E_6 \, \Thi{1} + 8 E_4^2 \, D\Thi{1}
+\Delta_\chi \left( \frac{1}{672}E_4 E_6 \, \Thi{1} + \frac{1}{28} E_4^2 \, D\Thi{1} \right) 
\ee
The $q$-expansions read
\bea
\Phi_1^{[1,1]}  &=&
-2+300 q+2496 q^{5/4}+217200 q^2+665600 q^{9/4}+10226400 q^3+21854400 q^{13/4}
+\dots\nn\\
\Phi_1^{[2,1]}  &=& 
-2+q^{1/4}+244 q+2856 q^{5/4}+203520 q^2+696569 q^{9/4}+10049280 q^3+22121160
   q^{13/4}
+ \dots\nn\\
\Phi_1^{[2,2]} & =& 
-2+2 \q^{1/4}+188 q+3216 q^{5/4}+189840 q^2+727538 q^{9/4}+9872160 q^3+22387920
   q^{13/4}
+\dots\nn\\
\eea
The generating series of base degree 1, genus 0 GW invariants are given by 
\be
\tilde h_1 = \frac{2ij E_4^2}{\eta^{24} (J-1728)}
\ee
leading to the base degree 1, genus 0 GV invariants
\bea
(i,j)=(1,1): && W_1=2+2496 q+ 1941264 q^2 +1327392512 q^3 + 861202986072 q^4+ \dots
\nn\\
(i,j)=(2,1): && W_1= 4 + 4752 q + 3603528 q^2 + 2445085664 q^3 + 1580414726844 q^4+ \dots
\nn\\
(i,j)=(2,2): && W_1=8 + 9024 q + 6677856 q^2 + 4501790208 q^3 + 2899431143088 q^4 + \dots\nn\\
\eea
The generating series of base degree 2, genus 0 GW invariants are given by 
\be
\tilde h_2 = \frac{P_{32}^{[i,j]}}{\eta^{72} (J-1728)^3}
\ee
where $P_{32}^{[i,j]}$ is a modular form of weight 32,
\bea
P_{32}^{[1,1]} &=& \frac{1}{43} E_4^2 \left( -89 E_4^6 +122  E_6^4- 53 E_4^3 E_6^2 \right)
\nn\\
P_{32}^{[2,1]} &=& \frac{1}{108} E_4^2 \left(-89 E_4^6 +87  E_6^4  -16 E_4^3 E_6^2\right) 
\nn\\
P_{32}^{[2,2]} &=& \frac{1}{27} E_4^2(-89 E_4^6 +63 E_6^4+21 E_4^3 E_6^2)
\eea
leading to the base degree 2, genus 0 GV invariants
\bea
(i,j)=(1,1): && W_2-\frac18 W_1(2\tau)= 223752 q^2 + 1327392512 q^3 + 2859010142112 q^4 +\dots
\nn\\
(i,j)=(2,1): &&  W_2-\frac18 W_1(2\tau)=252 q + 2204928 q^2 + 6905771456 q^3 + 12536489815488 q^4+\dots
\nn\\
(i,j)=(2,2): &&  W_2-\frac18 W_1(2\tau)= 3264 q + 14161056 q^2 + 33088733952 q^3 + 53279323863936 q^4+\dots
\nn\\
\eea
Similarly, for the base degree 3, genus 0 GW invariants are given by
\be
\tilde h_3=\frac{E_4{}^2 P_{48}^{[i,j]}}{\eta^{120} (J-1728)^5}
\ee
with
\bea
P_{48}^{[1,1]}&=&
\frac{20367 E_4{}^{12}-38052 E_4{}^9 E_6{}^2+18898 E_4{}^6 E_6{}^4-6260 E_4{}^3
   E_6{}^6+3895 E_6{}^8}{69984} \nn\\
P_{48}^{[2,1]}&=&   
   \frac{ 20367 E_4{}^{12}-37089 E_4{}^9 E_6{}^2+17404 E_4{}^6 E_6{}^4-2318 E_4{}^3
   E_6{}^6+1474 E_6{}^8}{8748} \\
P_{48}^{[2,2]}&=&    \frac{2\left(20367 E_4{}^{12}-36126 E_4{}^9 E_6{}^2+15559 E_4{}^6 E_6{}^4+328
   E_4{}^3 E_6{}^6-200 E_6{}^8\right)}{2187}   \nn
\eea
leading to the  base degree 3, genus 0 GV invariants
\bea
(i,j)=(1,1): && W_3-\frac1{27} W_1(3\tau)=  38637504 q^3 + 861202986072 q^4 + 4247105405354496 q^5 
+\dots
\nn\\
(i,j)=(2,1): &&  W_3-\frac1{27} W_1(3\tau)= 20520 q^2 + 2493967312 q^3 + 16303941022572 q^4
+\dots
\nn\\
(i,j)=(2,2): &&  W_3-\frac1{27} W_1(3\tau)= 6677856 q^2 + 58621711936 q^3 + 234194824987824 q^4
+\dots
\nn\\
\eea
For $(i,j)=(1,1)$, this agrees with the results in \cite{Candelas:1993dm,Henningson:1996jf}. 

\subsection{$m=2$}
Using \eqref{CTTT} we get 
\bea
(i,j)=(1,1): && C_{TTT}= 8 + 640 q + 80896 q^2 + 7787008 q^3 + 702799872 q^4 + \dots
\nn\\
(i,j)=(2,1): && C_{TTT}= 6 + 704 q + 78336 q^2 + 7857920 q^3 + 701009920 q^4 +\dots
\nn\\
(i,j)=(4,1): && C_{TTT}=5 + 640 q + 80896 q^2 + 7787008 q^3 + 702799872 q^4 + \dots
\nn\\
(i,j)=(2,2): && C_{TTT}= 4 + 768 q + 75776 q^2 + 7928832 q^3 + 699219968 q^4+\dots
\nn\\
(i,j)=(4,2): && C_{TTT}=3 + 704 q + 78336 q^2 + 7857920 q^3 + 701009920 q^4+ \dots
\nn\\
(i,j)=(4,4): && C_{TTT}=2 + 640 q + 80896 q^2 + 7787008 q^3 + 702799872 q^4 + \dots
\nn\\
\eea
Taking two further derivatives, we get a meromorphic modular form of weight 6  (generalizing  \cite{Antoniadis:1995cy} for $(i,j)=(1,1)$)
\be
C_{TTTTT} = \left[ \frac{128}{3} \frac{5 J_2^+ +768}{(J_2^+-256)^2} + \frac{16 \Delta_\chi }{21J_2^+} \right] 
E_2^{(2)} (E_4(\tau) - 4 E_4(2\tau)) 
\ee
where $\Delta_\chi=-4(i^2+j^2-2)-80(\tfrac{1}{i}+\tfrac{1}{j}-2)$.
By the GV/NL correspondence, this implies 
\bea
\Phi_2^{[i,j]} &=& -\frac{11}{9} E_4 E_6 \,\Thi{2} + 12 E_4^2 \, D\Thi{2}-32 E_6 D^2 \Thi{2} \nn\\
&&+\Delta_\chi \left( -\frac{1}{1512}E_4 E_6 \, \Thi{2} + \frac{1}{42} E_4^2 \, D\Thi{2} +\frac{4}{21}  D^2\Thi{2}\right) 
\eea
The $q$-expansions read
\bea
\Phi_2^{[1,1]}&=& \Phi_2^{[1,4]} =
-2+216 q+1280 q^{9/8}+10032 q^{3/2}+153900 q^2+546048 q^{17/8}+1280448 q^{5/2}
   +\dots
\nn\\
\Phi_2^{[2,1]}&=& \Phi_2^{[4,2]} =
-2+2q^{1/2}+188 q+1408 q^{9/8}+9656 q^{3/2}+152764 q^2+548224 q^{17/8}+1281168
   q^{5/2}
   +\dots
\nn\\
\Phi_2^{[2,2]} &=&
-2+4 \sqrt{q}+160 q+1536 q^{9/8}+9280 q^{3/2}+151628 q^2+550400 q^{17/8}+1281888
   q^{5/2}
   +\dots\nn\\
\eea
The generating series of base degree 1, genus 0 GW invariants are given by 
\be
\tilde h_1 = \frac{4ij (E_2^{(2)})^2}{\eta(\tau)^8\eta(2\tau)^8 (J_2^+ -256) } 
\ee
leading to the base degree 1, genus 0 GV invariants
\bea
(i,j)=(1,1): && W_1= 4 + 640 q + 72224 q^2 + 7539200 q^3 + 757561520 q^4 +\dots
\nn\\
(i,j)=(2,1): && W_1= 8 + 1184 q + 130192 q^2 + 13506752 q^3 + 1351951736 q^4 +\dots
\nn\\
(i,j)=(4,1): && W_1= 16 + 2144 q + 231888 q^2 + 23953120 q^3 + 2388434784 q^4 +\dots
\nn\\
(i,j)=(2,2): && W_1= 16 + 2176 q + 234176 q^2 + 24185856 q^3 + 2411703648 q^4 +\dots
\nn\\
(i,j)=(4,2): && W_1= 32 + 3904 q + 416736 q^2 + 42873024 q^3 + 4258889312 q^4\dots
\nn\\
(i,j)=(4,4): && W_1= 64 + 6912 q + 742784 q^2 + 75933184 q^3 + 7518494784 q^4 +\dots
\nn\\
\eea
The generating series of base degree 2, genus 0 GW invariants are given by 
\be
\tilde h_2 = \frac{P_{20}^{[i,j]}}{\eta(\tau)^{24}\eta(2\tau)^{24} (J_2^+ -256)^3} 
\ee
where $P_{20}^{[i,j]}$ is a Fricke-even modular form of weight 20,
\bea
P_{20}^{[1,1]} &=&-\frac{89 (E_2^{(2)})^{10}}{216}-\frac{31 (E_2^{(2)})^6
   (E_4^{(2,-)})^2}{2592}+\frac{41 (E_2^{(2)})^2 (E_4^{(2,-)})^4}{7776}
    \nn\\ 
P_{20}^{[2,1]} &=&-\frac{89 (E_2^{(2)})^{10}}{54}+\frac{55 (E_2^{(2)})^6
   (E_4^{(2,-)})^2}{1944}+\frac{229 (E_2^{(2)})^2 (E_4^{(2,-)})^4}{17496}
    \nn\\
P_{20}^{[4,1]} &=&-\frac{178 (E_2^{(2)})^{10}}{27}+\frac{43 (E_2^{(2)})^6 (E_4^{(2,-)})^2}{162}+\frac{5
   (E_2^{(2)})^2 (E_4^{(2,-)})^4}{162}
\\
P_{20}^{[2,2]} &=&-\frac{178 (E_2^{(2)})^{10}}{27}+\frac{203 (E_2^{(2)})^6
   (E_4^{(2,-)})^2}{486}+\frac{133 (E_2^{(2)})^2 (E_4^{(2,-)})^4}{4374} \nn
\\
P_{20}^{[4,2]} &=&-\frac{712 (E_2^{(2)})^{10}}{27}+\frac{554 (E_2^{(2)})^6
   (E_4^{(2,-)})^2}{243}+\frac{122 (E_2^{(2)})^2 (E_4^{(2,-)})^4}{2187}\nn
\\
P_{20}^{[4,4]} &=&\frac{104 (E_2^{(2)})^6 (E_4^{(2,-)})^2}{9}-\frac{2848
   (E_2^{(2)})^{10}}{27}
\nn
\eea
leading to the base degree 2, genus 0 GV invariants
\bea
(i,j)=(1,1): && W_2-\frac18 W_1(2\tau)=10032 q^2 + 7539200 q^3 + 2346819520 q^4+\dots
\nn\\
(i,j)=(2,1): && W_2-\frac18 W_1(2\tau)=
56 q + 82832 q^2 + 36973952 q^3 + 9855970048 q^4 +\dots
\nn\\
(i,j)=(4,1): && W_2-\frac18 W_1(2\tau)=120 q + 356368 q^2 + 144785584 q^3 + 36512550816 q^4+\dots
\nn\\
(i,j)=(2,2): && W_2-\frac18 W_1(2\tau)=768 q + 495360 q^2 + 168619008 q^3 + 40145749504 q^4+\dots
\nn\\
(i,j)=(4,2): && W_2-\frac18 W_1(2\tau)=3216 q + 2030176 q^2 + 644810720 q^3 + 146713202496 q^4+\dots
\nn\\
(i,j)=(4,4): && W_2-\frac18 W_1(2\tau)=14400 q + 8271360 q^2 + 2445747712 q^3 + 532817161216 q^4+\dots
\nn\\
\eea
The generating series of base degree 3, genus 0 GV invariants are given by 
\be
\tilde h_3 = \frac{P_{36}^{[i,j]}}{\eta(\tau)^{40}\eta(2\tau)^{40} (J_2^+ -256)^5} 
\ee
with
\bea
P_{36}^{[1,1]} &=&
\tfrac{2263 (E_2^{(2)})^{18}}{3888}-\tfrac{371 (E_2^{(2)})^{14}
   (E_4^{(2,-)})^2}{2916}+\tfrac{131 (E_2^{(2)})^{10} (E_4^{(2,-)})^4}{17496}-\tfrac{5
   (E_2^{(2)})^6 (E_4^{(2,-)})^6}{26244}+\tfrac{(E_2^{(2)})^2 (E_4^{(2,-)})^8}{104976}
 \nn\\ 
P_{36}^{[2,1]} &=&\tfrac{2263
   (E_2^{(2)})^{18}}{486}-\tfrac{4345 (E_2^{(2)})^{14} (E_4^{(2,-)})^2}{4374}+\tfrac{370
   (E_2^{(2)})^{10} (E_4^{(2,-)})^4}{6561}-\tfrac{107 (E_2^{(2)})^6
   (E_4^{(2,-)})^6}{177147}+\tfrac{29 (E_2^{(2)})^2 (E_4^{(2,-)})^8}{1594323}
\nn\\
P_{36}^{[4,1]} &=&\tfrac{9052 (E_2^{(2)})^{18}}{243}-\tfrac{5722
   (E_2^{(2)})^{14} (E_4^{(2,-)})^2}{729}+\tfrac{109 (E_2^{(2)})^{10}
   (E_4^{(2,-)})^4}{243}-\tfrac{23 (E_2^{(2)})^6 (E_4^{(2,-)})^6}{6561}
\nn\\
P_{36}^{[2,2]} &=&\tfrac{9052
   (E_2^{(2)})^{18}}{243}-\tfrac{16952 (E_2^{(2)})^{14} (E_4^{(2,-)})^2}{2187}+\tfrac{908
   (E_2^{(2)})^{10} (E_4^{(2,-)})^4}{2187}-\tfrac{128 (E_2^{(2)})^6
   (E_4^{(2,-)})^6}{177147}-\tfrac{176 (E_2^{(2)})^2 (E_4^{(2,-)})^8}{1594323}
 \nn
\\
P_{36}^{[4,2]} &=&\tfrac{72416
   (E_2^{(2)})^{18}}{243}-\tfrac{133904 (E_2^{(2)})^{14} (E_4^{(2,-)})^2}{2187}+\tfrac{21448
   (E_2^{(2)})^{10} (E_4^{(2,-)})^4}{6561}-\tfrac{1520 (E_2^{(2)})^6
   (E_4^{(2,-)})^6}{177147}-\tfrac{1372 (E_2^{(2)})^2 (E_4^{(2,-)})^8}{1594323}
\nn
\\
P_{36}^{[4,4]} &=&\tfrac{579328 (E_2^{(2)})^{18}}{243}-\tfrac{4352 (E_2^{(2)})^{14}
   (E_4^{(2,-)})^2}{9}+\tfrac{56000 (E_2^{(2)})^{10} (E_4^{(2,-)})^4}{2187}-\tfrac{320
   (E_2^{(2)})^6 (E_4^{(2,-)})^6}{2187}
\nn\\
\eea
leading to the  base degree 3, genus 0 GV invariants
\bea
(i,j)=(1,1): && W_3-\frac{1}{27} W_1(3\tau)=288384 q^3 + 757561520 q^4 + 520834042880 q^5 +\dots
\nn\\
(i,j)=(2,1): && W_3-\frac{1}{27} W_1(3\tau)=1104 q^2 + 13865248 q^3 + 12854281944 q^4 + 6201762710528 q^5+\dots
\nn\\
(i,j)=(4,1): && W_3-\frac{1}{27} W_1(3\tau)=-32 q + 14608 q^2 + 144051072 q^3 + 115675981232 q^4 
 + \dots
\nn\\
(i,j)=(2,2): && W_3-\frac{1}{27} W_1(3\tau)=234176 q^2 + 297281152 q^3 + 172102715744 q^4 
+\dots
\nn\\
(i,j)=(4,2): && W_3-\frac{1}{27} W_1(3\tau)=64 q + 2699616 q^2 + 2830008448 q^3 + 1466448925952 q^4 
+\dots
\nn\\
(i,j)=(4,4): && W_3-\frac{1}{27}W_1(3\tau)=6912 q + 31344000 q^2 + 26556152064 q^3 + 12305418469184 q^4 
+\dots
\nn\\
\eea

\subsection{$m=3$}
Using \eqref{CTTT} we get 
\bea
(i,j)=(1,1): && C_{TTT}=12 + 360 q + 21816 q^2 + 958104 q^3 + 38828088 q^4  + \dots
\nn\\
(i,j)=(2,1): && C_{TTT}=9 + 378 q + 21546 q^2 + 960390 q^3 + 38811690 q^4 +\dots
\nn\\
(i,j)=(3,1): && C_{TTT}=8 + 348 q + 21996 q^2 + 956580 q^3 + 38839020 q^4 + \dots
\nn\\
(i,j)=(2,2): && C_{TTT}=6 + 396 q + 21276 q^2 + 962676 q^3 + 38795292 q^4 +\dots
\nn\\
(i,j)=(3,2): && C_{TTT}=5 + 366 q + 21726 q^2 + 958866 q^3 + 38822622 q^4+ \dots
\nn\\
(i,j)=(3,3): && C_{TTT}=4 + 336 q + 22176 q^2 + 955056 q^3 + 38849952 q^4 + \dots
\nn\\
\eea
Taking two further derivatives, we get a meromorphic modular form of weight 6 
\be
C_{TTTTT} = \left[ \frac{9}{2} \frac{5 J_3^+ +432}{(J_3^+-108)^2} - \frac{3\Delta_\chi}{32 J_3^+(\tau)} \right] 
E_2^{(3)}( E_4(\tau) - 9 E_4(3\tau))
\ee
where $\Delta_\chi=-6(i^2+j^2-2)-24(\tfrac{1}{i}+\tfrac{1}{j}-2)$.
By the GV/NL correspondence, this implies 
\bea
\Phi_3^{[i,j]} &=& -\frac{49}{60} E_4 E_6 \,\Thi{3} 
+ \frac{47}{5} E_4^2 \, D\Thi{3}-48 E_6 D^2 \Thi{3} 
+\frac{576}{5} D^3 \Thi{3} 
\nn\\
&&+\Delta_\chi \left( -\frac{7}{11520}E_4 E_6 \, \Thi{3}
-\frac{13}{480} E_4^2 \, D\Thi{3} +\frac{1}{4}  D^2\Thi{3}
+\frac{6}{5}  D^3\Thi{3}\right) 
\eea
The $q$-expansions read
\bea
\Phi_3^{[1,1]} &=&
-2+8 q^{3/4}+180 q+720 q^{13/12}+5364 q^{4/3}+35280 q^{7/4}+125352 q^2+370224
   q^{25/12}
   +\dots
\nn\\
\Phi_3^{[2,1]} &=&
-2+14 q^{3/4}+168 q+756 q^{13/12}+5292 q^{4/3}+35220 q^{7/4}+125496 q^2+369684
   q^{25/12}+\dots
\nn\\
\Phi_3^{3,1} &=&
-2+4 q^{3/4}+188 q+696 q^{13/12}+5412 q^{4/3}+35320 q^{7/4}+125256 q^2+370584
   q^{25/12}+\dots
\nn\\
\Phi_3^{[2,2]} &=&
-2+20 q^{3/4}+156 q+792 q^{13/12}+5220 q^{4/3}+35160 q^{7/4}+125640 q^2+369144
   q^{25/12}+\dots
\nn\\
\Phi_3^{[3,2]} &=&
-2+10 q^{3/4}+176 q+732 q^{13/12}+5340 q^{4/3}+35260 q^{7/4}+125400 q^2+370044
   q^{25/12}+\dots
\nn\\
\Phi_3^{[3,3]} &=&
-2+196 q+672 q^{13/12}+5460 q^{4/3}+35360 q^{7/4}+125160 q^2+370944 q^{25/12}+1082328
   q^{7/3}
   +\dots
\nn\\
\eea
The generating series of  base degree 1, genus 0 GV invariants are given by 
\be
\tilde h_1 = \frac{-3ij E_2^{(3)}}{\eta(\tau)^6\eta(3\tau)^6 (J_3^+-108)}  
\ee
leading to 
\bea
(i,j)=(1,1): && W_1= 6 + 360 q + 17064 q^2 + 770280 q^3 + 33726420 q^4 + +\dots
\nn\\
(i,j)=(2,1): && W_1= 12 + 648 q + 30132 q^2 + 1353696 q^3 + 59046084 q^4+\dots
\nn\\
(i,j)=(3,1): && W_1= 18 + 900 q + 41778 q^2 + 1871784 q^3 + 81468792 q^4 + \dots
\nn\\
(i,j)=(2,2): && W_1= 24 + 1152 q + 53136 q^2 + 2377728 q^3 + 103323672 q^4 +\dots
\nn\\
(i,j)=(3,2): && W_1= 6 + 1584 q + 73728 q^2 + 3286224 q^3 + 142523712 q^4\dots
\nn\\
(i,j)=(3,3): && W_1= 54 + 2160 q + 102492 q^2 + 4539024 q^3 + 196580898 q^4+\dots
\nn\\
\eea
The generating series of base degree 2, genus 0 GW invariants are given by 
\be
 \tilde h_2 = \frac{P_{14}^{[i,j]}}{\eta(\tau)^{18}\eta(3\tau)^{18} (J_3^+ -108)^3} 
 \ee
where $P_{14}^{[i,j]}$ is a Fricke-odd modular form of weight 14,
\bea
P_{14}^{[1,1]} &=& \frac{5}{18432} (E_2^{(3)})^3 (2E_4-\frac52(E_2^{(3)})^2)^2 -\frac{15}{2} (E_2^{(3)})^4 \Xi_3 -\frac{5751}{2}(E_2^{(3)})  \Xi_3^2 
\nn\\
P_{14}^{[2,1]} &=& \frac{1}{1024}(E_2^{(3)})^3 (2E_4-\frac{375}{16} (E_2^{(3)})^2)^2 -\frac{375}{16} (E_2^{(3)})^4 \Xi_3 
-5832 (E_2^{(3)}) \Xi_3^2 
\nn\\
P_{14}^{[3,1]} &=& \frac{47}{18432} (E_2^{(3)})^3 (2E_4-\frac52(E_2^{(3)})^2)^2 -\frac{741}{16} (E_2^{(3)})^4 \Xi_3 
-\frac{15147}{2} (E_2^{(3)}) \Xi_3^2
 \nn\\
P_{14}^{[2,2]} &=& \frac{5}{4608} (E_2^{(3)})^3 (2E_4-\frac52(E_2^{(3)})^2)^2 -\frac{303}{4} (E_2^{(3)})^4 \Xi_3
-7776 (E_2^{(3)}) \Xi_3^2 
\nn\\
P_{14}^{[3,2]} &=& \frac{19}{9216} (E_2^{(3)})^3 (2E_4-\frac52(E_2^{(3)})^2)^2 -\frac{2415}{16} (E_2^{(3)})^4 \Xi_3 
-648\frac{5751}{2} (E_2^{(3)}) \Xi_3^2 
\nn\\
P_{14}^{[3,3]} &=& \frac{5}{2048} (E_2^{(3)})^3 (2E_4-\frac52(E_2^{(3)})^2)^2 -\frac{2403}{8} (E_2^{(3)})^4 \Xi_3 +\frac{64881}{2} (E_2^{(3)}) \Xi_3^2
\nn\\
\eea
leading to the  base degree 2, genus 0 GV invariants
\bea
(i,j)=(1,1): && W_2-\frac18 W_1(2\tau)=2682 q^2 + 770280 q^3 + 99533664 q^4
+\dots
\nn\\
(i,j)=(2,1): && W_2-\frac18 W_1(2\tau)=27 q + 19764 q^2 + 3622821 q^3 + 405431784 q^4
+\dots
\nn\\
(i,j)=(3,1): && W_2-\frac18 W_1(2\tau)=36 q + 46548 q^2 + 8009712 q^3 + 864795636 q^4
+\dots
\nn\\
(i,j)=(2,2): && W_2-\frac18 W_1(2\tau)=396 q + 112068 q^2 + 15951564 q^3 + 1602730872 q^4
+\dots
\nn\\
(i,j)=(3,2): && W_2-\frac18 W_1(2\tau)=909 q + 255960 q^2 + 34736049 q^3 + 3387935304 q^4
+\dots
\nn\\
(i,j)=(3,3): && W_2-\frac18 W_1(2\tau)=2160 q + 583470 q^2 + 75358080 q^3 + 7141800672 q^4+\dots
\nn\\
\eea
The generating series of base degree 3, genus 0 GW invariants are given by 
\be
\tilde h_3 = \frac{E_2^{(3)}  P_{24}^{[i,j]}}{\eta(\tau)^{30}\eta(3\tau)^{30} (J_3^+ -108)^5} 
\ee
with
\bea
P_{26}^{[1,1]} &=&  \frac{(E_2^{(3)})^{12}}{165888}
+\frac{\Delta _6 (E_2^{(3)})^9}{64}-\frac{271}{4}
   \Delta _6^2 (E_2^{(3)})^6-1647 \Delta _6^3 (E_2^{(3)})^3 -3519612 \Delta
   _6^4
\nn\\
P_{26}^{[2,1]} &=& \frac{(E_2^{(3)})^{12}}{147456}+\frac{275 \Delta _6
   (E_2^{(3)})^9}{3072}-\frac{8087}{16} \Delta _6^2 (E_2^{(3)})^6+48546 \Delta _6^3
   (E_2^{(3)})^3-1912896 \Delta
   _6^4
\nn\\
P_{26}^{[3,1]} &=& -\frac{13
   (E_2^{(3)})^{12}}{995328}+\frac{691 \Delta _6 (E_2^{(3)})^9}{2304}-\frac{26343}{16} \Delta
   _6^2 (E_2^{(3)})^6+\frac{415935}{2} \Delta _6^3 (E_2^{(3)})^3+5922396 \Delta _6^4
 \nn\\
P_{26}^{[2,2]} &=& \frac{(E_2^{(3)})^{12}}{41472}+\frac{43 \Delta _6 (E_2^{(3)})^9}{48}-\frac{15247}{4}
   \Delta _6^2 (E_2^{(3)})^6+516672 \Delta _6^3 (E_2^{(3)})^3+55240704 \Delta
   _6^4
\nn\\
P_{26}^{[3,2]} &=&\frac{127
   (E_2^{(3)})^{12}}{3981312}+\frac{29183 \Delta _6 (E_2^{(3)})^9}{9216}-\frac{207045}{16}
   \Delta _6^2 (E_2^{(3)})^6+1398870 \Delta _6^3 (E_2^{(3)})^3+64805184 \Delta _6^4
\nn\\
P_{26}^{[3,3]} &=& 
   \frac{(E_2^{(3)})^{12}}{18432}+\frac{1383 \Delta _6
   (E_2^{(3)})^9}{128}-\frac{724383}{16} \Delta _6^2 (E_2^{(3)})^6+2344464 \Delta _6^3
   (E_2^{(3)})^3-534511548 \Delta_6^4 
   \nn\\
\eea
leading to the  base degree 3, genus 0 GV invariants
\bea
(i,j)=(1,1): && W_3-\frac{1}{27} W_1(3\tau)=35472 q^3 + 33726420 q^4 + 9382024152 q^5 
+\dots
\nn\\
(i,j)=(2,1): && W_3-\frac{1}{27} W_1(3\tau)=324 q^2 + 1398216 q^3 + 529944336 q^4 + 106249923936 q^5
+\dots
\nn\\
(i,j)=(3,1): && W_3-\frac{1}{27} W_1(3\tau)=-4 q + 1512 q^2 + 5604204 q^3 + 1928672640 q^4 + 363480492960 q^5
+\dots
\nn\\
(i,j)=(2,2): && W_3-\frac{1}{27} W_1(3\tau)=53136 q^2 + 28024704 q^3 + 6746381496 q^4 + 1084701369600 q^5
+\dots
\nn\\
(i,j)=(3,2): && W_3-\frac{1}{27} W_1(3\tau)=16 q + 231336 q^2 + 106245024 q^3 + 23702767680 q^4 
+\dots
\nn\\
(i,j)=(3,3): && W_3-\frac{1}{27} W_1(3\tau)=336 q + 1011384 q^2 + 400151088 q^3 + 82704010218 q^4 
+\dots
\nn\\
\eea

\subsection{$m=4$}
Using \eqref{CTTT} we get 
\be
 C_{TTT}=8\left(\frac{1}{i}+\frac{1}{j}\right) + 256 q + 10240 q^2 + 283648 q^3 + 7159808 q^4  + \dots
\ee
\be
C_{TTTTT}= -32 \frac{J_4^+ + 64}{(J_4^+ - 64)^2} 
\left[ \theta_2^4(2\tau)-2 X_2(2\tau)\right]^2 \left[ 3\theta_2^4(2\tau)+2 X_2(2\tau)\right]
\ee
This implies that the generating series of NL invariants is independent of $(i,j)$\footnote{Indeed, $\Delta_\chi=-8(i^2+j^2-2)-48(\tfrac{1}{i}+\tfrac{1}{j}-2)$ vanishes 
for the relevant values $(1,1)$, $(1,2)$ and $(2,2)$.},
\bea
\Phi_4^{[i,j]} &=& -\frac{125}{243} E_4 E_6 \,\Thi{4} 
+ \frac{488}{81} E_4^2 \, D\Thi{4}-\frac{512}{9} E_6 D^2 \Thi{4} 
+\frac{2048}{9} D^3 \Thi{4} \nn \\
&=&-2+300 q+512 q^{17/16}+2496 q^{5/4}+20992 q^{25/16}+217200 q^2+290816 q^{33/16}\nn\\&&
+665600
   q^{9/4}+2286080 q^{41/16}+10226400 q^3
   +12441600 q^{49/16}+21854400 q^{13/4}+\dots
\eea
where $\Delta_\chi=0$ since all models have the same Euler number.
The generating series of base degree 1, genus 0 GW invariants are given by 
\be
\tilde h_1 = -\frac{8i j E_2^{(4,-)}(\tau)}{\eta(2\tau)^{12} (J_4^+-64)} 
\ee
leading to 
\bea
(i,j)=(1,1): && W_1=  8 + 256 q + 7232 q^2 + 197632 q^3 + 5263712 q^4 +\dots
\nn\\
(i,j)=(2,1): && W_1= 16 + 448 q + 12576 q^2 + 342144 q^3 + 9079536 q^4+\dots
\nn\\
(i,j)=(2,2): && W_1= 32 + 768 q + 21888 q^2 + 591872 q^3 + 15653568 q^4 +\dots
\nn\\
\eea
The generating series of base degree 2, genus 0 GW invariants are given by\footnote{Note that the eta product in the denominator is {\it not} the cube of the one appearing in $\tilde h_1$, unlike in the standard ansatz \eqref{htkansatz}.} 
\be
 \tilde h_2 = \frac{P_{12}^{[i,j]}}{\eta^8(\tau) \eta^{16}(2\tau) \eta^8(4\tau) (J_4^+-64)^3} 
 \ee
where $P_{12}^{[i,j]}$ is a Fricke-even modular form of weight 12,
\bea
P_{12}^{[1,1]} &=& \frac{1}{108} \left(-89 (E_2^{(4,-)})^6+15 (E_2^{(4,-)})^4 (E_2^{(4,+)})^2+54 (E_2^{(4,-)})^2
   (E_2^{(4,+)})^4\right)
   \nn\\
P_{12}^{[2,1]} &=&    
 \frac{1}{27} \left(-89 (E_2^{(4,-)})^6 + 52 (E_2^{(4,-)})^4 (E_2^{(4,+)})^2 + 19 (E_2^{(4,-)})^2 (E_2^{(4,+)})^4\right)
      \nn\\
P_{12}^{[2,2]} &=&       
-\frac{4}{27} \left(89 (E_2^{(4,-)})^6-89 (E_2^{(4,-)})^4 (E_2^{(4,+)})^2+5 (E_2^{(4,-)})^2
   (E_2^{(4,+)})^4\right)   
   \nn\\
\eea
leading to  base degree 2, genus 0 GV invariants
\bea
(i,j)=(1,1): && W_2-\frac18 W_1(2\tau)= 1248 q^2 + 197632 q^3 + 14953088 q^4+\dots
\nn\\
(i,j)=(2,1): &&  W_2-\frac18 W_1(2\tau)=16 q + 8464 q^2 + 899328 q^3 + 59453184 q^4+\dots
\nn\\
(i,j)=(2,2): &&  W_2-\frac18 W_1(2\tau)= 256 q + 46016 q^2 + 3851264 q^3 + 229545472 q^4+\dots
\nn\\
\eea
The generating series of base degree 3, genus 0 GW invariants are given by
\be
\tilde h_3 = \frac{(E_2^{(4,-)})^4  P_{20}^{[i,j]}}{\eta(\tau)^{8}\eta(2\tau)^{40} \eta(4\tau)^8 (J_4^+ -64)^5} 
\ee
with
\bea
P_{16}^{[1,1]} &=&
\tfrac{2263}{1944} (E_2^{(4,-)})^8-\tfrac{395}{162} (E_2^{(4,-)})^6 (E_2^{(4,+)})^2
+\tfrac{17}{12}   (E_2^{(4,-)})^4 (E_2^{(4,+)})^4 \nn\\
&& -\tfrac{13}{54} (E_2^{(4,-)})^2 (E_2^{(4,+)})^6
+\tfrac{7}{216}   (E_2^{(4,+)})^8
   \nn\\
P_{16}^{[2,1]} &=& \tfrac{2263}{243} (E_2^{(4,-)})^8-\tfrac{4633}{243} (E_2^{(4,-)})^6 (E_2^{(4,+)})^2
+\tfrac{884}{81}(E_2^{(4,-)})^4 (E_2^{(4,+)})^4 \nn\\
&& -\tfrac{286}{243} (E_2^{(4,-)})^2
   (E_2^{(4,+)})^6-\tfrac{14}{243} (E_2^{(4,+)})^8
      \nn\\
P_{16}^{[2,2]} &=&  \tfrac{18104}{243} (E_2^{(4,-)})^8-\tfrac{36208}{243} (E_2^{(4,-)})^6 (E_2^{(4,+)})^2+\tfrac{248}{3}
   (E_2^{(4,-)})^4 (E_2^{(4,+)})^4\nn\\
   &&-\tfrac{1984}{243} (E_2^{(4,-)})^2
   (E_2^{(4,+)})^6-\tfrac{64}{243}  (E_2^{(4,+)})^8
   \nn\\
\eea
leading to the  base degree 3, genus 0 GV invariants
\bea
(i,j)=(1,1): && W_3-\frac{1}{27} W_1(3\tau)= 10496 q^3 + 5263712 q^4 + 840087552 q^5 + 83742591808 q^6 + \dots
\nn\\
(i,j)=(2,1): &&  W_3-\frac{1}{27} W_1(3\tau)=160 q^2 + 355520 q^3 + 77860272 q^4 + 9142925312 q^5 +\dots
\nn\\
(i,j)=(2,2): &&  W_3-\frac{1}{27} W_1(3\tau)= 21888 q^2 + 6747904 q^3 + 952111808 q^4 + 90236788736 q^5 +\dots
\nn\\
\eea

\subsection{$m=5$}
Using \eqref{CTTT} we get 
\bea
(i,j)=(1,1): && C_{TTT}= 20 + 200 q + 6280 q^2 + 124400 q^3 + 2314120 q^4 + \dots
\nn\\
(i,j)=(2,1): && C_{TTT}= 15 + 190 q + 6310 q^2 + 124660 q^3 + 2311590 q^4 + \dots
\nn\\
(i,j)=(2,2): && C_{TTT}= 10 + 180 q + 6340 q^2 + 124920 q^3 + 2309060 q^4 + \dots
\nn\\
\eea
Taking two further derivatives, we get
\be
C_{TTTTT}= 5 E_2^{(5)} ( E_4(\tau)-25 E_4(5\tau))
 \left[\frac{9 (J_5^+) ^3+656 (J_5^+) ^2-2432 J_5^+ +15168}{24 \left((J_5^+) ^2-44 J_5^+ -16\right)^2} +(\Delta_\chi+4)
 \frac{8+J_5^+}{96(J_5^+)^2 } \right]
\ee
where $\Delta_\chi=-10(i^2+j^2-2)-40(\tfrac{1}{i}+\tfrac{1}{j}-2)$.
By the GV/NL correspondence, this implies 
\bea 
\Phi_5^{[i,j]} &=&-\tfrac{14497}{69984} E_4 E_6 \,\Thi{5} 
+ \tfrac{144001}{163296} E_4^2 \, D\Thi{5}
-\tfrac{3650}{567} E_6 D^2 \Thi{5} 
+ \tfrac{19900}{567} D^3 \Thi{5}
+ \tfrac{80000}{63} D^5 \Thi{5}\nn \\
&&+\Delta_\chi\left(
\tfrac{2261}{8398080} E_4 E_6 \,\Thi{5} 
+ \tfrac{48241}{19595520} E_4^2 \, D\Thi{5}
\tfrac{1525}{6804} D^2 \Thi{5} 
-\tfrac{11605}{3402} D^3 \Thi{5}
+\tfrac{2000}{189} D^5 \Thi{5}
\right)\nn\\
\eea
The $q$-expansions read
\bea
\Phi_5^{[1,1]} &=&
-2+20 q^{4/5}+148 q+400 q^{21/20}+1520 q^{6/5}+1136 q^{5/4}+9200 q^{29/20}+71640 q^{9/5}+97024
   q^2
   +\dots\nn\\
\Phi_5^{[2,1]} &=&
-2+10 q^{4/5}+158 q+380 q^{21/20}+1530 q^{6/5}+1156 q^{5/4}+9220 q^{29/20}+71800 q^{9/5}+96854
   q^2
   +\dots
\nn\\
\Phi_5^{[2,2]} &=&
-2+168 q+360 q^{21/20}+1540 q^{6/5}+1176 q^{5/4}+9240 q^{29/20}+71960 q^{9/5}+96684 q^2
+\dots\nn\\
\eea
The generating series of base degree 1, genus 0 GW invariants are given by 
\be
\tilde h_1 = \frac{5i j (E_2^{(5)})^2}{\eta(\tau)^{8}\eta(5\tau)^{8} [(J_5^+)^2 -44 J_5^+ -16]} 
\ee
leading to 
\bea
(i,j)=(1,1): && 
W_1=10 + 200 q + 4020 q^2 + 77600 q^3 + 1472510 q^4 + \dots
\nn\\
(i,j)=(2,1): && 
W_1=20 + 340 q + 6920 q^2 + 132740 q^3 + 2511600 q^4 +\dots
\nn\\
(i,j)=(2,2): && W_1= 40 + 560 q + 11960 q^2 + 226800 q^3 + 4281800 q^4+\dots
\eea
The generating series of base degree 2, genus 0 GW invariants are given by 
\be
 \tilde h_2 = \frac{5 (E_2^{(5)})^2  P_{16}^{[i,j]}}{\eta(\tau)^{24}\eta(5\tau)^{24} ((J_5^+)^2-44J_5^+-16)^3} 
 \ee
where $P_{16}^{[i,j]}$ is a Fricke-even modular form of weight 16,
\bea
P^{[1,1]}_{16}&=&\tfrac{576  \left(1101375 (E_2^{(5)})^6+172 (E_2^{(5)})^2
   (E_4^{(5,-)})^2\right)\Xi_5-901125 (E_2^{(5)})^8+394992 (E_2^{(5)})^4 (E_4^{(5,-)})^2+2128
   (E_4^{(5,-)})^4}{2^{24}3^7}
   \nn\\
P^{[2,1]}_{16}&=&\   \tfrac{ 576 \left(1101375 (E_2^{(5)})^6-1456 (E_2^{(5)})^2
   (E_4^{(5,-)})^2\right)\Xi_5 -901125 (E_2^{(5)})^8+397656 (E_2^{(5)})^4 (E_4^{(5,-)})^2+976
   (E_4^{(5,-)})^4}{2^{22}3^7}
   \nn\\
P^{[2,2]}_{16}&=&  \tfrac{ 192 \left(367125 (E_2^{(5)})^4-1028
   (E_4^{(5,-)})^2\right)\Xi_5 -100125 (E_2^{(5)})^6+44480 (E_2^{(5)})^2 (E_4^{(5,-)})^2}{2^{20}3^5} 
   \nn\\
\eea
where $E_4^{(5,-)}=E_4(\tau) - 25 E_4(5\tau)$ and $\Xi_5=\eta^4(\tau)\eta^4(5\tau)$.
This 
leads to the base degree 2, genus 0 GV invariants
\bea
(i,j)=(1,1): && W_2-\frac18 W_1(2\tau)=760 q^2 + 77600 q^3 + 4042740 q^4 + 158766880 q^5 +\dots
\nn\\
(i,j)=(2,1): &&  W_2-\frac18 W_1(2\tau)=10 q + 4780 q^2 + 343550 q^3 + 15771100 q^4+\dots
\nn\\
(i,j)=(2,2): &&  W_2-\frac18 W_1(2\tau)= 180 q + 25060 q^2 + 1438320 q^3 + 59793160 q^4+\dots
\nn\\
\eea
The generating series of base degree 3, genus 0 GV invariants are given by 
\be
\tilde h_3 = \frac{(E_2^{(5)})^{2} P_{32}^{[i,j]}}{\eta(\tau)^{40}\eta(5\tau)^{40} [(J_5^+)^2-44J_5^+-16]^5} 
\ee
with
\bea
P^{[1,1]}_{32}&=&-\frac{5}{2^{40}3^{13}}
 \left(-14151564421875 (E_2^{(5)})^{16}+6408303842250 (E_2^{(5)})^{12}
   (E_4^{(5,-)})^2 \right. \nn\\
   && -52791372216 (E_2^{(5)})^8 (E_4^{(5,-)})^4+11870208 (E_2^{(5)})^4
   (E_4^{(5,-)})^6
  \nn \\
 &&  +288 \Delta _4^5 (E_2^{(5)})^2 \left(34876259859375
   (E_2^{(5)})^{12}-417585537000 (E_2^{(5)})^8 (E_4^{(5,-)})^2 \right. 
  \nn \\ &&\left.  \left. +670396032 (E_2^{(5)})^4
   (E_4^{(5,-)})^4+827392 (E_4^{(5,-)})^6\right)+90112
   (E_4^{(5,-)})^8\right)
   \nn\\
P^{[1,2]}_{32}&= &
-\frac{5}{2^{37} 3^{13}}
 \left(-14151564421875 (E_2^{(5)})^{16}+6405905250000 (E_2^{(5)})^{12}
   (E_4^{(5,-)})^2 \right. \nn\\ 
   && -51737462424 (E_2^{(5)})^8 (E_4^{(5,-)})^4+17235360 (E_2^{(5)})^4
   (E_4^{(5,-)})^6 \nn\\
   && +288 \Delta _4^5 (E_2^{(5)})^2 \left(34876259859375
   (E_2^{(5)})^{12}-411674643000 (E_2^{(5)})^8 (E_4^{(5,-)})^2 \right. \nn \\
   && \left. \left. +679076064 (E_2^{(5)})^4
   (E_4^{(5,-)})^4+605888 (E_4^{(5,-)})^6\right)+87808
   (E_4^{(5,-)})^8\right)
  \nn\\
P^{[2,2]}_{32}&=&  
-\frac{5}{2^{34}3^9}
 (E_2^{(5)})^4 \left(-174710671875 (E_2^{(5)})^{12}+79055637750 (E_2^{(5)})^8
   (E_4^{(5,-)})^2  \right.\nn\\
   &&  -625683872 (E_2^{(5)})^4 (E_4^{(5,-)})^4 
    +96 \Delta _4^5 (E_2^{(5)})^2
   \left(1291713328125 (E_2^{(5)})^8
   \right.
   \nn\\
   && 
   \left. \left. -15028287000 (E_2^{(5)})^4 (E_4^{(5,-)})^2+25186448
   (E_4^{(5,-)})^4\right) +264416 (E_4^{(5,-)})^6\right)
   \nn\\
\eea
leading to 
\bea
(i,j)=(1,1): && W_3-\frac1{27} W_1(3\tau)=4600 q^3 + 1472510 q^4 + 158766880 q^5 + 10915610420 q^6 +\dots
\nn\\
(i,j)=(2,1): &&  W_3-\frac1{27} W_1(3\tau)=100 q^2 + 138700 q^3 + 20691180 q^4 + 1672278340 q^5+\dots
\nn\\
(i,j)=(2,2): &&  W_3-\frac1{27} W_1(3\tau)= 11960 q^2 + 2514480 q^3 + 244792240 q^4 + 16066988480 q^5+\dots
\nn\\
\eea

\subsection{$(m,i,j)=(6,1,1)$}
Using \eqref{CTTT} we get 
\be
C_{TTT}= 24 + 168 q + 4296 q^2 + 69720 q^3 + 1009608 q^4 +13989168 q^5+ \dots
\ee
\be
C_{TTTTT}= E_2^{6,+)} \left[ (E_2^{6,-)})^2 -\frac14  (E_2^{6,+)})^2 \right]
\frac{J_6^+  \left(7 (J_6^+) ^3+331 (J_6^+) ^2-331 J_6^+ -7\right)}{2 \left((J_6^+) ^2-34 J_6^++1\right)^2}
\ee
By the GV/NL correspondence, this implies 
\bea
\Phi_6^{[1,1]} &=&
-\frac{22}{9} E_4^2 \, D\Thi{6}
+\frac{44}{3} E_6 D^2 \Thi{6} 
-\frac{112}{5} D^3 \Thi{6}
+ \frac{9216}{5} D^5 \Thi{6}\nn \\
&=&
-2+140 q+672 q^{25/24}+1032 q^{7/6}+5152 q^{11/8}+5880 q^{3/2}+31416 q^{5/3}+88788 q^2\nn\\&&
 +431520
   q^{49/24}+380016 q^{13/6}+905856 q^{19/8}+736456 q^{5/2}+2730420 q^{8/3}+4169424
   q^3+\dots\nn\\
\eea
The generating series of base degree 1, genus 0 GW invariants are given by 
\be
\tilde h_1 = \frac{(E_2^{(6,-)}-5 E_2^{(6)})^2}
{48\eta(\tau)^4\eta(2\tau)^4 \eta(3\tau)^4\eta(6\tau)^4
 [(J_6^+)^2 -34 J_6^+ +1]} 
\ee
leading to 
\bea
W_1=12 + 168 q + 2604 q^2 + 39120 q^3 + 576384 q^4+ \dots
\eea
The generating series of base degree 2, genus 0 GW invariants are given by 
\be
 \tilde h_2 = \frac{P_{20}}{\eta^{12}(\tau) \eta^{12}(2\tau)\eta^{12}(3\tau)
 \eta^{12}(4\tau) ((J_6^+)^2-34J_6^+-1)^3} 
 \ee
with
\be
\begin{split}
P_{20}=& \frac{1}{2^{31} 3^{18}.19.59 }\left[
1119868701192394 (E_2^{(6)})^{10}-16721660451381924 (E_2^{(6)})^9
   {E_2^{(6,-)}} \right. \\
   & 
   +55171532155930641 (E_2^{(6)})^8 (E_2^{(6,-)})^2
   - 39299446415
   (E_2^{(6,-)})^{10} \\
   &
   +648 (E_2^{(6)})^5 \left(5415069069593
   (E_2^{(6,-)})^3 (E_2^{(6,+)})^2  -4044199392577 {E_2^{(6,-)}} (E_2^{(6,+)})^4\right)
   \\
 &   
   -843460158250 (E_2^{(6,-)})^8 (E_2^{(6,+)})^2-3458400589802 (E_2^{(6,-)})^6
   (E_2^{(6,+)})^4 \\
   & \left.
   -6029964794504 (E_2^{(6,-)})^4 (E_2^{(6,+)})^6-5054592585583 (E_2^{(6,-)})^2
   (E_2^{(6,+)})^8-1750651307606 (E_2^{(6,+)})^{10} \right]
   \end{split}
\ee
leading to 
 \be
 W_2-\frac18 W_1(2\tau)=516 q^2 + 39120 q^3 + 1543752 q^4 + 46475184 q^5 + 1195713912 q^6 + \dots
\ee
We have not determined the modular generating series of base degree 3, genus 0 GV invariants for this model.

\subsection{$(m,i,j)=(7,1,1)$}
Using \eqref{CTTT} we get 
\be
C_{TTT}= 28 + 140 q + 3388 q^2 + 43232 q^3 + 540988 q^4 + 6163640 q^5+ \dots
\ee
\be
C_{TTTTT}= 7 E_2^{(7)} ( E_4(\tau)-49 E_4(7\tau))
\frac{5 (J_7^+) ^5+274 (J_7^+) ^4-546 (J_7^+) ^3+12672 (J_7^+) ^2+6885 J_7^+ +4374}
{72 (J_7^+) ^2 \left((J_7^+) ^2-26  J_7^+ -27\right)^2}
\ee
By the GV/NL correspondence, this implies 
\bea
\Phi_7^{[1,1]} &=&
\tfrac{71111}{393660} E_4 E_6 \Thi{7}
-\tfrac{11518577}{2099520} E_4^2 \, D\Thi{7}
+\tfrac{34006}{729}E_6 D^2 \Thi{7} 
-\tfrac{563206}{3645} D^3 \Thi{7}
+ \tfrac{1075648}{405} D^5 \Thi{7}\nn \\
&=&
-2+56 q^{25/28}+136 q+280 q^{29/28}+812 q^{8/7}+2520 q^{9/7}+3192 q^{37/28}+16800
   q^{11/7}+23320 q^{7/4} \nn\\&& +97272 q^{53/28}+81920 q^2+194936 q^{57/28}+315728 q^{15/7}+582680
   q^{16/7}+677488 q^{65/28}
 +  \dots\nn\\
\eea
The generating series of base degree 1, genus 0 GW invariants are given by 
\be
\tilde h_1 = -\frac{7 E_2^{(7)}}{\eta(\tau)^{6}\eta(7\tau)^{6} [(J_7^+)^2 -26 J_7^+ -27]} 
\ee
leading to 
\bea
W_1=7 + 70 q + 931 q^2 + 11270 q^3 + 137298 q^4 + \dots
\eea
The generating series of base degree 2, genus 0 GW invariants are given by 
\be
 \tilde h_2 = \frac{P_{14}}
 {\eta(\tau)^{18}\eta(7\tau)^{18} [(J_7^+)^2 -26 J_7^+ -27]^3} 
\ee
with
\be
\begin{split}
P_{14}=& \frac{7 {E_2^{(7)}}}{2^{15} 3^{11} 5^3}
 \left[410257645 (E_2^{(7)})^6
-534174774 (E_2^{(7)})^4 {E_4^{(7,+)}}+255054520   (E_2^{(7)})^3 {E_6^{(7,+)}} \right. \\
& \left.  -50864760 {E_2^{(7)}} {E_4^{(7,-)}} {E_6^{(7,+)}}+14171760
   (E_6^{(7,+)})^2\right]
\end{split}
\ee
leading to 
\be
 W_2-\frac18 W_1(2\tau)=406 q^2 + 22540 q^3 + 713048 q^4 + 17372264 q^5 + 364190344 q^6 + \dots
\ee
The generating series of base degree 3, genus 0 GW invariants are given by 
\be
\tilde h_3 = \frac{P_{26}}{\eta(\tau)^{30}\eta(7\tau)^{30}  [(J_7^+)^2 -26 J_7^+ -27]^5} 
\ee
with
\be
\begin{split}
P_{26}=&
\frac{7 E_2^{(7)}}{2^{22} 3^{23} 5 } \left(49478895704035 (E_2^{(7)})^{12}-50479038842379 (E_2^{(7)})^{10}
   E_4^{(7,+)}+55888343232190 (E_2^{(7)})^9 E_6^{(7,-)} \right. \\
   &  -44136548857548 (E_2^{(7)})^7
   E_4^{(7,+)} E_6^{(7,-)}+7849405670040 (E_2^{(7)})^6 (E_6^{(7,-)})^2 \\
   &+3498388920432
   (E_2^{(7)})^4 E_4^{(7,+)} (E_6^{(7,-)})^2-607676554080 (E_2^{(7)})^3
   (E_6^{(7,-)})^3 \\
   &  \left. -49465111104 E_2^{(7)} E_4^{(7,+)} (E_6^{(7,-)})^3+4194840960
   (E_6^{(7,-)})^4\right)
   \end{split}
\ee
leading to 
\be
 W_3-\frac1{27} W_1(3\tau)=1596 q^3 + 274596 q^4 + 17372264 q^5 + 725242728 q^6+ \dots
\ee

\subsection{$(m,i,j)=(8,1,1)$}
Using \eqref{CTTT} we get 
\be
C_{TTT}= 32 + 128 q + 2560 q^2 + 31232 q^3 + 323584 q^4 + 3216128 q^5+ \dots
\ee
\be
C_{TTTTT}=  8 E_2^{(8,+)} \left[ (E_2^{(8,-)})^2 -\frac19  (E_2^{(8,+)})^2 \right]
\frac{(J_8^+) ^4+40 (J_8^+) ^3-640 J_8^+ -256}{3 \left((J_8^+) ^2-24 J_8^+ +16\right)^2}
\ee
By the GV/NL correspondence, this implies 
\bea
\Phi_8^{[1,1]} &=& \frac{385388}{1148175}\,E_4 E_6 \Thi{8}
-\frac{3022256}{382725}\,E_4^2 D\Thi{8}
+  \frac{851968}{8505}\,E_6 D^2\Thi{8}  -\frac{11763712}{42525}\, E_4 D^3\Thi{8} \nn\\
&&
+ \frac{16777216}{4725} D^5\Thi{8}+ \frac{1}{18}\,E_4 E_6 T_{2,4} 
-2 E_4^2 DT_{2,4}-16 E_6 D^2 T_{2,4}\nn\\
&=&
-1+108 q+128 q^{33/32}+640 q^{9/8}+1152 q^{41/32}+5016 q^{3/2}+6144 q^{49/32}+25728 q^{57/32} \nn\\
&&+76950 q^2+89088 q^{65/32}+273024
   q^{17/8}+267264 q^{73/32}+640224 q^{5/2}+\dots\nn\\
\eea
The generating series of base degree 1, genus 0 GW invariants are given by 
\be
\tilde h_1 = \frac{16 X_4(2\tau) (4 E_2^{(2)}(4\tau)+  E_2^{(8)}(\tau))}
{9\eta(2\tau)^8\eta(4\tau)^8 [(J_8^+)^2 -24 J_8^+ +16]}
\ee
leading to 
\bea
W_1=16 + 256 q + 3328 q^2 + 39936 q^3 + 452864 q^4 + \dots
\eea
The generating series of base degree 2, genus 0 GW invariants are given by 
\be
 \tilde h_2 = \frac{P_{20}}{ \eta^{24}(2\tau)\eta^{24}(4\tau)((J_8^+)^2-24J_8^+-16)^3} 
 \ee
with
\be
\begin{split}
P_{20}=& \tfrac{1}{2^{29} 3^{12}. 29 .41} \left[ -41961455973 (E_2^{(8)})^{10}
+471716259912 (E_2^{(8)})^9 {E_2^{(8,-)}}     \right. \\ 
& -1199633892840
   (E_2^{(8)})^8 (E_2^{(8,-)})^2  -88246779305184   (E_2^{(8,-)})^{10}\\
&  
   -3456 (E_2^{(8)})^5 \left(1555507834 (E_2^{(8,-)})^3
   (E_2^{(8,+)})^2-89432013 {E_2^{(8,-)}} (E_2^{(8,+)})^4\right)  \\
 &+60567037350768 (E_2^{(8,-)})^8 (E_2^{(8,+)})^2
   -17273815025328 (E_2^{(8,-)})^6
   (E_2^{(8,+)})^4  \\
      & \left.
   +2602795786488 (E_2^{(8,-)})^4 (E_2^{(8,+)})^6-213078241446 (E_2^{(8,-)})^2
   (E_2^{(8,+)})^8+7908818587 (E_2^{(8,+)})^{10}  \right]
   \end{split}
\ee
leading to 
\be
 W_2-\frac18 W_1(2\tau)=304 q^2 + 14848 q^3 + 393312 q^4 + 8109056 q^5 + 144154048 q^6 +  \dots
\ee

\subsection{$(m,i,j)=(9,1,1)$}
Using \eqref{CTTT} we get 
\be
C_{TTT}= 36 + 108 q + 2268 q^2 + 22464 q^3 + 216540 q^4 + 1876608 q^5+ \dots
\ee
\be
C_{TTTTT}= \frac{2187 E_2^{(9,+)}}{2}  \left[ 
(E_2^{(9,+)})^2 +16  (E_2^{(9,-)} )^2 -\frac{1}{108} (E_2^{(9)})^2 \right] 
\frac{J_9^+  \left((J_9^+) ^2+27\right)}{[(J_9^+) ^2-18 J_9^+)+27]^2}
\ee
By the GV/NL correspondence, this implies 
\bea
\Phi_9^{[1,1]} &=& -\tfrac{35669501}{513216},\,E_4 E_6 \Thi{9} 
+ \tfrac{220005401}{77760}\,E_4^2 D\Thi{9}
 -\tfrac{1479401}{135} \,E_6 D^2\Thi{9} 
+ \tfrac{5863382}{675}\, E_4 D^3\Thi{9} \nn\\
&& -\tfrac{35428544}{75}, D^5\Thi{9}
+\tfrac{198485}{10392624}, E_4 E_6 T_{3.3}
+  \tfrac{1505}{4374} E_4^2 DT_{3.3} \nn\\
&& -V_3 \left( \tfrac{392035}{62355744} E_4 E_6 \Thi{3}  + \tfrac{2345}{39366}, E_4^2 D\Thi{3}  +\tfrac{2170}{19683} E_6 D^2 \Thi{3} \right) \nn\\
&=&-1+150 q+108 q^{37/36}+270 q^{10/9}+1248 q^{5/4}+1944 q^{49/36}+3348 q^{13/9}+15012 q^{61/36} \nn\\ &&
+23814 q^{16/9}+108600 q^2+82944
   q^{73/36}+120852 q^{19/9}+332800 q^{9/4}+348732 q^{85/36} \nn\\
   &&+486324 q^{22/9}+1229904 q^{97/36}+1639872 q^{25/9}+5113200 q^3+\dots\nn\\
\eea
The generating series of base degree 1, genus 0 GW invariants are given by 
\be
\tilde h_1 = \frac{9E_2^{(9)}(\tau)^2}
{32\eta(3\tau)^{16} [(J_9^+)^2 -18 J_9^+ -27]} 
\ee
leading to 
\bea
W_1=18 + 108 q + 1134 q^2 + 10116 q^3 + 92718 q^4+\dots
\eea
The generating series of base degree 2, genus 0 GW invariants are given by 
\be
 \tilde h_2 = \frac{P_{20}}{ \eta^{48}(3\tau) ((J_9^+)^2-18J_9^+-27)^3} 
 \ee
with 
 \be
 \begin{split}
 P_{20} =& \frac{1}{2^{36} 3^3.11.19}
 \left[ -150209 (E_2^{(9)})^{10}+84359088 (E_2^{(9)})^9 {E_2^{(9,-)}}
 -1811538432 (E_2^{(9)})^8
   (E_2^{(9,-)})^2 \right. \\
   &  -71663616 {E_2^{(9)}}^5 \left(40884 (E_2^{(9,-)})^3 {E_2^{(9,+)}}^2+1781 
   {E_2^{(9,-)}}
   (E_2^{(9,+)})^4\right) \\
   & \left. -16779605704704 \left(12
   (E_2^{(9,-)})^2+(E_2^{(9,+)})^2\right)^5 \right]
      \end{split}
 \ee
 leading to 
\be
 W_2-\frac18 W_1(2\tau)= 270 q^2 + 10116 q^3 + 231012 q^4 + 4114152 q^5 + 63677754 q^6 +  \dots
\ee

\subsection{$(m,i,j)=(11,1,1)$}
Using \eqref{CTTT} we get, for the putative model $X_{11}^{[1,1]}$, 
\be
C_{TTT}= 44 + 88 q + 1672 q^2 + 14344 q^3 + 114312 q^4 + 814088 q^5+ \dots
\ee
By the GV/NL correspondence, this implies 
\bea
\Phi_{11}^{[1,1]} &=& -\tfrac{23693562426079}{764569164579840}\, E_4 E_6 \Thi{11}
-\tfrac{76956434703269}{101942555277312}\,E_4^2 D\Thi{11}
  -\tfrac{12172378220189}{884918014560}  \,E_6 D^2\Thi{11}
   \nn\\
&&+ \tfrac{32012336240599}{2212295036400} E_4 D^3\Thi{11} 
  -\tfrac{2751852346738}{15363159975} D^5\Thi{11} \nn\\
&&
+ \tfrac{71253827029}{6991944806400} E_4^2 V_{11}  D\Thi{1}
+   \tfrac{14295721}{7283275840} E_6 V_{11}  D^2\Thi{1}
 +\tfrac{779188993}{118353232400} E_4 V_{11}  D^3\Thi{1}
 \nn\\
&& +V_{11} \left( 
-\tfrac{9205661}{6991944806400} E_4^2 \Thi{1} +  \tfrac{61497}{3641637920} E_4 D^3\Thi{1} \right)
-\tfrac{71353991071}{136342923724800}E_4 V_{11} E_4 D\Thi{1} \nn\\
&=&
-1+66 q+88 q^{45/44}+198 q^{12/11}+264 q^{49/44}+528 q^{53/44}+924 q^{14/11}+1760
   q^{15/11}+3234 q^{16/11} \nn\\
   && +6512 q^{69/44}+9240 q^{7/4}+26730 q^{20/11}+30184 q^{81/44}+32472
   q^2+73392 q^{89/44}+99924 q^{23/11} \nn\\
   && + 110088 q^{93/44}+165264 q^{97/44}+219912
   q^{25/11}+319572 q^{26/11}+457468 q^{27/11}
   +\dots \nn\\
\eea
The generating series of base degree 1, genus 0 GW invariants are given by 
\be
\tilde h_1 = -\frac{22 [ E_2^{(11)}(\tau) -16 
\eta^2(\tau)\eta^2(11\tau) ] }
{10\eta^6(\tau)\eta^6(11\tau) [(J_{11}^+)^3-20 (J_{11}^+)^2 + 56 J_{11}^+-44]}
\ee
leading to 
\be
W_1= 22 + 88 q + 792 q^2 + 5720 q^3 + 42724 q^4 + 309496 q^5 + 2242900 q^6 
 +\dots
\ee
The generating series of base degree 2, genus 0 GW invariants are given by 
\be
 \tilde h_2 = \frac{P_{14}}{ \eta^{18}(\tau)\eta^{18}(11\tau) [(J_{11}^+)^3-(20J_{11}^+)^2+56 J_{11}^+-44]^3} 
 \ee
 with
\be
\begin{split}
P_{14}=&
\frac{11}{2^9 3^3 5^6} 
\left((E_2^{(11)})^7-424 (E_2^{(11)})^6 E_2^{(11,-)}+46928 (E_2^{(11)})^5 (E_2^{(11,-)})^2
\right. \\
& -643040
   (E_2^{(11)})^4 (E_2^{(11,-)})^3+21059200 (E_2^{(11)})^3 (E_2^{(11,-)})^4-226994176 (E_2^{(11)})^2
   (E_2^{(11,-)})^5\\
   & \left. -2586114816 E_2^{(11)} (E_2^{(11,-)})^6+13009661952 (E_2^{(11,-)})^7\right)
   \end{split}
   \ee
leading to 
\be
 W_2-\frac18 W_1(2\tau)=198 q^2 + 5720 q^3 + 103224 q^4 + 1465288 q^5 + 18237032 q^6 +\dots
\ee


\begin{thebibliography}{100}

\bibitem{Candelas:1985en}
P.~Candelas, G.~T. Horowitz, A.~Strominger, and E.~Witten, ``{Vacuum
  Configurations for Superstrings},'' {\em Nucl.Phys.} {\bf B258} (1985)
46--74.

\bibitem{Lerche:1989uy}
W.~Lerche, C.~Vafa, and N.~P. Warner, ``{Chiral Rings in N=2 Superconformal
  Theories},'' {\em Nucl. Phys. B} {\bf 324} (1989) 427--474.

\bibitem{MR1403918}
M.~Kontsevich, ``Homological algebra of mirror symmetry,'' in {\em Proceedings
  of the {I}nternational {C}ongress of {M}athematicians, {V}ol.\ 1, 2
  ({Z}\"urich, 1994)}, pp.~120--139.
\newblock Birkh\"auser, Basel, 1995.
\newblock \href{http://www.arXiv.org/abs/alg-geom/9411018}{{\tt
  alg-geom/9411018}}.

\bibitem{Candelas:1990rm}
P.~Candelas, X.~C. de~la Ossa, P.~S. Green, and L.~Parkes, ``{A pair of
  Calabi-Yau manifolds as an exactly soluble superconformal theory},'' {\em
  Nucl. Phys.} {\bf B359} (1991)
21--74.

\bibitem{Candelas:1993dm}
P.~Candelas, X.~De~La~Ossa, A.~Font, S.~H. Katz, and D.~R. Morrison, ``{Mirror
  symmetry for two parameter models. I},'' {\em Nucl. Phys.} {\bf B416} (1994)
  481--538,
\href{http://www.arXiv.org/abs/hep-th/9308083}{{\tt hep-th/9308083}}.

\bibitem{Candelas:1994hw}
P.~Candelas, A.~Font, S.~H. Katz, and D.~R. Morrison, ``{Mirror symmetry for
  two parameter models. 2},'' {\em Nucl. Phys.} {\bf B429} (1994) 626--674,
\href{http://www.arXiv.org/abs/hep-th/9403187}{{\tt hep-th/9403187}}.

\bibitem{hosono:1993qy}
S.~Hosono, A.~Klemm, S.~Theisen, and S.-T. Yau, ``{Mirror symmetry, mirror map
  and applications to Calabi-Yau hypersurfaces},'' {\em Commun. Math. Phys.}
  {\bf 167} (1995) 301--350,
\href{http://www.arXiv.org/abs/hep-th/9308122}{{\tt hep-th/9308122}}.

\bibitem{Green1987}
P.~Green and T.~H\"ubsch, ``{Calabi-Yau manifolds as complete intersections in
  products of complex projective spaces},'' {\em Communications in Mathematical
  Physics} {\bf 109} (Mar., 1987) 99--108.

\bibitem{Candelas1988}
P.~Candelas, A.~Dale, C.~L\"{u}tken, and R.~Schimmrigk, ``{Complete
  intersection Calabi-Yau manifolds},'' {\em Nuclear Physics B} {\bf 298}
  (Mar., 1988) 493--525.

\bibitem{Kreuzer:2000xy}
M.~Kreuzer and H.~Skarke, ``{Complete classification of reflexive polyhedra in
  four-dimensions},'' {\em Adv. Theor. Math. Phys.} {\bf 4} (2000) 1209--1230,
  \href{http://www.arXiv.org/abs/hep-th/0002240}{{\tt hep-th/0002240}}.

\bibitem{Batyrev:1993oya}
V.~V. Batyrev, ``{Dual polyhedra and mirror symmetry for Calabi-Yau
  hypersurfaces in toric varieties},'' {\em J. Alg. Geom.} {\bf 3} (1994)
  493--545, \href{http://www.arXiv.org/abs/alg-geom/9310003}{{\tt
  alg-geom/9310003}}.

\bibitem{Knapp:2021vkm}
J.~Knapp, E.~Scheidegger, and T.~Schimannek, ``{On genus one fibered Calabi-Yau
  threefolds with 5-sections},''
  \href{http://www.arXiv.org/abs/2107.05647}{{\tt 2107.05647}}.

\bibitem{klemm:2004km}
A.~Klemm, M.~Kreuzer, E.~Riegler, and E.~Scheidegger, ``{Topological string
  amplitudes, complete intersection Calabi-Yau spaces and threshold
  corrections},''
\href{http://www.arXiv.org/abs/hep-th/0410018}{{\tt hep-th/0410018}}.

\bibitem{doran2015families}
C.~F. Doran, A.~Harder, A.~Y. Novoseltsev, and A.~Thompson, ``{Families of
  lattice polarized K3 surfaces with monodromy},'' {\em International
  mathematics research notices} {\bf 2015} (2015), no.~23, 12265--12318,
  \href{http://www.arXiv.org/abs/1312.6434}{{\tt 1312.6434}}.

\bibitem{doran2016calabi}
C.~F. Doran, A.~Harder, A.~Y. Novoseltsev, and A.~Thompson, ``{Calabi--Yau
  threefolds fibred by mirror quartic K3 surfaces},'' {\em Advances in
  Mathematics} {\bf 298} (2016) 369--392.

\bibitem{doran2020calabi}
C.~F. Doran, A.~Harder, A.~Y. Novoseltsev, and A.~Thompson, ``{Calabi--Yau
  threefolds fibred by high rank lattice polarized K3 surfaces},'' {\em
  Mathematische Zeitschrift} {\bf 294} (2020) 783--815,
  \href{http://www.arXiv.org/abs/1701.03279}{{\tt 1701.03279}}.

\bibitem{kooistra2021threefolds}
R.~Kooistra and A.~Thompson, ``Threefolds fibred by mirror sextic double
  planes,'' {\em Canadian Journal of Mathematics} {\bf 73} (2021), no.~5,
  1305--1346.

\bibitem{Doran2019}
C.~F. Doran, A.~Harder, A.~Y. Novoseltsev, and A.~Thompson,
  ``Calabi{\textendash}yau threefolds fibred by high rank lattice polarized k3
  surfaces,'' {\em Mathematische Zeitschrift} {\bf 294} (Apr., 2019) 783--815,
  \href{http://www.arXiv.org/abs/1701.03279}{{\tt 1701.03279}}.

\bibitem{doran:2016uea}
C.~F. Doran, A.~Harder, and A.~Thompson, ``{Mirror symmetry, Tyurin
  degenerations and fibrations on Calabi-Yau manifolds},'' {\em Proc. Symp.
  Pure Math.} {\bf 96} (2017) 93--132,
  \href{http://www.arXiv.org/abs/1601.08110}{{\tt 1601.08110}}.

\bibitem{doran2019doran}
C.~F. Doran, J.~Kostiuk, and F.~You, ``{The Doran-Harder-Thompson conjecture
  for toric complete intersections},'' {\em Advances in Mathematics} {\bf 415}
  (2023) 108893.

\bibitem{Batyrev:1998kx}
V.~V. Batyrev, I.~Ciocan-Fontanine, B.~Kim, and D.~van Straten, ``{Conifold
  transitions and mirror symmetry for Calabi-Yau complete intersections in
  Grassmannians},'' {\em Nucl. Phys. B} {\bf 514} (1998) 640--666,
  \href{http://www.arXiv.org/abs/alg-geom/9710022}{{\tt alg-geom/9710022}}.

\bibitem{Haghighat:2008ut}
B.~Haghighat and A.~Klemm, ``{Topological Strings on Grassmannian Calabi-Yau
  manifolds},'' {\em JHEP} {\bf 01} (2009) 029,
  \href{http://www.arXiv.org/abs/0802.2908}{{\tt 0802.2908}}.

\bibitem{Borcea:1996mxz}
C.~Borcea, ``{K3 surfaces with involution and mirror pairs of Calabi-Yau
  manifolds},'' {\em AMS/IP Stud. Adv. Math.} {\bf 1} (1996) 717--743.

\bibitem{hofmann:2013}
J.~Hofmann, {\em Monodromy calculations for some differential equations}.
\newblock PhD thesis, Mainz, 2013.

\bibitem{Katz:2022lyl}
S.~Katz, A.~Klemm, T.~Schimannek, and E.~Sharpe, ``{Topological Strings on
  Non-commutative Resolutions},'' {\em Commun. Math. Phys.} {\bf 405} (2024),
  no.~3, 62, \href{http://www.arXiv.org/abs/2212.08655}{{\tt 2212.08655}}.

\bibitem{Ueda:2016wfa}
K.~Ueda and Y.~Yoshida, ``{Equivariant A-twisted GLSM and Gromov--Witten
  invariants of CY 3-folds in Grassmannians},'' {\em JHEP} {\bf 09} (2017) 128,
  \href{http://www.arXiv.org/abs/1602.02487}{{\tt 1602.02487}}.

\bibitem{Batyrev:2008rp}
V.~Batyrev and M.~Kreuzer, ``{Constructing new Calabi-Yau 3-folds and their
  mirrors via conifold transitions},'' {\em Adv. Theor. Math. Phys.} {\bf 14}
  (2010), no.~3, 879--898, \href{http://www.arXiv.org/abs/0802.3376}{{\tt
  0802.3376}}.

\bibitem{Bouchard:2016lfg}
V.~Bouchard, T.~Creutzig, D.-E. Diaconescu, C.~Doran, C.~Quigley, and
  A.~Sheshmani, ``{Vertical D4-D2-D0 Bound States on K3 Fibrations and
  Modularity},'' {\em Commun. Math. Phys.} {\bf 350} (2017), no.~3, 1069--1121,
\href{http://www.arXiv.org/abs/1601.04030}{{\tt 1601.04030}}.

\bibitem{maulik2007gromov}
D.~Maulik and R.~Pandharipande, ``{Gromov-Witten theory and Noether-Lefschetz
  theory},'' \href{http://www.arXiv.org/abs/0705.1653}{{\tt 0705.1653}}.

\bibitem{dolgachev:1996xw}
I.~V. Dolgachev, ``{Mirror symmetry for lattice polarized K3 surfaces},''
  \href{http://www.arXiv.org/abs/alg-geom/9502005}{{\tt alg-geom/9502005}}.

\bibitem{doran2001algebraic}
C.~F. Doran, ``Algebraic and geometric isomonodromic deformations,'' {\em
  Journal of Differential Geometry} {\bf 59} (2001), no.~1, 33--85.

\bibitem{nikulin1980integral}
V.~V. Nikulin, ``{Integral symmetric bilinear forms and some of their
  applications},'' {\em Mathematics of the USSR-Izvestiya} {\bf 14} (1980),
  no.~1, 103.

\bibitem{oguiso1993algebraic}
K.~Oguiso, ``On algebraic fiber space structures on a calabi-yau 3-fold,'' {\em
  International Journal of Mathematics} {\bf 4} (1993), no.~03, 439--465.

\bibitem{Berglund:1993ax}
P.~Berglund, P.~Candelas, X.~De~La~Ossa, A.~Font, T.~Hubsch, D.~Jancic, and
  F.~Quevedo, ``{Periods for Calabi-Yau and Landau-Ginzburg vacua},'' {\em
  Nucl. Phys. B} {\bf 419} (1994) 352--403,
  \href{http://www.arXiv.org/abs/hep-th/9308005}{{\tt hep-th/9308005}}.

\bibitem{Aspinwall:1994ay}
P.~S. Aspinwall, ``{The Moduli space of N=2 superconformal field theories},''
  in {\em {ICTP Summer School in High-energy Physics and Cosmology}},
  pp.~0352--401.
\newblock 11, 1994.
\newblock \href{http://www.arXiv.org/abs/hep-th/9412115}{{\tt hep-th/9412115}}.

\bibitem{Haghighat:2009nr}
B.~Haghighat and A.~Klemm, ``{Solving the Topological String on K3
  Fibrations},'' {\em JHEP} {\bf 01} (2010) 009,
  \href{http://www.arXiv.org/abs/0908.0336}{{\tt 0908.0336}}.

\bibitem{kachru:1995wm}
S.~Kachru and C.~Vafa, ``{Exact results for N=2 compactifications of heterotic
  strings},'' {\em Nucl. Phys.} {\bf B450} (1995) 69--89,
\href{http://www.arXiv.org/abs/hep-th/9505105}{{\tt hep-th/9505105}}.

\bibitem{kachru:1995fv}
S.~Kachru, A.~Klemm, W.~Lerche, P.~Mayr, and C.~Vafa, ``{Nonperturbative
  results on the point particle limit of N=2 heterotic string
  compactifications},'' {\em Nucl. Phys. B} {\bf 459} (1996) 537--558,
  \href{http://www.arXiv.org/abs/hep-th/9508155}{{\tt hep-th/9508155}}.

\bibitem{Kaplunovsky:1995tm}
V.~Kaplunovsky, J.~Louis, and S.~Theisen, ``{Aspects of duality in N=2 string
  vacua},'' {\em Phys. Lett. B} {\bf 357} (1995) 71--75,
  \href{http://www.arXiv.org/abs/hep-th/9506110}{{\tt hep-th/9506110}}.

\bibitem{Antoniadis:1995zn}
I.~Antoniadis, E.~Gava, K.~S. Narain, and T.~R. Taylor, ``{N=2 type II
  heterotic duality and higher derivative F terms},'' {\em Nucl. Phys. B} {\bf
  455} (1995), no.~1-2, 109--130,
  \href{http://www.arXiv.org/abs/hep-th/9507115}{{\tt hep-th/9507115}}.

\bibitem{klemm:1995tj}
A.~Klemm, W.~Lerche, and P.~Mayr, ``{K3 Fibrations and heterotic type II string
  duality},'' {\em Phys.Lett.} {\bf B357} (1995) 313--322,
\href{http://www.arXiv.org/abs/hep-th/9506112}{{\tt hep-th/9506112}}.

\bibitem{Klawer:2021ltm}
D.~Kl\"awer, ``{Modular curves and the refined distance conjecture},'' {\em
  JHEP} {\bf 12} (2021) 088, \href{http://www.arXiv.org/abs/2108.00021}{{\tt
  2108.00021}}.

\bibitem{Anderson:2015iia}
L.~B. Anderson, F.~Apruzzi, X.~Gao, J.~Gray, and S.-J. Lee, ``{A new
  construction of Calabi\textendash{}Yau manifolds: Generalized CICYs},'' {\em
  Nucl. Phys. B} {\bf 906} (2016) 441--496,
  \href{http://www.arXiv.org/abs/1507.03235}{{\tt 1507.03235}}.

\bibitem{Berglund:2016yqo}
P.~Berglund and T.~H\"ubsch, ``{On Calabi\textendash{}Yau generalized complete
  intersections from Hirzebruch varieties and novel $K3$-fibrations},'' {\em
  Adv. Theor. Math. Phys.} {\bf 22} (2018) 261--303,
  \href{http://www.arXiv.org/abs/1606.07420}{{\tt 1606.07420}}.

\bibitem{Doran:2005gu}
C.~F. Doran and J.~W. Morgan, ``{Mirror symmetry and integral variations of
  Hodge structure underlying one parameter families of Calabi-Yau
  threefolds},'' in {\em {Workshop on Calabi-Yau Varieties and Mirror
  Symmetry}}.
\newblock 5, 2005.
\newblock \href{http://www.arXiv.org/abs/math/0505272}{{\tt math/0505272}}.

\bibitem{Alexandrov:2018lgp}
S.~Alexandrov and B.~Pioline, ``{Black holes and higher depth mock modular
  forms},'' {\em Commun. Math. Phys.} {\bf 374} (2019), no.~2, 549--625,
\href{http://www.arXiv.org/abs/1808.08479}{{\tt 1808.08479}}.

\bibitem{Alexandrov:2019rth}
S.~Alexandrov, J.~Manschot, and B.~Pioline, ``{S-duality and refined BPS
  indices},'' {\em Commun. Math. Phys.} {\bf 380} (2020), no.~2, 755--810,
  \href{http://www.arXiv.org/abs/1910.03098}{{\tt 1910.03098}}.

\bibitem{Alexandrov:2023zjb}
S.~Alexandrov, S.~Feyzbakhsh, A.~Klemm, B.~Pioline, and T.~Schimannek,
  ``{Quantum geometry, stability and modularity},'' {\em Commun. Num. Theor.
  Phys.} {\bf 18} (2024), no.~1, 49--151,
  \href{http://www.arXiv.org/abs/2301.08066}{{\tt 2301.08066}}.

\bibitem{Alexandrov:2023ltz}
S.~Alexandrov, S.~Feyzbakhsh, A.~Klemm, and B.~Pioline, ``{Quantum geometry and
  mock modularity},'' \href{http://www.arXiv.org/abs/2312.12629}{{\tt
  2312.12629}}.

\bibitem{kudla1990intersection}
S.~S. Kudla and J.~J. Millson, ``Intersection numbers of cycles on locally
  symmetric spaces and fourier coefficients of holomorphic modular forms in
  several complex variables,'' {\em Publications Math{\'e}matiques de
  l'IH{\'E}S} {\bf 71} (1990) 121--172.

\bibitem{0919.11036}
R.~E. Borcherds, ``{Automorphic forms with singularities on Grassmannians.},''
  {\em Invent. Math.} {\bf 132} (1998), no.~3, 491--562.

\bibitem{Aspinwall:1996mn}
P.~S. Aspinwall, ``{K3 surfaces and string duality},'' in {\em {Theoretical
  Advanced Study Institute in Elementary Particle Physics (TASI 96): Fields,
  Strings, and Duality}}, pp.~421--540.
\newblock 11, 1996.
\newblock \href{http://www.arXiv.org/abs/hep-th/9611137}{{\tt hep-th/9611137}}.

\bibitem{Aspinwall:2000fd}
P.~S. Aspinwall, ``{Compactification, geometry and duality: N=2},'' in {\em
  {Theoretical Advanced Study Institute in Elementary Particle Physics (TASI
  99): Strings, Branes, and Gravity}}, pp.~723--805.
\newblock 1, 2000.
\newblock \href{http://www.arXiv.org/abs/hep-th/0001001}{{\tt hep-th/0001001}}.

\bibitem{Kawai:1996te}
T.~Kawai, ``{String duality and modular forms},'' {\em Phys. Lett. B} {\bf 397}
  (1997) 51--62, \href{http://www.arXiv.org/abs/hep-th/9607078}{{\tt
  hep-th/9607078}}.

\bibitem{Marino:1998pg}
M.~Marino and G.~W. Moore, ``{Counting higher genus curves in a Calabi-Yau
  manifold},'' {\em Nucl. Phys. B} {\bf 543} (1999) 592--614,
  \href{http://www.arXiv.org/abs/hep-th/9808131}{{\tt hep-th/9808131}}.

\bibitem{Enoki:2019deb}
Y.~Enoki and T.~Watari, ``{Modular forms as classification invariants of 4D $
  \mathcal{N} $ = 2 Heterotic-IIA dual vacua},'' {\em JHEP} {\bf 06} (2020)
  021, \href{http://www.arXiv.org/abs/1911.09934}{{\tt 1911.09934}}.

\bibitem{Vafa:1995gm}
C.~Vafa and E.~Witten, ``{Dual string pairs with N=1 and N=2 supersymmetry in
  four-dimensions},'' {\em Nucl. Phys. B Proc. Suppl.} {\bf 46} (1996)
  225--247, \href{http://www.arXiv.org/abs/hep-th/9507050}{{\tt
  hep-th/9507050}}.

\bibitem{Melnikov:2012cv}
I.~V. Melnikov, R.~Minasian, and S.~Theisen, ``{Heterotic flux backgrounds and
  their IIA duals},'' {\em JHEP} {\bf 07} (2014) 023,
  \href{http://www.arXiv.org/abs/1206.1417}{{\tt 1206.1417}}.

\bibitem{Israel:2023itj}
D.~Israel and Y.~Proto, ``{A worldsheet approach to \ensuremath{\mathscr{N}} =
  1 heterotic flux backgrounds},'' {\em JHEP} {\bf 06} (2023) 175,
  \href{http://www.arXiv.org/abs/2302.01889}{{\tt 2302.01889}}.

\bibitem{Israel:2023tjw}
D.~Israel, I.~V. Melnikov, R.~Minasian, and Y.~Proto, ``{Topology change and
  heterotic flux vacua},'' {\em JHEP} {\bf 06} (2024) 204,
  \href{http://www.arXiv.org/abs/2312.08923}{{\tt 2312.08923}}.

\bibitem{Angelantonj:2015rxa}
C.~Angelantonj, I.~Florakis, and B.~Pioline, ``{Threshold corrections,
  generalised prepotentials and Eichler integrals},'' {\em Nucl. Phys. B} {\bf
  897} (2015) 781--820, \href{http://www.arXiv.org/abs/1502.00007}{{\tt
  1502.00007}}.

\bibitem{Enoki:2021vid}
Y.~Enoki and T.~Watari, ``{Direct computation of monodromy matrices and
  classification of 4d $ \mathcal{N} $ = 2 heterotic-IIA dual vacua},'' {\em
  JHEP} {\bf 03} (2022) 059, \href{http://www.arXiv.org/abs/2111.01575}{{\tt
  2111.01575}}.

\bibitem{Enoki:2022cfc}
Y.~Enoki, Y.~Sato, and T.~Watari, ``{Direct computation of period polynomials
  and classification of K3-fibred Calabi--Yau threefolds},''
  \href{http://www.arXiv.org/abs/2212.07948}{{\tt 2212.07948}}.

\bibitem{Henningson:1996jf}
M.~Henningson and G.~W. Moore, ``{Counting curves with modular forms},'' {\em
  Nucl. Phys. B} {\bf 472} (1996) 518--528,
  \href{http://www.arXiv.org/abs/hep-th/9602154}{{\tt hep-th/9602154}}.

\bibitem{iskovskih1978fano1}
V.~A. Iskovskih, ``{Fano 3-folds. I},'' {\em Izvestiya Akademii Nauk SSSR,
  Seriya Matematicheskaya} {\bf 41} (1977) 516--562.

\bibitem{iskovskih1978fano2}
V.~A. Iskovskih, ``{Fano 3-folds. II},'' {\em Izvestiya Akademii Nauk SSSR,
  Seriya Matematicheskaya} {\bf 42} (1978), no.~3, 506--549.

\bibitem{DeBiase2021}
L.~De~Biase, E.~Fatighenti, and F.~Tanturri, ``{Fano 3-folds from homogeneous
  vector bundles over Grassmannians},'' {\em Revista Matemática Complutense}
  {\bf 35} (July, 2021) 649--710.

\bibitem{Bershadsky:1993cx}
M.~Bershadsky, S.~Cecotti, H.~Ooguri, and C.~Vafa, ``{Kodaira-Spencer theory of
  gravity and exact results for quantum string amplitudes},'' {\em Commun.
  Math. Phys.} {\bf 165} (1994) 311--428,
  \href{http://www.arXiv.org/abs/hep-th/9309140}{{\tt hep-th/9309140}}.

\bibitem{Huang:2006hq}
M.-x. Huang, A.~Klemm, and S.~Quackenbush, ``{Topological string theory on
  compact Calabi-Yau: Modularity and boundary conditions},'' {\em Lect. Notes
  Phys.} {\bf 757} (2009) 45--102,
  \href{http://www.arXiv.org/abs/hep-th/0612125}{{\tt hep-th/0612125}}.

\bibitem{Alim:2012ss}
M.~Alim and E.~Scheidegger, ``{Topological Strings on Elliptic Fibrations},''
  {\em Commun. Num. Theor. Phys.} {\bf 08} (2014) 729--800,
  \href{http://www.arXiv.org/abs/1205.1784}{{\tt 1205.1784}}.

\bibitem{Gopakumar:1998ii}
R.~Gopakumar and C.~Vafa, ``{M-theory and topological strings. I},''
\href{http://www.arXiv.org/abs/hep-th/9809187}{{\tt hep-th/9809187}}.

\bibitem{Gopakumar:1998jq}
R.~Gopakumar and C.~Vafa, ``{M-theory and topological strings. II},''
\href{http://www.arXiv.org/abs/hep-th/9812127}{{\tt hep-th/9812127}}.

\bibitem{alexeevengel23}
V.~Alexeev and P.~Engel, ``{Compact moduli of K3 surfaces},'' {\em Annals of
  Mathematics} {\bf 198} (2023) 727-- 789.

\bibitem{Debarre:2018mmi}
O.~Debarre, ``{Hyperk\"ahler manifolds},''
  \href{http://www.arXiv.org/abs/1810.02087}{{\tt 1810.02087}}.

\bibitem{doran1999picard}
C.~F. Doran~Jr, {\em {Picard-Fuchs uniformization and geometric isomonodromic
  deformations: Modularity and variation of the mirror map}}.
\newblock Harvard University, 1999.

\bibitem{doran2019geometric}
C.~Doran and J.~Kostiuk, ``Geometric variations of local systems and elliptic
  surfaces,'' {\em Israel Journal of Mathematics} (2023) 1--79,
  \href{http://www.arXiv.org/abs/1911.00402}{{\tt 1911.00402}}.

\bibitem{Lian:1994zv}
B.~H. Lian and S.-T. Yau, ``{Arithmetic properties of mirror map and quantum
  coupling},'' {\em Commun. Math. Phys.} {\bf 176} (1996) 163--192,
  \href{http://www.arXiv.org/abs/hep-th/9411234}{{\tt hep-th/9411234}}.

\bibitem{golyshev2016proof}
V.~V. Golyshev and D.~Zagier, ``Proof of the gamma conjecture for fano 3-folds
  of picard rank 1,'' {\em Izvestiya: Mathematics} {\bf 80} (2016), no.~1, 24.

\bibitem{eguchi1997gravitational}
T.~Eguchi, K.~Hori, and C.-S. Xiong, ``Gravitational quantum cohomology,'' {\em
  International Journal of Modern Physics A} {\bf 12} (1997), no.~09,
  1743--1782.

\bibitem{harderthesis2016}
A.~Harder, {\em The Geometry of Landau-Ginzburg Models}.
\newblock PhD thesis, University of Alberta, 2016.

\bibitem{giovthom2024}
L.~Giovenzana and A.~Thompson, ``Degenerations and fibrations of k3 surfaces:
  Lattice polarisations and mirror symmetry,''
  \href{http://www.arXiv.org/abs/2405.12009}{{\tt 2405.12009}}.

\bibitem{doran2021degenerations}
C.~F. Doran, J.~Kostiuk, and F.~You, ``{Degenerations, fibrations and higher
  rank Landau-Ginzburg models},'' {\em arXiv preprint arXiv:2112.12891} (2021).

\bibitem{Doran2016}
C.~Doran, A.~Harder, A.~Novoseltsev, and A.~Thompson, ``Calabi{\textendash}yau
  threefolds fibred by mirror quartic k3 surfaces,'' {\em Advances in
  Mathematics} {\bf 298} (Aug., 2016) 369--392,
  \href{http://www.arXiv.org/abs/1501.04019}{{\tt 1501.04019}}.

\bibitem{doranthompMCSseq}
C.~Doran and A.~Thompson, ``{The Mirror Clemens-Schmid Sequence},''
  \href{http://www.arXiv.org/abs/2109.04849}{{\tt 2109.04849}}.

\bibitem{pichonpharabod2024}
E.~Pichon-Pharabod, ``A semi-numerical algorithm for the homology lattice and
  periods of complex elliptic surfaces over $\mathbb{P}^1$,''
  \href{http://www.arXiv.org/abs/2401.05131}{{\tt 2401.05131}}.

\bibitem{hosono:1994ax}
S.~Hosono, A.~Klemm, S.~Theisen, and S.-T. Yau, ``{Mirror symmetry, mirror map
  and applications to complete intersection Calabi-Yau spaces},'' {\em Nucl.
  Phys. B} {\bf 433} (1995) 501--554,
  \href{http://www.arXiv.org/abs/hep-th/9406055}{{\tt hep-th/9406055}}.

\bibitem{Brodie:2021toe}
C.~Brodie, A.~Constantin, A.~Lukas, and F.~Ruehle, ``{Flops for complete
  intersection Calabi-Yau threefolds},'' {\em J. Geom. Phys.} {\bf 186} (2023)
  104767, \href{http://www.arXiv.org/abs/2112.12106}{{\tt 2112.12106}}.

\bibitem{Batyrev:1994pg}
V.~V. Batyrev and L.~A. Borisov, ``{On Calabi-Yau complete intersections in
  toric varieties},'' in {\em Higher Dimensional Complex Varieties}.
\newblock {DE} {GRUYTER}, 12, 1994.
\newblock \href{http://www.arXiv.org/abs/alg-geom/9412017}{{\tt
  alg-geom/9412017}}.

\bibitem{sturmfels1996grobner}
B.~Sturmfels, {\em {Grobner Bases and Convex Polytopes}}.
\newblock Memoirs of the American Mathematical Society. American Mathematical
  Society, 1996.

\bibitem{Coates2018}
T.~Coates, S.~Galkin, A.~Kasprzyk, and A.~Strangeway, ``{Quantum Periods for
  Certain Four-Dimensional Fano Manifolds},'' {\em Experimental Mathematics}
  {\bf 29} (Aug., 2018) 183--221,
  \href{http://www.arXiv.org/abs/1406.4891}{{\tt 1406.4891}}.

\bibitem{Parshin2010-pn}
A.~N. Parshin and I.~R. Shafarevich, eds., {\em Algebraic geometry {V}}.
\newblock Encyclopaedia of Mathematical Sciences. Springer, Berlin, Germany,
  Dec., 2010.

\bibitem{Coates}
T.~Coates, A.~Corti, S.~Galkin, V.~Golyshev, and A.~Kasprzyk, {\em {Mirror
  Symmetry and Fano Manifolds}}, pp.~285--300.
\newblock European Mathematical Society Publishing House.

\bibitem{Coates_2022}
T.~Coates and A.~M. Kasprzyk, ``{Databases of quantum periods for Fano
  manifolds},'' {\em Scientific Data} {\bf 9} (Apr., 2022).

\bibitem{Coates2007}
T.~Coates and A.~Givental, ``{Quantum Riemann-Roch, Lefschetz and Serre},''
  {\em Annals of Mathematics} {\bf 165} (Jan., 2007) 15--53.

\bibitem{Coates2016}
T.~Coates, A.~Corti, S.~Galkin, and A.~Kasprzyk, ``{Quantum periods for
  3-dimensional Fano manifolds},'' {\em Geometry \& Topology} {\bf 20} (Feb.,
  2016) 103--256, \href{http://www.arXiv.org/abs/1303.3288}{{\tt 1303.3288}}.

\bibitem{Greene:1995hu}
B.~R. Greene, D.~R. Morrison, and A.~Strominger, ``{Black hole condensation and
  the unification of string vacua},'' {\em Nucl. Phys. B} {\bf 451} (1995)
  109--120, \href{http://www.arXiv.org/abs/hep-th/9504145}{{\tt
  hep-th/9504145}}.

\bibitem{Chialva:2007sv}
D.~Chialva, U.~H. Danielsson, N.~Johansson, M.~Larfors, and M.~Vonk,
  ``{Deforming, revolving and resolving - New paths in the string theory
  landscape},'' {\em JHEP} {\bf 02} (2008) 016,
  \href{http://www.arXiv.org/abs/0710.0620}{{\tt 0710.0620}}.

\bibitem{Grimm:2008ed}
T.~W. Grimm and A.~Klemm, ``{U(1) Mediation of Flux Supersymmetry Breaking},''
  {\em JHEP} {\bf 10} (2008) 077,
  \href{http://www.arXiv.org/abs/0805.3361}{{\tt 0805.3361}}.

\bibitem{Gendler:2022ztv}
N.~Gendler, B.~Heidenreich, L.~McAllister, J.~Moritz, and T.~Rudelius,
  ``{Moduli space reconstruction and Weak Gravity},'' {\em JHEP} {\bf 12}
  (2023) 134, \href{http://www.arXiv.org/abs/2212.10573}{{\tt 2212.10573}}.

\bibitem{Greene:1996dh}
B.~R. Greene, D.~R. Morrison, and C.~Vafa, ``{A Geometric realization of
  confinement},'' {\em Nucl. Phys. B} {\bf 481} (1996) 513--538,
  \href{http://www.arXiv.org/abs/hep-th/9608039}{{\tt hep-th/9608039}}.

\bibitem{Addington:2013gpa}
N.~Addington and P.~S. Aspinwall, ``{Categories of Massless D-Branes and del
  Pezzo Surfaces},'' {\em JHEP} {\bf 07} (2013) 176,
  \href{http://www.arXiv.org/abs/1305.5767}{{\tt 1305.5767}}.

\bibitem{Aspinwall:2014vea}
P.~S. Aspinwall, ``{Exoflops in Two Dimensions},'' {\em JHEP} {\bf 07} (2015)
  104, \href{http://www.arXiv.org/abs/1412.0612}{{\tt 1412.0612}}.

\bibitem{Candelas:1987kf}
P.~Candelas, A.~M. Dale, C.~A. Lutken, and R.~Schimmrigk, ``{Complete
  Intersection Calabi-Yau Manifolds},'' {\em Nucl. Phys. B} {\bf 298} (1988)
  493.

\bibitem{Green:1987cr}
P.~S. Green, T.~Hubsch, and C.~A. Lutken, ``{All Hodge Numbers of All Complete
  Intersection Calabi-Yau Manifolds},'' {\em Class. Quant. Grav.} {\bf 6}
  (1989) 105--124.

\bibitem{Gerhardus:2016iot}
A.~Gerhardus and H.~Jockers, ``{Quantum periods of Calabi\textendash{}Yau
  fourfolds},'' {\em Nucl. Phys. B} {\bf 913} (2016) 425--474,
  \href{http://www.arXiv.org/abs/1604.05325}{{\tt 1604.05325}}.

\bibitem{Cota:2019cjx}
C.~F. Cota, A.~Klemm, and T.~Schimannek, ``{Topological strings on genus one
  fibered Calabi-Yau 3-folds and string dualities},'' {\em JHEP} {\bf 11}
  (2019) 170, \href{http://www.arXiv.org/abs/1910.01988}{{\tt 1910.01988}}.

\bibitem{iritani2009integral}
H.~Iritani, ``An integral structure in quantum cohomology and mirror symmetry
  for toric orbifolds,'' {\em Advances in Mathematics} {\bf 222} (2009), no.~3,
  1016--1079.

\bibitem{Iritani:2023ngp}
H.~Iritani, ``{Gamma classes and quantum cohomology},'' in {\em {International
  Congress of Mathematicians}}.
\newblock 7, 2023.
\newblock \href{http://www.arXiv.org/abs/2307.15938}{{\tt 2307.15938}}.

\bibitem{Orlov1997}
D.~O. Orlov, ``Equivalences of derived categories and k3 surfaces,'' {\em
  Journal of Mathematical Sciences} {\bf 84} (May, 1997) 1361--1381.

\bibitem{Berglund:1997eb}
P.~Berglund, M.~Henningson, and N.~Wyllard, ``{Special geometry and automorphic
  forms},'' {\em Nucl. Phys. B} {\bf 503} (1997) 256--276,
  \href{http://www.arXiv.org/abs/hep-th/9703195}{{\tt hep-th/9703195}}.

\bibitem{Haghighat:2015qdq}
B.~Haghighat, H.~Movasati, and S.-T. Yau, ``{Calabi-Yau modular forms in limit:
  Elliptic Fibrations},'' {\em Commun. Num. Theor. Phys.} {\bf 11} (2017)
  879--912, \href{http://www.arXiv.org/abs/1511.01310}{{\tt 1511.01310}}.

\bibitem{Hull:1994ys}
C.~M. Hull and P.~K. Townsend, ``Unity of superstring dualities,'' {\em Nucl.
  Phys.} {\bf B438} (1995) 109--137,
  \href{http://www.arXiv.org/abs/hep-th/9410167}{{\tt hep-th/9410167}}.

\bibitem{Sen:1995ff}
A.~Sen and C.~Vafa, ``{Dual pairs of type II string compactification},'' {\em
  Nucl. Phys. B} {\bf 455} (1995) 165--187,
  \href{http://www.arXiv.org/abs/hep-th/9508064}{{\tt hep-th/9508064}}.

\bibitem{deWit:1995dmj}
B.~de~Wit, V.~Kaplunovsky, J.~Louis, and D.~Lust, ``{Perturbative couplings of
  vector multiplets in N=2 heterotic string vacua},'' {\em Nucl. Phys. B} {\bf
  451} (1995) 53--95, \href{http://www.arXiv.org/abs/hep-th/9504006}{{\tt
  hep-th/9504006}}.

\bibitem{Antoniadis:1995ct}
I.~Antoniadis, S.~Ferrara, E.~Gava, K.~Narain, and T.~Taylor, ``{Perturbative
  prepotential and monodromies in N=2 heterotic superstring},'' {\em
  Nucl.Phys.} {\bf B447} (1995) 35--61,
\href{http://www.arXiv.org/abs/hep-th/9504034}{{\tt hep-th/9504034}}.

\bibitem{Antoniadis:1995cy}
I.~Antoniadis and H.~Partouche, ``{Exact monodromy group of N=2 heterotic
  superstring},'' {\em Nucl. Phys. B} {\bf 460} (1996) 470--488,
  \href{http://www.arXiv.org/abs/hep-th/9509009}{{\tt hep-th/9509009}}.

\bibitem{Klemm2010}
A.~Klemm, D.~Maulik, R.~Pandharipande, and E.~Scheidegger, ``Noether-lefschetz
  theory and the yau-zaslow conjecture,'' {\em Journal of the American
  Mathematical Society} {\bf 23} (June, 2010) 1013--1040,
  \href{http://www.arXiv.org/abs/0807.2477}{{\tt 0807.2477}}.

\bibitem{katz:1999xq}
S.~H. Katz, A.~Klemm, and C.~Vafa, ``{M theory, topological strings and
  spinning black holes},'' {\em Adv. Theor. Math. Phys.} {\bf 3} (1999)
  1445--1537, \href{http://www.arXiv.org/abs/hep-th/9910181}{{\tt
  hep-th/9910181}}.

\bibitem{Maulik:2010jw}
D.~Maulik, R.~Pandharipande, R.~P. Thomas, and A.~Pixton, ``{Curves on K3
  surfaces and modular forms},'' {\em J. Topol.} {\bf 3} (2010) 937--996,
  \href{http://www.arXiv.org/abs/1001.2719}{{\tt 1001.2719}}.

\bibitem{Harvey:1995fq}
J.~A. Harvey and G.~W. Moore, ``{Algebras, BPS States, and Strings},'' {\em
  Nucl. Phys.} {\bf B463} (1996) 315--368,
\href{http://www.arXiv.org/abs/hep-th/9510182}{{\tt hep-th/9510182}}.

\bibitem{gw-dt}
D.~Maulik, N.~Nekrasov, A.~Okounkov, and R.~Pandharipande, ``Gromov-{W}itten
  theory and {D}onaldson-{T}homas theory. {I},'' {\em Compos. Math.} {\bf 142}
  (2006), no.~5, 1263--1285.

\bibitem{gw-dt2}
D.~Maulik, N.~Nekrasov, A.~Okounkov, and R.~Pandharipande, ``Gromov-{W}itten
  theory and {D}onaldson-{T}homas theory. {II},'' {\em Compos. Math.} {\bf 142}
  (2006), no.~5, 1286--1304.

\bibitem{Gholampour:2013mka}
A.~Gholampour, A.~Sheshmani, and Y.~Toda, ``{Stable pairs on nodal $K3$
  fibrations},'' {\em Int. Math. Res. Not.} {\bf 2018} (2018), no.~17,
  5297--5346, \href{http://www.arXiv.org/abs/1308.4722}{{\tt 1308.4722}}.

\bibitem{Alexandrov:2020qpb}
S.~Alexandrov and S.~Nampuri, ``{Refinement and modularity of immortal
  dyons},'' {\em JHEP} {\bf 01} (2021) 147,
  \href{http://www.arXiv.org/abs/2009.01172}{{\tt 2009.01172}}.

\bibitem{Diaconescu:2007bf}
E.~Diaconescu and G.~W. Moore, ``{Crossing the wall: Branes versus bundles},''
  {\em Adv. Theor. Math. Phys.} {\bf 14} (2010), no.~6, 1621--1650,
  \href{http://www.arXiv.org/abs/0706.3193}{{\tt 0706.3193}}.

\bibitem{Batyrev2000}
V.~V. Batyrev, I.~Ciocan-Fontanine, B.~Kim, and D.~Straten, ``{Mirror symmetry
  and toric degenerations of partial flag manifolds},'' {\em Acta Mathematica}
  {\bf 184} (2000), no.~1, 1--39.

\bibitem{Paule2003}
P.~Paule and C.~Schneider, ``{Computer proofs of a new family of harmonic
  number identities},'' {\em Advances in Applied Mathematics} {\bf 31} (Aug.,
  2003) 359--378.

\bibitem{almkvist2005tables}
G.~Almkvist, C.~Van~Enckevort, D.~Van~Straten, and W.~Zudilin, ``{Tables of
  Calabi-Yau equations},'' {\em arXiv preprint math/0507430} (2005).

\bibitem{cooper2015hypergeometric}
S.~Cooper, J.~Ge, and D.~Ye, ``{Hypergeometric transformation formulas of
  degrees 3, 7, 11 and 23},'' {\em Journal of Mathematical Analysis and
  Applications} {\bf 421} (2015), no.~2, 1358--1376.

\bibitem{skoruppa1987jacobi}
N.-P. Skoruppa and D.~Zagier, ``{Jacobi forms and a certain space of modular
  forms},'' {\em Inventiones Mathematicae} {\bf 94} (1988) 113--146.

\bibitem{conway1979monstrous}
J.~H. Conway and S.~P. Norton, ``Monstrous moonshine,'' {\em Bulletin of the
  London Mathematical Society} {\bf 11} (1979), no.~3, 308--339.

\bibitem{vanHoeij}
M.~van Hoeij, ``{Parametrization of the modular curve $X_0(N)$ for $N\leq 37$
  },'' 2012.

\bibitem{skoruppa1989developments}
N.-P. Skoruppa, ``{Developments in the theory of Jacobi forms},''.

\bibitem{Alexandrov:2022pgd}
S.~Alexandrov, N.~Gaddam, J.~Manschot, and B.~Pioline, ``{Modular bootstrap for
  D4-D2-D0 indices on compact Calabi-Yau threefolds},''
  \href{http://www.arXiv.org/abs/2204.02207}{{\tt 2204.02207}}.

\bibitem{Cheng:2016klu}
M.~C.~N. Cheng and J.~F.~R. Duncan, ``{Optimal Mock Jacobi Theta Functions},''
  {\em Adv. Math.} {\bf 372} (2020) 107284,
  \href{http://www.arXiv.org/abs/1605.04480}{{\tt 1605.04480}}.

\bibitem{skoruppa1988explicit}
N.-P. Skoruppa, ``{Explicit formulas for the Fourier coefficients of Jacobi and
  elliptic modular forms},''.

\bibitem{skoruppa1991heegner}
N.-P. Skoruppa, ``{Heegner cycles, modular forms and Jacobi forms},'' {\em
  Journal de th{\'e}orie des nombres de Bordeaux} {\bf 3} (1991), no.~1,
  93--116.

\bibitem{zbMATH06251204}
K.~{Bringmann}, B.~{Kane}, and M.~{Viazovska}, ``{Theta lifts and local Maass
  forms.},'' {\em {Math. Res. Lett.}} {\bf 20} (2013), no.~2, 213--234.

\bibitem{bringmann2014locally}
K.~Bringmann, B.~Kane, and W.~Kohnen, ``{Locally harmonic Maass forms and the
  kernel of the Shintani lift},'' {\em International Mathematics Research
  Notices} (2014) rnu024, \href{http://www.arXiv.org/abs/1206.1100}{{\tt
  1206.1100}}.

\end{thebibliography}

\providecommand{\href}[2]{#2}\begingroup\raggedright\endgroup

\end{document}